\documentclass[11pt,twoside]{report}
\usepackage{epsf}
\usepackage{ruthesis}

\newcommand{\met}{\,/\!\!\!\!E_{T}}
\newcommand{\et}{E_T}
\newcommand{\pt}{p_T}
\newcommand{\zprime}{Z^\prime}

\begin{document}

\phd
\title{\bf A Search for New Physics with High Mass Tau Pairs in 
           Proton--Antiproton Collisions at
           \mbox{$\bf \sqrt{\rm s}$ = 1.96 TeV} at CDF}
\author{Zongru Wan}
\campus{New Brunswick}
\program{Physics and Astronomy}
\director{Professor John Conway}
\approvals{5}
\copyrightpage
\submissionmonth{May}
\submissionyear{2005}
\figurespage
\tablespage
\abstract{We present the results of a search for new particles
decaying to tau pairs using the data corresponding to
an integrated luminosity of 195 pb$^{-1}$ collected
from March 2002 to September 2003 with the CDF
detector at the Tevatron.  Hypothetical particles,
such as $\zprime$ and MSSM Higgs bosons can 
potentially produce the tau pair final state.  We 
discuss the method of tau identification, and show 
the signal acceptance versus new particle mass.  The 
low-mass region, dominated by $Z\to\tau\tau$, is used 
as a control region.  In the high-mass region, we 
expect $2.8\pm0.5$ events from known background sources, 
and observe $4$ events in the data sample.  Thus no 
significant excess is observed, and we set upper 
limits on the cross section times branching ratio as 
a function of the masses of heavy scalar and vector 
particles.

}  

\beforepreface 


\acknowledgements{I would like to thank all of the CDF collaborators
and the Fermilab staffs for making CDF an excellent
experiment.  High energy physics is amazing, which 
I learned by being a part of the CDF 
collaboration, learning how the experiment runs, 
and systematically analysing the data. 

I thank my advisor John Conway for bringing me to 
CDF, recommending this interesting thesis topic, 
his clear goals on this search, genuinely valuable 
guidance on physics, constant encouragement, 
elegant presentations on statistics, good humor on 
my English, and coming often to CDF.  He patiently 
read through the drafts of this thesis and helped 
to make it a more complete work.  His sharp physics 
insight and nice presentation style will certainly 
benefit my career in years to come. 

It is my pleasure to thank everybody directly 
involved in this analysis: Anton Anastassov my mentor 
at CDF and Amit Lath my advisor at Rutgers for their 
inspiring inputs and generous help during all of these 
years, Dongwook Jang my fellow graduate student and 
good friend for sharing the techniques,
and the collaborators in the Tau group for their 
important supports. I also thank John Zhou and Aron 
Soha for the very useful techniques, and Pieter 
Jacques and John Doroshenko for keeping the great 
hex farm running. 

I am grateful to the conveners of the Tau group Fedor 
Ratnikov and Teruki Kamon, the convener of the Lepton 
plus Track group Alexei Safonov, the conveners of the 
VEGY group Kaori Maeshima, Rocio Vilar, and Chris Hays, 
and the conveners of the Exotics group Stephan Lammel 
and Beate Heinemann.  They gave me numerous 
opportunities to present my work and offered valuable 
advices that came up during the discussions. 

I thank Teruki Kamon, M\"{u}ge Karag\"{o}z \"{U}nel,
and Ronan McNulty for being great godparents
of the paper on this thesis topic.

For my colleagues at Rutgers: the weekly group meeting 
has been one of the most important parts of my education. 
For my professors Tom Devlin, Sunil Somalwar, Terry 
Watts, and Steve Worm, I am grateful for their great 
advices based on deep understanding and wide experience 
on physics analysis. For my fellow graduate students: 
Paul DiTuro, Sourabh Dube and Jared Yamaoka, I thank 
them for showing me the great opportunities and 
challenges in their Higgs and SUSY searches and for the 
fun time.  For my office mate Pete McNamara, I thank 
him for demonstrating me a mature understanding of 
statistics, offering nice suggestions on an astrophysics 
term paper, and recommending fun movies.  For John Zhou, 
it has been my good luck to work with him. He greatly 
improved my English in this thesis, provided many useful 
comments and suggestions on my analysis, and shared the 
cheerful time to learn his SUSY search.

I appreciate the members of my thesis committee over 
the years: Amit Lath, Ronald Gilman, John Bronzan, 
Jolie Cizewski, Ron Ransome, and 
Chris Tully, for their overview of my progress and 
for the many useful comments and suggestions that 
have improved my presentation and the thesis.

I thank Nancy DeHaan, Kathy DiMeo, Jennifer 
Fernandez--Villa, Phyllis Ginsberg, Carol Picciolo 
and Marie Tamas for their administrative efforts.

I also thank Tom Devlin for his recommendation on 
phenomenology books and experience on accelerators. 
And I thank Vincent Wu my good friend at Fermilab
Beam Division for teaching me the concepts of
accelerators.

For the theorists at Fermilab, their lectures and 
papers are very inspiring and helpful sources, and 
I thank Marcela Carena, Alejandro Daleo, Bogdan 
Dobrescu, Stephen Mrenna and Tim Tait for their very 
useful suggestions. 

A special thank goes to 
Ming-Tsung Ho my good friend at Rutgers for patiently 
showing me the power of explicitly writing down the 
equations step-by-step. Another special thank goes to 
Willis Sakumoto my colleague at CDF for the relaxed 
discussions on how to calculate cross sections 
during lunch times.

I thank my family for their constant support and 
encouragement.  I thank my wife Meihua Zhu for her 
unconditional love and support and for bringing much 
happiness into my life.}



\afterpreface

%


\chapter{Introduction}
\label{cha:Intro}

The Standard Model (SM) combines the electroweak 
theory together with Quantum Chromodynamics (QCD) 
of strong interactions and shows good agreement with 
collider experiments.  However the SM 
does not include gravity and 
is expected to be an effective 
low-energy theory.

The Fermilab Tevatron is currently the high energy 
frontier of particle physics and delivers 
proton-antiproton collisions at high luminosity.

The Run II of the Collider Dectector at Fermilab (CDF)
continues the precision measurements 
of hadron collider physics and the search for new 
physics at and above the electroweak scale. 
With the precision capability at the energy frontier, we 
can attack the open questions of high energy physics
from many complementary directions, including: the properties 
of top quark, the precision electroweak measurements, 
e.g. mass of the $W$ boson,
the direct searches for new phenomena, the tests of 
perturbative QCD at Next-to-Leading-Order and large 
$Q^2$, and the constraint of the CKM matrix with high 
statistics of the B decays.  

This thesis is about a direct
search for new particles decaying to tau pairs.  
The evidence for such new particles is 
that at accessible energies the events with tau pairs 
deviate clearly and significantly from the SM 
prediction.

\begin{figure}
   \begin{center}
      \parbox{5.5in}{\epsfxsize=\hsize\epsffile{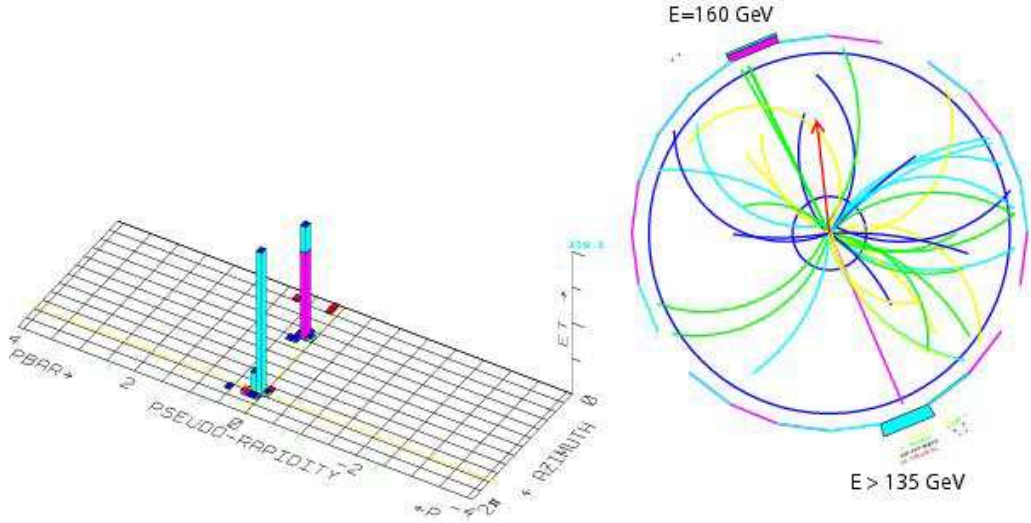}}
   \end{center}
   \caption[Run I high-mass di-tau candidate event]
           {Run I high-mass di-tau candidate event.
            The left plot shows energy measurement in calorimeters
            and the event is very clean. The right plot shows
            the display in the transverse plane. The three-prong
            identified tau object has energy 160 GeV. The one-prong
            identified tau object has energy at least 135 GeV.
            There is also a significant missing transverse energy
            indicating significant neutrinos.
            The scale of the invariant mass of the two tau objects
            and the neutrinos is above 300 GeV/$c^2$.}
   \label{fig:Intro}
\end{figure}

In Run I CDF recorded an unusual event in which 
there were two very high energy $\tau\to h\nu$ 
candidates nearly back-to-back in direction.  
Figure~\ref{fig:Intro} shows a display of the event.  
This event was recorded in the data sample from the
missing transverse energy trigger, and was 
noticed in the context of the Run I charged Higgs 
search~\cite{CDFnote:3546}.  In Run I, 
{\em a posteriori}, it was not possible to estimate 
a probability for observing such an event, though 
less than about 0.1 such events were expected from 
backgrounds, including $Z/\gamma^*\to\tau\tau$ 
Drell-Yan ($q\bar{q}\to Z/\gamma^*\to l^+l^-$).

Various new physics processes can lead to very 
high-mass tau pairs, for example, the new vector
boson $\zprime\to\tau\tau$ predicted in the 
extension to the Standard Model by adding a 
new U(1) gauge symmetry and the pseudoscalar Higgs
boson $A\to\tau\tau$ predicted in the minimum
supersymmetric extension of the Standard Model (MSSM).
The known backgrounds are from 
the high-mass tail of Drell-Yan processes 
(mainly $Z/\gamma^*\to\tau\tau$) as well as jet$\to\tau$ 
fakes from $W$+jets, QCD di-jet, and mutli-jet 
events. 

In this analysis we search for such signal 
processes by performing a counting experiment.  
We select events with $e+\tau_h$, $\mu+\tau_h$, 
and $\tau_h+\tau_h$ (here, ``$\tau_h$'' means a 
$\tau$ hadronic decay).  We construct an 
invariant mass which we call $m_{vis}$ using the 
four-vector sum of the lepton, the tau, and the missing transverse energy
vector (ignoring in the latter the $z$ component).
The region which has $m_{vis}>120$ GeV/$c^2$ is
defined as the signal region, while the region
which has $m_{vis}<120$ GeV/$c^2$ is retained as 
a control region.  We 
perform a blind analysis in the signal region, i.e., 
we do not look at the data in the signal region until we have 
precisely estimated the backgrounds.
If there is a significant excess 
over the known backgrounds, we have 
discovered new physics; otherwise, we set 
limits on the possible signal rates.

The thesis is organized as follows:  
theorectical models including the SM, 
extensions of the SM, and high-mass 
tau pair phenomenology are described
in Chapter~\ref{cha:theory}.
The experimental appratus including the Fermilab 
Accelerator and CDF detector is introduced 
in Chapter~\ref{cha:apparatus}.
We discuss the logic behind the analysis
in Chapter~\ref{cha:Search_Strategy}.
Particle identifications for tau, electron and 
muon, and the study of missing transverse energy 
are discussed in detail
in Chapter~\ref{cha:PID}.
The data samples and event selection are discussed
in Chapter~\ref{cha:event}.
The low-mass control region background estimate,
uncertainties, and the observed events are discussed
in Chapter~\ref{cha:control}.
The high-mass signal region, signal acceptance,
background estimate, and uncertainties are 
discussed in Chapter~\ref{cha:signal}.
The results of the observed events after opening
the box, and the method to extract limit are 
discussed 
in Chapter~\ref{cha:results}.
Finally, the conclusion is presented
in Chapter~\ref{cha:conclusions}.



\chapter{Theoretical Model}
\label{cha:theory}

The goal of elementary particle physics is to 
answer the following fundamental questions:
\begin{itemize}
\item What is the world made of? 
\item How do the parts interact?
\end{itemize}
The Standard Model (SM)~\cite{Weinberg:1967tq}
of particle physics 
is a beautiful theory which attempts to
\textit{find the simplest model} 
that quantitatively answer these questions.  
The thousands of cross sections and decay 
widths listed in the Particle Data Group (PDG)~\cite{Eidelman:2004wy}, 
and all of the data from collider 
experiments, are calculable and explained in 
the framework of the SM, which is the bedrock
of our understanding of Nature.  

Building on the success of the SM, ambitious attempts 
have been made to extend it.  
This thesis is concerned about a direct search for
new particles decaying to two taus.  The phenomenology
of tau pairs, namely the production rates of intermediate
bosons and the branching ratio of their decays to tau 
pairs, in the framework of the SM and some of the 
extensions will be presented in this chapter.


\section{The Standard Model}
\label{sec:theory_SM}

The SM elementary particles include the fermion
matter particles and the force carriers.
There are three generations of fermion matter 
particles: leptons and quarks. The second and 
third generations have the same quantum numbers 
of the first generation, but with heavier masses.  
The masses of the leptons and quarks are listed in 
Table~\ref{tab:LeptonsQuarks}.
The force carriers include the gluon for the strong 
interaction, and the photon, the W and Z vector bosons 
for the electroweak interaction.  The masses of the 
force carriers are listed in Table~\ref{tab:ForceCarriers}.
The Higgs boson predicted in the SM is a fundamental
scalar particle and has special interaction strength
proportional to the mass of the elementary particles.
Since it is not discovered yet, it is not
listed in Table~\ref{tab:ForceCarriers}.

\begin{table}
   \begin{center}
      \begin{tabular}{|c|lc|c|} \hline
         Generation & Particle          &              & Mass [GeV/$c^2$] \\ \hline \hline
         I          & electron neutrino & $\nu_e$      & 0                \\
                    & electron          & $e$          & 0.00051          \\
                    & up quark          & $u$          & 0.002 to 0.004   \\
                    & down quark        & $d$          & 0.004 to 0.008   \\ \hline
         II         & muon neutrino     & $\nu_{\mu}$  & 0                \\
                    & muon              & $\mu$        & 0.106            \\
                    & charm quark       & $c$          & 1.15 to 1.35     \\
                    & strange quark     & $s$          & 0.08 to 0.13     \\ \hline
         III        & tau neutrino      & $\nu_{\tau}$ & 0                \\
                    & tau               & $\tau$       & 1.777            \\
                    & top quark         & $t$          & 174.3 $\pm$ 5.1  \\
                    & bottom quark      & $b$          & 4.1 to 4.4       \\ \hline
      \end{tabular}
      \caption[Leptons and quarks in the SM]
              {Three generations of leptons and quarks in the Standard
               Model and their masses.}
      \label{tab:LeptonsQuarks}
   \end{center}
\end{table}

\begin{table}
   \begin{center}
      \begin{tabular}{|l|lc|c|} \hline
         Force           & Carrier &           & Mass [GeV/$c^2$]     \\ \hline \hline
         electromagnetic & photon  & $\gamma$  & 0                    \\
         charged weak    & W boson & $W^{\pm}$ & 80.425  $\pm$ 0.038  \\
         neutral weak    & Z boson & $Z^0$     & 91.1876 $\pm$ 0.0021 \\
         strong          & gluon   & $g$       & 0                    \\ \hline
      \end{tabular}
      \caption[Force carriers in the SM]
              {Force carriers in the Standard Model and their masses.}
      \label{tab:ForceCarriers}
   \end{center}
\end{table}

\begin{figure}
   \begin{center}
      \parbox{5.5in}{\epsfxsize=\hsize\epsffile{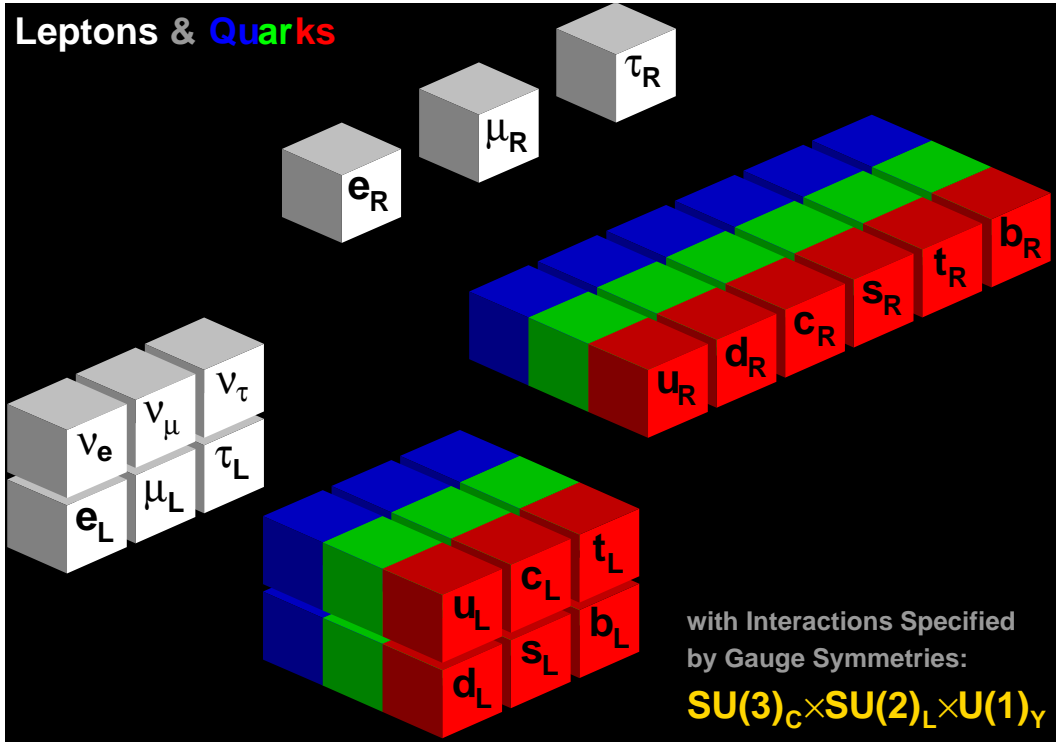}}
      \caption[SU(3)$_C$$\times$SU(2)$_L$$\times$U(1)$_Y$ gauge symmetries]
              {SU(3)$_C$$\times$SU(2)$_L$$\times$U(1)$_Y$
               gauge symmetries of fermions in the Standard Model.
               Quarks have three color degrees-of-freedom, while
               leptons are colorless. Left-handed fermions 
               are arranged in SU(2) weak isospin doublets and 
               right-handed fermions are arranged in SU(2) 
               singlets. Each fermion also has
               U(1) weak hyper-charge. The interactions 
               are uniquely specified by the gauge symmetries.}
      \label{fig:GaugeSymmetries}
   \end{center}
\end{figure}

The SU(3)$_C$$\times$SU(2)$_L$$\times$U(1)$_Y$
structure of the leptons and quarks is shown in 
Fig.~\ref{fig:GaugeSymmetries}. 
The quarks are arranged in triplets with respect to the 
color gauge group SU(3)$_C$, with indices as red ($r$), 
green ($g$), and blue ($b$).
\begin{equation}
   q = \left( \begin{array}{c}
                 q_r \\
                 q_g \\
                 q_b
              \end{array}
       \right)
\end{equation}
The left- and right-handed fermions have different
transformation properties under the weak isospin
group SU(2)$_L$.  The left-handed fermions
are arranged in doublets, and the right-handed
fermions are arranged in singlets. There is no 
right-handed neutrino in the SM.  
\begin{equation}
   \begin{array}{rccccccccc}
      \mbox{Leptons:}
               & \left(\begin{array}{c} \nu_e      \\ e    \end{array}\right)_L
               & \left(\begin{array}{c} \nu_{\mu}  \\ \mu  \end{array}\right)_L
               & \left(\begin{array}{c} \nu_{\tau} \\ \tau \end{array}\right)_L
               &
               & e_R
               & 
               & \mu_R
               &
               & \tau_R \\
      \mbox{Quarks:}
               & \left(\begin{array}{c} u          \\ d    \end{array}\right)_L
               & \left(\begin{array}{c} c          \\ s    \end{array}\right)_L
               & \left(\begin{array}{c} t          \\ b    \end{array}\right)_L
               & u_R
               & d_R
               & c_R
               & s_R  
               & t_R
               & b_R   
   \end{array}
\end{equation}
Table~\ref{tab:QuantumNumbers}
lists the transformation properties, i.e., the quantum
numbers, of the fermions of the first generation under 
the gauge groups.  The hypercharge of U(1)$_Y$ is 
related to the electric charge by $Q = T_L^3 + \frac{Y}{2}$.
The assignments of the quantum numbers to the second and third 
generations are the same.  
A brief review about how this structure emerges is given in 
Appendix~\ref{cha:app_structure}.

\begin{table}
   \begin{center}
      \begin{tabular}{|c|c|c|c|c|} \hline
                 &  $Q$ & $T_L^3$ &  $Y$ &   $C$   \\ \hline \hline
         $\nu_e$ &   0  &   1/2   &   -1 &    0    \\
         $e_L$   &  -1  &  -1/2   &   -1 &    0    \\ \hline
         $e_R$   &  -1  &    0    &   -2 &    0    \\ \hline
         $u_L$   &  2/3 &   1/2   &  1/3 & $r,g,b$ \\ 
         $d_L$   & -1/3 &  -1/2   &  1/3 & $r,g,b$ \\ \hline 
         $u_R$   &  2/3 &    0    &  4/3 & $r,g,b$ \\ \hline 
         $d_R$   & -1/3 &    0    & -2/3 & $r,g,b$ \\ \hline
      \end{tabular}
      \caption[Quantum numbers of the fermions]
              {Quantum numbers of the fermions.}
      \label{tab:QuantumNumbers}
   \end{center}
\end{table}

The interactions are uniquely specified by the 
SU(3)$_C$$\times$SU(2)$_L$$\times$U(1)$_Y$ 
gauge symmetries.  All of the gauge bosons 
and fermions acquire mass by the 
Higgs mechanism~\cite{Higgs:1964ia}.
It introduces an extra Higgs boson, and its physical
vacuum is spontaneously broken in the field space
of the Higgs potential.
The quark states in charged weak interactions mediated by
$W^{\pm}$ bosons are not the physical states, but rather 
a quantum superposition of the physical states, described 
by the CKM (Cabbibo-Kobayashi-Maskawa) 
matrix~\cite{Cabibbo:1963yz}.
\begin{equation}
   \left( \begin{array}{c}
             d \\
             s \\
             b
          \end{array}
   \right)_{\mbox{weak}}
   = 
   \left( \begin{array}{ccc}
             V_{ud} & V_{us} & V_{ub} \\
             V_{cd} & V_{cs} & V_{cb} \\
             V_{td} & V_{ts} & V_{tb} 
          \end{array}
   \right)
   \left( \begin{array}{c}
             d \\
             s \\
             b
          \end{array}
   \right)_{\mbox{mass}}
\end{equation}
The topic of this thesis is mostly related
to the fermion couplings.  The couplings to 
fermions in the SM are listed in 
Table~\ref{tab:CouplingsToFermions}.
A very detailed review with explicit derivations
on these topics starting from the gauge symmetry 
to the couplings to the fermions in the SM
is given in Appendix~\ref{cha:app_gs_ssb}.

\begin{table}
   \begin{center}
      \begin{tabular}{|ll|c|c|} \hline
                               &
                               & Left Coupling 
                               & Right Coupling \\ \hline \hline
         Higgs                 &
         $H \to f\bar{f}$      & $\frac{m_f}{v}$ 
                               & $\frac{m_f}{v}$ \\ \hline
         Strong                &
         $g \to q\bar{q}$      & $\frac{g_3}{2}\lambda^a$
                               & $\frac{g_3}{2}\lambda^a$ \\ \hline
         EM                    &
         $\gamma \to f\bar{f}$ & $eQ_f$ 
                               & $eQ_f$ \\ \hline
         Weak                  &
         $Z^0 \to f\bar{f}$    &
            $\frac{g_2}{\cos\theta_W}(T^3_f - \sin^2\theta_W Q_f)$ 
                               & 
            $-\frac{g_2}{\cos\theta_W}\sin^2\theta_WQ_f$ \\ 
                               &
         $W^{\pm}\to l\nu_l$   & $\frac{g_2}{\sqrt{2}}$ 
                               & 0 \\ 
                               &
         $W^{\pm}\to qq'$      & $V_{qq'}\frac{g_2}{\sqrt{2}}$ 
                               & 0 \\ \hline
      \end{tabular}
      \caption[Couplings to fermions in the SM]
              {Couplings to fermions in the Standard Model.}
      \label{tab:CouplingsToFermions}
   \end{center}
\end{table}

In spite of its tremendous success in explaining 
collider results, there are still many 
unexplained aspects in the SM.   The set of group 
representations and hypercharge it requires are quite 
bizarre, and there are 18 free parameters which must be 
input from experiment: 3 gauge couplings (usually traded 
as $e$, $\sin^2\theta_W$ and $g_3$), 2 Higgs potential 
couplings (usually traded as $m_Z$ and $m_H$), 9 fermion 
masses, and 4 CKM mixing parameters.  Do particle masses 
really originate from a Higgs field?  Can all the particle 
interactions be unified in a simple gauge group? What is 
the origin of the CKM matrix?   The ultimate ``theory of 
everything'' should explain all of these parameters.  The 
imaginary goal, for example, is probably to express 
everything in terms of the Planck constant $\hbar$, the 
speed of light $c$, the mathematical constant $\pi$, and 
without any free parameters.  That would be an amazing accomplishment. There 
are still many things to do in particle physics in the 
direction to \emph{find the simplest model} and many 
exciting challenges are ahead!


\section{Extensions to the Standard Model}
\label{sec:theory_BSM}

One interesting extension to the SM is to add a new U(1) gauge 
group.  This predicts a new 
$Z'$ gauge boson~\cite{Carena:2004xs}
at high energy scale.  
We will use the $Z'$ as our model to calculate the signal 
acceptance for any kind of new vector boson.

\begin{figure}
   \begin{center}
      \parbox{3.7in}{\epsfxsize=\hsize\epsffile{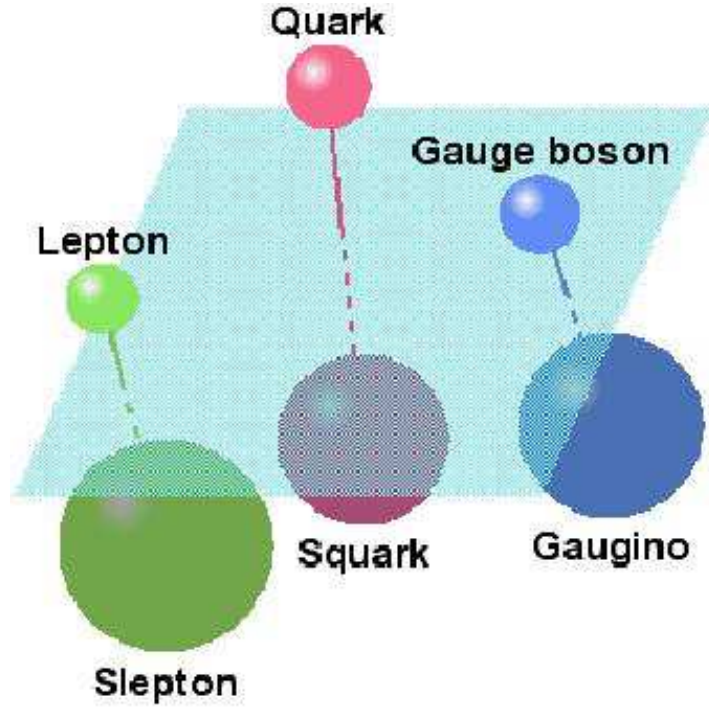}}
      \caption[Particles in the Supersymmetry Theory]
              {Particles in the Supersymmetry Theory.}
      \label{fig:Supersymmetry}
   \end{center}
\end{figure}

Another interesting extension is 
supersymmetry~\cite{Wess:1974tw}, 
which is 
motivated by the desire to unify fermions and bosons, 
shown in Fig.~\ref{fig:Supersymmetry}. 
For each fermion 
(lepton and quark) it predicts a bosonic super partner 
(slepton and squark), and for each gauge boson it predicts 
a fermionic super partner (gaugino).  
There is a divergence
from scalar contributions to radiative corrections 
for the Higgs mass in the SM, while the new fermion loops 
appearing in supersymmetry have a negative sign 
relative to the scalar contributions, thus cancel the 
divergence.  
We will use the pseudoscalar Higgs particle $A$, one of the 
Higgs particles predicted in 
the minimal supersymmetric extension of the Standard Model 
(MSSM)~\cite{Nilles:1983ge}
as our model to 
calculate the signal acceptance for any kind of new scalar 
boson.


\section{High Mass Tau Pairs}
\label{sec:theory_tt}

\begin{figure}
   \begin{center}
      \parbox{4.5in}{\epsfxsize=\hsize\epsffile{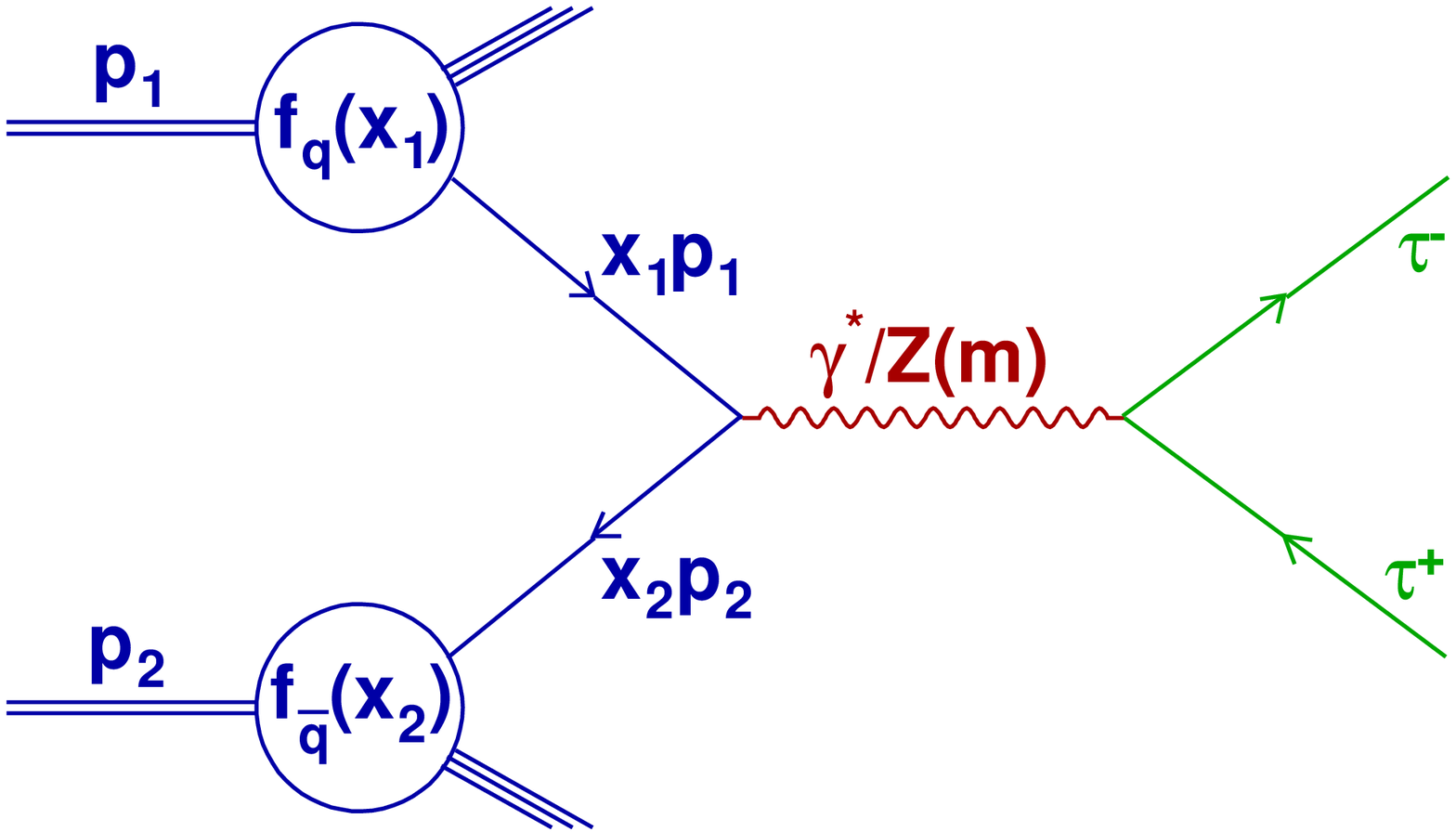}}
      \caption[Tau pair production 
               $p\bar{p}\to\gamma^*/Z\to\tau\tau$ 
               in the SM]
              {Tau pair production
               $p\bar{p}\to\gamma^*/Z\to\tau\tau$
               in the Standard Model.}
      \label{fig:TauTau_1}
   \vspace{0.5in}
      \parbox{5.5in}{\epsfxsize=\hsize\epsffile{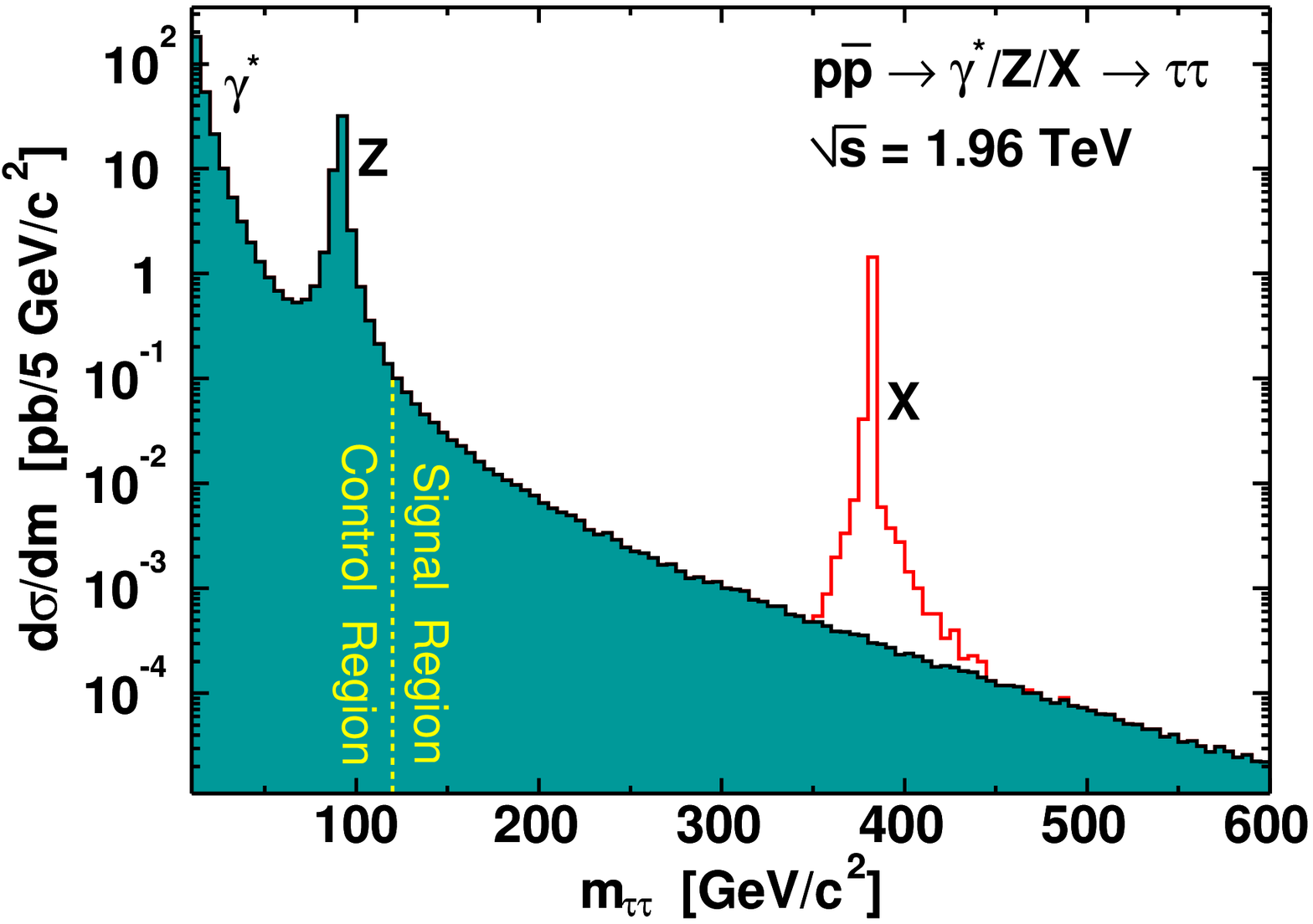}}
      \caption[High mass tau pair search]
              {High mass tau pair search.
               Low-mass region including the $Z$ peak is the control
               region. High-mass region is the signal region.
               The high-mass tail of the Drell-Yan process is
               the main background of this search. The signature
               of new particles is a sigficant deviation from
               the known backgrounds, such as $X$ shown in this plot.}
      \label{fig:TauTau_3}
   \end{center}
\end{figure}

At the Tevatron, the
tau pair production in the SM is through the Drell-Yan process,
$p\bar{p}\to\gamma^*/Z\to\tau\tau$,
as shown in Fig.~\ref{fig:TauTau_1}.  
The center-of-mass energy of 
$p\bar{p}$ collisions at the Tevatron is 1.96 TeV.  At the parton 
level, one incoming quark from a proton and the other anti-quark from 
an anti-proton collide via an intermediate boson which decays to two outgoing 
taus.  The details about how to calculate cross sections are shown in 
Appendix~\ref{cha:HowToCalculateXSec} and the mass spectrum of the
final two taus is shown in Fig.~\ref{fig:TauTau_3}.  We perform 
a direct search for new hypothetical particle in high mass region by 
its decay to two taus $X\to\tau\tau$.  The low mass region of the SM 
processes $\gamma^*/Z\to\tau\tau$ is the control region 
and its high mass Drell-Yan tail is the major background for 
this search.

\begin{figure}
   \begin{center}
      \parbox{3.0in}{\epsfxsize=\hsize\epsffile{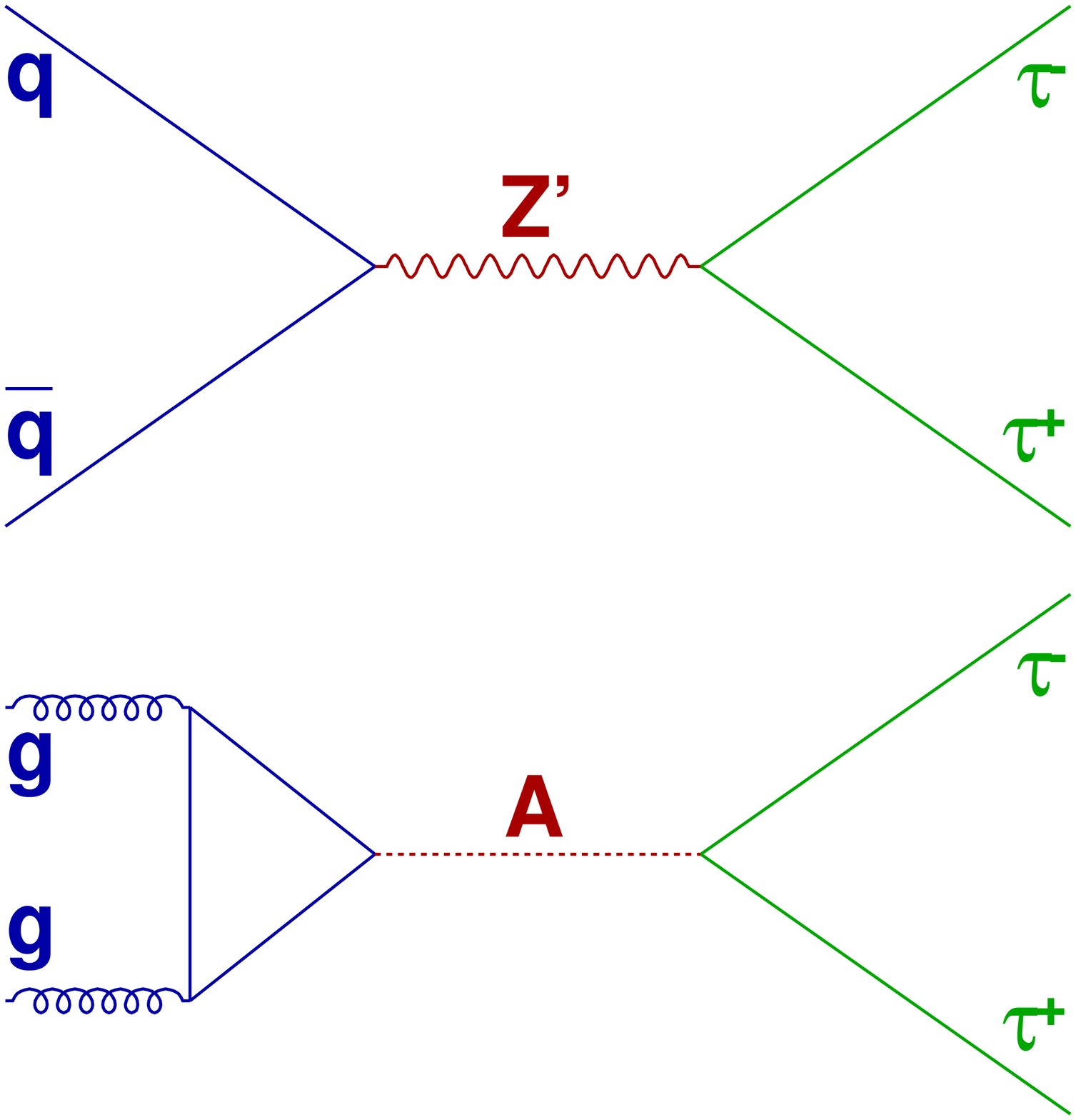}}
      \caption[Tree-level Feynman diagrams for 
               $Z'\to\tau\tau$ and $A\to\tau\tau$]
              {Tree-level Feynman diagrams for the productions
               at $p\bar{p}$ collider and decays of $Z'$ predicted 
               in U(1) extension and pseudoscalar $A$ predicted in 
               minimum supersymmetric extension of the Standard
               Model.}
      \label{fig:TauTau_2}
   \vspace{0.5in}
      \parbox{5.5in}{\epsfxsize=\hsize\epsffile{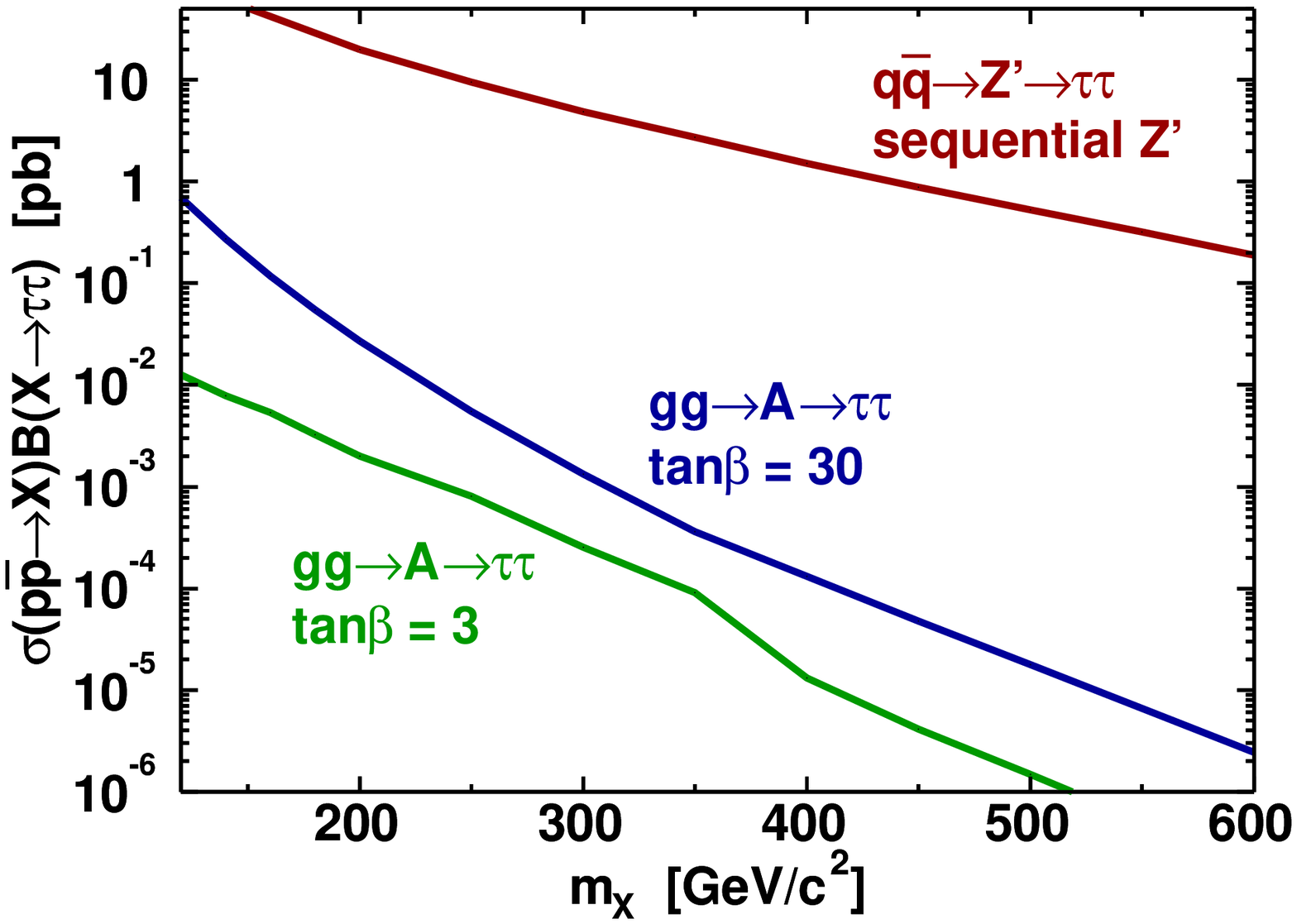}}
      \caption[Theoretical signal
               $\sigma(p\bar{p}\to X)\cdot\mbox{B}(X\to\tau\tau)$]
              {Theoretical signal 
               $\sigma(p\bar{p}\to X)\cdot\mbox{B}(X\to\tau\tau)$.}
      \label{fig:TauTau_4}
   \end{center}
\end{figure}

The two extensions described above are 
shown in Fig.~\ref{fig:TauTau_2}.  For U(1) extension, we consider 
the simplest model with the same interactions as the $Z$ boson in the 
SM, called the sequential $Z'$, and the only unknown parameter is 
the mass of the new gauge boson.  The MSSM requires two Higgs doublets 
and the ratio of the two Higgs expectation values is defined as 
$\tan\beta$, which is undetermined and should be treated as a free 
parameter.  Thus the $A$ boson is governed by one more free parameter 
in addition to its mass.

The couplings to fermions in the SM are listed in 
Table~\ref{tab:CouplingsToFermions}.  For each mass point of 
the sequential $Z'$, we can use the same couplings to fermions
as the $Z$ boson in the SM and repeat the procedure to calculate 
the cross section.  
The leading order cross section $\sigma_0$ is subject to a
correction $K$ factor~\cite{Barger:1997}
such that the corrected cross section
$\sigma 
 = (1 + \mbox{correction}) \times \sigma_0 
 = K \times \sigma_0$.
Including the $K$ factor, the predicted cross section 
versus mass for the sequential $Z'$ is shown in 
Fig.~\ref{fig:TauTau_4}.

The SM requires one Higgs doublet with a coupling of the SM 
Higgs boson to fermions as $m_f/v$, where $m_f$ is the fermion
mass and $v$ is the vacuum expectation value of the SM Higgs 
boson, about 246 GeV.  Therefore Higgs boson prefers to couple to the
fermions in the heaviest generation.  In the MSSM, at large 
$\tan\beta$, the coupling of $A\to\tau\tau$ and 
$A\to b\bar{b}$ are enhanced to $m_f\tan\beta/v$, whereas 
the coupling of $A\to t\bar{t}$ is suppressed to 
$m_t\cot\beta/v$ when the top quark is kinematically available, 
i.e. $m_A > 2 m_t \approx 350$ GeV/$c^2$.  We use the programs 
{\textsc HIGLU}~\cite{Spira:1995mt} and 
{\textsc HDECAY}~\cite{Djouadi:1997yw} 
to calculate the next-to-leading-order cross 
section of $gg\to A\to\tau\tau$.  They are also shown in 
Fig.~\ref{fig:TauTau_4}.

\chapter{The Tevatron Accelerator and the CDF Detector}
\label{cha:apparatus}

Fermilab is the home of the highest energy 
particle accelerator in the world, the Tevatron.  
The center-of-mass energy of proton-antiproton
($p\bar{p}$) collision is $\sqrt{s}=1.96$ TeV.  
We shall describe the Tevatron accelerator and 
the Collider Detector at Fermilab (CDF) in this 
chapter.  


\section{Fermilab's Accelerator Chain}
\label{sec:accelerator}

Protons and antiprotons have equal and opposite electric 
charge.  The advantage of $p\bar{p}$ collider is that $p$ and
$\bar{p}$ travel in opposite 
directions through the magnets and a $p\bar{p}$ collider 
can be built with one ring of magnets instead of two.  
The disadvantage is that it is difficult 
to produce and accumulate $\bar{p}$ at a high efficiency.

\begin{figure}
   \begin{center}
      \parbox{4.0in}{\epsfxsize=\hsize\epsffile{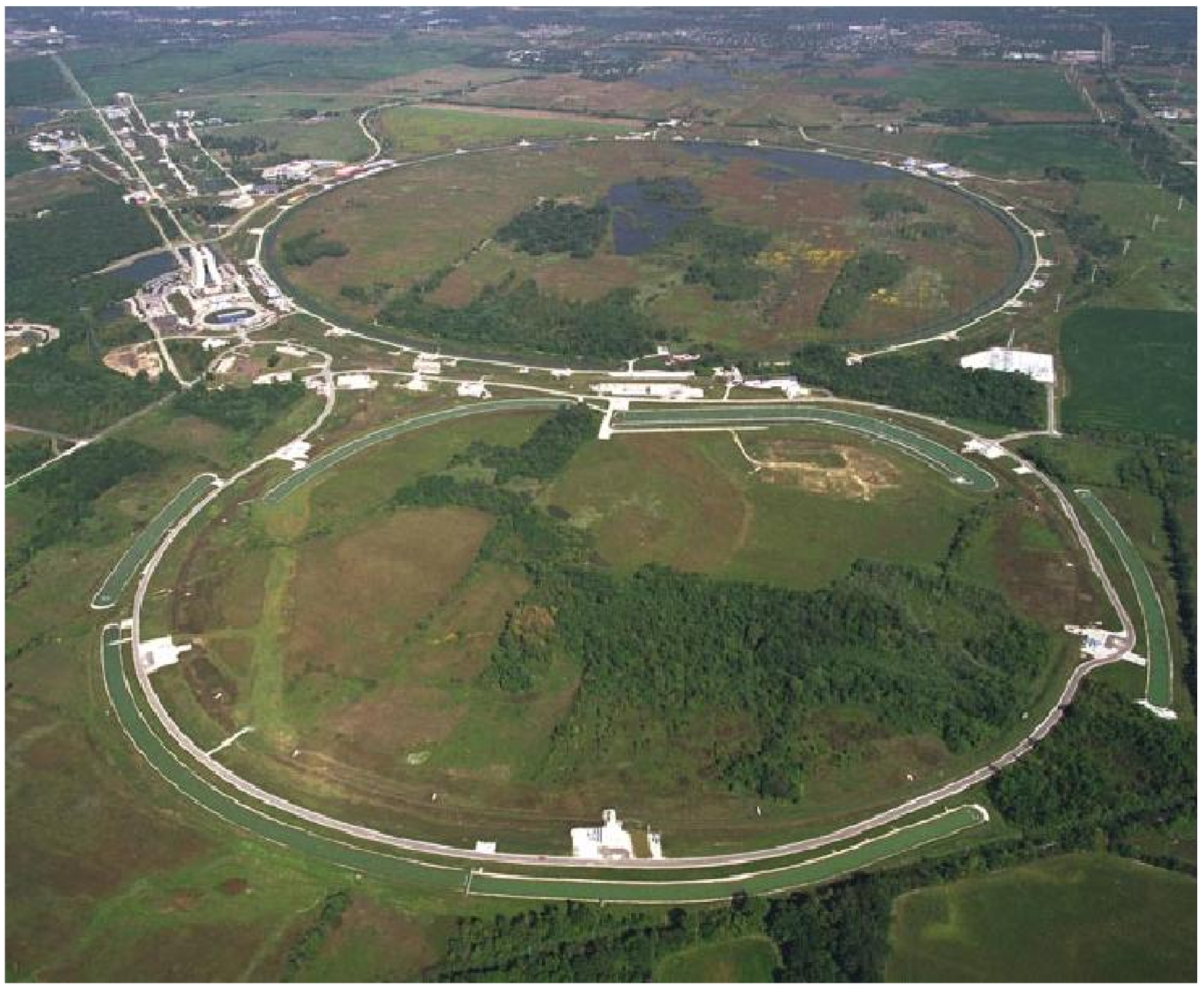}}
      \caption[Aerial view of Fermilab]
              {Aerial view of Fermilab showing 
               the Main Injector in the foreground, 
               the Tevatron collider ring and the 
               fixed target facilities in the background.}
      \label{fig:Ferimlab}
   \vspace{0.5in}
      \parbox{4.4in}{\epsfxsize=\hsize\epsffile{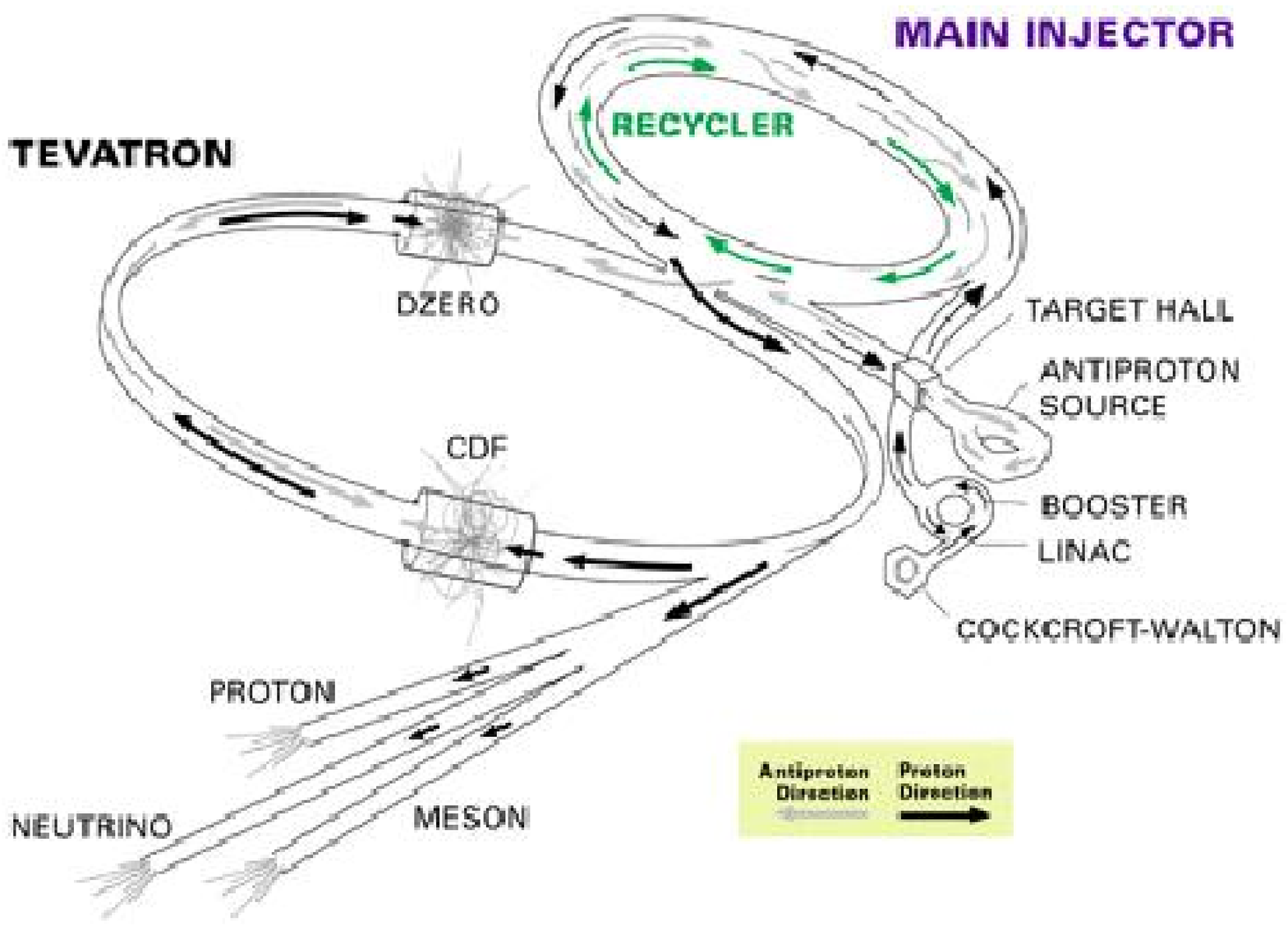}}
      \caption[Fermilab's accelerator chain]
              {Fermilab's accelerator chain consists of
               the 8 GeV proton source, the 8 GeV anti-proton
               source, the Main Injector, the Recycler for
               recycling the precious anti-protons,
               and the Tevatron. The Main Injector accelerates
               protons and anti-protons to 150 GeV. The 
               Tevatron ramps up their energies to 980 GeV.
               The center-of-mass energy of $p\bar{p}$
               collision is thus 1.96 TeV. The linear
               accelerators for the fixed targed experiments
               are also shown.}
      \label{fig:accelerator}
   \end{center}
\end{figure}

The aerial view of Fermilab is shown in 
Fig.~\ref{fig:Ferimlab}.  The Fermilab's accelerator 
chain is shown in Fig.~\ref{fig:accelerator}.  It 
consists of the Proton/Antiproton Sources (8 GeV), the Main 
Injector (150 GeV), the Recycler, and the Tevatron (980 GeV).  

The Proton Source includes the Cockcroft-Walton, the Linear 
Accelerator (Linac), and the Booster.  The Cockcroft-Walton 
uses DC power to accelerate H$^-$ ions to 750 KeV.  
The Linac uses Radio Frequency (RF) power to 
accelerate H$^-$ ions to 400 MeV.  The electrons are 
stripped off and the bare protons are injected into 
the Booster.  The Booster uses RF cavities to accelerate 
protons to 8 GeV.

The Anti-proton Source includes the Target Station, 
the Debuncher and the Accumulator.  A bunched beam of 120 GeV 
protons from the Main Injector hits a Nickel Target to 
make anti-protons and other particles as well.  The 
particles are focused with a lithium lens and filtered 
through a pulsed magnet acting as a charge-mass 
spectrometer to select anti-protons.  The antiproton 
beam is bunched since the beam from the Main Injector 
is bunched and the antiprotons have a wide range of 
energies, positions and angles.  The transverse spread of the beam out of 
the Target Station is ``hot'', in terms analogous to 
temperature.  Both RF and stochastic cooling systems 
are used in the momentum stacking process.   
The Debuncher exchanges the large energy spread and narrow 
time spread into a narrow energy spread and large time 
spread.  The Accumulator stacks successive pulses of 
antiprotons from the Debuncher over several hours or 
days.  For every million protons that hit the target, 
only about twenty 8 GeV anti-protons finally get stacked
into the Accumulator.

Protons at 8 GeV from the Booster are injected into
the Main Injector.  They are accelerated to 120 GeV 
for fixed target experiments or 150 GeV for injection 
into the Tevatron.  Antiprotons at 8 GeV from either 
the Accumulator or the Recycler are accelerated to 
150 GeV in the Main Injector and then injected into 
the Tevatron.

The Recycler is placed directly above the Main 
Injector beamline, near the ceiling.  One role of 
the Recycler is a post-Accumulator ring.  Another 
role, and by far the leading factor in the luminosity 
increase, is to act as a recycler for the precious 
antiprotons left over at the end of Tevatron stores.  
It is a ring of 
steel cases 
holding bricks of ``refrigerator'' magnets (the 
same permanent magnet used in home refrigerators).  
Permanent magnets do not need power supplies, 
cooling water systems, or electrical safety systems.  
The Recycler is a highly reliable storage ring 
for antiprotons.

The Tevatron was the world's first superconducting 
synchrotron.  A magnet with superconducting coils 
has no electrical resistance, and consumes minimal electrical 
power,  except that is needed to keep the magnets cold.  
The particles of a beam are guided around the closed 
path by dipole magnetic field.  The radius of the 
circle is 1000 meters.  As the beam energy is ramped 
up by RF cavities from 150 GeV to 980 GeV, the bending 
magnetic field and the RF frequency must be 
synchronized to keep the particles in the ring and 
this enables a stable longitudinal motion.  
The stability of the transverse motion is achieved with 
a series quadrupole magnets with alternating gradient.

\begin{figure}
   \begin{center}
      \parbox{1.5in}{\epsfxsize=\hsize\epsffile{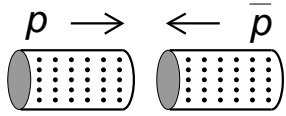}}
      \caption[$p\bar{p}$ collision]
              {$p\bar{p}$ collision.}
      \label{fig:ppbar}
   \end{center}
\end{figure}

\begin{figure}
   \begin{center}
      \parbox{5.0in}{\epsfxsize=\hsize\epsffile{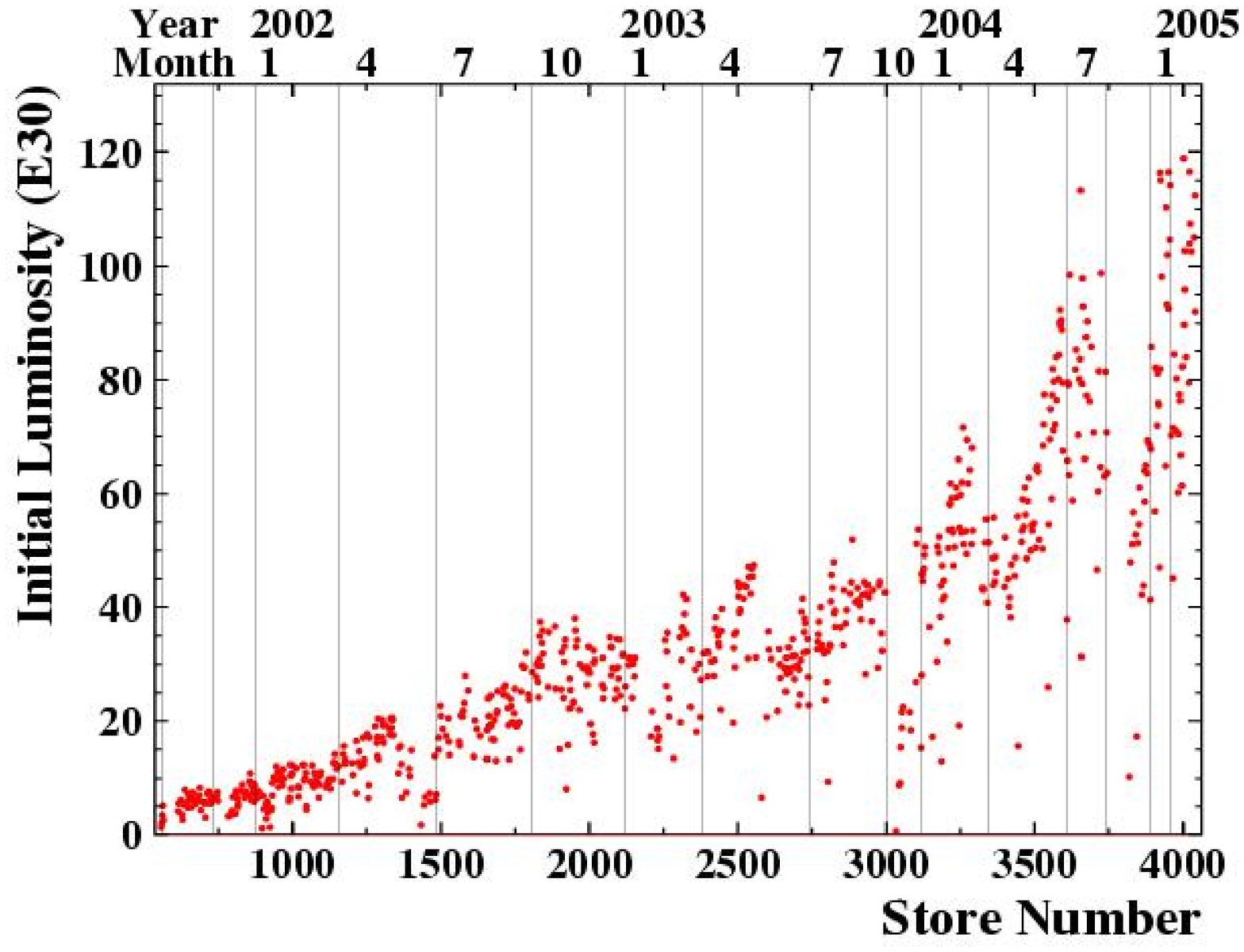}}
      \caption[Run II instantaneous initial luminosity]
              {Run II instantaneous initial luminosity.}
      \label{fig:store_lum}
   \vspace{0.5in}
      \parbox{5.0in}{\epsfxsize=\hsize\epsffile{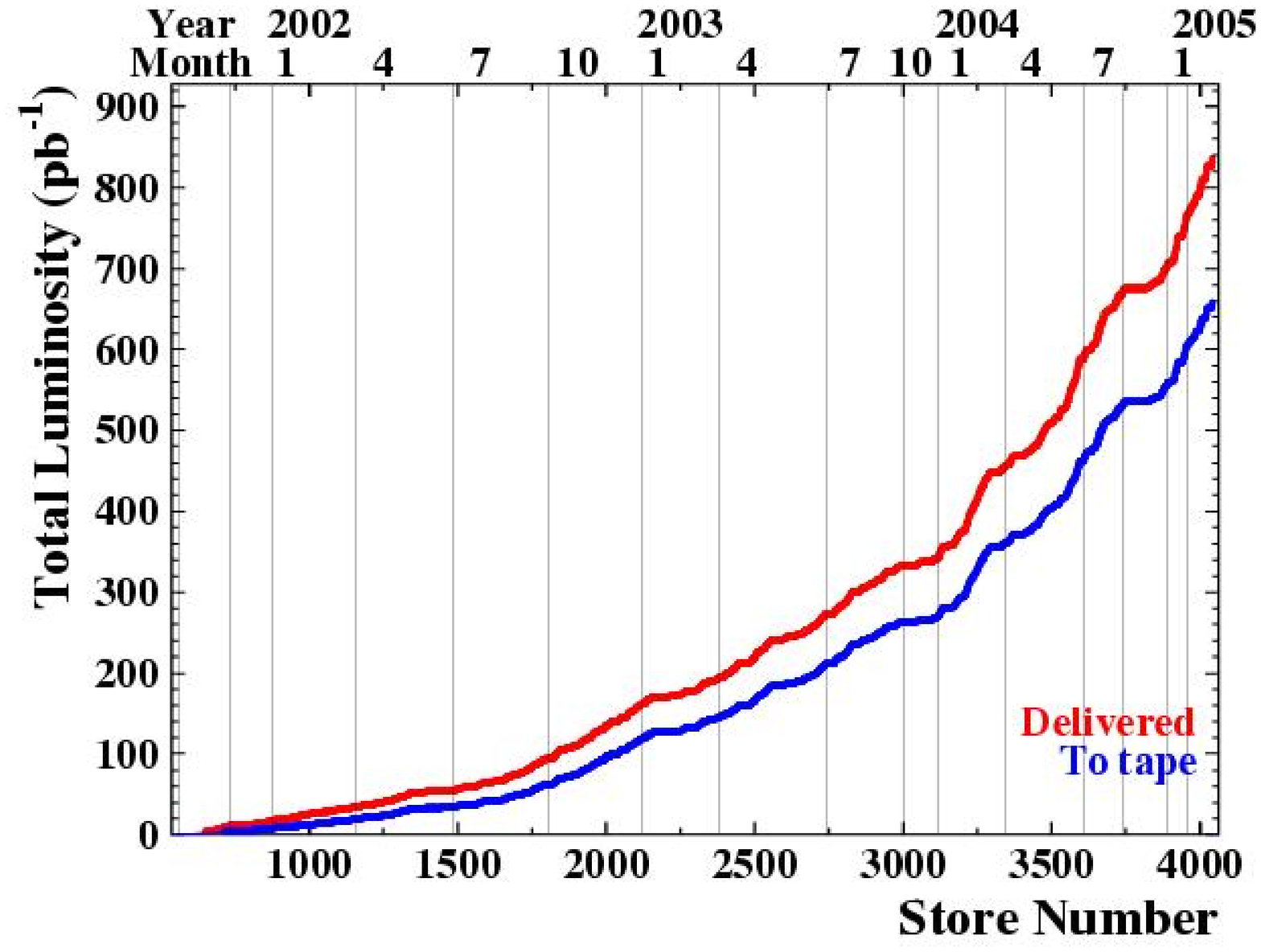}}
      \caption[Run II integrated luminosity]
              {Run II integrated luminosity.}
      \label{fig:store_tot}
   \end{center}
\end{figure}

Luminosity is a measure of the chance that a proton
will collide with an antiproton.  To achieve high 
luminosity we place as many particles as possible 
into as small a collision region as possible.  At the 
interaction point, the two beams of $p$ and 
$\bar{p}$ are brought together 
by special quadrupole magnets called Low Beta 
magnets, shown in Fig.~\ref{fig:ppbar}.
The current status (at the writing of the thesis) 
of the luminosity is shown in Fig.~\ref{fig:store_lum}, 
and the integrated luminosity delivered and to tape is 
shown in Fig.~\ref{fig:store_tot}. 

The design value for the peak instantaneous luminosity 
during Run II is $2\times10^{32}$ cm$^{-2}$s$^{-1}$.
Typically a year allows 10$^7$ seconds of running at 
the peak instantaneous luminosity.  This is about one third 
of the actual number of seconds in a year, which accounts 
both for the drop in luminosity and for a normal amount of 
down-time.  Using the conversion constant 
$1 \mbox{ fb} = 10^{-39} \mbox{ cm}^2$, the design value
corresponds to an integrated luminosity about 2 fb$^{-1}$
per year.  
Ultimately it is hoped that an integrated luminosity of
8$-$10 fb$^{-1}$ can be attained in Run II.
The total number of events $N$ in a scattering process 
is proportional to the luminosity and the cross section 
$\sigma$ of the process,
\begin{equation}
   N = L\sigma
\end{equation}
We can get a rough sense of the reach for new physics and 
the challenge of enhancing signal and suppressing 
background by considering the following examples.
At a center-of-mass energy of 1.96 TeV, we have
\begin{eqnarray}
   \sigma(p\bar{p}\to          \mbox{anything}) & \approx & 75 \mbox{ mb} \\
   \sigma(p\bar{p}\to t\bar{t}+\mbox{anything}) & \approx & 6  \mbox{ pb} \\
   \sigma(p\bar{p}\to hZ      +\mbox{anything}) & \approx & 75 \mbox{ fb} 
\end{eqnarray}

\newpage


\section{The CDF Dectector}
\label{sec:detector}

\begin{figure}
   \begin{center}
      \parbox{5.2in}{\epsfxsize=\hsize\epsffile{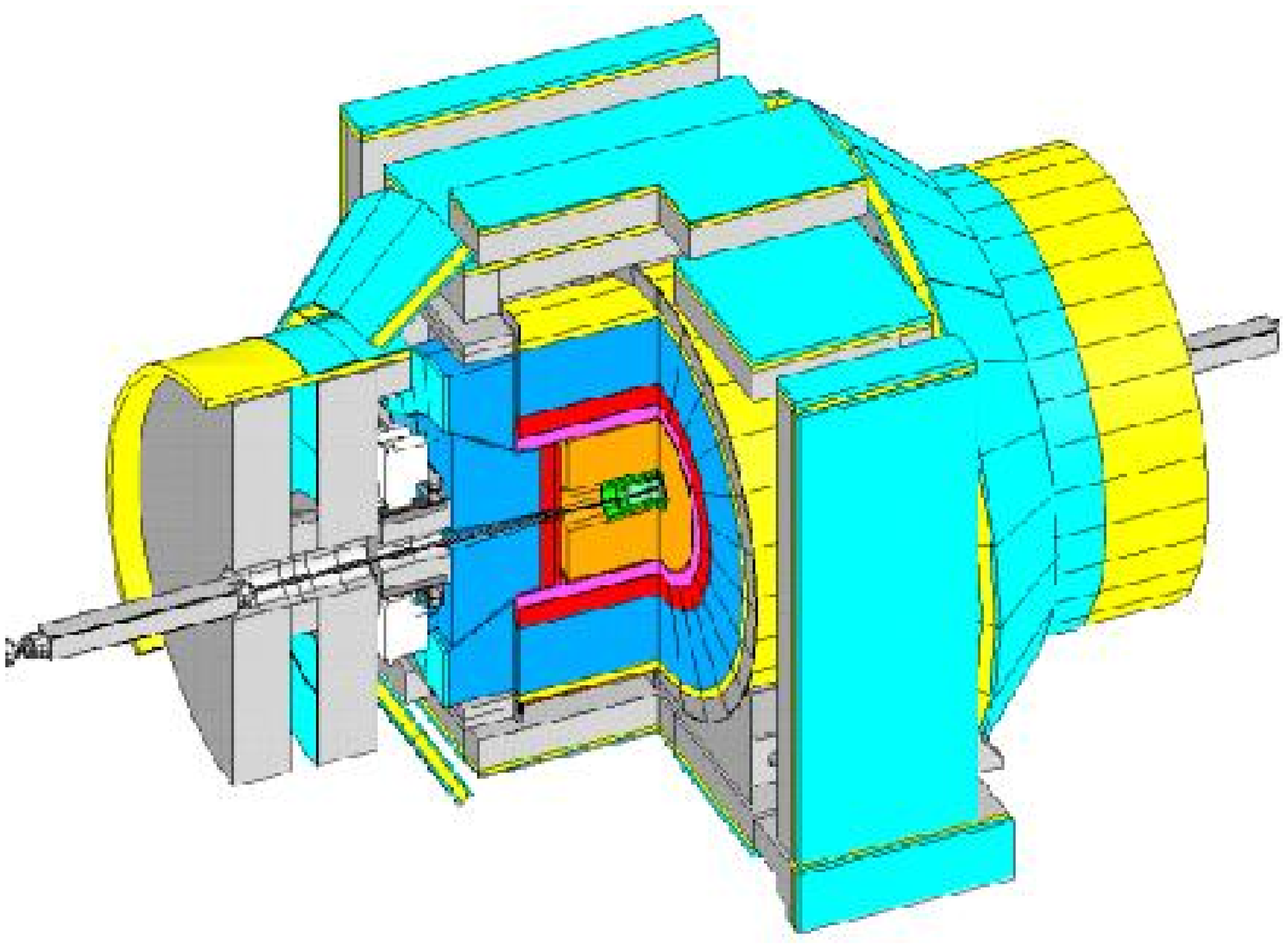}}
      \caption[Solid cutaway view of CDF II detector]
              {Solid cutaway view of CDF II detector.}
      \label{fig:cdfiso}
   \vspace{0.5in}
      \parbox{4.5in}{\epsfxsize=\hsize\epsffile{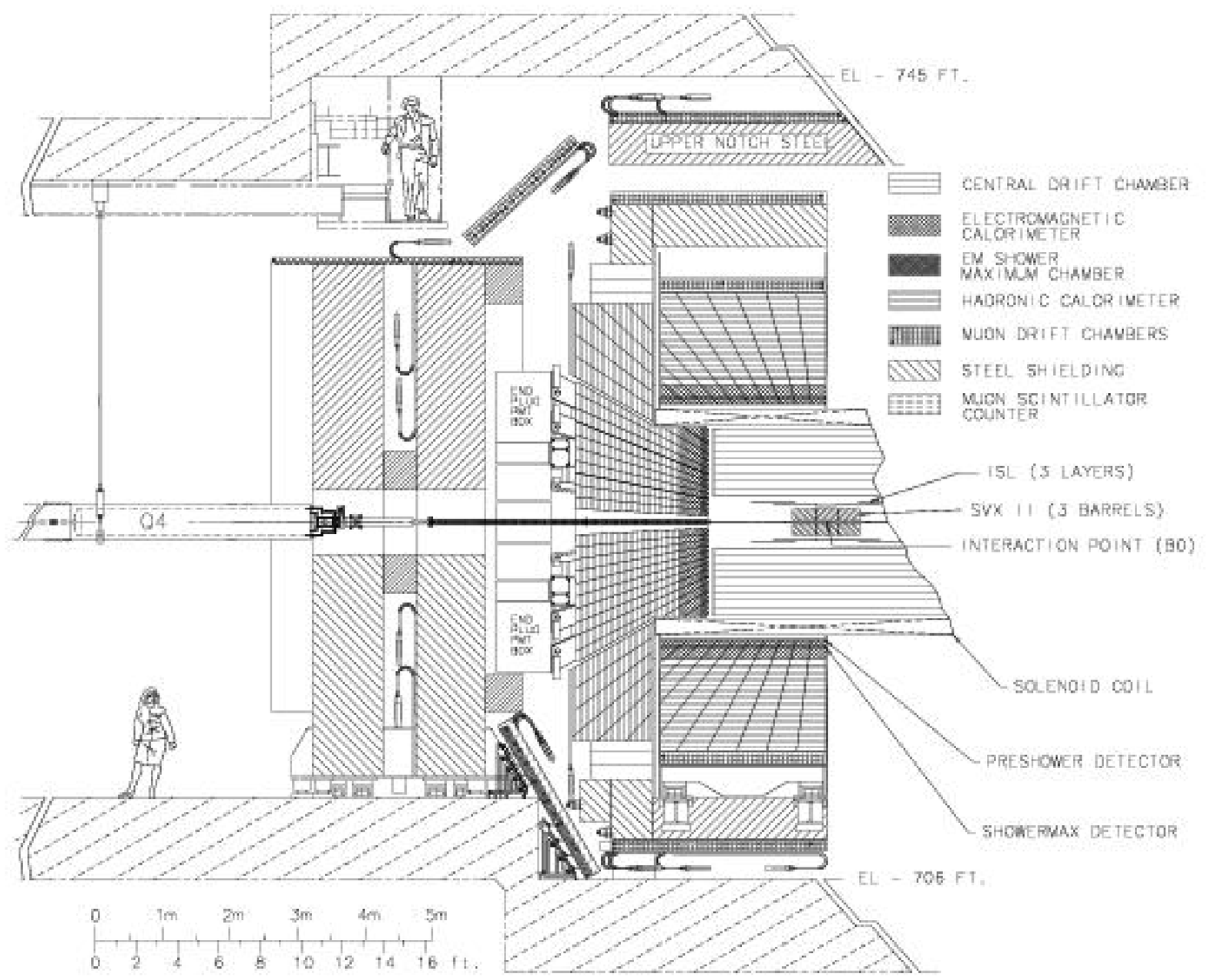}}
      \caption[Elevation view of CDF II detector]
              {Elevation view of CDF II detector.}
      \label{fig:cdfelev}
   \end{center}
\end{figure}

The CDF detector~\cite{Acosta:2004yw} is 
cylindrically symmetric around the beamline. 
A solid cutaway view is shown in Fig.~\ref{fig:cdfiso}, 
and an elevation view is shown in Fig.~\ref{fig:cdfelev}. 
It is a general-purpose solenoidal detector
with tracking system, calorimetry and muon detecion.
The tracking system is contained in a superconducting 
solenoid, 1.5~m in radius and 4.8~m in length. 
The magnetic field is 1.4~T, parallel to
the beamline.  The calorimetry and muon system are 
outside the solenoid.  These sub-systems will be
described in more details below.

\subsection{CDF Coordinate System}
\label{subsec:detector_xyz}

The origin of the CDF detector is its geometric
center.  The luminous region of the beam at the 
interaction point has Gaussian profiles with 
$(\sigma_x, \; \sigma_y, \; \sigma_z)_{beam}
 \approx 
 (0.003,    \; 0.003,    \; 30)$ cm. 
The $p\bar{p}$ collision point is not 
necessarily at the origin.

The CDF detector uses a right-handed coordinate
system.  The horizontal direction pointing out of 
the ring of the Tevatron is the positive $x$-axis.  
The vertical direction pointing upwards is the 
positive $y$-axis.  The proton beam direction 
pointing to the east is the positive $z$-axis.  

A spherical coordinate system is also used.  
The radius $r$ is measured from the center of 
the beamline.  The polar angle $\theta$ is 
taken from the positive $z$-axis.  The 
azimuthal angle $\phi$ is taken anti-clockwise 
from the positive $x$-axis.  

At a $p\bar{p}$ collider, the production of any 
process starts from a parton-parton interaction  
which has an unknown boost along the $z$-axis, 
but no significant momentum in the plane 
perpendicular to the $z$-axis, i.e. the 
transverse plane.  This makes the transverse 
plane 
an important plane in $p\bar{p}$ collision.   
Momentum conservation requires the vector sum of 
the transverse energy and momentum of all of the 
final particles to be zero.  The transverse 
energy $\et$ and transverse momentum $\pt$ are 
defined by
\begin{eqnarray}
   \et & = & E\sin\theta \\
   \pt & = & p\sin\theta 
\end{eqnarray}

Hard $p\bar{p}$ head-on collisions produce
significant momentum in the transverse plane.
The CDF detector has been optimized to measure 
these events.  On the other hand, the soft 
collisions such as elastic or diffractive 
interactions or minimum-bias events, and 
by-products from the spectator quarks from hard 
collisions, have most of their energy directed 
along the beampipe, and will not be measured by 
the detector.

Pseudorapidity $\eta$ is used by 
high energy physicists and is defined as
\begin{equation}
   \eta = -\ln\tan\frac{\theta}{2}
\end{equation}

Consider occupancy in a sample of large 
amount of $p\bar{p}$ collision events.  
Typically, particles in a $p\bar{p}$ 
collision event tend to be more in the forward 
and backward regions than in the central 
region because there is usually a boost 
along the $z$-axis, which could be shown 
in $\theta$ occupancy of the particles 
of the events in the sample.  Now we 
transform $\theta$ to $\eta$.  The derivative 
of $\eta$ is
\begin{equation}
   d\eta = -\frac{d\theta}{\sin\theta}
   \label{eq:eta_derivative}
\end{equation} 
A constant $\eta$ slice corresponds to 
variant $\theta$ slice which is smaller 
in the forward and backward regions 
than in the central region.  This can 
make the $\eta$ occupancy more uniform than 
$\theta$ occupancy.  For example, 
calorimeters are constructed in $\eta$ 
slices, instead of $\theta$ slices.  


\subsection{Tracking}
\label{subsec:detector_trk}

\begin{figure}
   \begin{center}
      \parbox{4.7in}{\epsfxsize=\hsize\epsffile{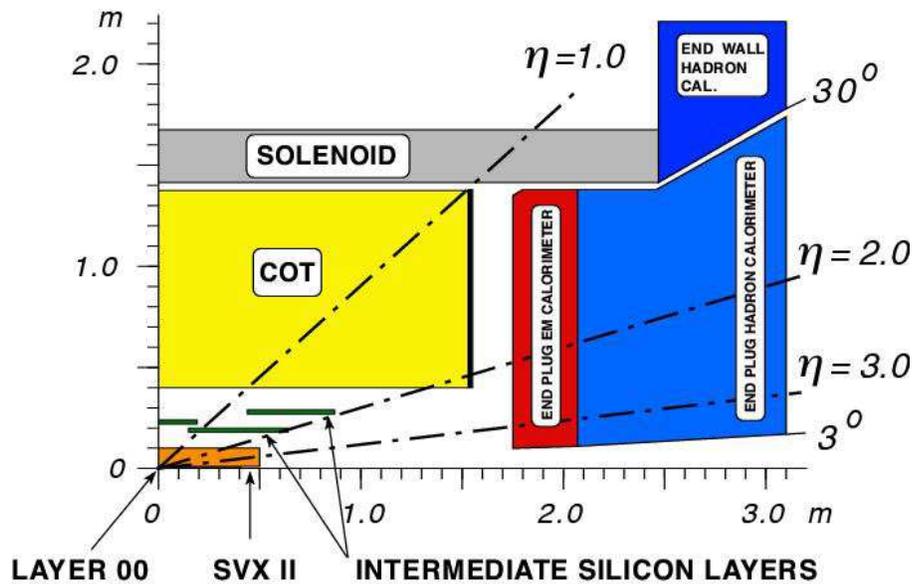}}
      \caption[CDF II tracking volume]
              {CDF II tracking volume.}
      \label{fig:cdfii_tracker_quad}
   \end{center}
\end{figure}

The tracking volume is surrounded by the solenoid 
magnet and the endplug calorimeters as shown in 
Fig.~\ref{fig:cdfii_tracker_quad}.  The tracking
system records the paths of charged particles 
produced in the $p\bar{p}$ collisions.  It consists of a 
silicon microstrip system~\cite{Sill:2000zz} with 
radius from $r=1.5$ to 28 cm and $|\eta|<2$, and 
an open-cell wire drift chamber called central 
outer tracker (COT)~\cite{Affolder:2003ep} with 
radius from $r=40$ to 137 cm and $|\eta|<1$.

The silicon microstrip is made from Si with a p-n junction.  When 
p-type semiconductors and n-type semiconductors are 
brought together to form a p-n junction, migration
of holes and electrons leaves a region of net charge
of opposite sign on each side, called the depletion 
region (depleted of free charge carriers).  The 
p-n junction can be made at the surface of a 
silicon wafer with the bulk being n-type (or the 
opposite way).  By applying a reverse-bias voltage 
we can increase the depletion region to the full volume of the device.  
A charged 
particle moves through this depletion region, 
creates electron-hole pairs which drift and are
collected at the surfaces.  
This induces a signal on metal strips deposited
on the surface, connected to readout amplifiers.

\begin{figure}
\begin{center}
   \begin{minipage}[b]{2.5in}
   \begin{center}
      \parbox{2.5in}{\epsfxsize=\hsize\epsffile{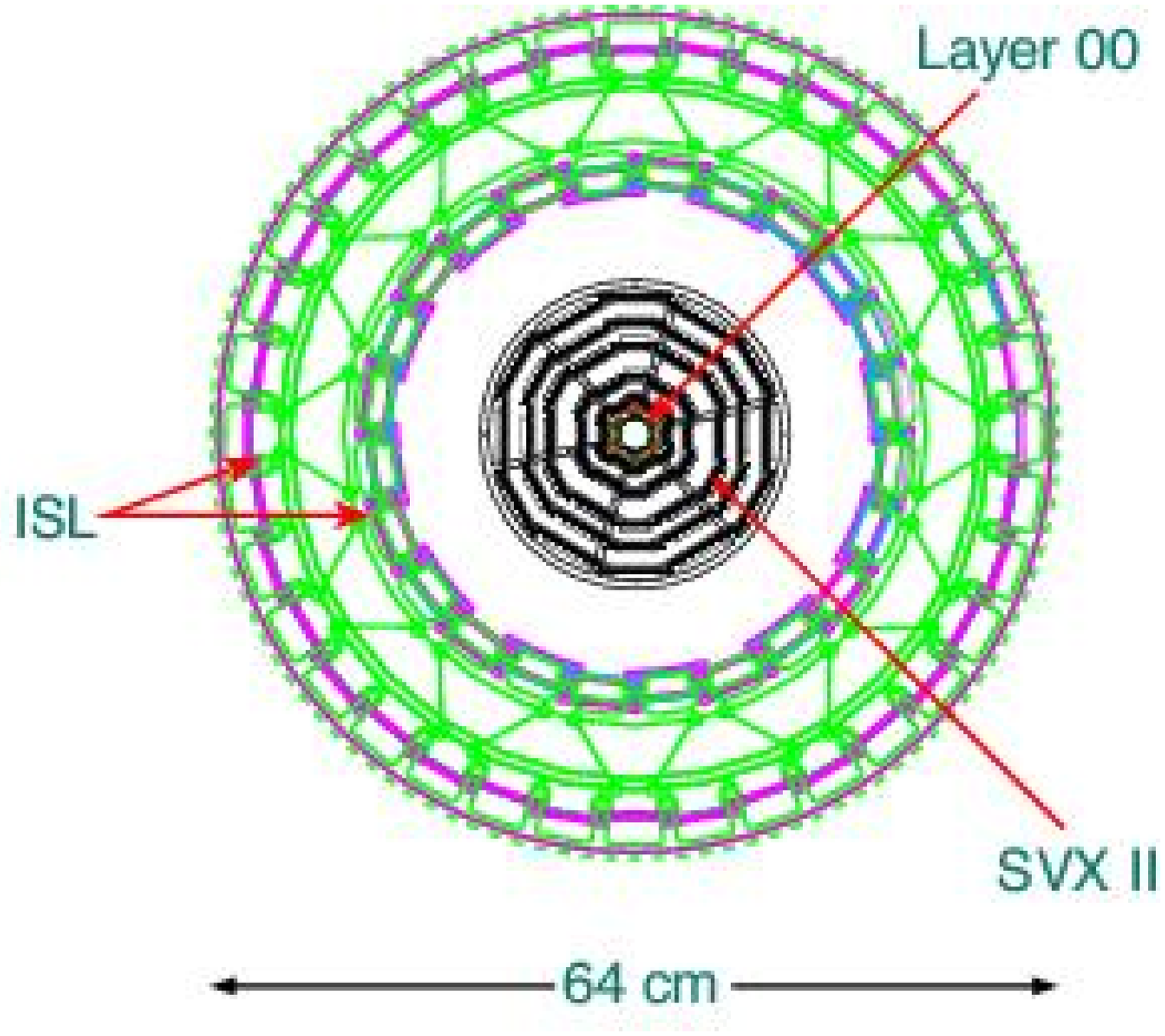}}
      \caption[Silicon system]
              {Silicon system.}
      \label{fig:cdf_silicon_endview}
   \end{center}
   \end{minipage}
\hspace{0.4in} 
   \begin{minipage}[b]{2.5in}
   \begin{center}
      \parbox{2.5in}{\epsfxsize=\hsize\epsffile{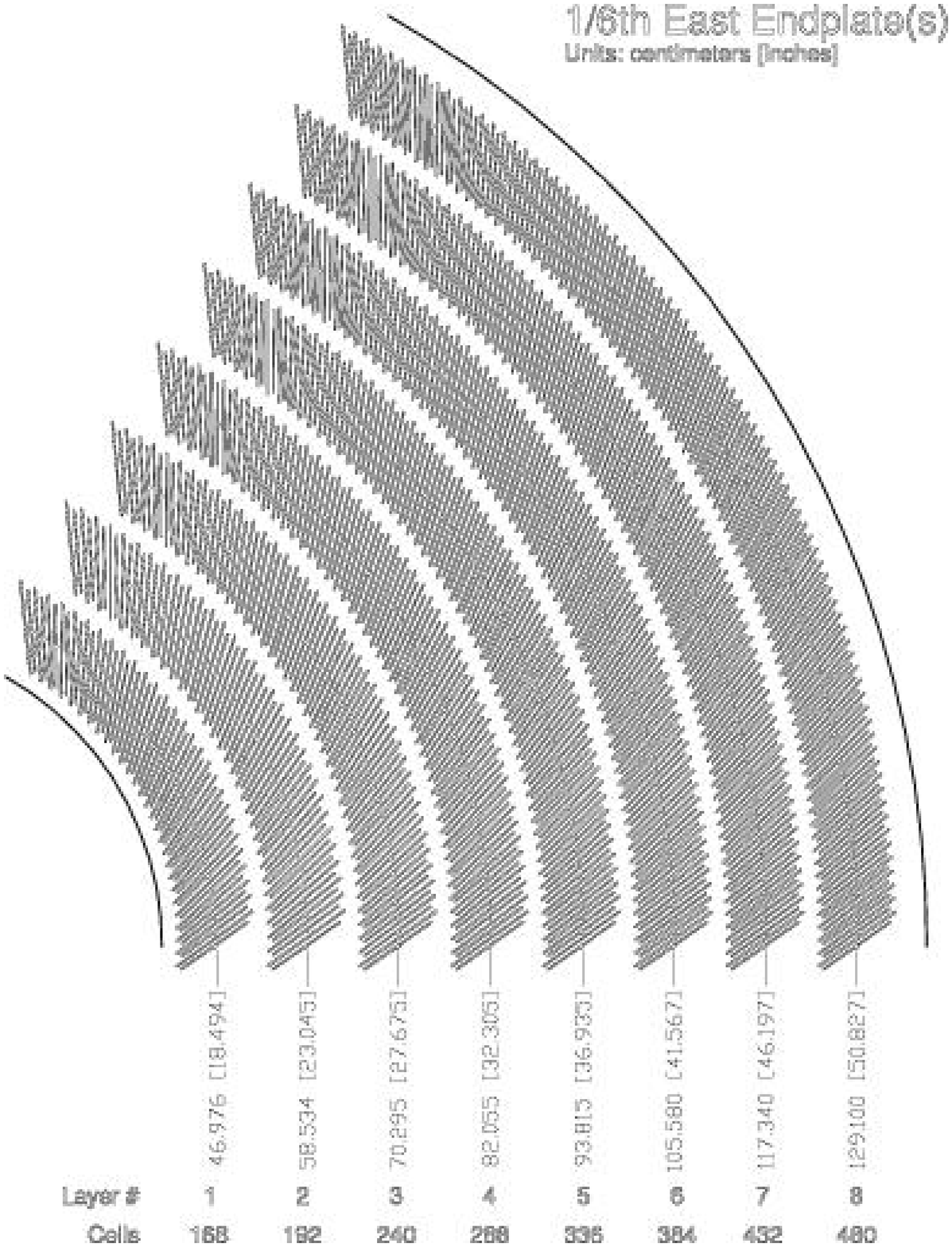}}
      \caption[COT superlayers]
              {COT superlayers.}
      \label{fig:cot_plate}
   \end{center}
   \end{minipage}
\end{center}
\end{figure}

The silicon microstrip detector consists of
three components: 
the Layer~00, the Silicon VerteX detector II (SVX~II), and 
the Intermediate Silicon Layers (ISL).
An end 
view is shown in Fig.~\ref{fig:cdf_silicon_endview}.
Layer~00 is physically mounted on and supported by 
the beam pipe.
The sensors are single-sided p-in-n 
silicon and have a pitch of 25~$\mu$m.
The next five layers compose the SVX II and are 
double-sided detectors.  The axial side of each 
layer is used for $r$-$\phi$ measurements and the 
sensors have a strip pitch of about 60~$\mu$m.  The 
stereo side of each layer is used for $r$-$z$ 
measurements.  Both 90$^{\circ}$ and small-angle 
stereo sensors are used in the pattern 
(90, 90, $-$1.2, 90, +1.2) degrees and have a 
strip pitch of (141, 125.5, 60, 141, 60)~$\mu$m from 
the innermost to outermost layers.  
The two outer layers compose the ISL and are
double-sided detectors with a strip pitch of
112~$\mu$m on both the axial and the 
1.2$^{\circ}$ stereo sides. 
This entire system allows charged particle 
track reconstruction 
in three dimensions.  The impact parameter 
resolution of SVX~II~+~ISL is 40~$\mu$m 
including 30~$\mu$m contribution from the 
beamline.  The~$z_0$ resolution of SVX~II~+~ISL 
is 70~$\mu$m.  

The COT is arranged in 8 superlayers shown in
Fig.~\ref{fig:cot_plate}.  The superlayers are
alternately axial and $\pm$2$^{\circ}$ stereo, 
four axial layers for $r$-$\phi$ measurement and  
four stereo layers for $r$-$z$ measurement.  Within 
each superlayer are cells which are tilted about 
30$^{\circ}$ to the radial direction to compensate 
for the Lorentz angle of the drifting charged 
particles due to the solenoid magnet field.  Each 
cell consists of 12 layers of sense wires, thus 
total 8$\times$12 = 96 measurements per track.

The COT is filled with a mixture of 
argone:ethane~=~50:50 which determines the drift 
velocity $v$.  A charged particle enters 
gas, ionizes gas and produces electrons.   
There is an electric field around each sense wire.  
In the low 
electric field region, the ionization electrons 
drift toward the sense wire.  In the high electric field 
region within a few radii of the sense wire, there 
is an avalanche multiplication of charges by 
electron-atom collision.  A signal is 
induced via the motion of electrons.  By measuring 
the drift time $t$ (the  arrival time of ``first'' 
electrons) at sense wire relative to collision time 
$t_0$, we can calculate the distance of the hit 
$D = v \Delta t$.  

A track is formed from a series of hits, fit to a helix.
We can measure the curvature of a track $C = 1/R$ and 
then calculate transverse momentum $\pt = 0.3$$R$$B$, 
with $\pt$, $R$ and 
$B$ in the units GeV/$c$, m, and T, respectively.  
The hit position resolution is approximately 140 
$\mu$m and the momentum resolution 
$\sigma(\pt)/\pt^2$ = 0.0015~(GeV/$c$)$^{-1}$.  


\subsection{Calorimetry}
\label{subsec:detector_calo}

\begin{figure}
   \begin{center}
      \parbox{5.5in}{\epsfxsize=\hsize\epsffile{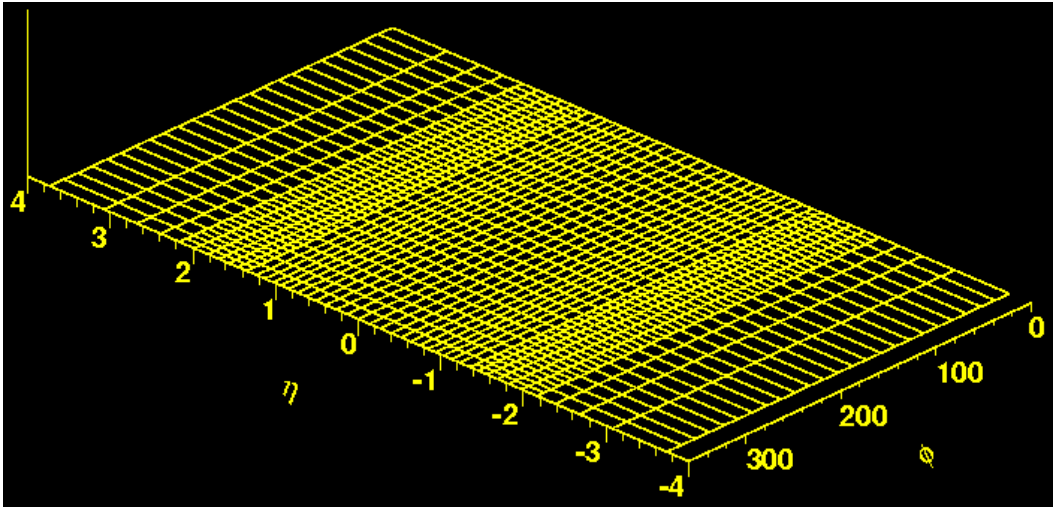}}
      \caption[Calorimeter tower segmentation in $\eta-\phi$ space]
              {Calorimeter tower segmentation in $\eta-\phi$ space.}
      \label{fig:calor_tower_segementation}
   \end{center}
\end{figure}

The CDF electromagnetic and hadronic sampling calorimeters
surround the tracking system and measure the energy flow 
of interacting particles up to $|\eta|<3.64$.  They are
segmented in $\eta$ and $\phi$ with a projective ``tower'' 
geometry, shown in Fig.~\ref{fig:calor_tower_segementation}.

The energy measurement is done by sampling 
calorimeters 
which are absorber and 
sampling scintillator sandwich with phototude readout.  
When interacting with the absorber, electrons lose energy by 
ionization and bremsstrahlung, and photons lose energy 
by the photoelectric effect, Compton scattering and pair 
production.  Both electrons and photons develop 
electromagnetic shower cascades.  The size of the 
longitudinal shower cascade grows only logarithmically 
with energy.  A very useful cascade parameter is 
the radiation length $X_0$ which is the mean distance 
for the $e^{\pm}$ to lose all but 1/e of its energy. 
For example, for a 10 GeV electron in 
lead glass, the maximum electromagnetic shower is at 
about 6$X_0$ and the 95\% containment depth is at about 16$X_0$.
Hadrons lose energy by nuclear interaction cascades which
can have charged pions, protons, kaons, neutrons, neutral 
pions, neutrinos, soft photons, muons, etc. It is much more 
complicated than an electromagnetic cascade and thus results in a large 
fluctuation in energy measurement.  In analogy to $X_0$, 
a hadronic interaction length $\lambda$ can be defined.  
Hadronic showers are much longer than the electromagnetic 
ones.

\begin{figure}
\begin{center}
   \begin{minipage}[b]{2.8in}
   \begin{center}
      \parbox{2.2in}{\epsfxsize=\hsize\epsffile{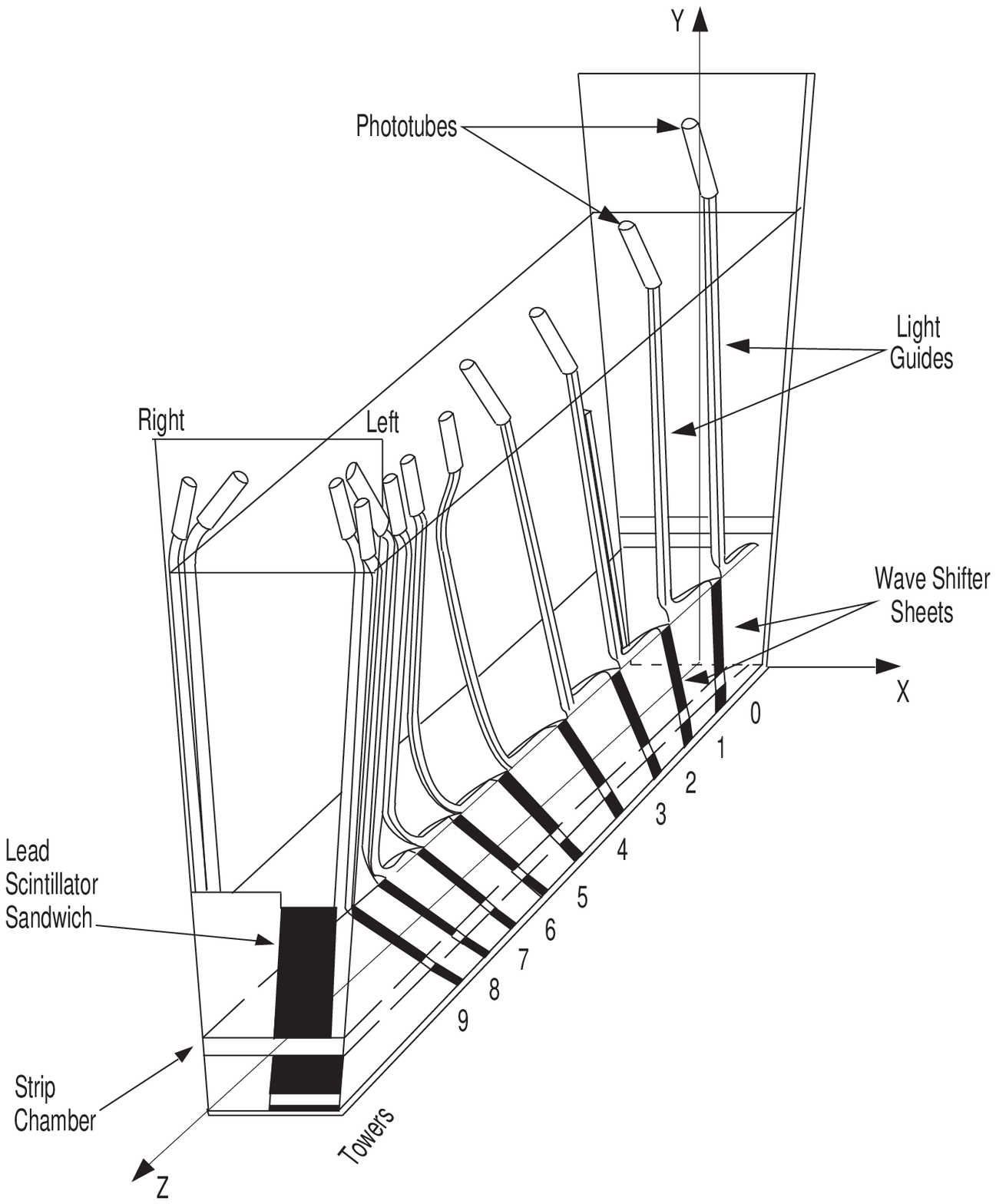}}
      \caption[CEM/CES/CHA wedge]
              {CEM/CES/CHA wedge.}
      \label{fig:CCAL}
   \end{center}
   \end{minipage}
\hspace{0.4in}
   \begin{minipage}[b]{2.2in}
   \begin{center}
      \parbox{2.2in}{\epsfxsize=\hsize\epsffile{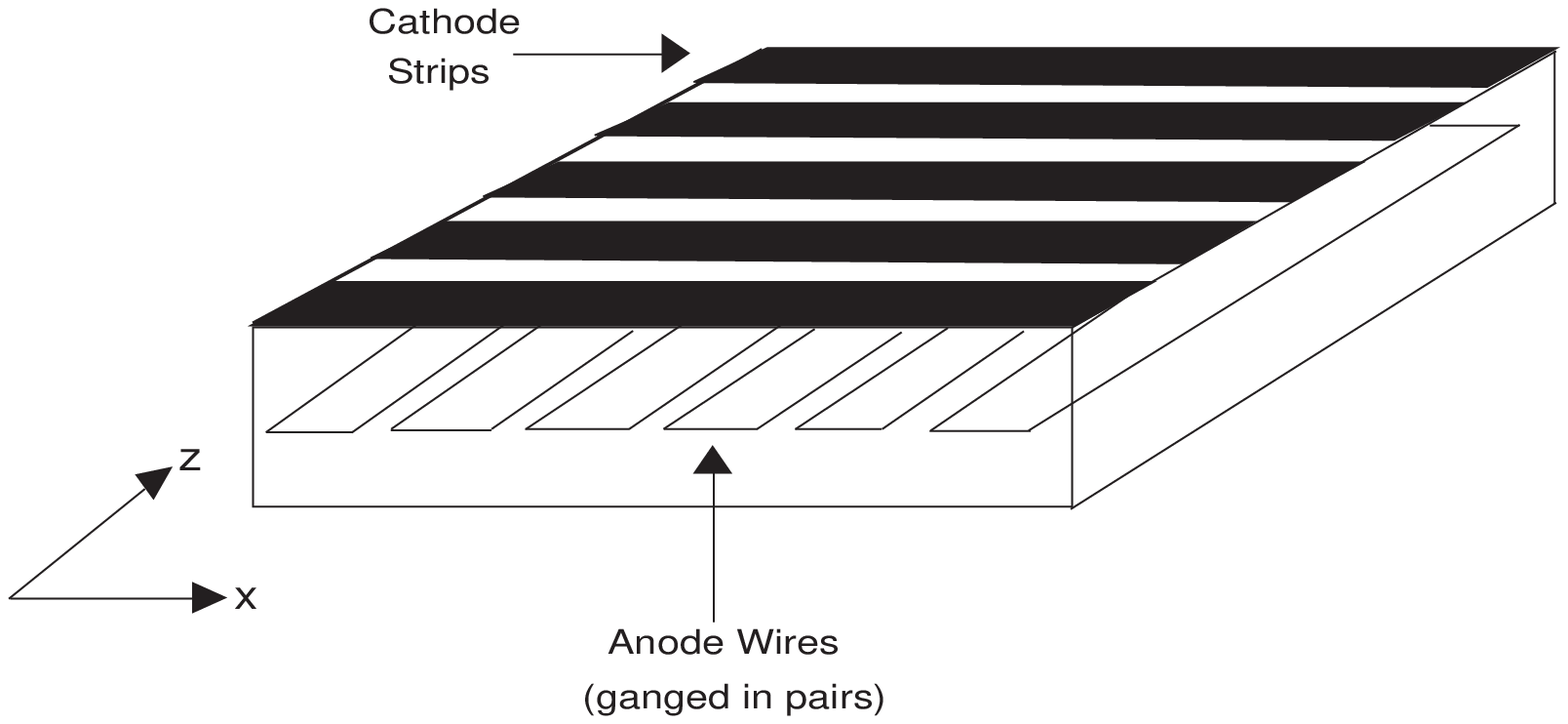}}
      \caption[CES strip and wire]
              {CES strip and wire.}
      \label{fig:CES}
   \end{center}
   \end{minipage}
\\
\vspace{0.3in}
   \begin{minipage}[b]{2.8in}
   \begin{center}
      \parbox{2.8in}{\epsfxsize=\hsize\epsffile{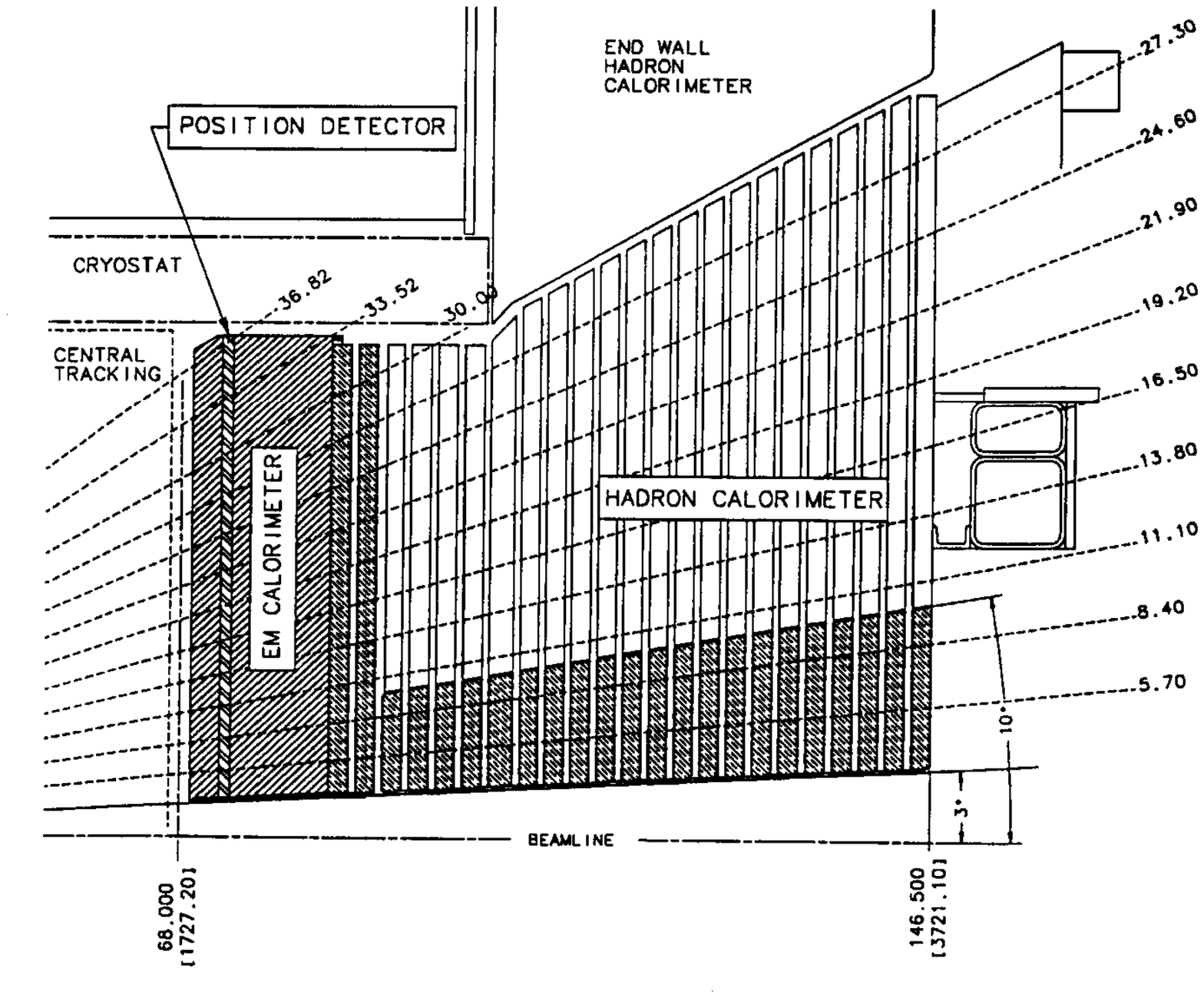}}
      \caption[PEM/PES/PHA layout]
              {PEM/PES/PHA layout.}
      \label{fig:PCAL}
   \end{center}
   \end{minipage}
\hspace{0.4in}
   \begin{minipage}[b]{2.2in}
   \begin{center}
      \parbox{1.8in}{\epsfxsize=\hsize\epsffile{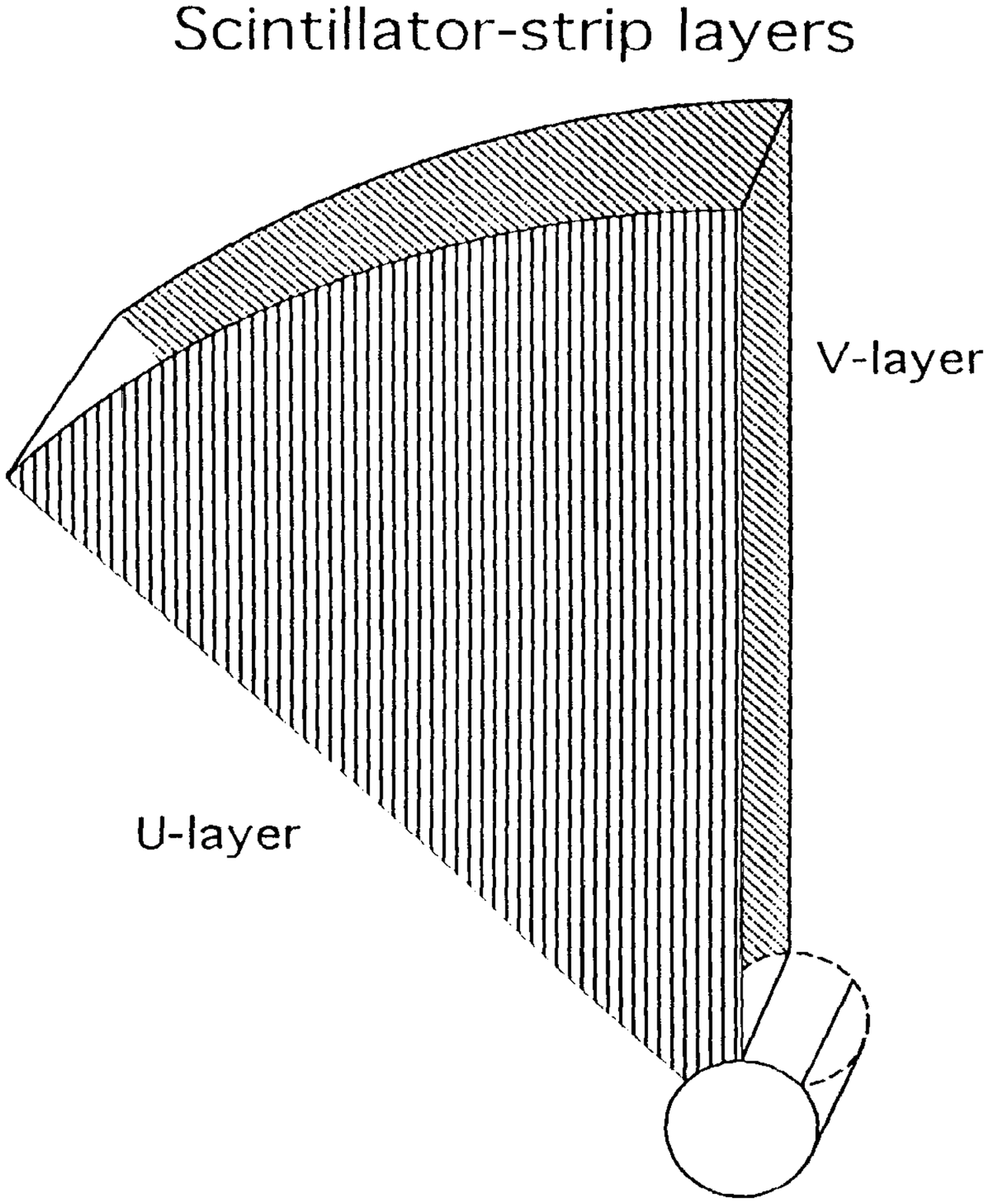}}
      \caption[PES U and V layers]
              {PES U and V layers.}
      \label{fig:PES}
   \end{center}
   \end{minipage}
\end{center}
\end{figure}

The central calorimeters consist of the central 
electromagnetic calorimeter (CEM)~\cite{Balka:1987ty},
the central hadronic calorimeter (CHA)~\cite{Bertolucci:1987zn}, 
and the end wall hadronic calorimeter (WHA).
At approximately 6$X_0$ in depth in the CEM, at which 
electromagnetic showers typically reach the maximum 
in their shower profile,
is the central shower maximum detector 
(CES).  The CEM and CHA are constructed in wedges which 
span 15$^{\circ}$ in azimuth and extend about 250~cm in 
the positive and negative $z$ direction, shown in 
Fig.~\ref{fig:CCAL}.  There are thus 24 wedges on both 
the $+z$ and $-z$ sides of the detector, for a total of 
48.  A wedge contains ten towers, each of which covers 
a range 0.11 in pseudorapidity.  Thus each tower subtends 
$0.11\times15^{\circ}$ in $\eta\times\phi$.  CEM covers 
$0<|\eta|<1.1$, CHA covers $0<|\eta|<0.9$, and WHA covers 
$0.7<|\eta|<1.3$.  

The CEM uses lead sheets interspersed with polysterene 
scintillator as the active medium and employs phototube 
readout, approximately 19$X_0$ in depth, and has an
energy resolution $13.5\%/\sqrt{\et}\oplus2\%$, where
$\oplus$ denotes addition in quadrature.
The CES uses proportional strip and wire counters in a 
fine-grained array, as shown in Fig.~\ref{fig:CES}, to 
provide precise position (about 2~mm resolution) and 
shape information for electromagnetic cascades.
The CHA and WHA use steel absorber interspersed with 
acrylic scintillator as the active medium.  They are approximately 
4.5$\lambda$ in depth, and have an energy resolution of
$75\%/\sqrt{\et}\oplus3\%$.

The plug calorimeters consist of the plug electromagnetic 
calorimeter (PEM)~\cite{Albrow:2001jw}, and the plug 
hadronic calorimeter (PHA).  At approximately 6$X_0$ in 
depth in PEM is the plug shower maximum detector (PES).
Fig.~\ref{fig:PCAL} shows the layout of the detector and 
coverage in polar angle $36.8^{\circ}>\theta>3^{\circ}$ 
($1.1<|\eta|<3.64$).  Each plug wedge spans 15$^{\circ}$ 
in azimuth, however in the range 
$36.8^{\circ}>\theta>13.8^{\circ}$ ($1.1<|\eta|<2.11$)
the segmentation in azimuth is doubled and each tower spans
only 7.5$^{\circ}$.

The PEM is a lead-scintillator sampling
calorimeter.  It is approximately 21$X_0$ in depth, and has an
energy resolution of $16\%/\sqrt{E}\oplus1\%$.
The PES consists of two layers
of scintillating strips: U and V layers offset from the
radial direction by $+22.5^{\circ}$ and $-22.5^{\circ}$ 
respectively, as shown in Fig.~\ref{fig:PES}.  The position 
resolution of the PES is about 1~mm. 
The PHA is a steel-scintillator sampling
calorimeter.  It is approximately 7$\lambda$ in depth, and
has an energy resolution of $74\%/\sqrt{E}\oplus4\%$.

\subsection{Muon Chambers}
\label{subsec:detector_muon}

The muon chambers are situated outside the calorimeters.
In addition to the calorimeters, the magnet return 
yoke and additional steel shielding are used to
stop electrons, photons and hadrons from entering 
the muon chambers.  The muon is a minimum ionizing 
particle which loses very little energy in detector 
materials.  The muon's lifetime is long enough to allow
it to pass through all the detector components,
reach the muon chambers, and decay outside.  

\begin{figure}
   \begin{center}
      \parbox{5.0in}{\epsfxsize=\hsize\epsffile{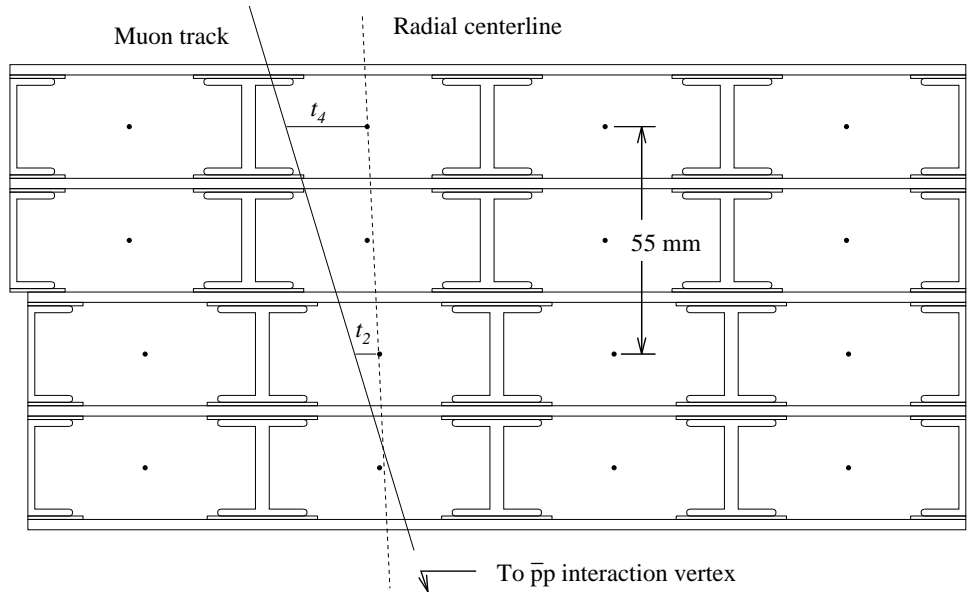}}
      \caption[Muon stub matching to a track]
              {Muon stub matching to a track.}
      \label{fig:CMU_tower}
   \end{center}
\end{figure}

A muon chamber contains a stacked array of drift 
tubes and operates with a gas mixture of 
argon:ethane~=~50:50.  The basic drift principle 
is the same as that of the COT, but the COT is a 
multi-wire chamber, while at the center of a muon
drift tube there is only a single sense wire.  The
sense wire is connected to a positive high voltage 
while the wall of the tube is connected to a 
negative high voltage to produce a roughly uniform 
time-to-distance relationship throughout the tube. 
The drift time of a single hit gives the distance
to the sense wire, and the charge division at each 
end of a sense wire can in principle be used to 
measure the longitudinal coordinate along the 
sense wire.  The hits in the muon chamber are linked 
together to form a short track segment called a 
muon stub.
If a muon stub is 
matched to an extrapolated track, a muon is 
reconstructed.  This is shown in Fig.~\ref{fig:CMU_tower}.

\begin{figure}
   \begin{center}
      \parbox{5.5in}{\epsfxsize=\hsize\epsffile{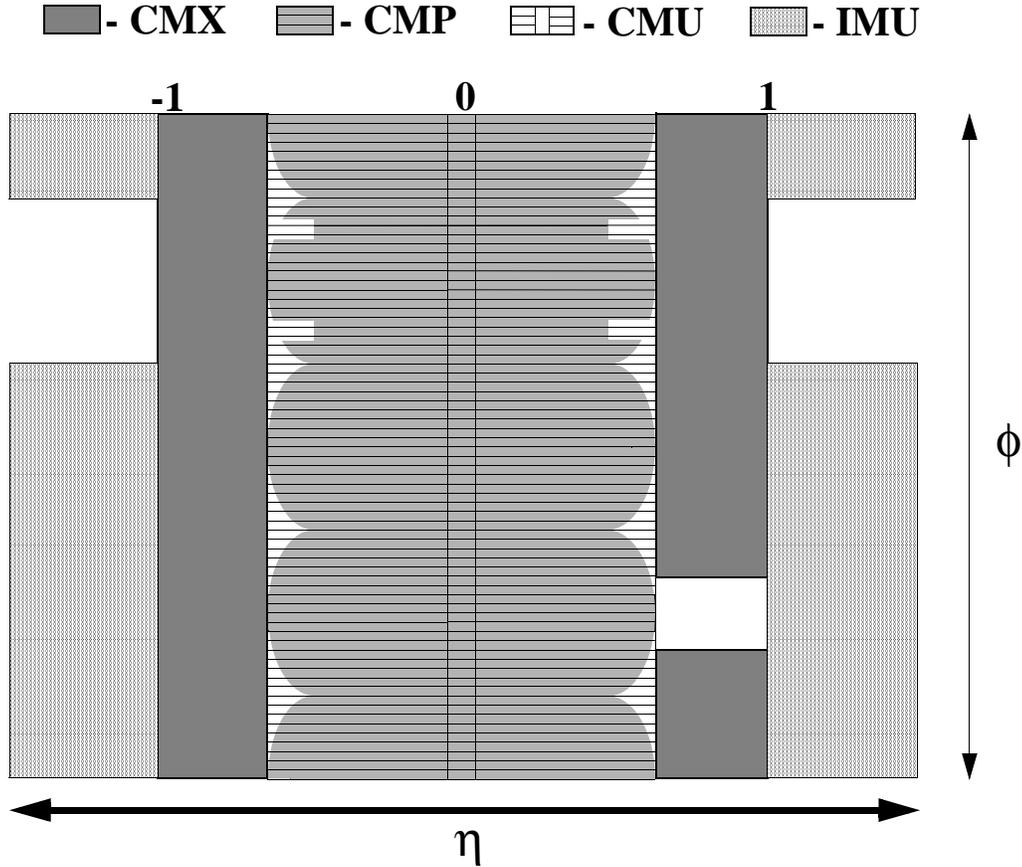}}
      \caption[Muon coverage in $\eta$ and $\phi$]
              {Muon coverage in $\eta$ and $\phi$.}
      \label{fig:newetaphimu}
   \end{center}
\end{figure}

There are four independent muon detectors: the 
central muon detector (CMU)~\cite{Ascoli:1987av}, 
the central muon upgrade (CMP), the central
muon extension (CMX), and the intermediate muon 
detector (IMU).  The muon coverage in $\eta-\phi$ 
space is shown in Fig.~\ref{fig:newetaphimu}.

The CMU is behind the central hadronic calorimeter
and has four layers of cylindrical drift chambers.
The CMP is behind an additional 60 cm of shielding 
steel outside the magnet return yoke.  It consists of 
a second set of four layers with a fixed length in 
$z$ and forms a box around the central detector.
Its psuedorapidity coverage thus varies with the 
azimuth.  A layer of scintillation counters (the
CSP) is installed on the outside surface of the 
CMP.  The CMU and CMP each covers $|\eta|<0.6$.  
The maximum drift time of the CMU is longer than 
the $p\bar{p}$ bunch crossing separation.  This 
can cause an ambiguity in the Level 1 trigger 
(described in the next section) 
about which bunch the muon belongs to.  By requiring CMP 
confirmation, this ambiguity is resolved by 
the CSP scintillators.

The CMX has eight layers and covers $0.6<|\eta|<1.0$.
A layer of scintillation counters (the CSX) is 
installed on both the inside and the outside 
surfaces of the CMX.  No additional steel was 
added for this detector because the large angle 
through the hadron calorimeter, magnet yoke,
and steel of the detector end support structure
provides more absorber material than in the 
central muon detectors.  The azimuthal coverage 
of CMX has a 30$^{\circ}$ gap for the solenoid 
refrigerator.

The IMU consists of barrel chambers (the BMU) 
and scintillation counters (the BSU), and covers 
the region $1.0<|\eta|<1.5$.  


\section{Trigger and Data Acquisition System}
\label{sec:trgdaq}

The trigger system has a three-level architecture:
level 1 (L1), level 2 (L2), and level 3 (L3).
The data volume is reduced at each level which 
allows more refined filtering at subsequent levels
with minimal deadtime.  The trigger needs to be fast 
and accurate to record as many interesting events as 
possible, while rejecting uninteresting events.

\begin{figure}
   \begin{center}
      \parbox{4.0in}{\epsfxsize=\hsize\epsffile{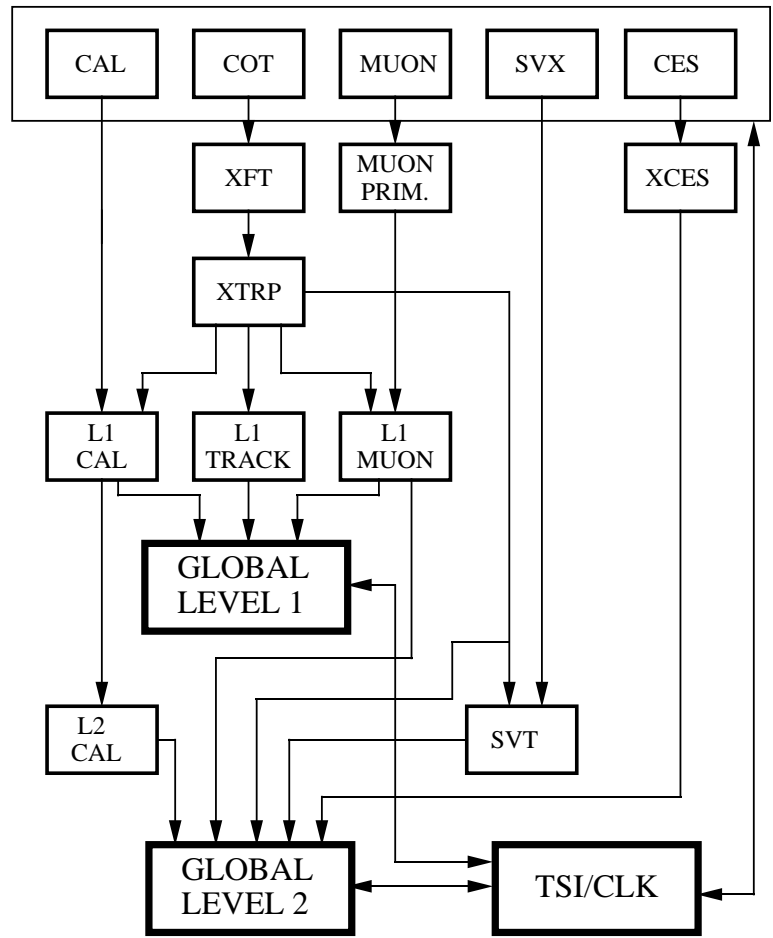}}
      \caption[Trigger system block diagram]
              {Trigger system block diagram.}
      \label{fig:trigger_system}
   \end{center}
\end{figure}

Each sub-detector generates primitives that we can 
``cut'' on.  The trigger system block diagram is 
shown in Fig.~\ref{fig:trigger_system}.  
The available trigger primitives at L1 are 
\begin{itemize}
\item XFT tracks, with $\phi$ and $\pt$ 
      provided by the eXtreme Fast Tracker using 
      the hits in the axial layers of the COT,
\item electrons, based on XFT and HAD/EM which is
      the ratio of the hadronic energy and the
      electromagnetic energy of a calorimeter 
      tower,
\item photons, based on HAD/EM ratio,
\item jets, based on EM+HAD,
\item muons, based on muon hits and XFT, and
\item missing $\et$ and sum $\et$ which are
      the negative of the vector sum and the 
      scalar sum of the energies of all of 
      the calorimeter towers, respectively.
\end{itemize}
The available trigger primitives at L2 are
\begin{itemize}
\item SVT, the Silicon Vertex Tracker
      trigger based on the track impact 
      parameter of displaced tracks,
\item jet clusters,
\item isolated clusters, and
\item EM ShowerMax which is the strip and 
      wire clusters in the CES.
\end{itemize}

There are two important factors for trigger design:
the time between beam crossing and $\bar{N}$, the 
average number of overlapping interactions in a 
given beam crossing.

      We can have many bunches in the Tevatron
      to enhance the luminosity.  Since the radius of the ring is 
      1000 m, a proton (or an anti-proton) at a speed 
      very close to the speed of light circulates the 
      ring once every 20 $\mu$s.  
      To accomodate 36 bunches, the maximum bunch 
      separation allowed is about 600 ns, and the 
      Run IIa configuration is 396 ns. 
      The 
      bunch separation defines an overall time constant 
      for signal integration, data acquisition and 
      triggering.  

\begin{figure}
   \begin{center}
      \parbox{5.0in}{\epsfxsize=\hsize\epsffile{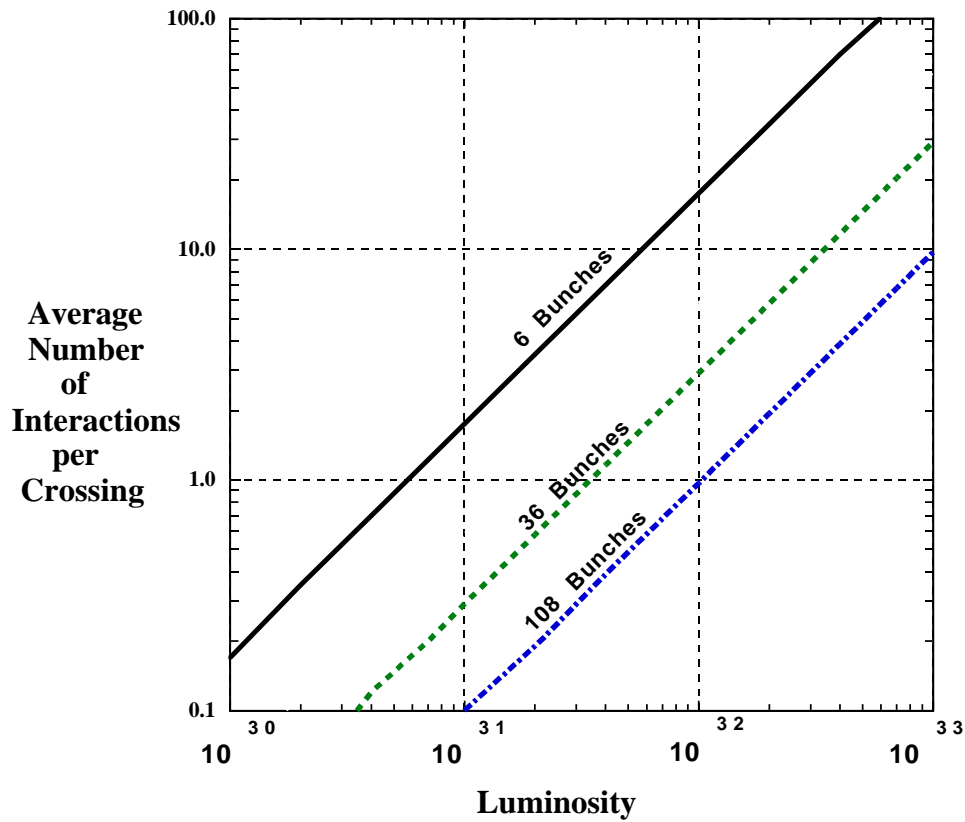}}
      \caption[Average number of interactions per crossing]
              {Average number of interactions per crossing
               for various bunches, as a function of 
               instantaneous luminosity.}
      \label{fig:nbar_noctc}
   \end{center}
\end{figure}

      Another key design input is the average number of 
      overlapping interactions $\bar{N}$, which is 
      shown as a function of luminosity and the number
      of bunches in 
      Fig.~\ref{fig:nbar_noctc}~\cite{Blair:1996kx}.  
      For example, with 36 bunches, $\bar{N}$ is about 
      1 at $3\times31$ cm$^{-2}$s$^{-1}$ and 
      about 10 at $4\times32$ cm$^{-2}$s$^{-1}$.  The 
      trigger with fast axial tracking at L1 can handle 
      the former environment, but cannot handle the 
      latter environment because of the presence of too many fake tracks.  
      To be able to handle $4\times32$ cm$^{-2}$s$^{-1}$ 
      we would need
      108 bunches and even that seems not enough, thus 
      we will also need to upgrade the trigger to include, 
      for example, stereo tracking at L1 to suppress fake 
      tracks.

\begin{figure}
   \begin{center}
      \parbox{5.0in}{\epsfxsize=\hsize\epsffile{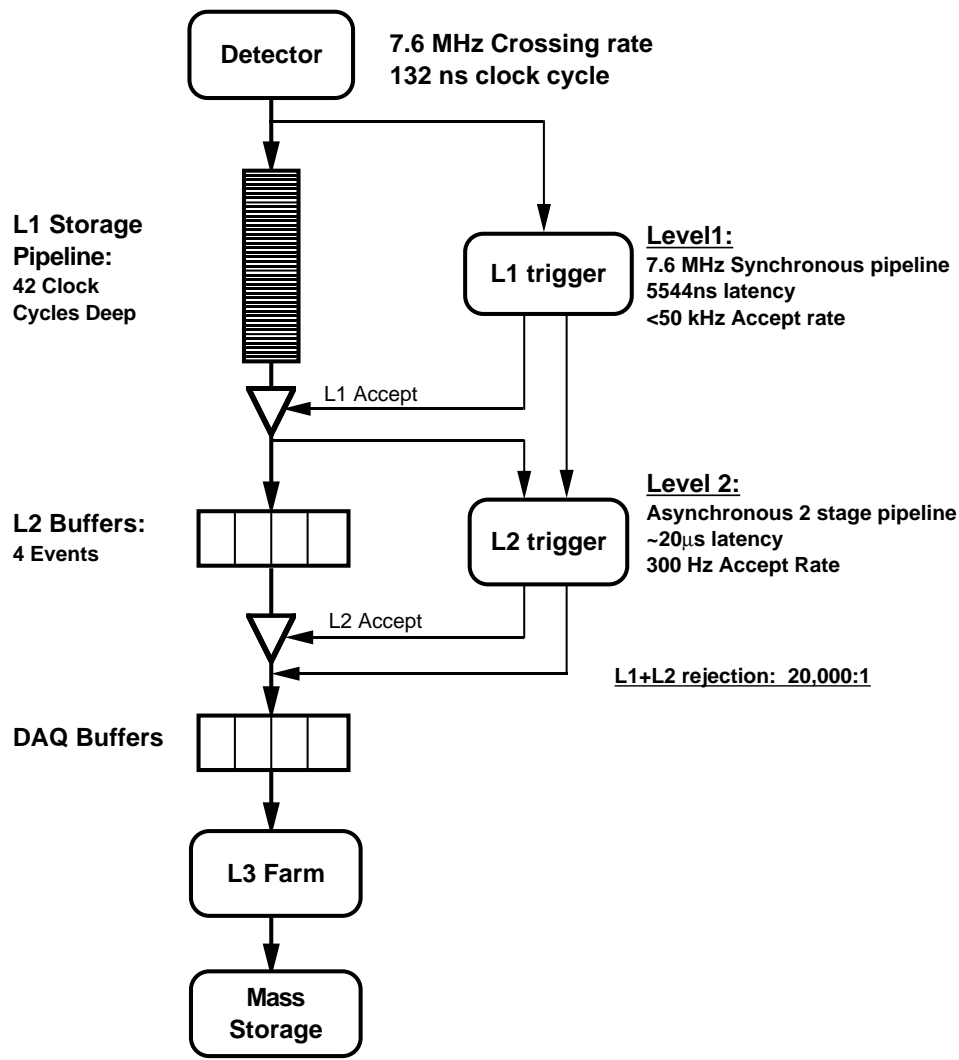}}
      \caption[Data flow of ``deadtimeless'' trigger and data acquisition]
              {Data flow of ``deadtimeless'' trigger and data acquisition.}
      \label{fig:cdf_dataflow}
   \end{center}
\end{figure}

The data flow in the trigger system is constrained 
by the processing time, i.e. how fast a decision 
can be made to clear events at each level and the 
tape writing speed for permanant storage at the 
end of the triggering process.  The implementation 
needs a sufficient buffer while filtering because 
any overflow means deadtime.  The ``deadtimeless'' 
design for 132 ns crossing is shown in 
Fig.~\ref{fig:cdf_dataflow}. 

      The L1 decision occurs at a fixed time 
      about 5.5 $\mu$s after beam collision.
      L1 is a synchronous hardware trigger.
      To process one event every 132 ns, each 
      detector element is pipelined to have 
      local data buffering for 42 beam crossings.
      The L1 accept rate is less than 50 KHz which
      is limited by the L2 processing time.

      The L2 decision time is about 20 $\mu$s.
      L2 is a combination of hardware and 
      software triggers and is asynchronous.
      If an event is accepted by L1, the
      front-end electronics moves the data
      to one of the four onboard L2 buffers.
      This is sufficient to process a L1 
      50 KHz accept rate and to average 
      out the rate fluctuations.
      The L2 accept rate is about 300 Hz which
      is limited by the speed of the 
      event-builder in L3.

      L3 is purely a software trigger consisting
      of the event builder running on a large PC 
      farm.  The event builder assembles
      event fragments from L1 and L2 into
      complete events, and then the PC farm
      runs a version of the full offline
      reconstruction code.  This means that
      fully reconstructed three-dimensional
      tracks are available to the trigger 
      decision.  The L3 accept rate is about 75 Hz 
      which is limited by tape writing speed 
      for permanent storage.

Once an event passes L3 it is delivered to the 
data-logger sub-system which sends the event out
to permanent storage for offline reprocess, 
and to online monitors which verify the entire 
detector and trigger systems are working properly.

The data used in this 
analysis were collected from March 2002 to 
September 2003.  It was for 396 ns with 36 
bunches and for luminosity about $3\times31$ cm$^{-2}$s$^{-1}$.  
This means that the trigger (designed for 132 ns) 
was sufficiently capable to handle the timing
of bunch crossing with no need to worry about 
multiple interactions in this environment.  



\chapter{Search Strategy}
\label{cha:Search_Strategy}

This chapter describes the overall logic of the high-mass 
tau tau search.  There are three steps:
\begin{enumerate}
\item Use $W\to\tau\nu$ events to cross check the $\tau$
      identification efficiency.

\item Use $Z\to\tau\tau$ events to study the low-mass 
      control region with $m_{vis}<$ 120 GeV/$c^2$.

\item Examine the high-mass signal region with
      $m_{vis}>$ 120 GeV/$c^2$ 
      for evidence of an excess signalling new physics.
\end{enumerate}

\vspace{0.2in}
\noindent{\bf Tau Hadronic Decays}
\vspace{0.1in}

The dominant decays of $\tau$'s are into leptons or 
into either one or three charged hadrons, shown in 
Table~\ref{tab:strategy_1}.  The following
short-hand notations for $\tau$ and its decays
are used,
\begin{eqnarray}
   \tau_e     & & \tau\to e\bar{\nu}\nu \\ 
   \tau_{\mu} & & \tau\to\mu\bar{\nu}\nu \\
   \tau_h     & & \tau\to\mbox{hadrons}~\nu 
\end{eqnarray}
The leptonic decays cannot be distinguished from 
prompt leptons.  So tau identification requires a hadronic 
tau decay only, with a mass less than 
\begin{equation}
   m(\tau) = 1.777 \mbox{ GeV/}c^2
\end{equation}
The net charge of the charged tracks is $\pm$1.  But 
we will not cut on charge because for very high energy 
taus there is an ambiguity of charge sign for very 
straight tracks. 

\begin{table}
   \begin{center}
      \begin{tabular}{|l|l|r|} \hline
         Decay Mode       & Final Particles                    & BR     \\ \hline \hline
         Leptonic         & $e^- \bar{\nu}_e \nu_{\tau}$       & 17.8\% \\
                          & $\mu^- \bar{\nu}_{\mu} \nu_{\tau}$ & 17.4\% \\ \hline
         Hadronic 1-prong & $\pi^- \nu_{\tau}$                 & 11.1\% \\
                          & $\pi^- \pi^0 \nu_{\tau}$           & 25.4\% \\
                          & $\pi^- 2\pi^0 \nu_{\tau}$          &  9.2\% \\
                          & $\pi^- 3\pi^0 \nu_{\tau}$          &  1.1\% \\
                          & $K^- \nu_{\tau}$                   &  0.7\% \\
                          & $K^- \pi^0 \nu_{\tau}$             &  0.5\% \\ \hline
         Hadronic 3-prong & $2\pi^- \pi^+ \nu_{\tau}$          &  9.5\% \\
                          & $2\pi^- \pi^+ \pi^0 \nu_{\tau}$    &  4.4\% \\ \hline
      \end{tabular}
   \end{center}
   \caption[Tau dominant decay modes]
           {Tau dominant decay modes and branching ratios.}
   \label{tab:strategy_1}
\end{table}

The characteristic signature of hadronically decaying 
taus is the track multiplicity distribution 
with an excess in the 1- and 3-track bins.  The excess, 
about 2:1 in these bins, is related to the tau hadronic 
branching ratios to one or three charged pions.  Quark 
or gluon jets from QCD processes tend not to have such 
low charged track multiplicity, but have a broader 
distribution peaking at higher multiplicities (3-5 
charged tracks).  Other final particles, namely photons,
electrons, and muons have mainly 0, 1, or 1 tracks, respectively, 
which are different from tau hadronic decays too.  
Seeing the tau's characteristic track multiplicity signature is 
a very important 
indication that backgrounds are under control.

Since $\sigma\cdot B(W\to\tau\nu)$ is about ten times
larger than $\sigma\cdot B(Z\to\tau\tau)$~\cite{Acosta:2004uq}
we will use $W\to\tau\nu$ events to cross check the 
tau identification efficiency.

\vspace{0.2in}
\noindent{\bf Di-Tau Visible Mass}
\vspace{0.1in}

There are six final states for tau pairs, shown
in Table~\ref{tab:strategy_2}.  $\tau_e\tau_e$ and 
$\tau_{\mu}\tau_{\mu}$ modes cannot be distinguished 
from the prompt $ee$ or the prompt $\mu\mu$, 
respectively.  $\tau_e\tau_{\mu}$ mode has a special 
signature, but its branching ratio is small and its
final particles tend to have low energy.  
For this analysis, we will look for three golden final 
states with at least one hadronic decay.

\begin{table}
   \begin{center}
      \begin{tabular}{|c|r|} \hline
         Final States           &   BR \\ \hline \hline
         $\tau_e\tau_h$         & 22\% \\
         $\tau_{\mu}\tau_h$     & 22\% \\
         $\tau_h\tau_h$         & 41\% \\
         $\tau_e\tau_{\mu}$     &  3\% \\
         $\tau_e\tau_e$         &  6\% \\
         $\tau_{\mu}\tau_{\mu}$ &  6\% \\ \hline
      \end{tabular}
   \end{center}
   \caption[Tau pair final states]
           {Tau pair final states and their branching ratios.}
   \label{tab:strategy_2}
\end{table}

The high-mass tau pair search will be based on just
counting the number of events with some specified set
of cuts.  It is desirable to measure for some variable
a distribution which agrees with the Standard Model in
some range, but deviates from it in another, thus giving a
more convincing signal while also providing an estimate
of the new particle's mass scale. 
 
There are at least two missing neutrinos in the golden
final states, and therefore six unknown momentum components. 
With only two constraints from the two components of 
the missing transverse energy 
and the two constraints from two tau masses, there is at 
least a 2-fold ambiguity. It is not possible to 
reconstruct the tau pair invariant mass in 
general. 

The mass of the sum of the two tau's visible momentum and 
the missing transverse energy $\met$ with its $z$-component 
set to zero is called the visible mass,
\begin{equation}
   m_{vis} = m(\tau^1_{vis} + \tau^2_{vis} + \met)
\end{equation}

The invariant mass of the irreducible $Z\to\tau\tau$ 
background peaks at $m(Z)\approx$ 91 GeV/$c^2$.  The 
visible mass distribution will be broadened and peak at 
somewhere less than 91 GeV/$c^2$.  We will study the
sample with
$m_{vis} < 120$ GeV/$c^2$ 
for $Z\to\tau\tau$ cross check.
After all of the cuts, we want the control sample to 
be dominated by $Z\to\tau\tau$ background, with jet 
background under control and other backgrounds
negligible.  A successful cross check between data 
and MC in the low-mass region will give us confidence 
to go further to the high-mass region.

\newpage

\vspace{0.2in}
\noindent{\bf Blind Analysis}
\vspace{0.1in}

If a new particle with high mass exists and the 
statistics are sufficient, it will show up in the 
high-mass signal region.  The strategy we choose is 
a blind analysis.  The data sample with
$m_{vis} > 120$ GeV/$c^2$
will be put aside until all selection criteria are
fixed and all backgrounds are determined.  The principle 
of a blind analysis is to avoid human bias.  If 
the selection cuts are decided by the distributions 
of high mass region in the real data sample, there 
will be a strong bias 
and the probabilities 
calculated are meaningless.  
Given good 
understanding of backgrounds, 
there will be two possibilities after examining the
data in the signal region.  Either one will observe
a number of events statistically consistent with the
expected background rate, or there will be an excess
signalling new physics.



\chapter{Particle Identification and Missing Transverse Energy}
\label{cha:PID}

\begin{figure}
   \begin{center}
      \parbox{4.3in}{\epsfxsize=\hsize\epsffile{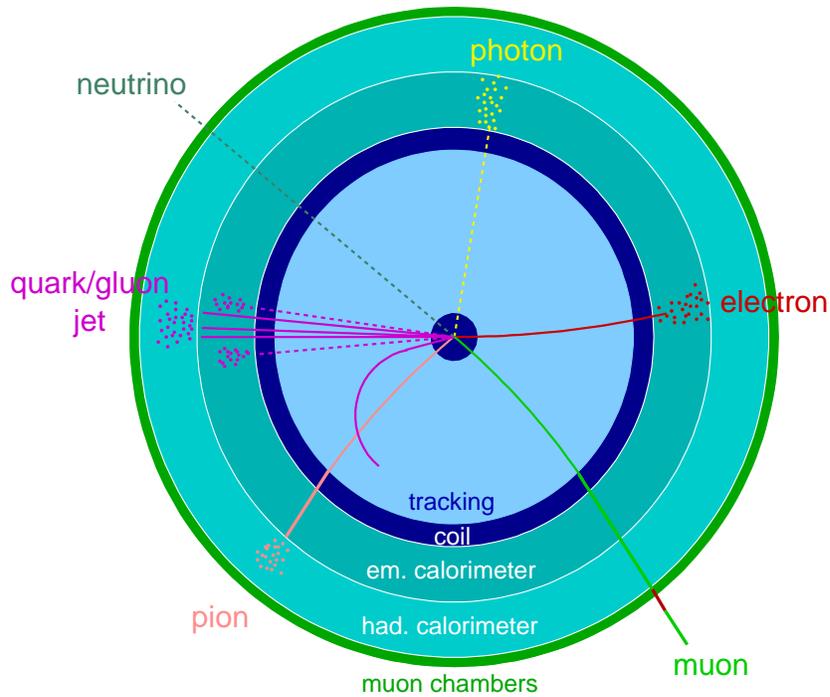}}
      \caption[Particle identification and missing transverse energy]
              {Patterns for identifying photon,
               electron, muon, charged hadron, and jet.
               Neutrino induces missing transverse energy.}
      \label{fig:PID}
   \end{center}
\end{figure}

High energy $p\bar{p}$ collisions can produce a large number of 
particles.  As illustrated in Fig.~\ref{fig:PID}, the CDF detector with 
its tracking system, calorimeter and muon chambers can identify the 
following particles by the following patterns:
\begin{itemize}
\item photon:  cascade showering in electromagnetic calorimeter,
      but no associated charged tracks;

\item electron: a track, and cascade showering in 
      electromagnetic calorimeter;

\item muon: a track, minimum ionization energy 
      deposit in calorimeter, and hits in muon chambers;

\item jet: an object which cannot be identified as an isolated 
      photon, or an isolated electron, or an isolated muon is 
      identified as a jet; 

\item missing transverse energy ($\met$): an imbalance of transverse 
      energy in the whole calorimeter.  
\end{itemize}
The final particles and the $\met$ are 
reconstructed by CDF II offline programs.

\section{Monte Carlo Simulation}
\label{sec:MC}

Often we need to predict the output in the detector 
including the final reconstructed particles and the $\met$ of 
a particular interesting process and compare 
with data.  Usually the phase space of an event
of the $p\bar{p}$ collision is too 
complicated to be calculated analytically.
In this case Monte Carlo (MC) simulation is used.
It has become a powerful tool used in 
many research areas including high energy 
physics.

A well-known MC example is the Buffon's Needle.
It involves dropping a needle on a lined sheet 
of paper and determining the probability of the 
needle crossing one of the lines on the page. 
The remarkable result is that the probability 
is directly related to the value of the 
mathematical $\pi$.  Suppose the length of the 
needle is one unit and the distance between 
the lines is also one unit.  There are two 
variables, the angle $\theta$ at which the 
needle falls and the distance $D$ from the 
center of the needle to the closest line.
$\theta$ can vary from 0$^{\circ}$ to 180$^{\circ}$ 
and is measured against a line parallel to the 
lines on the paper.  $D$ can never be more than 
half the distance between the lines.  The needle 
will hit the line if $D\le\frac{1}{2}\sin\theta$.  
How often does this occur?  The probability 
$\mathcal{P}$ is $2/\pi$ by integrating over 
$\theta$.  With a computer, we can 
generate a large sample of random needle
drops.  The probability $\mathcal{P}$ can 
be simply taken as the number of hits divided 
by the number of drops, yielding
$\pi=2/\mathcal{P}$.

Here we discuss the basic techniques of MC simulation.
For a one-dimensional integral, we can choose $n$ numbers
$x_i$ randomly with probability density uniform on the
interval from $a$ to $b$, and for each $x_i$ evaluate
the function $f(x_i)$.  The sum of these function values,
divided by $n$, will converge to the expectation of the
function $f$.
\begin{equation}
   \int_a^bf(x)dx 
   =       (b-a)\langle f(x)\rangle 
   \approx (b-a)\frac{1}{n}\sum_{i=1}^nf(x_i) 
   =       (b-a)\overline{f_n}
\end{equation}

The central limit theorem tells us that the sum
of a large number of independent random variables
is always normally distributed (i.e. a Gaussian
distribution), no matter how the individual 
random variables are distributed.  To understand
this, we can test with uniformly distributed 
random variable $x_1$, $x_2$, $x_3$, $x_4$, 
(a) $x_1$ is a uniform distribution; 
(b) $x_1+x_2$ is a triangle distribution; 
(c) $x_1+x_2+x_3$ is already close to a Gaussian 
    distribution; 
(d) $x_1+x_2+x_3+x_4$ is almost like the exact
    Gaussian distribution.
Applying this theorem, we know the MC  
method is particularly useful as we can also
calculate an error on the estimate by computing
the standard deviation,
\begin{equation}
   \langle f(x)\rangle 
   = \overline{f_n} \pm \frac{\sigma_n}{\sqrt{n}}
\end{equation} 
where 
$\sigma_n = (\overline{f_n^2} - \overline{f_n}^2)^{1/2}$ 
and
$\overline{f_n^2} = \frac{1}{n}\sum_{i=1}^nf^2(x_i)$.
The convergence for numerically evaluating the 
integral goes as $1/\sqrt{n}$ with the number of 
function evaluation, $n$.  
And obviously if the distribution $f(x)$ is flatter, then the
$\sigma_n$ is smaller for the same number of events in a
sample generated. If there is a peak in the distribution
such as the distribution of a resonance production, it is 
better to transform that variable to 
some other variable with a flatter distribution in order 
to converge faster.

The generalisation to multi-dimensional integrals
$\int\!f(x,y,z,...)dxdydz...$ is straightforward.  
We can choose $n$ numbers of grid $(x,y,z,...)$ 
randomly with probability density uniform on the 
multi-dimensional phase space, and for each grid
evaluate the function $f(x,y,z,...)$.  The sum of 
these function values, divided by $n$, will 
converge to the expectation of the function $f$.
A nice feature is that it will always converge as 
$1/\sqrt{n}$, even for very high dimensional 
integrals.  This can make the performance of the 
MC method on multi-dimensional integrals 
very efficient.  

In high energy physics, an event occurs with a
probability in the phase space of the kinematic
variables.  A MC simulation generates a large
number of random events according to the probability
described by a model.
With a large sample, we can get the predictions 
of the model by looking at the distributions of the 
kinematic variables and the derived variables, and 
the correlations among the variables.  By confronting
the predictions with real data, it is possible to tell
if a model describes Nature correctly.

For this analysis, 
we use PYTHIA 6.215 program~\cite{Sjostrand:2000wi}
with CTEQ5L parton density functions (PDF's)~\cite{Lai:1999wy}
to generate the large samples of the processes of 
$p\bar{p}$ collision, such as 
$p\bar{p}\to\gamma^*/Z\to\tau\tau$, 
$p\bar{p}\to Z'\to\tau\tau$,
$p\bar{p}\to A\to\tau\tau$,
and use TAUOLA 2.6~\cite{Was:2000st} to simulate tau decays.
We use GEANT 3~\cite{GEANT3:1993} to simulate the 
response to the final particles in the CDF II detector.  


\section{Tau Identification}
\label{sec:TauId}

\begin{figure}
   \begin{center}
      \parbox{3.2in}{\epsfxsize=\hsize\epsffile{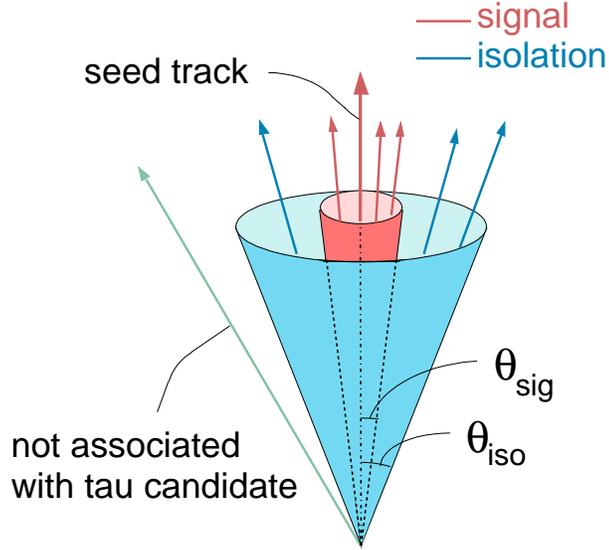}}
      \caption[Illustration of tau isolation cone definitions]
              {Illustration of tau isolation cone definitions.}
      \label{fig:TauId_iso}
   \end{center}
\end{figure}

Tau leptons decay predominantly into charged and neutral pions 
and suffer from large backgrounds from jet production.  Hadronic 
tau decays appear in the detector as narrow isolated jets.  The 
most powerful cut to suppress the jet background is in fact isolation,
requiring no other tracks or $\pi^0$s near the tau cone. 
To do this we define a signal cone and an isolation cone around 
the direction of the seed track (the track with the highest $\pt$)
and then require that there is no track 
or $\pi^0$ between the signal cone and the isolation cone.  This is 
shown in Fig.~\ref{fig:TauId_iso}.

\subsection{Cone Size Definition}
\label{subsec:TauId_cone}

There are two useful cone size definitions.
One is to construct a cone in $\Delta R$ defined 
below which has relativity invariance under a
boost along the $z$-axis.  The other 
is to construct a cone in three-dimensional separation 
angle, $\alpha$, which has geometry invariance.
Below we discuss
why $\Delta R$ is chosen as cone size definition 
for jet identification and why $\alpha$ is chosen 
as cone size definition for hadronic tau 
identification.

We start with the discussion on relativity invariance.
For a particle under a boost $\beta = v/c$ along 
the $z$-axis and $\gamma = (1-\beta^2)^{-1/2}$, 
its four-momentum $(p_x, \; p_y, \; p_z, \; E)$ 
is transformed to
\begin{equation}
   \left(
      \begin{array}{cccc}
         1 & 0 &           0 &           0 \\
         0 & 1 &           0 &           0 \\
         0 & 0 &      \gamma & \beta\gamma \\
         0 & 0 & \beta\gamma &      \gamma
      \end{array}
   \right)
   \left(
      \begin{array}{c}
         p_x  \\
         p_y  \\
         p_z  \\
         E
      \end{array}
   \right)
   =
   \left(
      \begin{array}{c}
         p_x                \\
         p_y                \\
         \gamma(p_z+\beta E) \\
         \gamma(\beta p_z+E)
      \end{array}
   \right)
\end{equation}
The $p_x$ and $p_y$ components in the transverse 
plane are not changed, while the $p_z$ component 
and the energy are changed.  
Rapidity is defined by
\begin{equation}
   y = \frac{1}{2}\ln\frac{E+p_z}{E-p_z}
\end{equation}
Using 
$\tanh^{-1}\beta
 = \frac{1}{2}
   \ln
   \frac{1+\beta}{1-\beta}$,
it is easy to check that rapidity has a nice 
additive property under the boost along 
the $z$-axis,
\begin{equation}
   y \to y + \tanh^{-1}\beta
\end{equation}
For ultra-relativistic particle with 
$p\gg m$, we have 
$p_z/E 
 \approx p_z/p 
  = \cos\theta$.
Using
$\cos\theta
  = (1-\tan^2\frac{\theta}{2})/(1+\tan^2\frac{\theta}{2})$,
the rapidity is well approximated by 
pseudorapidity~$\eta$, 
\begin{equation}
   \eta = -\ln\tan\frac{\theta}{2}
\end{equation}

Particles in a jet deposits energy in the calorimeter towers.
For the traditional cone jet
algorithm, we can call the tower with 
$\et$ above a seed threshold as the seed 
(abbreviated as $s$), and the other 
towers with $\et$ above a shoulder 
threshold as shoulders (abbreviated as
$h$).  To identify a jet, we can put the 
seed at the center and make a cone 
starting at a reconstructed interaction 
vertex point and around the seed to 
include the shoulders.  Since the 
transverse components of a particle's 
four-momentum are not changed under the 
unknown boost $\beta$ of the parton-parton 
system along the $z$-axis, $\phi$ is not 
changed.  For an ultra-relativistic 
particle, $\eta$ is a good approximation 
of its rapidity.  We have
\begin{equation}
   \begin{array}[c]{llllllll}
      \phi_{s} & \to & \phi_{s}, & & &
      \phi_{h} & \to & \phi_{h} 
      \\
      \eta_{s} & \to & \eta_{s} + \tanh^{-1}\beta, & & &
      \eta_{h} & \to & \eta_{h} + \tanh^{-1}\beta
   \end{array}
\end{equation}
The separations in $\phi$ and $\eta$ are 
not changed under the unknown boost along
the $z$-axis, 
\begin{equation}
   \begin{array}[c]{ccc}
      \Delta\phi = \phi_{h} - \phi_{s} & \to & \Delta\phi \\
      \Delta\eta = \eta_{h} - \eta_{s} & \to & \Delta\eta
   \end{array}
\end{equation}
Therefore the separation in $\Delta R$ which 
is constructed in the combination of 
$\Delta\phi$ and $\Delta\eta$ is not 
changed under the unknown boost along
the $z$-axis,  
\begin{equation}
   \begin{array}[c]{ccc}
      \Delta R = \sqrt{(\Delta\eta)^2+(\Delta\phi)^2}& \to & \Delta R
   \end{array}
\end{equation}
Given the $\et$ and the configuration 
(shape) of a jet, whatever the magnitude of
the boost along the $z$-axis of the 
parton-parton system is, or, equivalently, whatever the 
direction of the seed of the jet is, we can 
use the same cone to include or exclude a 
tower
into the jet by calculating 
its separation in $\Delta R$ to the seed.  
Thus $\Delta R$ is a very useful shape 
variable for jet identification.  

It also makes sence that there is a strong
correlation between the two variables $\et$ 
and $\Delta R$: a higher $\et$ should give 
a smaller cone in $\Delta R$ to include all 
of the final particles, e.g. of a jet.  It 
is very common that there are hundreds of 
final particles after a $p\bar{p}$ collision.  
The problem is that the energy of a jet in 
real data cannot be measured before a cone 
is actually constructed, otherwise there is 
no constraint to tell which 
tower
should be included or excluded.  
Jet identification usually starts with a large 
and constant cone around a seed.  The towers 
with significant energy in the cluster may 
or may not be contiguous.  The energy of the 
jet is determined afterwards by summing up 
the energies of all of the towers in the 
cluster.  

Now consider hadronic tau identification
with a narrow cone and 
small number of final particles. 
The situation is quite different from
jet identification.  Since there 
are only a small amount of final particles, 
each final particle has significant energy.  
And since all of the final particles are in 
a narrow cone, they make a narrow and 
contiguous cluster with significant energy 
in each tower.  This constraint of a narrow 
and contiguous cluster with significant 
energy in each tower tells us that we can 
determine energy first, and then construct 
a narrow cone to include or exclude charged 
particles reconstructed in the tracking 
system and/or neutral $\pi^0$s reconstructed 
in the shower maximum detector which is 
inside the electromagnetic calorimeter.

The question now is: is $\Delta R$ a good
choice of cone size definition for hadronic
tau identification?

\begin{figure}
   \begin{center}
      \parbox{4.5in}{\epsfxsize=\hsize\epsffile{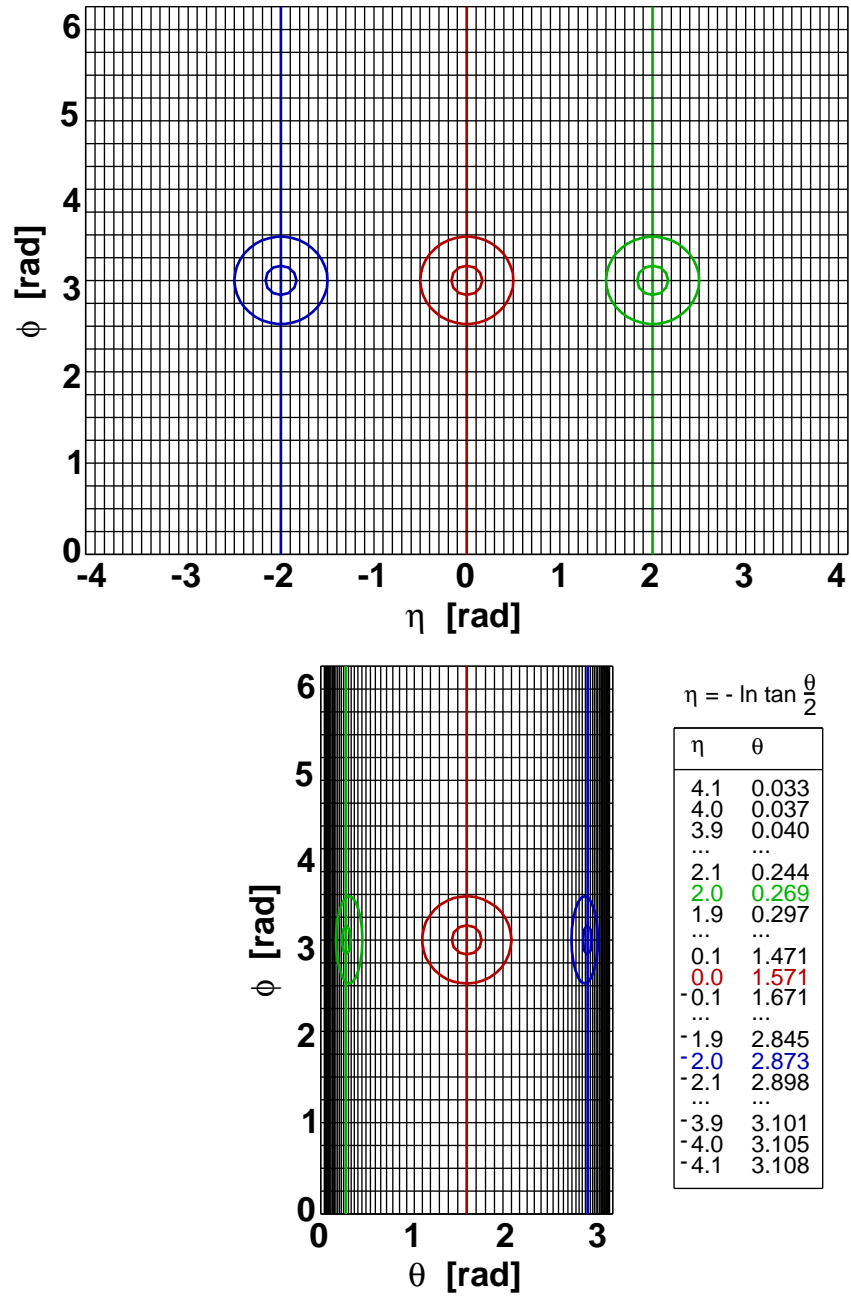}}
      \caption[Lack of geometry invariance in $\Delta R$ cone]
              {Lack of geometry invariance in $\Delta R$ cone.}
      \label{fig:TauId_cone}
   \end{center}
\end{figure}

A $\Delta R$ cone has a relativity invariance
under a boost along the $z$-axis.
However a $\Delta R$ cone 
does not have geometry
invariance.  
What does a constant $\Delta R$ imply in geometry?  The top 
plot of Fig.~\ref{fig:TauId_cone} shows three constant isolation 
annulus at different $\eta$ in a uniform $\eta$-$\phi$ space; 
the bottom plot shows the same three isolation annulus in a 
uniform $\theta$-$\phi$ space after using the function 
$\eta = -\ln\tan\frac{\theta}{2}$ to map $\eta$ slices to $\theta$ 
slices.  In the central region, the isolation annulus is almost 
unchanged; outside the central region, they are severely squeezed,
thus $\Delta\eta$ doesn't have geometry invariance.  
$\Delta\phi$ doesn't have geometry invariance either.
Think of one step at the Equator of the Earth and another
step at the North Pole of the Earth, the former is a tiny
one in $\Delta\phi$ while the latter is a giant one in
$\Delta\phi$.
A constant $\Delta R$ cone 
with relativity invariance is not expected 
to be a constant cone with geometry 
invariance. 

Instead of $\et$ and $\Delta R$, we can use energy 
$E$ and three-dimensional separation angle $\alpha$
to construct a cone for hadronic tau identification.  
There are two reasons.

      First, consider a rotation of a solid cone; 
      the geometry invariance of a three-dimensional 
      separation angle $\alpha$ is easy to 
      visualize.  
      The unknown 
      boost of the parton-parton system along the 
      $z$-axis
      doesn't affect the 
      energy measurement of the hadronic tau 
      identification at all.  Under the known high 
      energy boost, the final particles are flying 
      together in a narrow cone.  In one case the 
      boost is to the central region, and in another 
      case the boost is to somewhere forward or 
      backward.  Are these two cones geometrically 
      invariant?  The answer is yes. 

      Second, the correlation of $E$ and $\alpha$ 
      is very strong.  The case with the simplest 
      phase space of final particles is calculable, 
      see Appendix~\ref{cha:Alpha_Gamma}. 
      Comparing with a constant cone, a variable 
      cone determined by this correlation can give 
      extra power to suppress the jet 
      background for hadronic tau identification.  
This is described by the ``shrinking'' cone algorithm 
for hadronic tau identification below.


\subsection{The ``Shrinking'' Cone}
\label{subsec:TauId_shrinking}

As shown in Fig.~\ref{fig:TauId_iso},
tau isolation cone, i.e., the outer cone, is a 
constant 30$^{\circ}$ (0.525~rad) cone.  
For a particle with definite mass like 
tau, the bigger the energy, the smaller the 
separation angle of its decay daughters, 
hence a smaller signal cone which is the 
inner cone in Fig.~\ref{fig:TauId_iso}.

The tau resonctruction algorithm~\cite{CDFnote:6252}
starts with a seed tower with
$\et>6$~GeV.  It adds all of the adjacent shoulder 
towers with $\et>1$~GeV to make a calorimeter cluster.
The cluster is required to be narrow, i.e., the number 
of towers~$\leq6$.  The visible energy, denoted as
$E_{vis}$, of the final particles of tau haronic decays 
is measured by the energy of the calorimeter cluster, 
denoted as $E^{\tau\;obj}_{cluster}$.  Then the 
algorithm asks a seed track with $\pt>4.5$~GeV/$c$ to 
match with the cluster.  The matched seed track is a 
track with the highest $\pt$ in the neighbor of the 
calorimeter cluster.  The tau signal cone is 
constructed around the direction of the seed track.  
The other tracks with $\pt>1$~GeV/$c$, and the $\pi^0$s 
with $\pt>1$~GeV/$c$ which are reconstructed by the 
strip and wire clusters in the CES detector, are 
included in the tau candidate if they are inside the 
tau signal cone.  The size of the tau signal cone is 
determined by $E_{vis}$.

The phase space of tau hadronic decays is
very complicated and the energy dependence
of the signal cone cannot easily be calculated
analytically.  We use a large MC sample
of $p\bar{p}\to Z\to\tau\tau$ to get this
correlation.

\begin{figure}
   \begin{center}
      \parbox{4.9in}{\epsfxsize=\hsize\epsffile{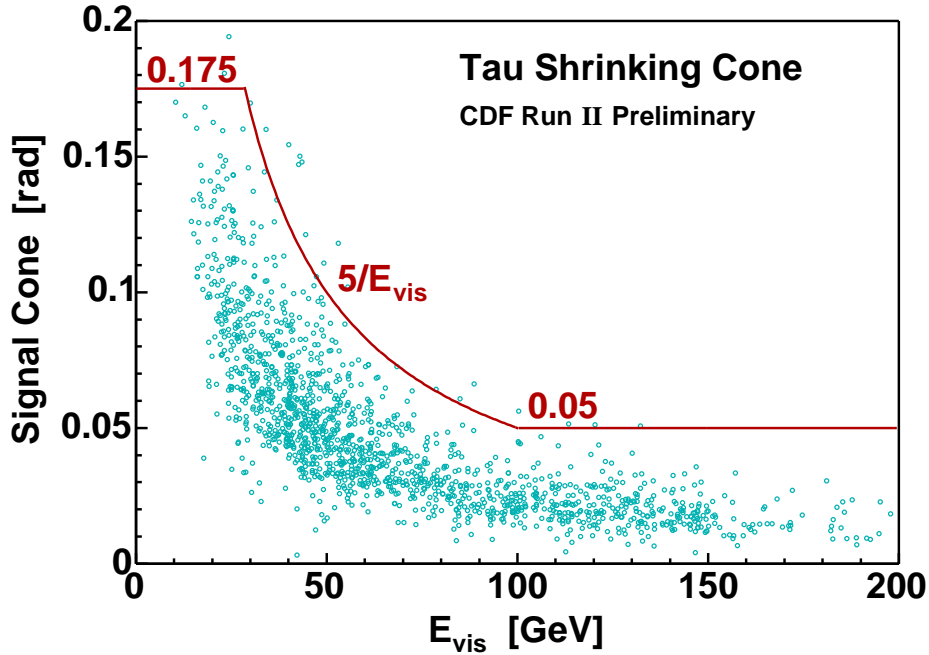}}
      \caption[Tau ``shrinking'' signal cone as a function of energy]
              {Distribution of maximum angle between
               tau decay products and tau seed track as a function of
               tau visible decay product energy.  The red line indicates
               the half-width of the "shrinking" tau signal cone as a
               function of energy.}
      \label{fig:TauId_shr1}
   \end{center}
\end{figure}

The concept of tau shrinking signal cone at generation 
level (without underlying track or $\pi^0$) is shown in 
Fig.~\ref{fig:TauId_shr1}.  The cone starts out 
at a constant 10$^{\circ}$, and then, if the 
quantity (5~rad)/$E_{vis}$ is less than 10$^{\circ}$ 
we use this angle, unless it is less than 50~mrad.

\begin{figure}
   \begin{center}
      \parbox{5.5in}{\epsfxsize=\hsize\epsffile{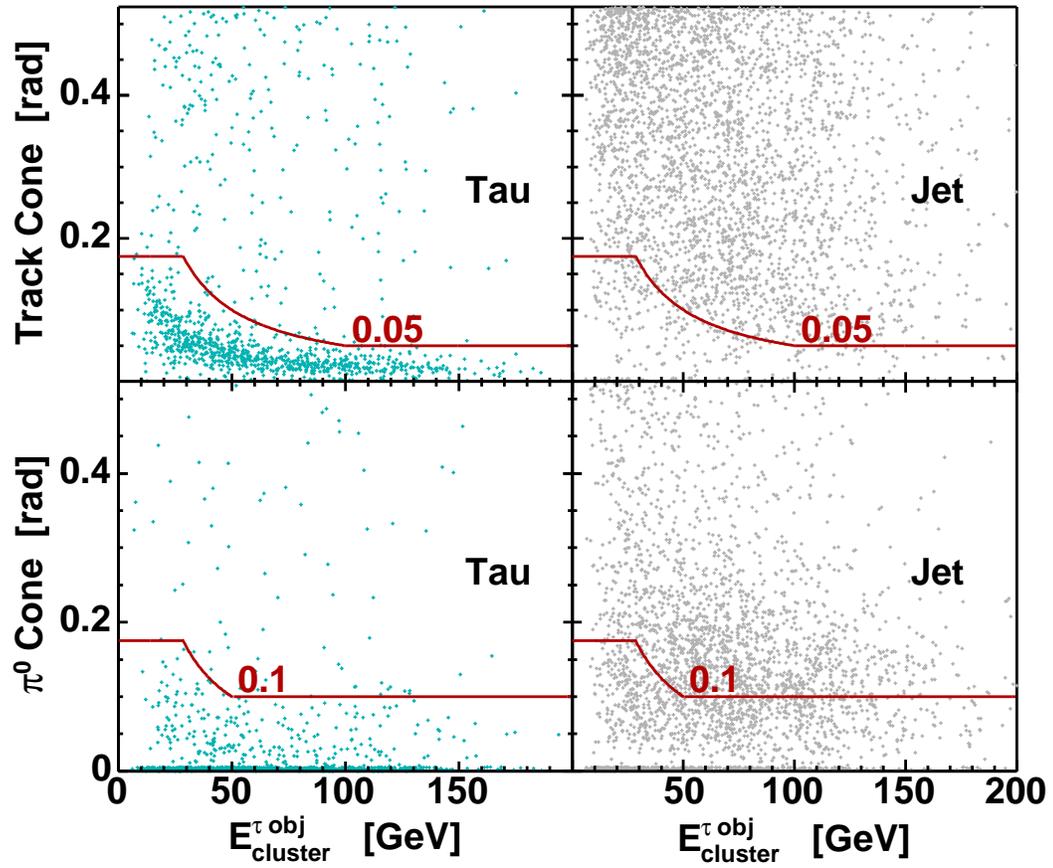}}
      \caption[Tau track/$\pi^0$ ``shrinking'' signal cone]
              {Due to different reconstruction 
               resolutions, the minimum cone sizes
               of the ``shrinking'' cone for 
               track and $\pi^0$ are 0.05 and
               0.1 radian, respectively,
               shown in the left two plots for
               tau. The right two plots show
               how the ``shrinking'' cone looks when
               applied to jets reconstructed as
               tau objects.}
      \label{fig:TauId_shr2}
   \end{center}
\end{figure}

For reconstructed tracks a cone defined as that
shown in Fig.~\ref{fig:TauId_shr1} 
is efficient and selective against jet backgrounds.  
However, for $\pi^0$s, the 
reconstructed angle can, at large visible energies, 
be larger than 50~mrad.  Thus we relax the minimum 
to 100~mrad.  With underlying track or $\pi^0$, the 
shrinking cone is shown in the left two plots of 
Fig.~\ref{fig:TauId_shr2}.  Inside the tau isolation 
cone (the outer 0.525~rad cone), the separation angle 
between the farthest track/$\pi^0$ and the seed track 
is ploted.  A tau object between the tau isolation 
cone and the shrinking signal cone is non-isolated 
and will be removed by isolation cut.   
The right two plots of Fig.~\ref{fig:TauId_shr2} 
show how the shrinking cone looks when applied 
to jets reconstructed as tau objects.  Comparing with
a constant signal cone, the shrinking signal cone, a 
natural consequence of the tau's relativistic boost, 
dramatically helps to reduce jet background in the 
high mass search.

\subsection{Tau Identification Cuts}
\label{subsec:TauId_cuts}

Now we can put the seed track in the center of the cone and 
include in the tau candidate all tracks and $\pi^0$s 
whose direction is within the ``shrinking'' signal cone. 
Table~\ref{tab:TauId_cuts} 
shows the list of tau identification cuts
using the information about calorimeter cluster,
seed track, shoulder tracks/$\pi^0$s of the tau 
candidate.  
The $\pt$(tracks + $\pi^0$s)
threshold is not listed because it is not an 
identification cut and it should be chosen by 
looking at the trigger cuts applied and by comparing 
tau identification efficiency with the jet$\to\tau$ 
misidentification rate.  
We do not cut on charge 
because there is an ambiguity in the charge for high 
$\pt$ tracks; we do not cut on
track multiplicity either because we will check 
track multiplicity to see hadronic tau signature.

\begin{table}
   \begin{center}
      \begin{tabular}{|l|l|l|l|} \hline
         Variable                   & Cut                    & Note                       & Denominator       \\ \hline \hline
         $|\eta_{det}|$             & $<$1                   & central calorimeter        &                   \\
         $|z_{loc}|$                & 9$<|z_{loc}|<$230 cm   & fiducial ShowerMax         &                   \\
         $\xi$                      & $>$0.2                 & electron removal           & $D_{\xi}$         \\
         $p_T^{seed}$               & $>$6 GeV/$c$           & seed track $\pt$           &                   \\
         10$^\circ$ track isolation & constant cone          & weaker than shrinking      & $D_{trkIso10Deg}$ \\
         m(tracks)                  & $<$1.8 GeV/$c^2$       & weaker than vis. mass      & $D_{trkMass}$     \\
         $|z_0|$                    & $<$60 cm               & vertex $z$                 &                   \\
         $|d_0|$                    & $<$0.2 cm              & impact prameter            &                   \\
         seed track ax. seg.        & $\ge$3$\times$7        & COT axial segments         &                   \\
         seed track st. seg.        & $\ge$3$\times$7        & COT stereo segments        &                   \\
         track isolation            & shrinking track cone   & shoulder tracks            &                   \\
         $\pi^0$ isolation          & shrinking $\pi^0$ cone & shoulder $\pi^0$s          &                   \\
         $E^{em}_{iso}$             & $<$2 GeV               & EM cal. isolation          &                   \\
         m(tracks + $\pi^0$s)       & $<$1.8 GeV/$c^2$       & visible mass               & Numerator         \\ \hline
      \end{tabular}
      \caption[Tau identification cuts]
              {Tau identification cuts.}
      \label{tab:TauId_cuts}
   \end{center} 
\end{table}

\vspace{0.2in}
\noindent{\bf Electron Removal}
\vspace{0.1in}

\begin{figure}
   \begin{center}
      \parbox{3.4in}{\epsfxsize=\hsize\epsffile{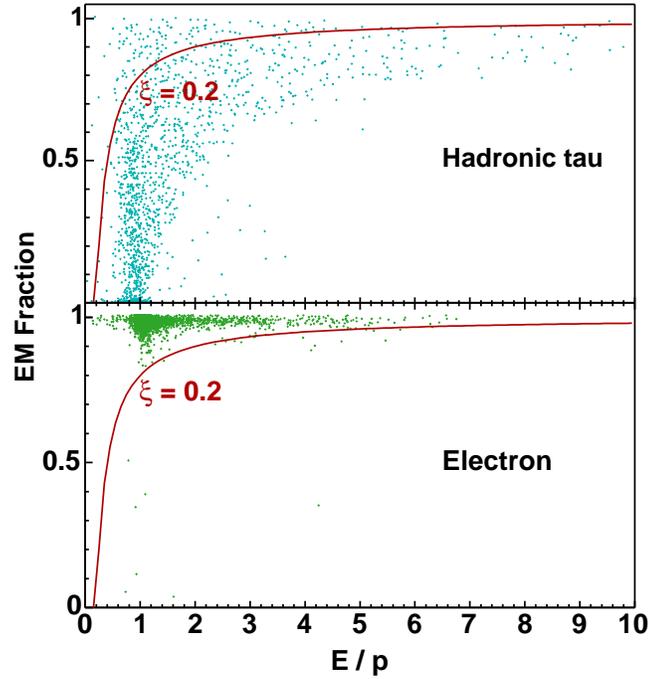}}
      \caption[Electron removal in tau identification]
              {Distributions of EM fraction ($E_{em}/E$) vs.
               $E/p$ for hadronic tau and electron.
               $\xi>0.2$ is used to remove electron.}
      \label{fig:TauId_cuts_1}
   \end{center}
\end{figure}

Using the requirements discussed above, electrons can be 
reconstructed as hadronic tau objects if they have a narrow
calorimeter cluster and a high $\pt$ seed track.
To remove electrons we demand that the tau be consistent with 
having only pions in the final state.  
We define the variable $\xi$ as 
\begin{equation}
       \xi \equiv \frac{E}{p}(1 - \frac{E_{em}}{E}) = \frac{E_{had}}{p}
\end{equation}
Fig.~\ref{fig:TauId_cuts_1} 
shows the tau object EM fraction ($E_{em}/E$)
versus E/p.  The top plot is for hadronic taus reconstructed 
as tau objects, and the bottom plot is for electrons 
reconstructed as tau objects.  For an ideal hadronic tau 
and a perfect calorimeter, $\xi$ = 1.  For an ideal electron, 
$\xi$ = 0.  However, the calorimeter is not perfect and there 
can be a large background from $Z\to ee$ events.  To remove 
this background we use a very tight cut, $\xi>$ 0.2.  The 
remaining background is discussed below.

\vspace{0.2in}
\noindent{\bf EM Calorimeter Isolation}
\vspace{0.1in}

\begin{figure}
   \begin{center}
      \parbox{5.2in}{\epsfxsize=\hsize\epsffile{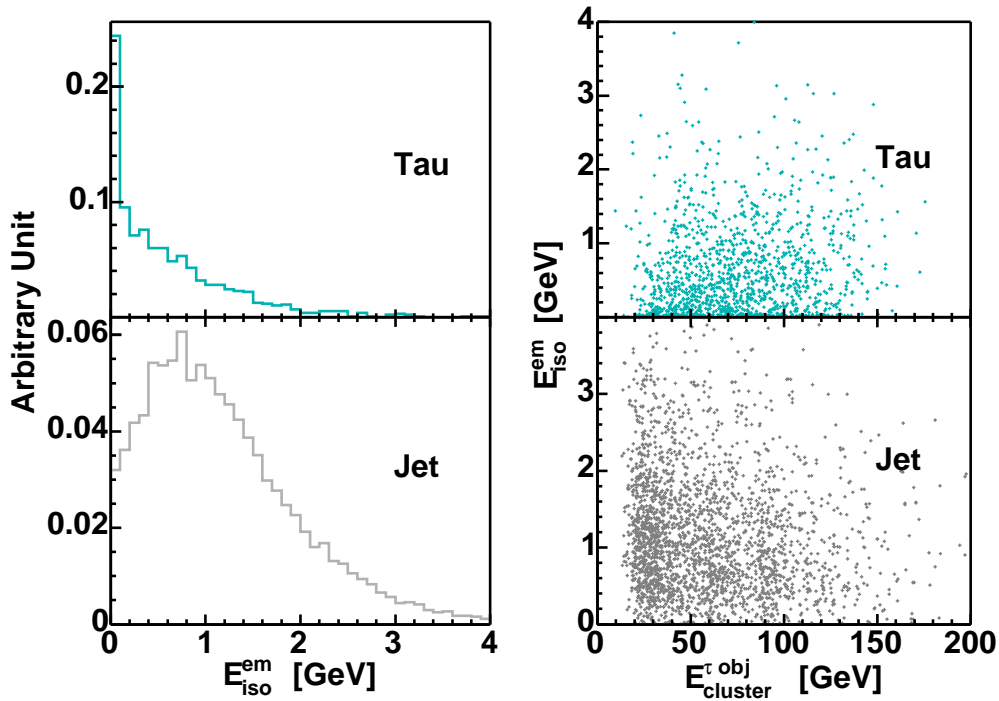}}
      \caption[EM calorimeter isolation in tau identification]
              {Disbutions of EM calorimeter isolation
               for tau and jet, and distributions of
               EM calorimeter isolation vs. energy
               of reconstructed tau object for tau 
               and jet.}
      \label{fig:TauId_cuts_2}
   \end{center}
\end{figure}

The motivation for the EM calorimeter isolation cut is due to
$\pi^0$ reconstruction inefficiency, 
for example, some CES clusters are not reconstructed as 
$\pi^0$s if a track is nearby.  This affects the power 
of the $\pi^0$ isolation requirement.  We add an EM 
calorimeter isolation cut to deal with the remaining jet
background.  We calculate the EM energy in a 
$\Delta R = 0.4$ cone around the seed track, summing over 
all EM towers which are not members of the tau cluster. 
Here $\Delta R$ is used to calculate the distance between 
the centroid of a calorimeter tower and the seed track 
because the calorimeter tower segmentation is fixed
in $\eta\times\phi$ space, namely $0.11\times15^{\circ}$
around the central region.
Since the EM calorimeter isolation cut is strongly correlated with 
other isolation cuts, its marginal distribution is shown 
in Fig.~\ref{fig:TauId_cuts_2}.  The EM cal. isolation 
energy versus cluster energy plots show that we do not
need to use a relative (fractional) cut, which is 
necessary if for high energy tau objects there is 
significant energy leakage outside tau cluster.  We 
instead choose an abosolute cut, $E^{em}_{iso}<2$ GeV.

\vspace{0.2in}
\noindent{\bf Object Uniqueness}
\vspace{0.1in}

Though not listed in the summary table of tau 
identification cuts, we note that all reconstructed 
objects in the event are required to be unique.
Thus we only apply the tau identification cuts to 
objects not already reconstructed as a photon, 
electron, or muon.  In practice, we require that a tau
object be 30$^{\circ}$ away from any identified
photon, electron, or muon.

\vspace{0.2in}
\noindent{\bf Denominators}
\vspace{0.1in}

For various subsequent studies presented here we 
will use specific subsets of the tau identification
cuts listed in the summary table.  The cuts are in 
cumulative order which is important for 
calculating rates and efficiencies.  There are three 
different denominators in Table~\ref{tab:TauId_cuts} 
corresponding to three different relative rates, 
which will be applied on different data samples with 
consistent denominators later. 


\subsection{Tau Identification Efficiency}
\label{subsec:TauId_efficiency}

Table~\ref{tab:TauId_efficiency} 
shows the procedure to measure the tau 
identification efficiency, using different samples.  
For all of the generated taus, we pick those taus decaying
hadronically, and consider the central ones in the
pseudorapidity range $|\eta|<1$ which are able to be 
reconstructed as tau object, called CdfTau in the table.  
We require 
the seed track of the generated tau to match with the seed track of a
reconstructed tau object
within 0.2 radian.  
Then we apply the tau identification cuts on the reconstructed tau objects
and calculate tau identification efficiency.

\begin{figure}
   \begin{center}
      \parbox{5.4in}{\epsfxsize=\hsize\epsffile{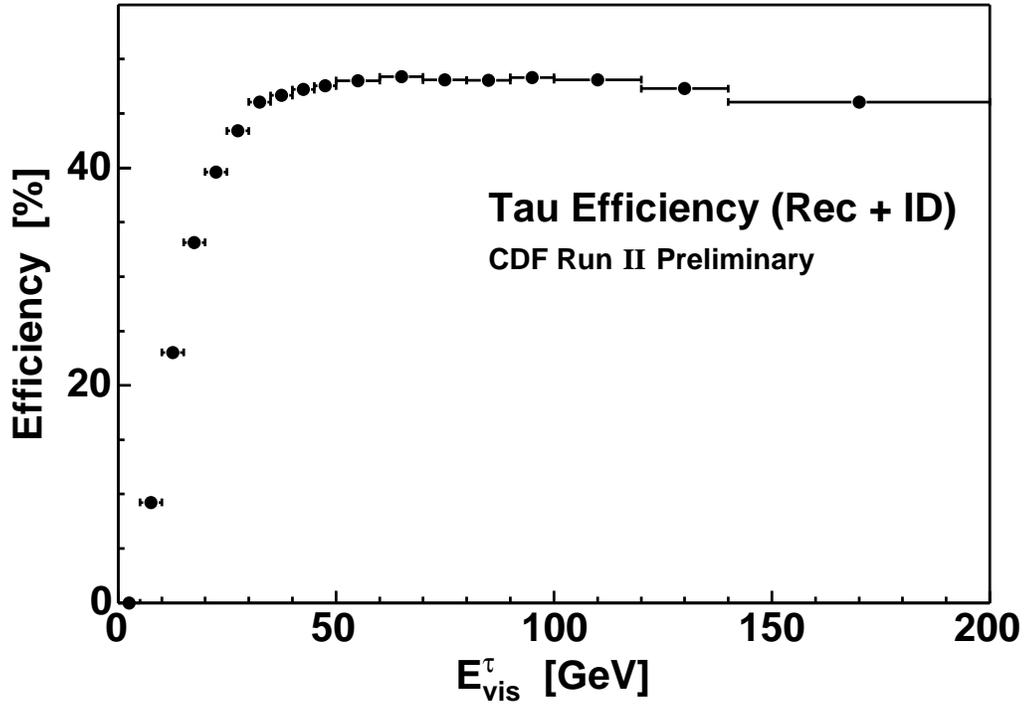}}
      \caption[Tau identification efficiency vs. energy]
              {Tau identification efficiency vs. tau visible energy.}
      \label{fig:TauId_efficiency}
   \end{center}
\end{figure}

Fig.~\ref{fig:TauId_efficiency} shows the absolute
tau identification efficiency, which includes the effects of 
both reconstruction and identification, vs. tau visible energy, 
using the $Z'$ sample which has a lot of high energy taus.

\subsection{Jet$\to\tau$ Misidentification Rate}
\label{subsec:TauId_jet}

Table~\ref{tab:TauId_jet} 
shows the procedure to measure the jet$\to\tau$
misidentification rate, using four different jet 
samples called JET20, JET50, JET70, and JET100 
samples collected with different trigger thesholds.
The L1 tower $\et$, L2 cluster $\et$ and L3 jet 
$\et$ trigger thresholds in the unit of GeV for 
a triggered jet in each jet sample are
\begin{itemize}
\item JET20: 5, 15, 20
\item JET50: 5, 40, 50
\item JET70: 10, 60, 70
\item JET100: 10, 90, 100
\end{itemize}
We use the central jets with $|\eta|<1$ which may be
reconstructed as tau object, called CdfTau in the table.  
We require the central jet to match with a reconstructed tau 
object by requiring that they share the seed tower of 
the reconstructed tau object.
Then we apply the tau identification cuts on the reconstructed tau objects
and calculate jet$\to\tau$ misidentification rate.

\begin{figure}
   \begin{center}
      \parbox{5.4in}{\epsfxsize=\hsize\epsffile{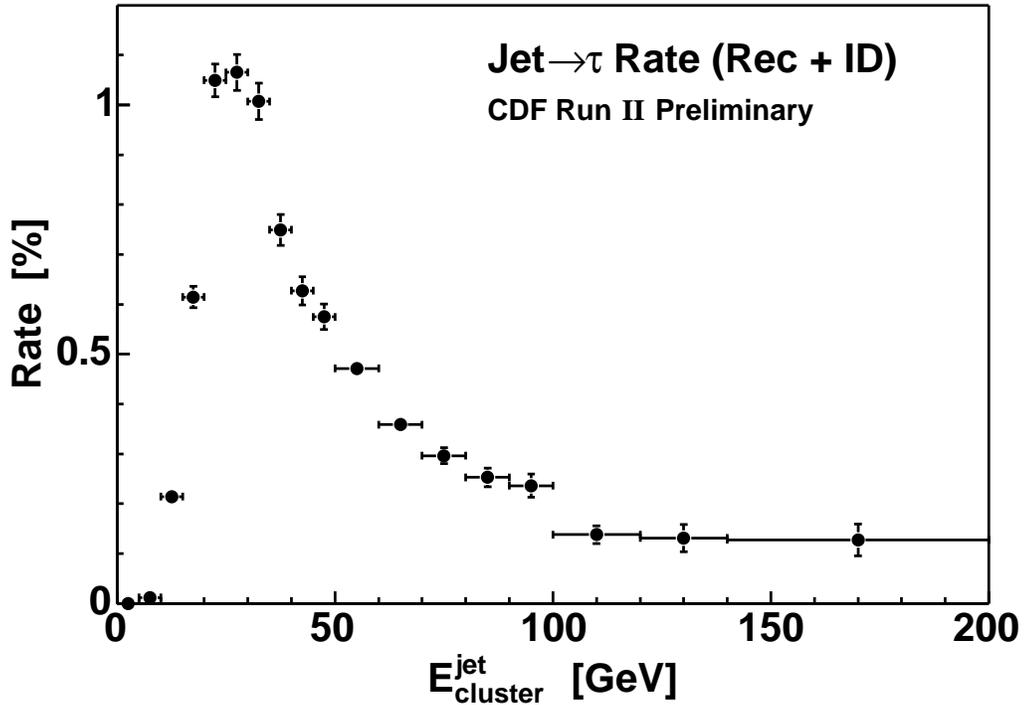}}
      \caption[Jet$\to\tau$ misidentification rate vs. energy]
              {Jet$\to\tau$ misidentification rate vs. energy, 
               using JET50 sample.}
      \label{fig:TauId_jet_1}
   \end{center}
\end{figure}

Fig.~\ref{fig:TauId_jet_1} shows the absolute 
jet$\to\tau$ misidentification rate, which includes the effects of 
both reconstruction and identification, vs. jet cluster
energy, using JET50 sample.

\begin{table}
   \begin{center}
      \begin{tabular}{|l|r|r|r|l|} \hline
         Procedure                      & $W\to\tau\nu$ & $Z\to\tau\tau$ & $Z'\to\tau\tau$ & Denominator       \\ \hline \hline
         event                          &        491513 &         492000 &         1200000 &                   \\ \hline
         tau hadronic                   &        319357 &         637889 &         1554159 &                   \\
         tau central                    &        150984 &         275330 &          898102 & $D_{absolute}$    \\
         tau match CdfTau               &         86325 &         165495 &          800262 & $D_{CdfTau}$      \\ \hline
         $|\eta_{det}|<1$               &         85899 &         164722 &          797705 &                   \\
         $9<|z_{loc}|<230$ cm           &         82240 &         157748 &          758403 &                   \\
         $\xi>0.2$                      &         65854 &         127403 &          663845 & $D_{\xi}$         \\
         $p_T^{seed}>6$ GeV/$c$         &         60960 &         119451 &          651328 &                   \\
         10$^\circ$ track isolation     &         50309 &          98717 &          540485 & $D_{trkIso10Deg}$ \\
         m(tracks) $<1.8$ GeV/$c^2$     &         50141 &          98355 &          532190 & $D_{trkMass}$     \\
         $|z_0|<60$ cm                  &         48659 &          95333 &          515239 &                   \\
         $|d_0|<0.2$ cm                 &         47975 &          93969 &          506453 &                   \\
         seed track ax. seg. $\ge3\times7$ &      47822 &          93657 &          501965 &                   \\
         seed track st. seg. $\ge3\times7$ &      47312 &          92666 &          494069 &                   \\
         track isolation (shrinking)    &         47112 &          92042 &          475017 &                   \\
         $\pi^0$ isolation (shrinking)  &         45687 &          89148 &          451129 &                   \\
         $E^{em}_{iso}<2$ GeV           &         43981 &          85910 &          428641 &                   \\
         m(tracks + $\pi^0$s) $<1.8$ GeV/$c^2$ &  43155 &          84218 &          404105 & Numerator         \\ \hline
      \end{tabular}
      \caption[Tau identification efficiency measurement]
              {Number of events for tau identification efficiency measurement.}
      \label{tab:TauId_efficiency}
   \end{center}
\end{table}

\begin{table}
   \begin{center}
      \begin{tabular}{|l|r|r|r|r|l|} \hline
         Procedure                      &    JET20 &   JET50 &   JET70 &  JET100 & Denominator       \\ \hline \hline
         event                          &  7696880 & 1951396 &  910618 & 1137840 &                   \\
         event goodrun                  &  4309784 & 1213104 &  556961 &  697231 &                   \\ \hline
         jet non-triggered              & 21957203 & 6071557 & 2641643 & 2935801 &                   \\
         jet central                    &  8214991 & 2480232 & 1127376 & 1321840 & $D_{absolute}$    \\
         jet match CdfTau               &   653680 &  425086 &  189148 &  201530 & $D_{CdfTau}$      \\ \hline
         $|\eta_{det}|<1$               &   643190 &  416560 &  184996 &  196651 &                   \\
         $9<|z_{loc}|<230$ cm           &   611401 &  393222 &  174474 &  184980 &                   \\
         $\xi>0.2$                      &   521326 &  354504 &  159320 &  169441 & $D_{\xi}$         \\
         $p_T^{seed}>6$ GeV/$c$         &   414966 &  315384 &  145124 &  156391 &                   \\
         10$^\circ$ track isolation     &   105846 &   74425 &   36231 &   42727 & $D_{trkIso10Deg}$ \\
         m(tracks) $<1.8$ GeV/$c^2$     &    92475 &   63616 &   31865 &   37709 & $D_{trkMass}$     \\
         $|z_0|<60$ cm                  &    85754 &   56951 &   28146 &   32747 &                   \\
         $|d_0|<0.2$ cm                 &    79889 &   51829 &   25391 &   28994 &                   \\
         seed track ax. seg. $\ge3\times7$ & 78500 &   50043 &   24293 &   27474 &                   \\
         seed track st. seg. $\ge3\times7$ & 71926 &   42754 &   20058 &   21828 &                   \\
         track isolation (shrinking)    &    64489 &   20679 &    7475 &    7293 &                   \\
         $\pi^0$ isolation (shrinking)  &    50886 &   13910 &    5025 &    4965 &                   \\
         $E^{em}_{iso}<2$ GeV           &    41749 &   11132 &    4073 &    3969 &                   \\
         m(tracks + $\pi^0$s) $<1.8$ GeV/$c^2$  
                                        &    35314 &    7965 &    2879 &    2792 & Numerator         \\ \hline 
      \end{tabular}
      \caption[Jet$\to\tau$ misidentification rate measurement]
              {Number of events for jet$\to\tau$ misidentification rate
               measurement.}
      \label{tab:TauId_jet}
   \end{center}
\end{table}

\vspace{0.2in}
\noindent{\bf Discrepancies}
\vspace{0.1in}

\begin{figure}
   \begin{center}
      \parbox{5.5in}{\epsfxsize=\hsize\epsffile{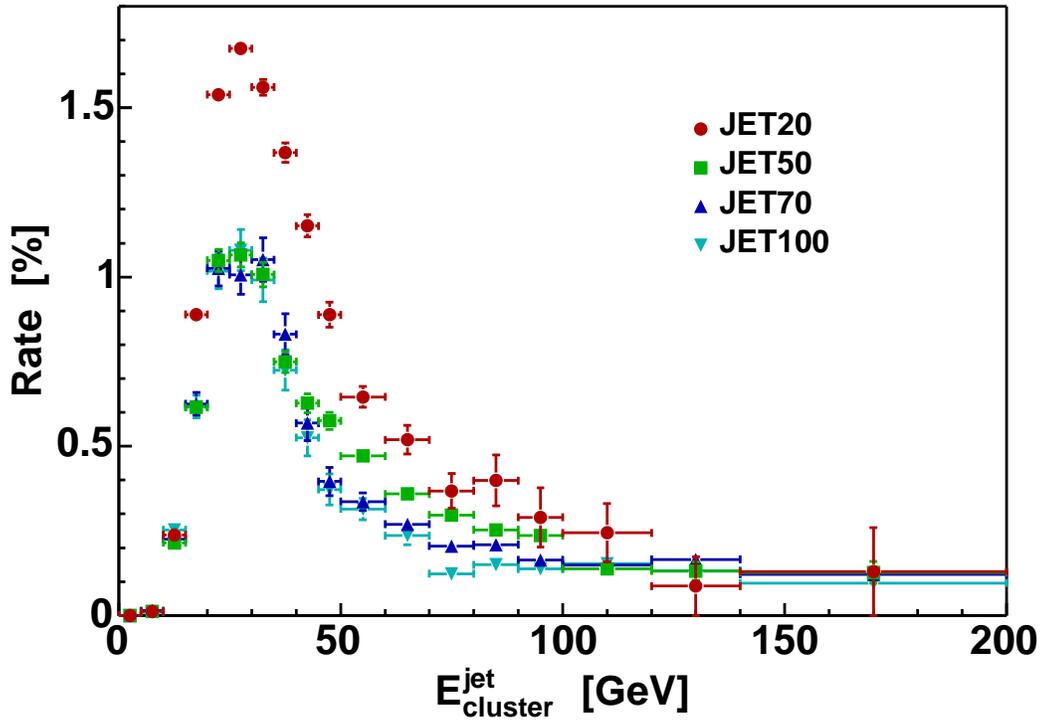}}
      \caption[Discrepancies of jet$\to\tau$ misidentification rates]
              {Discrepancies of jet$\to\tau$ misidentification rates 
               in JET samples.}
      \label{fig:TauId_jet_2}
   \end{center}
\end{figure}

To try to minimize trigger bias, we use 
non-triggered jet only.  Based on the L1 tower $\et$, 
L2 cluster $\et$ and L3 jet $\et$ trigger thresholds
in each sample, we find all of the 
jets which can satisfy the trigger requirements.  
The choice of the triggered jets in an event in the 
case of zero, one or more than one jet 
satisfying trigger requirements are 
\begin{itemize}
\item If zero, throw away the event
\item If only one, choose that jet
\item If more than one, do not choose any as triggered
\end{itemize}
Non-triggered jets are just the jets not chosen as 
the triggered jet.  Even after trying to minimize 
trigger bias by using non-triggered jet only, 
there are still discrepancies among jet$\to\tau$ 
misidentification rates obtained from different 
jet samples, shown in Fig.~\ref{fig:TauId_jet_2}. 

\vspace{0.2in}
\noindent{\bf Two-Dimensional Parametrization}
\vspace{0.1in}

\begin{figure}
   \begin{center}
      \parbox{5.2in}{\epsfxsize=\hsize\epsffile{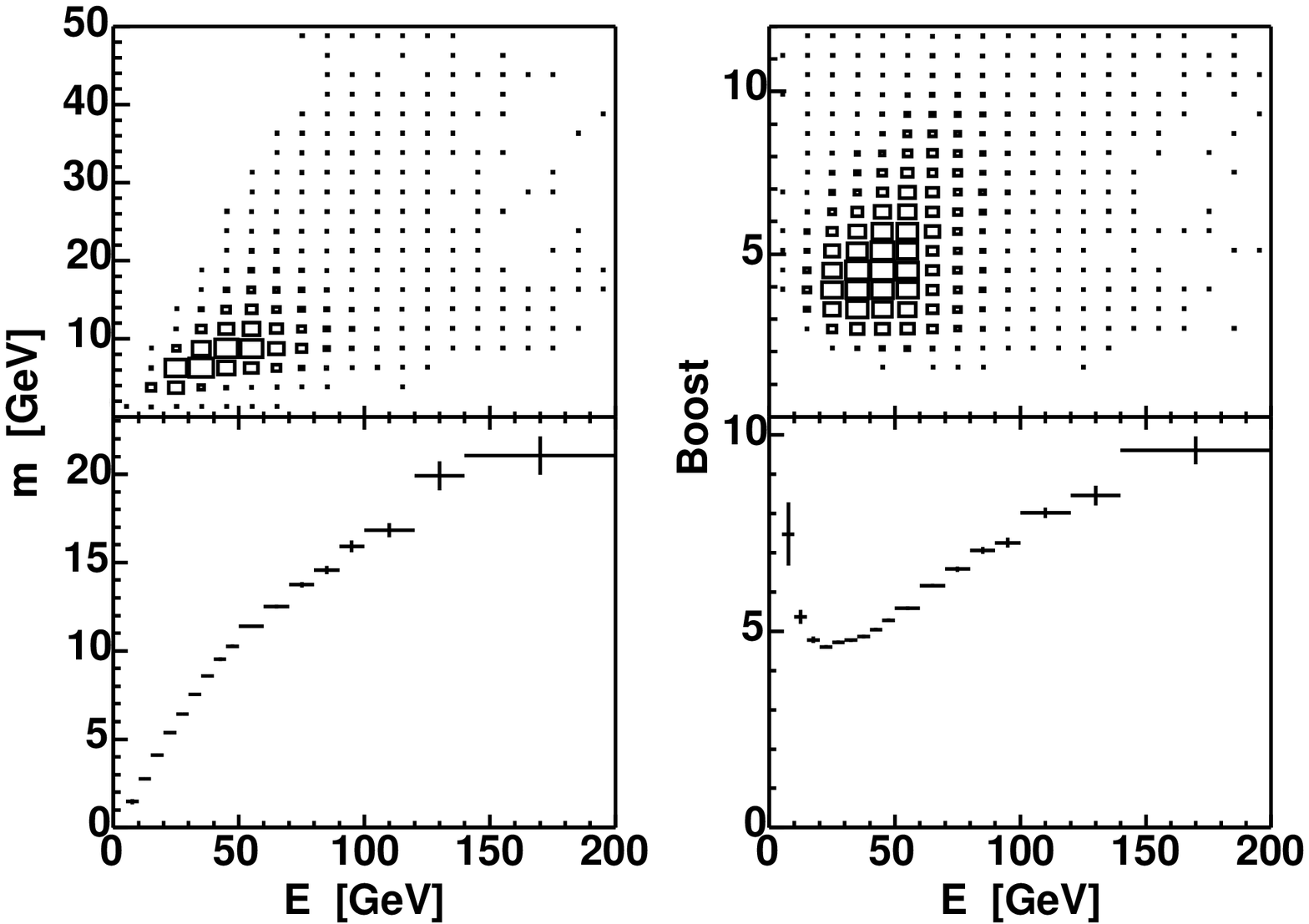}}
      \caption[Jet cluster: mass vs. energy, and boost vs. energy]
              {Distributions of jet cluster: mass vs. energy, 
               and boost vs. energy, in JET50 sample.}
      \label{fig:TauId_jet_3}
   \vspace{0.5in}
      \parbox{5.2in}{\epsfxsize=\hsize\epsffile{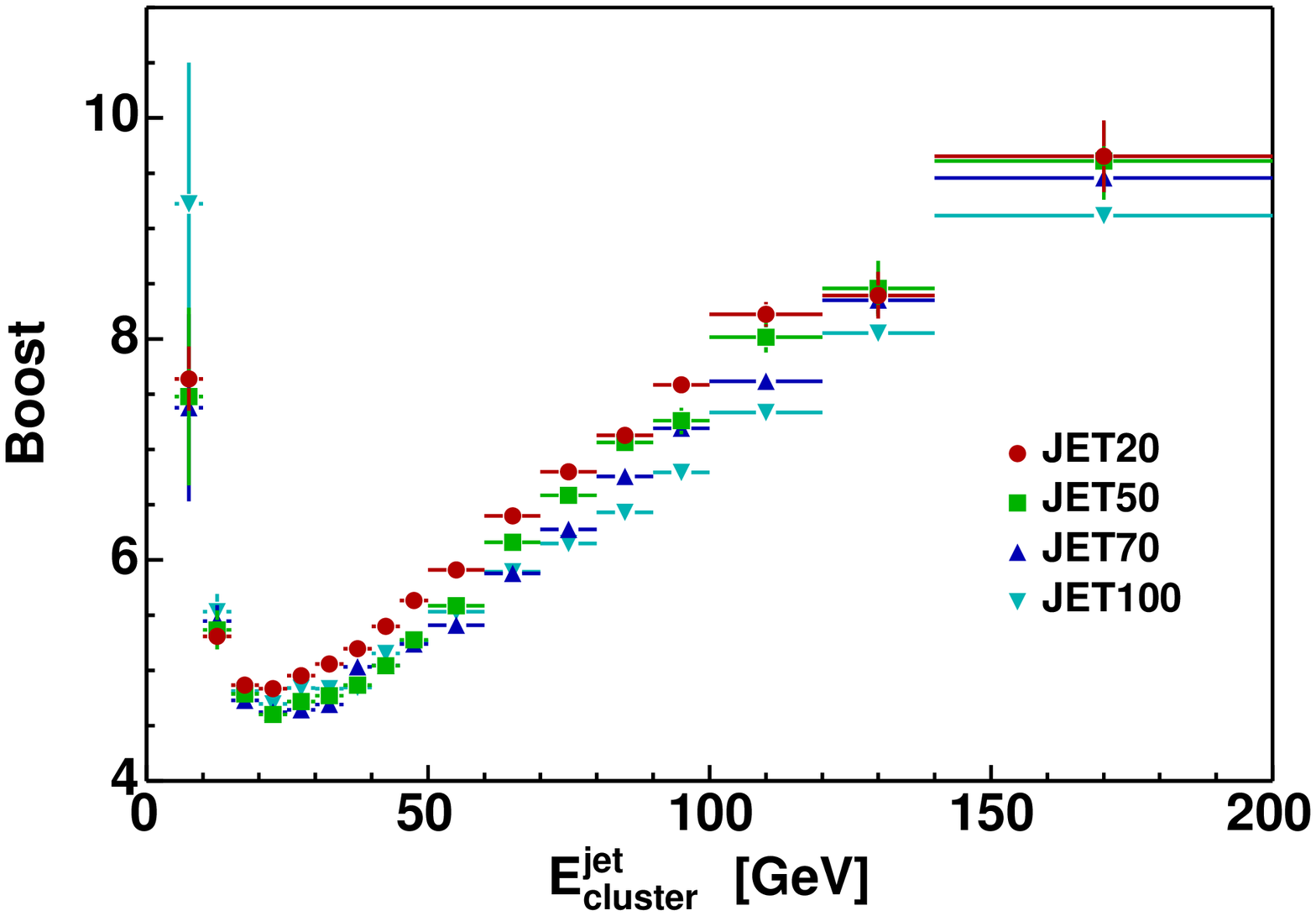}}
      \caption[Jet cluster: boost vs. energy in various samples]
              {Profiles of jet cluster boost vs. 
               cluster energy in JET samples.}
      \label{fig:TauId_jet_4}
   \end{center}
\end{figure}

There is no doubt that the jet$\to\tau$ 
misidentification rate has a very strong dependence
on energy because the tau isolation annulus is a function 
of energy.  To resolve the discrepancies among the jet$\to\tau$ rates, 
we add another parameter to make a two-dimensional parametrization. 
The second parameter should not be correlated strongly with energy, 
otherwise adding another parameter is meaningless.  
Given the final particles, the transverse size of a jet depends 
on its boost: jets with a bigger boost have smaller size and  
smaller size jets have higher probability to
survive tau identification.  The relativistic boost 
$\gamma$ is
\begin{eqnarray}
   \gamma = \frac{E}{m} 
\end{eqnarray}
where $E$ is the energy of the jet which can be
measured by its cluster energy in calorimeter, and $m$ 
is the invariant mass of its final particles.  The mass $m$ is
not easy to measure because some of the final particles 
can be neutral and leave no track in tracking system.  
We use cluster mass, which treats each tower in the
cluster as a massless photon and sums up the photons, 
as an approximation of $m$.  The cluster mass has
a strong correlation with energy, while the cluster
boost does not.  This is shown in 
Fig.~\ref{fig:TauId_jet_3}.  We choose cluster boost as
the second parameter.

In the one-dimensional jet$\to\tau$ misidentification 
rate what we see is the average over all of the bins 
of cluster boost.  Given the energy of a jet, the 
average cluster boost is different in JET samples, 
shown in Fig.~\ref{fig:TauId_jet_4}.  

\begin{figure}
   \begin{center}
      \parbox{5.5in}{\epsfxsize=\hsize\epsffile{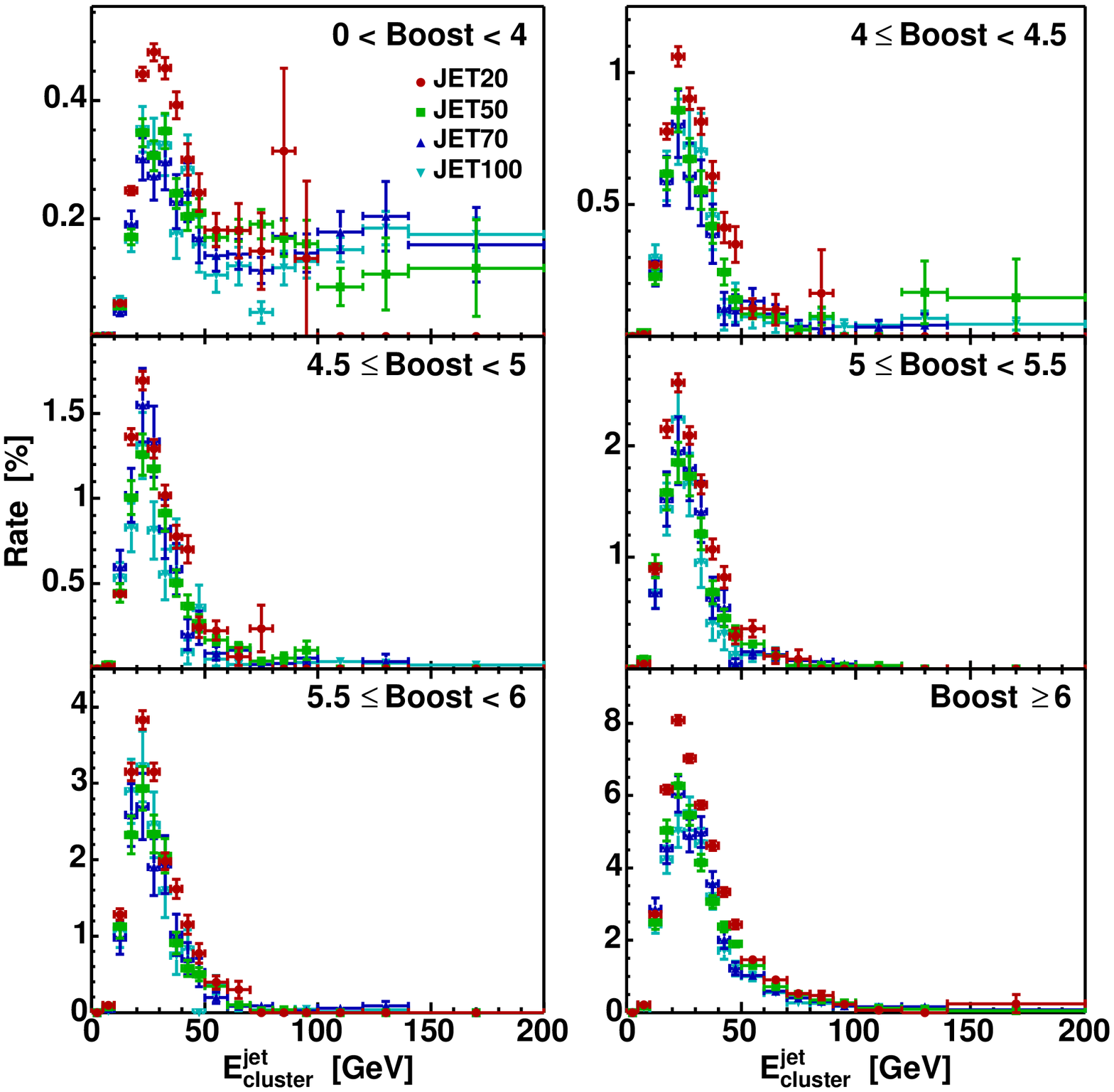}}
      \caption[Jet$\to\tau$ misidentification rate vs. energy, in boost slices]
              {Jet$\to\tau$ misidentification rate vs. energy 
               in JET samples and in jet cluster 
               boost slices.}
      \label{fig:TauId_jet_5}
   \end{center}
\end{figure}

Now we plot the jet$\to\tau$ misidentification rate 
vs. energy, in each boost slice, shown in
Fig.~\ref{fig:TauId_jet_5}.  With the new 
two-dimensional parametrization, the overall 
discrepancy drops down to about 20\%.  Since the 
discrepancies are not totally resolved, there are other 
unknown effects.

\subsection{Jet$\to\tau$ Background Estimate}
\label{subsec:TauId_jetBg}

After applying the full set of tau identification cuts, 
there will be some jet background left because of the 
huge production rate of jets in $p\bar{p}$ 
collisions.  The jet$\to\tau$ misidentification rate  
and tau identification efficiency are very useful for
estimating jet background.

To estimate the jet background, the starting point is not jets, or
tau candiates, but tau candidates with at least electron
removal, with a very tight $\xi>$ 0.2 cut applied.  Muons 
usually cannot have enough energy to make a tau cluster
in the calorimeter.  We have two general equations,
\begin{eqnarray}
   \mbox{Before full tau ID:} &   & \tilde{N} =  \tilde{N}^{\tau} +  \tilde{N}^{jet} \\
    \mbox{After full tau ID:} &   &        N  =         N^{\tau}  +         N^{jet}                 
                                              = e\tilde{N}^{\tau} + f\tilde{N}^{jet} 
\end{eqnarray}
where $f$ is jet$\to\tau$ misidentification rate and 
$e$ is tau identification efficiency.  Both are
relative in a sense that they are relative to the starting point chosen
as ``Before full tau ID''.  The solution is
\begin{equation}
   N^{jet} = \frac{f}{e-f}(e\tilde{N} - N).
\end{equation}

\begin{figure}
   \begin{center}
      \parbox{5.5in}{\epsfxsize=\hsize\epsffile{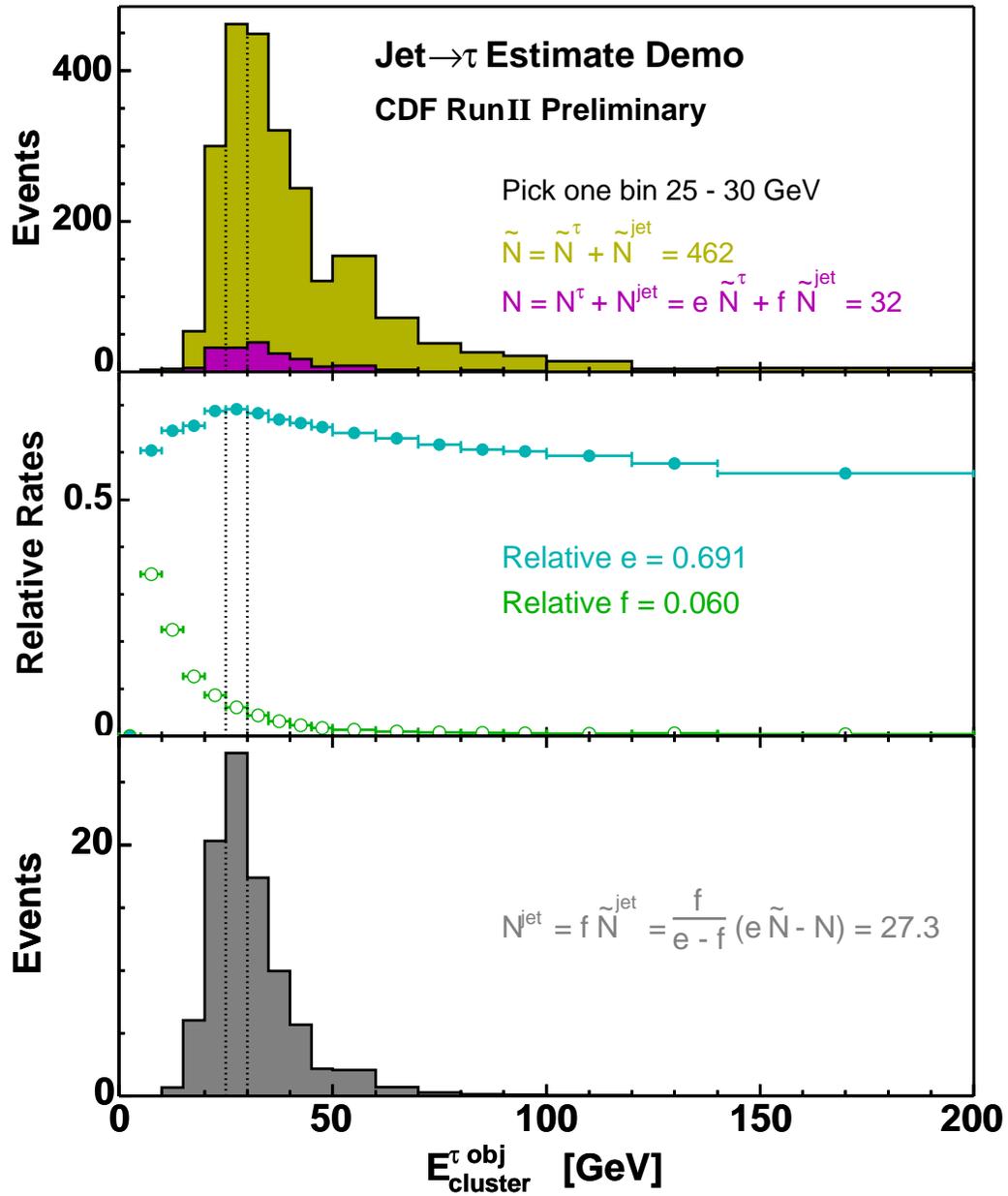}}
      \caption[Demonstration of estimating jet$\to\tau$ misidentification]
              {Demonstration of estimating jet$\to\tau$ misidentification.}
      \label{fig:TauId_jetBg}
   \end{center}
\end{figure}

Fig.~\ref{fig:TauId_jetBg} 
is a demonstration of picking one bin and using the
formula to estimate jet background.  This is only 
an example
because the parametrization 
of the relative rates is a one-dimensional function of 
energy.  
For the jet$\to\tau$ misidentification rate there is a 
better parametrization, i.e., the two-dimensional function of 
energy and boost.

\vspace{0.2in}
\noindent{\bf Implementation}
\vspace{0.1in}

The actual implementation is done on an event-by-event basis. For a tau 
object in an event under consideration, the knowns are: 
$\tilde{N}$ = 1, $e$, $f$ and whether this tau object passes the 
full set of the tau identification cuts.  If it does,
$N$ = 1; otherwise, $N$ = 0.  For the two cases, the weight 
to be a jet is estimated as
\begin{eqnarray}
   \mbox{If not passing the full tau ID cuts:} & & \omega^{jet} = \frac{f}{e - f}(e - 0) 
   \label{eq:weight_jet_notpassing} \\
       \mbox{If passing the full tau ID cuts:} & & \omega^{jet} = \frac{f}{e - f}(e - 1)  
   \label{eq:weight_jet_passing}    
\end{eqnarray}
In terms of coding, it means the rest of full tau 
identification cuts are replaced by the weight 
$\omega^{jet}$.  We sum up the weights of all the events 
in the sample, and get the jet background estimate 
$N^{jet}$, 
\begin{equation}
   N^{jet} = \sum \omega^{jet}
   \label{eq:sum_weights_jet}
\end{equation}

\newpage

\vspace{0.2in}
\noindent{\bf Special Case}
\vspace{0.1in}

This method actually needs both the jet$\to\tau$ 
misidentification rate $f$ and the tau identification 
efficiency $e$.    The main idea is to remove the
contribution from any real tau signal in jet 
background estimate.

The special case is that if we start with 
a jet-dominated sample and $f$ is much smaller than $e$, then
we can suppress signal by replacing tau identification 
cuts with the jet$\to\tau$ misidentification rate,
\begin{equation}
   N^{jet} = f\tilde{N}^{jet} \approx f\tilde{N} \;\;\;\; (f \ll e)
\end{equation}


\section{Tau Scale Factor Using $W\to\tau\nu$}
\label{sec:TauSF}

In this section, we apply tau
identification cuts to select hadronic taus
in $W\to\tau\nu$ events, estimate jet$\to\tau$ 
misidentification background, study tau identification scale 
factor and compare tau distributions in data 
and MC simulation.


\subsection{Data/MC Scale Factor}
\label{subsec:TauSF_chart}

\begin{figure}
   \begin{center}
      \parbox{5.5in}{\epsfxsize=\hsize\epsffile{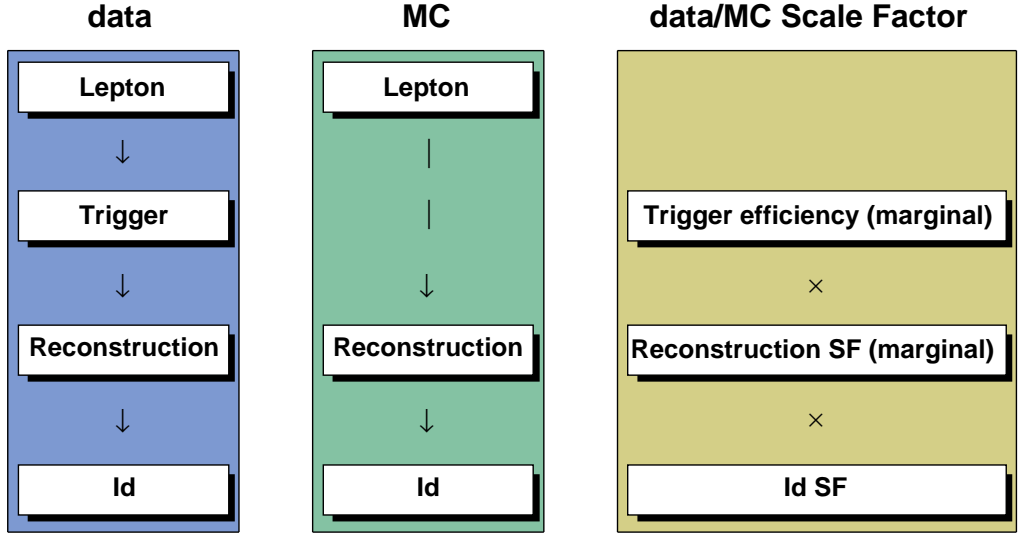}}
      \caption[Lepton in data and in MC, and lepton data/MC scale factors]
              {Lepton in data and in MC, and lepton data/MC scale factors.}
      \label{fig:TauSF_chart}
   \end{center}
\end{figure}

The scale factor for a set of cuts 
quantifies and corrects for
the difference between data and MC simulation.  
It should be multiplied on MC to get the 
scaled efficiency consistent with the 
efficiency in data.
Fig.~\ref{fig:TauSF_chart} 
shows lepton flow in data and in MC, and 
lepton data/MC scale factors.

\vspace{0.2in}
\noindent{\bf Ratio of Efficiencies}
\vspace{0.1in}

A data/MC scale factor is defined as the ratio
of efficiencies,
\begin{equation}
   f_{data/MC} = \frac{\epsilon_{data}}
                      {\epsilon_{MC}}
\end{equation}
where $\epsilon_{MC}$ is the efficiency in MC 
which is straightforward to obtain because the MC 
simulation has the true information of particle identity,
and $\epsilon_{data}$ is the efficiency in data, 
which can be a challenge to measure.  

In the electron or muon case, we can use 
electron or muon pairs from the Z boson 
peak, which gives us a pure sample with negligible 
background in real data.  This is so reliable that we can use 
it as ``standard candle'' to calibrate detector 
and even measure luminosity.  We select 
one leg to satisfy the trigger requirements in 
data, and ask whether the second leg passes the set of 
cuts, and thereby get the efficiency in data.  

\vspace{0.2in}
\noindent{\bf Ratio of Numbers}
\vspace{0.1in}

Due to the missing energy from the neutrino in 
tau decays, the tau pair mass at the Z boson peak is 
severely broadened.  Instead, we will use 
$W\to\tau\nu$ to select a relatively clean tau 
sample.  There is no second leg to get efficiency 
data/MC. We use the method of absolute number 
data/MC,
\begin{equation}
    f_{data/MC} = \frac{n_{data}}
                       {n_{MC}}
\end{equation}
where $n_{MC}$ is the absolute number of 
$W\to\tau\nu$ events in MC normalized to the 
luminosity of data, and $n_{data}$ is the number 
of $W\to\tau\nu$ events observed in the data after 
subtracting backgrounds.  


\subsection{$W\to\tau\nu$ Selection}
\label{subsec:TauSF_WTauNu}

We select $W\to\tau\nu$ events by using a data sample
from the TAU\_MET trigger which requires:
 \begin{itemize}
\item level 1 trigger (L1) $\met>$ 25 GeV
\item level 3 trigger (L3) tau $\et>$ 20 GeV
\end{itemize}
where 
(a) L1 $\met$ is based on
    a tower threshold of 1 GeV for a fast calculation; 
(b) for L3 tau, the cuts $|\eta_{det}|<$ 1, 
    10$^{\circ}$ track isolation and m(tracks) 
    $<$ 2 GeV/$c^2$ are applied in the trigger.

\begin{figure}
   \begin{center}
      \parbox{3.3in}{\epsfxsize=\hsize\epsffile{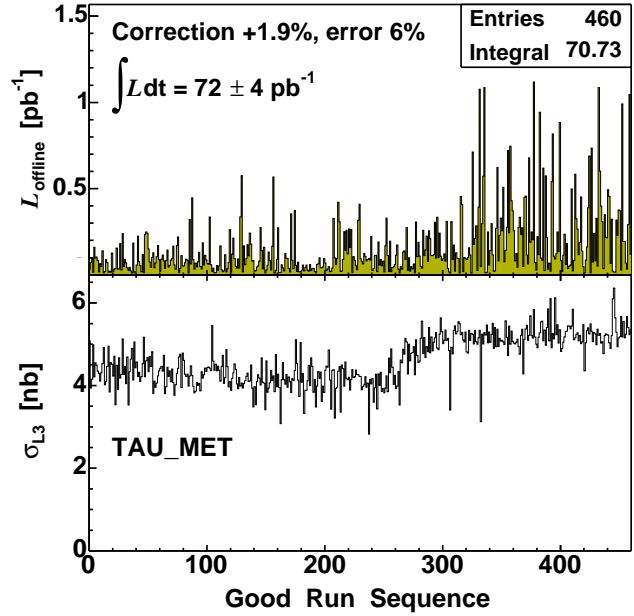}}
      \caption[Offline luminosity and L3 cross section of the TAU\_MET trigger]
              {Distributions of offline luminosity vs.
               good run sequence and L3 cross 
               section vs. good run sequence, 
               in the data sample from TAU\_MET trigger.}
      \label{fig:TauSF_WTauNu_1}
   \end{center}
\end{figure}

The top plot of 
Fig.~\ref{fig:TauSF_WTauNu_1} shows that the 
integrated luminosity of the good runs is 
72 $\pm$ 4 pb$^{-1}$, and the bottom plot
shows the L3 cross section is reasonablly
flat (no sudden drop to zero), thus all
of the good runs are present in the data
file. 

The offline selection cuts are: 
\begin{itemize}
\item Monojet
\item $\met>$ 30 GeV
\item Tau $\pt$(tracks + $\pi^0$s) $>$ 25 GeV/$c$
\end{itemize}
where 
(a) monojet selection requires 
      one central cone 0.7 jet 
         with $|\eta_{det}|<$ 1 
         and $\et>$ 25 GeV,
      no other jets 
         with $\et>$ 5 GeV 
         anywhere;
(b) offline $\met$ is obtained from the vector 
    sum of $\et$ for towers with $\et>0.1$ GeV;
(c) in addition to tau $\pt$ threshold, the 
    whole set of tau identification cuts under 
    study will be applied on the offline tau 
    candidates.

The monojet cut dramatically helps clean up
the data sample.  But, to get the estimated 
$n_{MC}$ of $W\to\tau\nu$ events, we need to 
study the monojet cut and the L1 $\met>$ 25 GeV 
trigger efficiency for monojet-type events.

\vspace{0.2in}
\noindent{\bf Monojet Selection}
\vspace{0.1in}

The monojet selection essentially requires there is no other 
underlying jet with $\et>$ 5 GeV.  We select 
$Z\to\mu\mu$ events, count the number of 
cone 0.7 jets with $\et>$ 5 GeV,
no $\eta$ cut, and 0.7 radian in $\Delta$R 
away from muons.  

The $Z\to\mu\mu$ selection cuts are:
(a) cosmic veto~\cite{CDFnote:6089},
(b) one tight muon and one track with $\pt>$ 20 GeV/$c$,
(c) opposite charges,
(d) track $|z_0(1)-z_0(2)|<$ 4 cm, and
(e) $80<m_{\mu\mu}<100$ GeV/$c^2$.
We require one tight muon and one track, instead 
of two tight muons to get higher statistics.
The track is required to be of minimum ionisation 
particle (MIP) type.
Both the tight muon and the track requires tau-like 
track isolation which is to mimic the isolated tau 
in $W\to\tau\nu$ events.  

We use a data sample from a trigger
designed to select ``muon plus track'' events
which have $\mu$ with $\pt>8$ GeV/$c$ plus another
charged track with $\pt>5$ GeV/$c$.
We select 5799 events with negligible background which is 
confirmed by the negligible number of same-charge 
muon pair events.
There are 2152 events in the zero jet bin.  The fraction 
of zero jet events in the data is 
2152/5799 = 0.371.

We use about 500K MC events.
46297 events survived after the same selection cuts as in data.  
There are 20149 events in the zero jet bin.  The
fraction of zero jet events in the MC is
20149/46297 = 0.435.

\begin{figure}
   \begin{center}
      \parbox{5.3in}{\epsfxsize=\hsize\epsffile{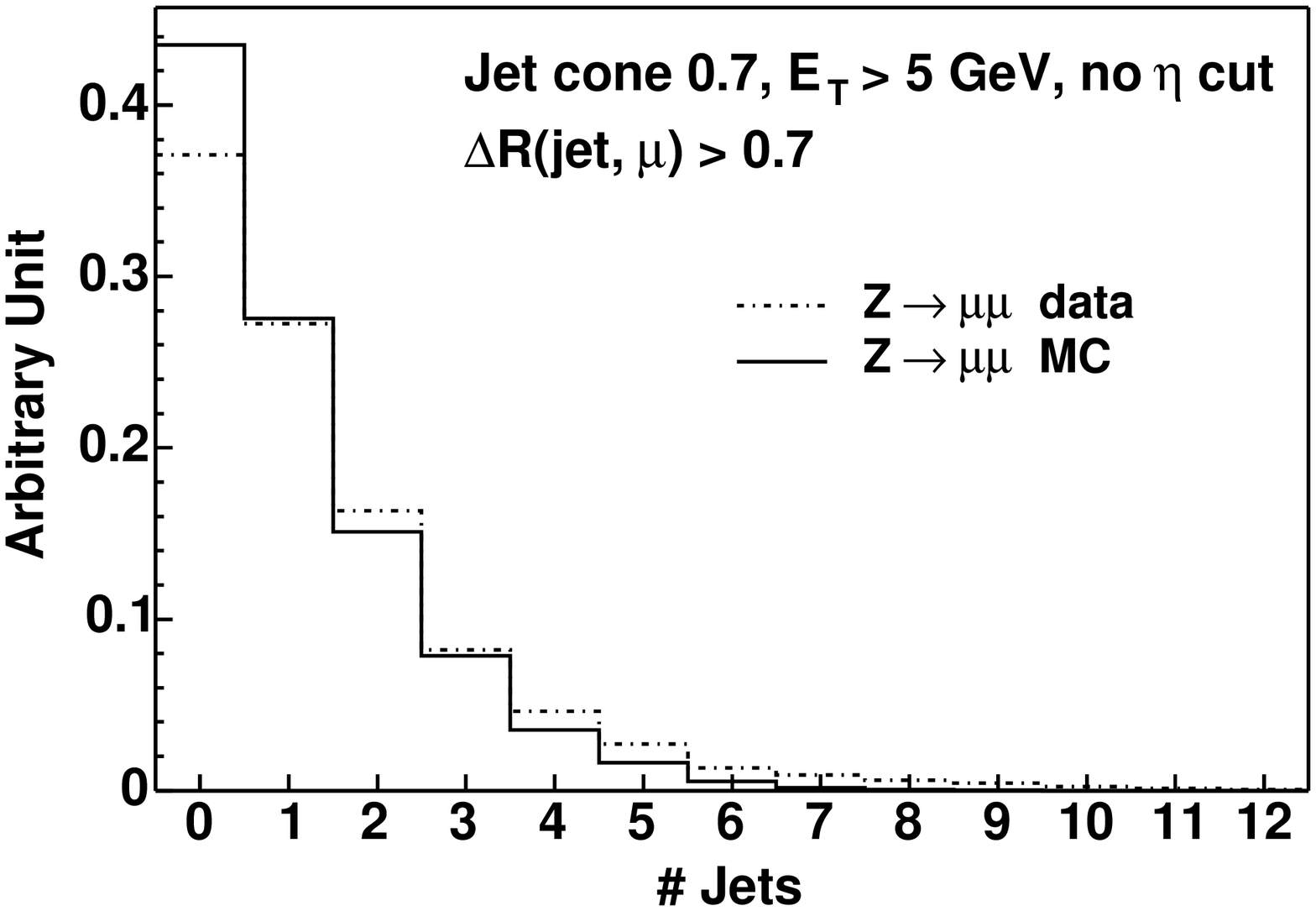}}
      \caption[Number of jets in $Z\to\mu\mu$ data and MC]
              {Distributions of the number of jets 
               in $Z\to\mu\mu$ data and MC.}
      \label{fig:TauSF_WTauNu_2}
   \vspace{0.5in}
      \parbox{5.3in}{\epsfxsize=\hsize\epsffile{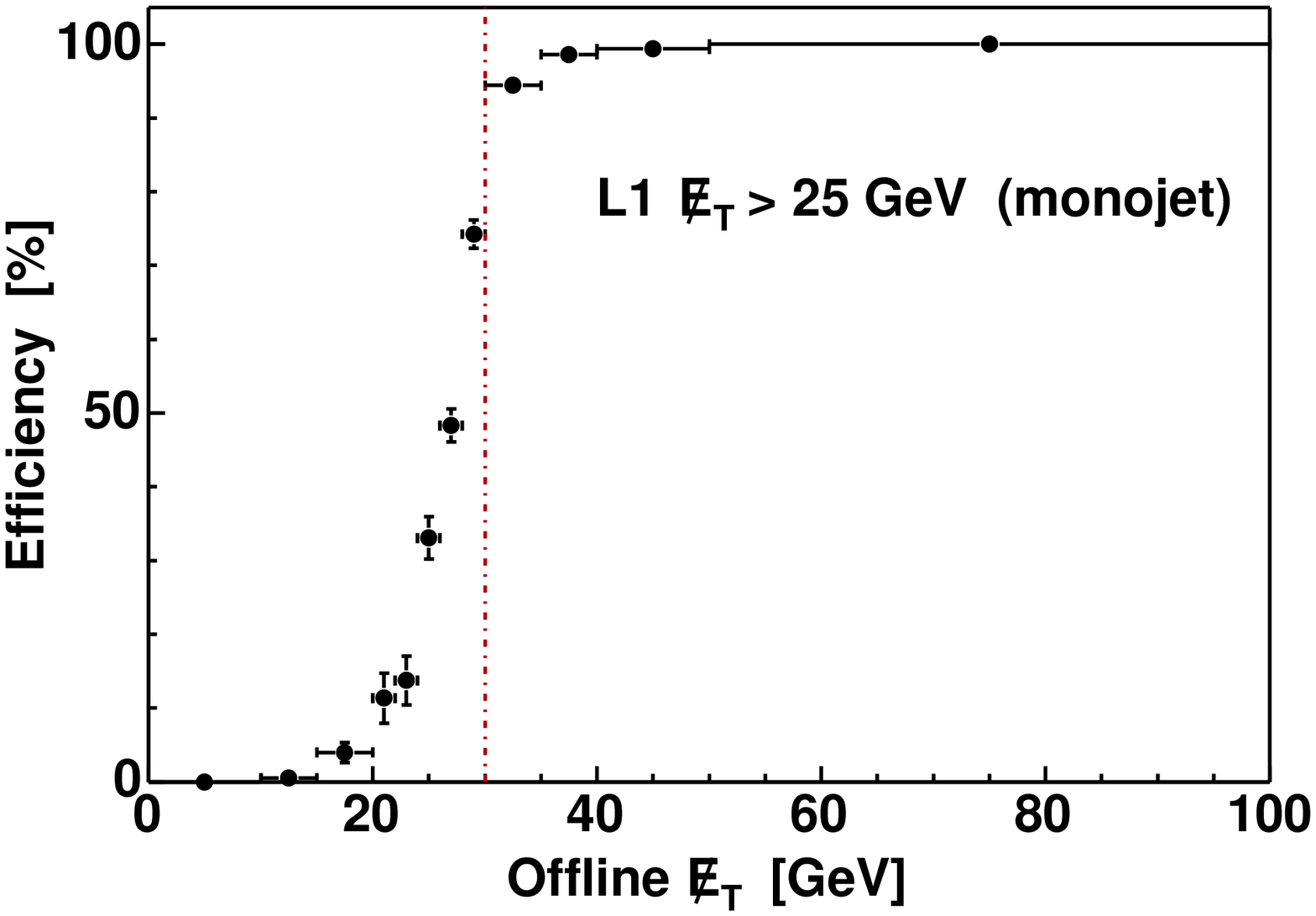}}
      \caption[L1 $\met>$ 25 GeV trigger efficiency vs. offline $\met$ for monojet event]
              {L1 $\met>$ 25 GeV trigger efficiency vs. offline $\met$ for monojet event.}
      \label{fig:TauSF_WTauNu_3}
   \end{center}
\end{figure}

The number of jets distribution in data and in MC 
are shown in Fig.~\ref{fig:TauSF_WTauNu_2}.
So $W\to\tau\nu$ monojet data/MC scale factor is 
\begin{equation}
   f^{monojet}_{data/MC} = \frac{2152 / 5799}{20149 / 46297}
                         = \frac{0.371}{0.435}
                         = 0.85 \pm 0.02
\end{equation}
The uncertainty is statistical only.

\vspace{0.2in}
\noindent{\bf L1 $\met>$ 25 GeV}
\vspace{0.1in}

The TAU\_MET trigger triggers directly on tau objects, and so there 
is no marginal trigger efficiency from TAU side. 
But there is marginal trigger efficiency from
MET side: L1 $\met$ uses a 1 GeV tower 
threshold, and offline $\met$ uses a 0.1
GeV tower threshold.  

We use JET20 data to study this trigger 
efficiency.  The event topology is monojet-like, 
since here that is what we are interested in.  
The L1 $\met>$ 25 GeV trigger efficiency vs 
offline $\met$ for monojet event is shown in 
Fig.~\ref{fig:TauSF_WTauNu_3}. 
It is a slow turn-on due to a large tower 
threshold.  An offline $\met>$ 30 GeV 
cut is not fully efficient.

\subsection{Tau Scale Factor}
\label{subsec:TauSF_sf}

After all of the above, we count 
the absolute number of $W\to\tau\nu$ events 
$n_{data}$ and $n_{MC}$ for total integrated 
luminosity 72 pb$^{-1}$.  Their ratio will be 
the tau scale factor.
\begin{itemize} 
\item To get $n_{data}$, we will use the data sample
      from the TAU\_MET trigger.  We apply the
      offline cuts to get the observed number 
      of $W\to\tau\nu$ candidates, and subtract 
      various backgrounds.  

\item To get $n_{MC}$, we will use  
      $W\to\tau\nu$ MC simulation.  We apply the
      offline cuts, multiply the number of accepted events
      by the monojet scale factor and  
      the trigger efficiency, and 
      normalize to 72~pb$^{-1}$.
\end{itemize}

The main source of backgrounds are $W\to e\nu$,
$W\to\mu\nu$, $Z/\gamma^*\to\tau\tau$, and jet 
background.

We will use MC simulation to get 
      $W\to e\nu$, $W\to\mu\nu$, and
      $Z/\gamma^*\to\tau\tau$ backgrounds. 
      We apply the offline cuts, multiply the number of
      accepted events by the monojet 
      scale factor and the trigger 
      efficiency, and normalize to 72~pb$^{-1}$.  
      For the normalization in MC, 
      $\sigma\cdot B(W\to l\nu)$ is 2700~pb~\cite{Acosta:2004uq},
      and $\sigma\cdot B(Z/\gamma^*\to l l)$ is 
      326~pb with $m_{Z/\gamma^*}>30$~GeV/$c^2$,
      which is obtained from the measured value 250~pb~\cite{Acosta:2004uq} 
      at the Z boson mass peak with $66<m_{Z/\gamma^*}<116$~GeV/$c^2$ and normalizing
      the $Z/\gamma^*\to l l$ mass spectrum generated by PYTHIA.

The jet background will be estimated directly 
      from the data by applying the relative
      jet$\to\tau$ misidentification rate and
      the relative tau identification efficiency.
      Since the cuts $|\eta_{det}|<1$, 
      10$^{\circ}$ track isolation and m(tracks)~$<2$~GeV/$c^2$ 
      are applied in the trigger,
      we use the relative rates up to the 
      denominator $D_{trkMass}$. Then we just follow
      the implementation described in 
      section~\ref{subsec:TauId_jetBg}.

Table~\ref{tab:TauSF_sf_1} shows the procedure
to estimate the contributions from signal and
backgrounds estimated from MC, the jet
background estimated from data, and the observed
number of events in data.

\begin{table}
   \begin{center}
      \begin{tabular}{|l|c|c|c|c||c|c|} \hline
                        & $W\to\tau\nu$ 
                        & $W\to e\nu$ 
                        & $W\to\mu\nu$ 
                        & $Z/\gamma^*\to\tau\tau$ 
                        & \multicolumn{2}{c|}{etau08} \\ \cline{6-7}
                        & \multicolumn{1}{c|}{signal}
                        & \multicolumn{1}{c|}{bkgd}
                        & \multicolumn{1}{c|}{bkgd}
                        & \multicolumn{1}{c||}{bkgd}
                        & \multicolumn{1}{c|}{jet bkgd}
                        & \multicolumn{1}{c|}{observed} \\ \hline
         event          & 491513 
                        & 1480550 
                        & 760457 
                        & 492000 
                        & \multicolumn{2}{c|}{3747680} \\ \hline
         hadronic tau   & 319517 
                        & N/A
                        & N/A
                        & N/A
                        & \multicolumn{2}{c|}{TAU\_MET 342164} \\
         monojet        & 11368 
                        & 192806 
                        & 7557 
                        & 7311 
                        & \multicolumn{2}{c|}{23818} \\
         $\met>$ 30 GeV & 4874 
                        & 154256 
                        & 4535 
                        & 3149 
                        & \multicolumn{2}{c|}{17490} \\
         tau ID         & 1982 
                        & 319 
                        & 130 
                        & 1230 
                        & \multicolumn{2}{c|}{$D_{trkMass}$ 1519} \\ \hline
         monojet SF     & 1684.7 
                        & 271.2 
                        & 110.5 
                        & 1045.5 
                        & \multicolumn{1}{c|}{$\sum\omega^{jet}$}
                        & \multicolumn{1}{c|}{tau ID} \\  \cline{6-7}
         trigger eff.   & 1622.1 
                        & 267.0 
                        & 107.6 
                        & 1012.3 
                        & 81.8
                        & 814 \\
         normalized     & 638.8 
                        & 34.9 
                        & 27.4 
                        & 48.3 
                        & 81.8
                        & 814 \\ \hline
      \end{tabular}
      \caption[Estimate $W\to\tau\nu$ events]
              {Expected number of events for the signal,
               backgrounds and observed number of $W\to\tau\nu$ events.}
      \label{tab:TauSF_sf_1}
   \end{center}
\end{table}

The uncertainties include
\begin{itemize}
\item statistical uncertainty,

\item monojet scale factor: 2\%,

\item luminosity: 6\%~\cite{Klimenko:2003if}, and

\item $\sigma\cdot B(W\to l\nu)$ and
      $\sigma\cdot B(Z/\gamma^*\to l l)$, 
      2\%, aside from luminosity uncertainty~\cite{Acosta:2004uq}.
\end{itemize}

Since there are discrepancies among the
      jet$\to\tau$ misidentification rates 
      obtained from different jet samples, we
      use the average jet$\to\tau$ 
      misidentification rate to get a central
      value of 81.8 events.  The estimates using the
      individual jet$\to\tau$ 
      misidentification rate from JET20,
      JET50, JET70, and JET100 samples are
      90.3, 67.1, 72.8, and 66.3, respectively.
      We take the biggest difference as the
      the uncertainty for jet background:
      $|$(66.3-81.8)/81.8$|$ = 18.9\%.

The numbers and the uncertainties of each 
channel are summarized in 
Table~\ref{tab:TauSF_sf_2}.
We now arrive at the tau scale factor as follows:
\begin{eqnarray}
   f^{\tau}_{data/MC} & = & \frac{n_{data}}
                                {n_{MC}} \nonumber \\
                     & = & \frac{n_{obs.} - n_{WZ\;bgs} - n_{jet\;bg}}
                                {n_{sig.}} \nonumber \\
                     & = & \frac{814 - (34.9 + 27.4 + 48.3) - 81.8}
                                {638.8} \nonumber \\
                     & = & 0.97\pm0.10
\end{eqnarray}
with statistical uncertainty and all of the
systematic uncertainties.

\begin{table}
   \begin{center}
      \begin{tabular}{|c|c|} \hline
         $W\to\tau\nu$           & $638.8\pm42.7$   \\
         $W\to e\nu$             &  $34.9\pm2.9$    \\
         $W\to\mu\nu$            &  $27.4\pm3.0$    \\
         $Z/\gamma^*\to\tau\tau$ &  $48.3\pm3.3$    \\
         Jet$\to\tau$            &  $81.8\pm15.5$   \\ \hline
         Expected                & $831.2\pm45.7$   \\ \hline
         Observed                &       814        \\ \hline
      \end{tabular}
      \caption[Expected and observed $W\to\tau\nu$ events]
              {Cross check for the numbers of $W\to\tau\nu$ events.}
      \label{tab:TauSF_sf_2}
   \end{center}
\end{table}



\begin{figure}
   \begin{center}
      \parbox{5.7in}{\epsfxsize=\hsize\epsffile{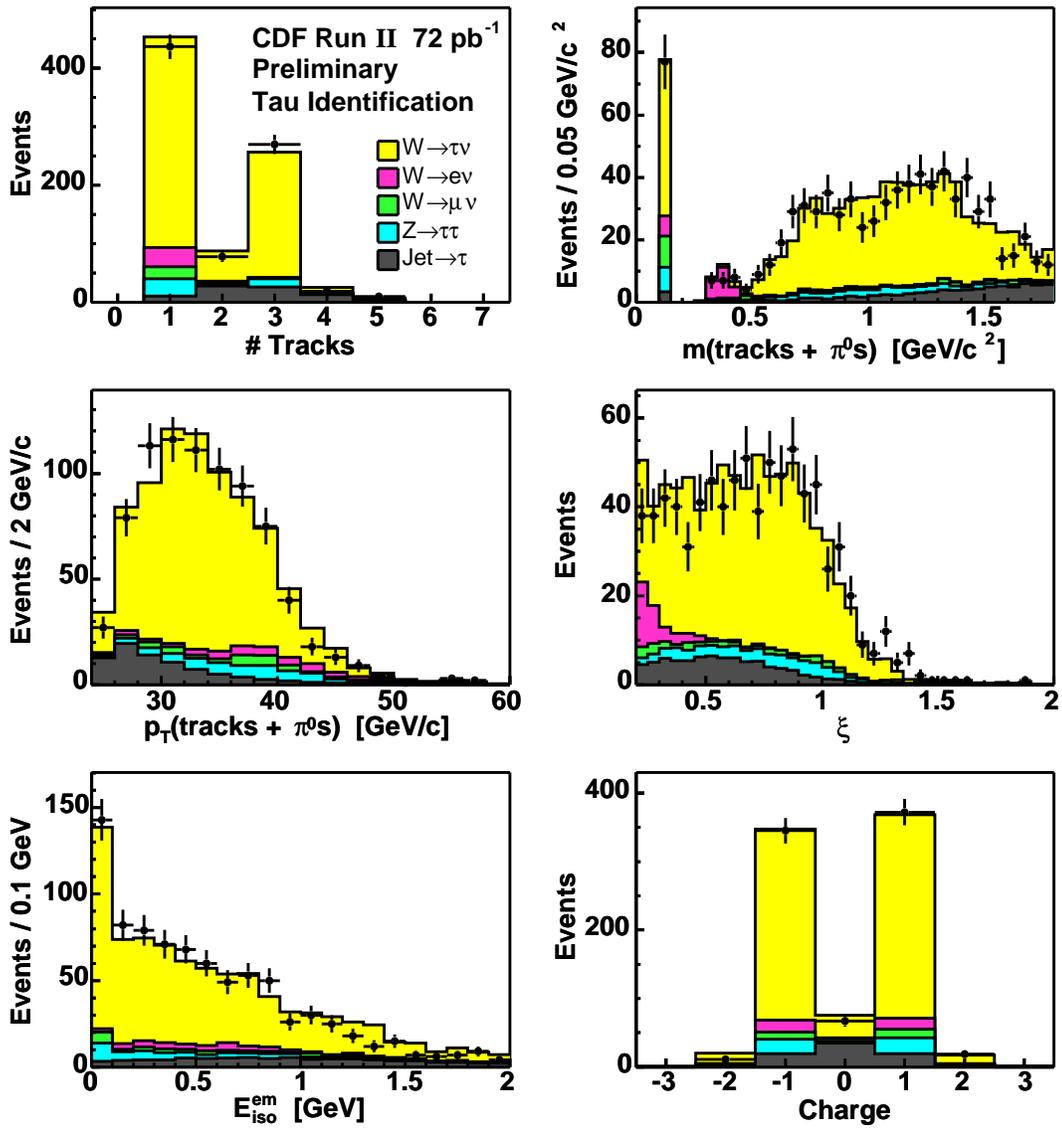}}
      \caption[Distributions of tau variables using $W\to\tau\nu$ events]
              {Distributions of hadronic tau identification using 
               $W\to\tau\nu$ events for data (points) and predicted 
               backgrounds (histograms).}
      \label{fig:TauSF_TAUMET}
   \end{center}
\end{figure}

Lastly we put signal and background together, and show the $W\to\tau\nu$
kinematic distributions in data and MC in Fig.~\ref{fig:TauSF_TAUMET}.  
The agreement between data and MC is very good.

\section{Electron Identification}
\label{sec:EleId}

Identification of electrons is based on the energy
it deposits in the calorimeter, its track in the COT,
and its position in the CES.
The central electron reconstruction algorithm~\cite{CDFnote:5456} 
starts with clusters in the CEM detector. The electromagnetic
towers are ordered in $\et$ and the highest $\et$ tower 
that has not yet been clustered is taken as a seed.  The 
available shoulder towers are added to the cluster if they 
are adjacent in $\eta$ to the seed, and the clusters are 
restricted to two towers.  The default threshods for seed 
towers and shoulder towers are 3.0 and 0.1 GeV, respectively.  
For the leading electrons used in our analysis with 
$e + \tau_h$ channel, the threshods for seed towers and 
shoulder towers are set to be 8 and 7.5 GeV, respectively.
Then we associate tracks with the candidate cluster.  For 
all of the tracks associated, the one with highest $\pt$ 
is chosen as the matched one. The CES strip and wire clusters are 
associated with the CEM cluster if they are reconstructed 
in the same wedge.  The ``best-matching'' CES cluster is 
the one seeded by the matched track.


\subsection{Electron Identification Cuts}
\label{subsec:EleId_cuts}

The electron identification~\cite{CDFnote:6580} cuts, and 
the conversion veto~\cite{CDFnote:6250} cuts to remove electrons
from photon conversion, are listed in Table~\ref{tab:EleId_cuts}.
The $\et$ and $\pt$ thresholds are not listed because 
they depend on the process and trigger sample.  The probe 
electron must be a fiducial CEM electron and pass the 
vertex $z$ cut.  

\begin{table}
   \begin{center}
      \begin{tabular}{|l|l|l|l|} \hline
         Variable         & Cut                         & Note                   & Denominator \\ \hline \hline
         region           & ==0                         & CEM                    &             \\
         fiducial         & ==1                         & fiducial $X_{CES}$, 
                                                                   $Z_{CES}$     &             \\
         $|z_0|$          & $<$60 cm                    & vertex $z$             & Probe       \\
         track ax. seg.   & $\ge$3$\times$7             & COT axial segments     &             \\
         track st. seg.   & $\ge$3$\times$7             & COT stereo segments    &             \\
         cal. isolation   & $<$0.1                      & cone 0.4               &             \\
         $E_{had}/E_{em}$ & $<$0.055+0.00045$\times$$E$ & had./em.               &             \\
         $E/p$            & $<$4 (for $\et<$100 GeV)    & cal./track with brem.  &             \\
         $L_{shr}$        & $<$0.2                      & lateral shower profile &             \\
         $|\Delta X|$     & $<$3 cm                     & $X_{track} - X_{CES}$  &             \\
         $|\Delta Z|$     & $<$5 cm                     & $Z_{track} - Z_{CES}$  &             \\
         conversion veto  & $|\Delta XY|$$<$0.2 cm, and & separation, and        &             \\   
                          & $|\Delta\cot\theta|$$<$0.04 & parallel               & Numerator   \\ \hline
      \end{tabular}
      \caption[Electron identification cuts]
              {Electron identification cuts.}
      \label{tab:EleId_cuts}
   \end{center}
\end{table}


\subsection{Electron Scale Factor}
\label{subsec:EleSF}

The electron identification scale factor is the ratio 
of the efficiency in data/MC.  The data sample is
from the TAU\_ELE trigger which requires an electron 
with $\et>$ 8 GeV, $\pt>$ 8 GeV/$c$ and an isolated track 
with $\pt>$ 5 GeV/$c$. We study the electron scale factor 
versus $\et$. 
\begin{itemize}
\item For medium-$\et$ (between 5 GeV and 20 GeV) electrons, 
      the MC uses electrons 
      from $Z\to\tau\tau\to eX$, and in the real data we use
      the second leg after selecting 
      $\Upsilon\to ee$.  
      We require the probe electrons have 
      $\et>$ 5 GeV and $\pt>$ 5 GeV/$c$ in both the real data and the MC.

\item For high-$\et$ (above 20 GeV) electron, the MC uses electrons 
      from $Z\to ee$, and in the real data we use the second leg 
      after selecting $Z\to ee$. 
      We require the probe electrons have 
      $\et>$ 20 GeV and $\pt>$ 10 GeV/$c$ in both the real data and the MC.
\end{itemize}
The procedure to select $\Upsilon\to ee$ events is:
\begin{itemize}
\item Require a tight electron with $\et>$ 8 
      GeV, $\pt>$ 8 GeV/$c$ which are the trigger 
      requirements and the electron 
      identification cuts.

\item Require a probe electron with $\et>$ 5 
      GeV, $\pt>$ 5 GeV/$c$.

\item Same-sign pair will be used later for fitting 
      the slope of background and opposite-sign pair
      will be used later for fitting signal + 
      background.

\item Require the invariance mass of the $ee$ 
      pair to lie in the range (0, 20) GeV/$c^2$.
\end{itemize}
The procedure to select $Z\to ee$ events is:
\begin{itemize}
\item Require a tight electron with $\et>$ 20
      GeV, $\pt>$ 10 GeV/$c$ and the electron 
      identification cuts. 

\item Require a probe electron with $\et>$ 20
      GeV, $\pt>$ 10 GeV/$c$.

\item Require opposite sign.

\item Require the invariance mass of the $ee$ 
      pair to lie in the range (75, 105) GeV/$c^2$.
\end{itemize}
The procedure to select the second leg is:
\begin{itemize}
\item Require exactly one $\Upsilon$ or Z boson.

\item If there is one tight electron, the probe electron 
      is the second leg.

\item If there are two tight electrons, both are used as 
      second leg.
\end{itemize}
Then we apply the set of electron identification 
cuts under study on the second leg electrons in 
data, and on the probe electrons in the MC.
The result of the procedure is shown in 
Table~\ref{tab:EleSF_1}.

\begin{table}
   \begin{center}
      \begin{tabular}{|l|c|c||l|c|c|l|} \hline
                    & \multicolumn{2}{c||}{data}
                    &
                    & \multicolumn{2}{c|}{MC}
                    & 
           \\ \cline{2-3} \cline{5-6}
                    & $\Upsilon\to ee$
                    & \multicolumn{1}{c||}{$Z\to ee$}
                    &
                    & $Z\to\tau_e\tau_x$
                    & $Z\to ee$
                    &
           \\ \cline{1-6}
         event      & \multicolumn{2}{c||}{11922805}
                    & event
                    & 492000
                    & 398665
                    &
           \\ \cline{1-6}
         good run   & \multicolumn{2}{c||}{9103020}
                    &
                    &
                    &
                    &
           \\
         triggered  & \multicolumn{2}{c||}{5575584}
                    &
                    &
                    &
                    &
           \\
         unique     & \multicolumn{2}{c||}{5310963}
                    &
                    &
                    &
                    &
           \\ \cline{1-6}
         process                           & 10373 &   4534 & electron& 175515 & 797330 &       \\ \cline{1-6}
         second leg                        & 10770 &   7973 & match   &  36733 & 204680 & Probe \\ \hline \hline
         track ax. seg. $\ge3\times7$      & 10687 &   7946 & same    &  36665 & 204266 &       \\
         track st. seg. $\ge3\times7$      & 10165 &   7721 & same    &  36448 & 202907 &       \\
         cal. isolation  $<0.1$            &  2797 &   7484 & same    &  32094 & 197065 &       \\
         $E_{had}/E_{em}<0.055$+0.00045$E$ &  2553 &   7427 & same    &  31290 & 194746 &       \\
         $E/p<4$ (for $\et<100$ GeV)       &  2551 &   7379 & same    &  31187 & 194133 &       \\
         $L_{shr}<0.2$                     &  2331 &   7318 & same    &  30240 & 188653 &       \\
         $|\Delta X|<3$ cm                 &  2304 &   7198 & same    &  30028 & 186360 &       \\
         $|\Delta Z|<5$ cm                 &  2292 &   7189 & same    &  29976 & 186222 &       \\
         conversion veto                   &  2249 &   6878 & same    &  29714 & 181449 & Id    \\ \hline
      \end{tabular}
      \caption[Electron identification efficiency measurement]
              {Number of events for electron identification efficiency
               measurement.}
      \label{tab:EleSF_1}
   \end{center}
\end{table}

For the $Z\to ee$ selection in the real data, 
the backgrounds in the sample with a tight electron plus 
a probe electron and in the sample with two tight electrons  
are both negligible which is confirmed by the negligible 
number of same-sign events in these two samples.  

For the $\Upsilon\to ee$ in the real data,
the backgrounds in the sample with tight electron plus 
probe electron and in the sample with two tight electrons  
are both significant.
The same-sign samples provide the shapes of the invariant 
mass distribution of the backgrounds,
which are taken as the slopes of linear backgrounds.
Then in the opposite sign samples we fit the invariant
mass distributions by the ``Crystal Ball'' function~\cite{Skwarnicki:1986}
plus a linear background.  The $\Upsilon\to ee$ invariant 
mass distribution has a Bremsstrahlung tail at lower mass 
side where at least one of the electrons radiates.  
The ``Crystal Ball'' line-shape serves to model this 
Gaussian core with a power-law tail.
The yield of signal is obtained by the entries in the histogram
subtracted by the integral of the linear background.

\begin{figure}
   \begin{center}
      \parbox{3.9in}{\epsfxsize=\hsize\epsffile{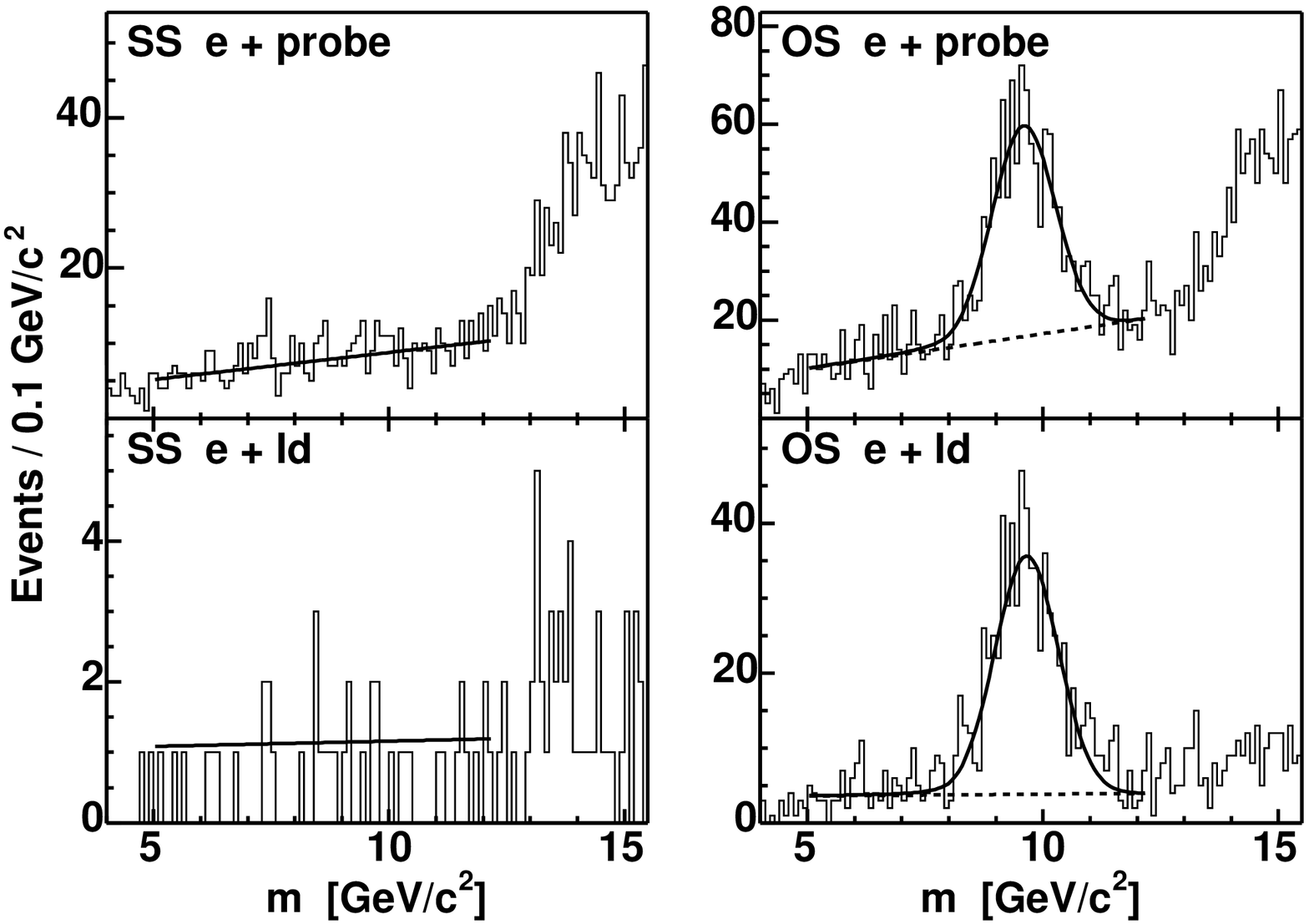}}
      \caption[$\Upsilon\to ee$ in data]
              {Distributions of the invariant mass
               of $\Upsilon\to ee$ for medium $\et$
               electron identification 
               efficiency measurement in data. 
               Same-sign samples provide
               the slopes of the linear backgrounds.
               The numbers of $\Upsilon\to ee$
               signal events are obtained 
               from fitting the histograms with 
               the ``Crystal Ball''
               function plus a linear background
               in the oppisite-sign samples.}
      \label{fig:EleSF_1}
   \vspace{0.5in}
      \parbox{3.9in}{\epsfxsize=\hsize\epsffile{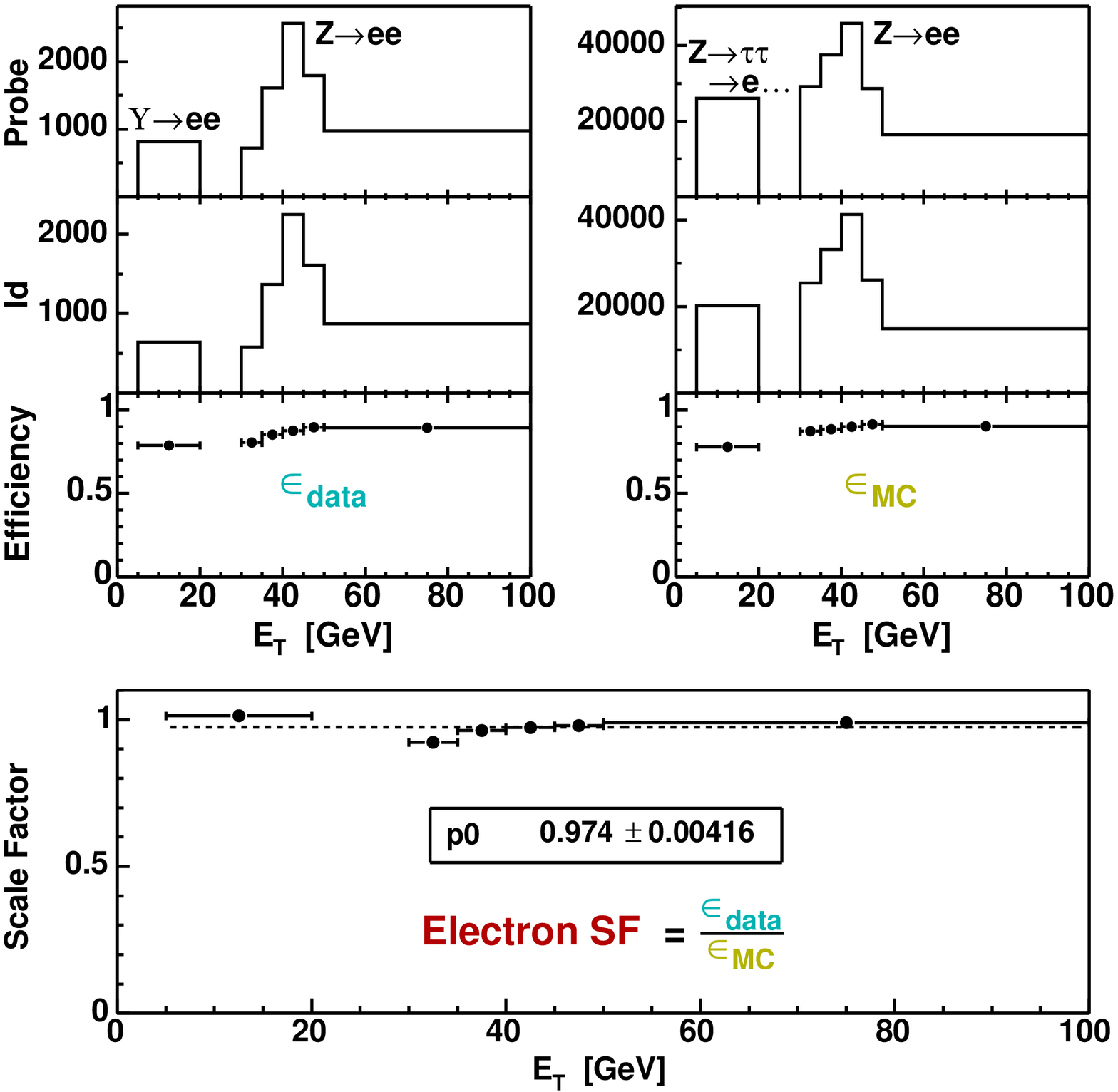}}
      \caption[Electron scale factor vs. $\et$]
              {Electron scale factor vs. $\et$. This is 
               obtained from dividing the efficiency in 
               data by the efficiency in MC.}
      \label{fig:EleSF_2}
   \end{center}
\end{figure}

Up to this point, all of the $\Upsilon\to ee$ candidates 
in the mass window (0,~20)~GeV/$c^2$ are accepted.  
We then subtract background, as shown in 
Fig.~\ref{fig:EleSF_1}.
The plot only shows the mass window (4,~15.5)~GeV/$c^2$. 
The fit is performed in the mass window (5,~12.2)~GeV/$c^2$. 
The fitting result is N(e + probe) = 818.0, 
N(e + Id) = 644.4, efficiency = 78.8\%. 

Now we put everything together to get the electron 
scale factor vs. $\et$ and perform a fit in $\et$.
This is shown in Fig.~\ref{fig:EleSF_2}.
Data: the medium $\et$ electrons (5, 20) GeV are
      from the second leg of $\Upsilon\to ee$; 
      the high $\et$ electrons (30, 100) GeV are from 
      the second leg of $Z\to ee$;  there is 
      a gap (20, 30) GeV which has very low 
      statistics and is not used.
MC:   the medium $\et$ electrons (5, 20) GeV are 
      from the probe electrons of 
      $Z\to\tau\tau\to eX$; 
      the high $\et$ electrons  
      (30, 100) GeV are from the probe 
      electrons of $Z\to ee$.
In each $\et$ bin, the efficiency in data 
      divided by the efficiency in MC gives 
      scale factor in that $\et$ bin.
For all of the $\et$ bins, the scale 
      factor is flat.  A fit by a polynomial 
      of degree 0, which is exactly the same 
      as the weighted average, gives a 
      scale factor $0.974\pm0.004$. 

There are two bins (45, 50) GeV and 
      (50, 100) GeV with efficiency close to 100\%, 
      in data and MC. 
      The binomial uncertainty in this case is 
      always close to zero and underestimated. 
      This propagates to the scale factors in 
      those two $\et$ bins, and finally 
      propagates to the weighted average.
There is also uncertainty in the (5, 20)
      GeV bin due to $\Upsilon\to ee$ background 
      subtraction. This uncertainty is not
      estimated.
We assign a conservative 4\% uncertainty for 
electron scale factor~\cite{CDFnote:7288}:  
\begin{equation}
   f^{e}_{data/MC} = 0.97 \pm 0.04
\end{equation}

\section{Muon Identification}
\label{sec:MuId}

Muon reconstruction~\cite{CDFnote:5870} uses information from tracking,
the calorimeters and the muon chambers.  The momentum is
measured by the curvature of the muon trajectory 
bent by magnetic field in tracking system.
Muons behave as minimum ionizing particles and 
they are the only charged particle that can
travel through the large amount of material in
calorimeter with a very small 
energy loss.  Muons are not stable, but they are 
so long lived that they can reach the muon
chamber, leave hits there, and continue to travel and 
decay outside the detector.  These features
allow a rather simple and clean muon 
identification.


\subsection{Muon Identification Cuts}
\label{subsec:MuId_cuts}

The muon identification cuts~\cite{CDFnote:6825} are listed in 
Table~\ref{tab:MuId_cuts}.
We use COT-only tracks and add the beam constraint 
to the track.  The $\et$ and $\pt$ thresholds are not 
listed because they depend on the process and 
trigger.  
For data/MC scale factor studies, we require the 
track to be fiducial which means that the track is 
headed in a direction that will lead it to hit 
enough chambers for a stub to be reconstructed. 
All three subdetectors CMU, CMP, and CMX which
are used in this analysis require 3 hits in 3 
different layers for a stub to be reconstructed.  
And we will study two kinds of data/MC scale 
factors:
\begin{itemize}
\item Muon identification scale factor.
      A fiducial stub muon and vertex $z$ cut
      are required for the probe muon for this study,
      called ``Probe (Id)'' in Table~\ref{tab:MuId_cuts}.

\item Marginal muon reconstruction scale 
      factor.  We require a fiducial track and a stubless muon,
      which also has the information of energy 
      loss in calorimeter, for the probe muon for this study,
      called ``Probe (Rec)'' in Table~\ref{tab:MuId_cuts}.  
      It is not necessary 
      to have hits in the muon chambers.  The vertex $z$ cut, 
      calorimeter isolation cut, EM energy cut, and 
      hadronic energy cut are required.  Then 
      we check if this track has a muon stub.
      The default track $\pt$ threshold to 
      make a stubless muon is 10 GeV;
      we lower it to 5 GeV to allow more medium
      $\pt$ stubless muons.
\end{itemize}

\begin{table}
   \begin{center}
      \begin{tabular}{|l|l|l|l|} \hline
         Variable                  & Cut                                       & Note             & Probe              \\ \hline \hline
         $|z_0|$                   & $<$60 cm                                  & vertex $z$       & Probe (Id)         \\
         cal. isolation            & $<$0.1                                    & cone 0.4         &                    \\
         $E_{em}$                  & $<$2+max(0,($p$$-$100)$\times$0.0115) GeV & EM energy        &                    \\
         $E_{had}$                 & $<$6+max(0,($p$$-$100)$\times$0.028) GeV  & had. energy      & Probe (rec.)       \\ 
         $|d_0|$                   & $<$0.2 cm                                 & impact parameter &                    \\ 
         track ax. seg.            & $\ge$3$\times$7                           & COT ax. seg.     &                    \\
         track st. seg.            & $\ge$3$\times$7                           & COT st. seg.     &                    \\
         $|\Delta x_{\mbox{CMU}}|$ & $<$3 cm (for CMUP)                        & $x_{\mbox{track}} - x_{\mbox{CMU}}$ & \\
         $|\Delta x_{\mbox{CMP}}|$ & $<$5 cm (for CMUP)                        & $x_{\mbox{track}} - x_{\mbox{CMP}}$ & \\
         $|\Delta x_{\mbox{CMX}}|$ & $<$6 cm (for CMX)                         & $x_{\mbox{track}} - x_{\mbox{CMX}}$ & \\
         $\rho_{\mbox{COT}}$       & $>$140 cm (for CMX)                       & COT exit radius  &                    \\ \hline 
      \end{tabular}
      \caption[Muon identification cuts]
              {Muon identification cuts.}
      \label{tab:MuId_cuts}
   \end{center}
\end{table}


\subsection{Muon Scale Factor}
\label{subsec:MuSF}

The muon identification scale factor is the ratio 
of the identification efficiency in real data to that in MC.
The muon marginal reconstruction scale factor is 
the ratio of the marginal reconstruction
efficiency in data/MC.  
The data sample is from the TAU\_CMU trigger 
which requires a CMUP muon with 
$\pt>$ 8 GeV/$c$ and an isolated track with $\pt>$ 5 
GeV/$c$. (A CMUP muon is required to have stubs in
both CMU and CMP). We study the muon scale factors versus muon $\pt$.  
\begin{itemize}
\item For medium $\pt$ (between 5 and 20 GeV/$c$) muons, the MC uses muon from 
      $Z\to\tau\tau\to\mu X$, data uses 
      the second leg after selecting 
      $\Upsilon\to\mu\mu$. 
      We require the probe muons have 
      $\pt>$ 5 GeV/$c$ in both the real data and the MC.

\item For high $\pt$ (above 20 GeV/$c$) muons, the MC uses muon from 
      $Z\to\mu\mu$, and for the data we use 
      the second leg after selecting 
      $Z\to\mu\mu$.  
      We require the probe muons have 
      $\pt>$ 20 GeV/$c$ in both the real data and the MC.
\end{itemize}
The procedure to select $\Upsilon\to\mu\mu$ events is:
\begin{itemize}
\item Cosmic veto~\cite{CDFnote:6089}.

\item Require a tight CMUP muon with $\pt>$ 
      8 GeV/$c$ which are trigger requirements 
      and the CMUP muon identification cuts.

\item Require a probe muon with $\pt>$ 5  
      GeV/$c$.

\item Require $|z_0(1)-z_0(2)|<$ 4 cm.

\item Require opposite sign.

\item Mass window (7, 13) GeV/$c^2$. We 
      will use side band for background 
      subtraction.
\end{itemize}
The procedure to select $Z\to\mu\mu$ events is:
\begin{itemize}
\item Cosmic veto.

\item Require a tight CMUP muon with 
      $\pt>$ 20 GeV/$c$ and the CMUP 
      muon identification cuts. 

\item Require a probe muon with 
      $\pt>$ 20 GeV/$c$.

\item Require $|z_0(1)-z_0(2)|<$ 4 cm.

\item Require opposite sign.  The
      negligible number of same sign 
      events confirms that background 
      is negligible.

\item Mass window (80, 100) GeV/$c^2$.
\end{itemize}
The procedure to select the second leg is:
\begin{itemize}
\item Require exactly one $\Upsilon$ or 
      Z boson.

\item If one tight muon, the probe 
      muon is the second leg.

\item If two tight muons, both are used 
      as second leg.
\end{itemize}

\vspace{0.2in}
\noindent{\bf Muon Identification Scale Factor}
\vspace{0.1in}

In the muon identification scale factor study, 
we apply the set of muon identification 
cuts under study on the second leg muons in 
data, and on the probe muons in the MC.
Table~\ref{tab:MuSF_1} shows the
procedure in data and Table~\ref{tab:MuSF_2}
shows the procedure in MC. 

\begin{table}
   \begin{center}
      \begin{tabular}{|l|c|c|c|c|l|} \hline
                        & \multicolumn{2}{c|}{$\Upsilon\to\mu\mu$}
                        & \multicolumn{2}{c|}{$Z\to\mu\mu$}
                        &
           \\ \cline{1-5}
         event          & \multicolumn{4}{c|}{11922805}
                        &
           \\ \cline{1-5}
         good run       & \multicolumn{4}{c|}{9103020}
                        &
           \\
         triggered      & \multicolumn{4}{c|}{1881529}
                        &
           \\
         unique         & \multicolumn{4}{c|}{1800059}
                        &
           \\ \cline{1-5}
         process        & \multicolumn{2}{c|}{971}
                        & \multicolumn{2}{c|}{2025}
                        &
           \\ \cline{1-5}
         second leg     & \multicolumn{2}{c|}{1047}
                        & \multicolumn{2}{c|}{2805}
                        &
           \\ \cline{2-5}
                                                          & CMUP & CMX & CMUP & CMX &        \\ \cline{2-5}
                                                          &  762 & 285 & 1820 & 985 & Probe  \\ \hline \hline 
         $|d_0|<0.2$ cm                                   &  758 & 283 & 1816 & 982 &        \\
         track ax. seg. $\ge3$$\times$7                   &  756 & 283 & 1814 & 978 &        \\
         track st. seg. $\ge3$$\times$7                   &  739 & 280 & 1777 & 955 &        \\
         cal. isolation $<0.1$                            &  527 & 194 & 1750 & 947 &        \\
         $E_{em}<$ 2+max(0,($p$$-$100)$\times$0.0115) GeV &  525 & 191 & 1700 & 929 &        \\
         $E_{had}<$ 6+max(0,($p$$-$100)$\times$0.028) GeV &  524 & 191 & 1670 & 907 &        \\
         $|\Delta x_{\mbox{CMU}}|<3$ cm
        ($|\Delta x_{\mbox{CMX}}|<5$ cm)                  &  427 & 129 & 1590 & 877 &        \\
         $|\Delta x_{\mbox{CMP}}|<6$ cm 
        ($\rho_{\mbox{COT}}>140$ cm)                      &  287 & 113 & 1560 & 753 & Id     \\ \hline
      \end{tabular}
      \caption[Muon identification efficiency measurement in data]
              {Number of events for muon identification efficiency 
               measurement in data.}
      \label{tab:MuSF_1}
   \end{center}
\end{table}

\begin{table}
   \begin{center}
      \begin{tabular}{|l|c|c|c|c|l|} \hline
                        & \multicolumn{2}{c|}{$Z\to\tau_{\mu}\tau_x$}
                        & \multicolumn{2}{c|}{$Z\to\mu\mu$}
                        &
           \\ \cline{1-5}
         event          & \multicolumn{2}{c|}{492000}
                        & \multicolumn{2}{c|}{405291}
                        &
           \\ \cline{1-5}
         muon           & \multicolumn{2}{c|}{170596}
                        & \multicolumn{2}{c|}{810582}
                        &
           \\ \cline{1-5}
         match          & \multicolumn{2}{c|}{32388}
                        & \multicolumn{2}{c|}{170382}
                        &
           \\ \cline{2-5}
                                                          &  CMUP  &  CMX  &  CMUP  &  CMX  &        \\ \cline{2-5}
                                                          &  20516 & 11872 & 107481 & 62901 & Probe  \\ \hline \hline 
         $|d_0|<0.2$ cm                                   &  20499 & 11827 & 107410 & 62692 &        \\
         track ax. seg. $\ge3$$\times$7                   &  20490 & 11732 & 107348 & 62184 &        \\
         track st. seg. $\ge3$$\times$7                   &  20417 & 11601 & 106985 & 61463 &        \\
         cal. isolation  $<0.1$                           &  18361 & 10448 & 104247 & 59929 &        \\
         $E_{em}<$ 2+max(0,($p$$-$100)$\times$0.0115) GeV &  18055 & 10261 & 100115 & 57680 &        \\
         $E_{had}<$ 6+max(0,($p$$-$100)$\times$0.028) GeV &  17871 & 10080 &  97906 & 55799 &        \\
         $|\Delta x_{\mbox{CMU}}|<3$ cm
        ($|\Delta x_{\mbox{CMX}}|<5$ cm)                  &  16799 &  8994 &  97707 & 55617 &        \\
         $|\Delta x_{\mbox{CMP}}|<6$ cm
        ($\rho_{\mbox{COT}}>140$ cm)                      &  14198 &  7419 &  96788 & 46059 & Id     \\ \hline
      \end{tabular}
      \caption[Muon identification efficiency measurement in MC]
              {Number of events for muon identification efficiency
               measurement in MC.}
      \label{tab:MuSF_2}
   \end{center}
\end{table}

\begin{figure}
   \begin{center}
      \parbox{3.9in}{\epsfxsize=\hsize\epsffile{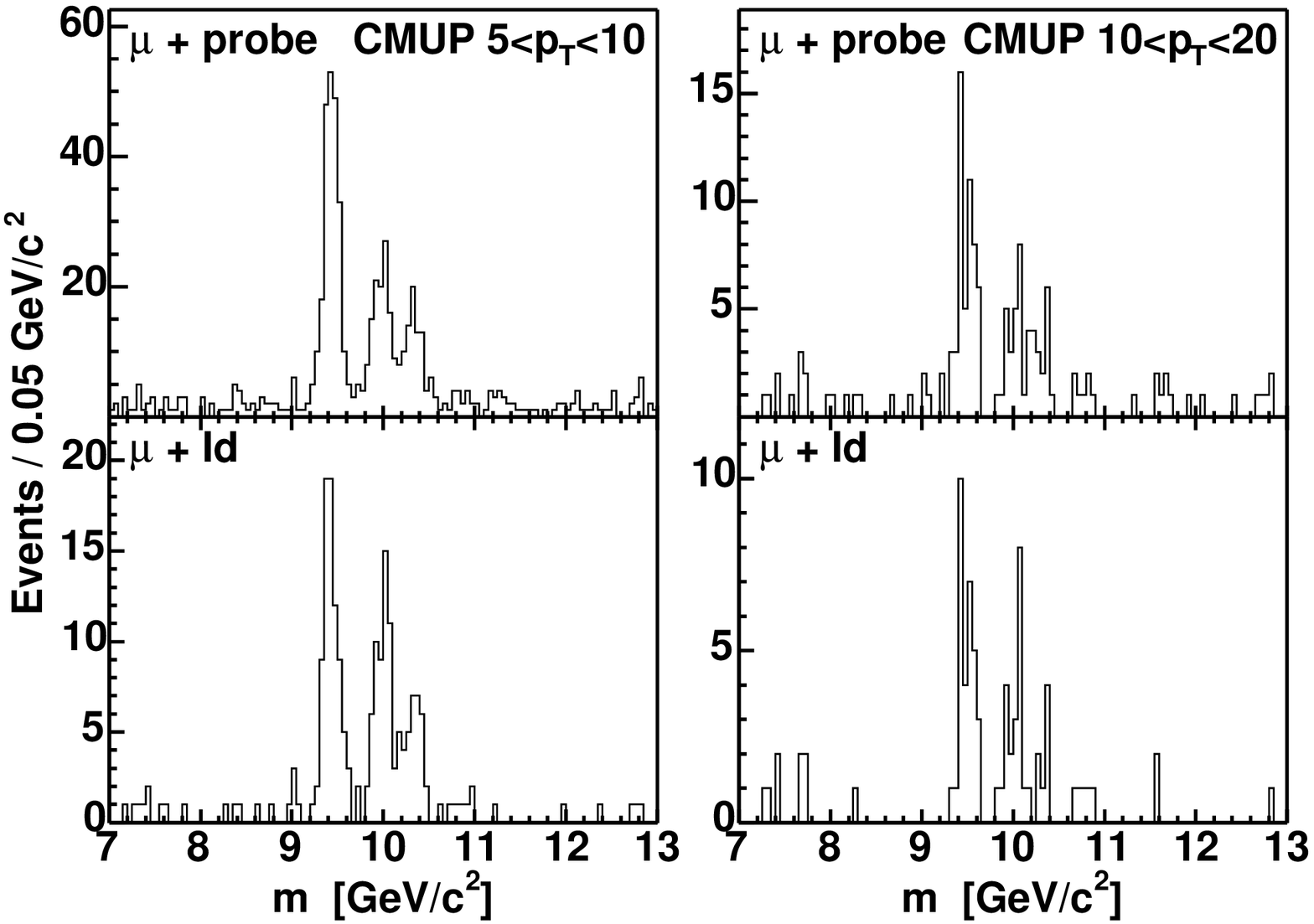}}
      \caption[$\Upsilon\to\mu\mu$ for CMUP muon identification efficiency measurement]
              {Distributions of the invariant mass
               of $\Upsilon\to\mu\mu$ for medium $\pt$
               CMUP muon identification 
               efficiency measurement in data.
               The three peaks are signature
               of $\Upsilon\to\mu\mu$.
               The left two plots are for CMUP
               muons with $5<\pt<10$ GeV/$c$.
               The right two plots are for CMUP
               muons with $10<\pt<20$ GeV/$c$.
               Side-band method is used for 
               background subtractions.}
      \label{fig:MuSF_1}
   \vspace{0.5in}
      \parbox{3.9in}{\epsfxsize=\hsize\epsffile{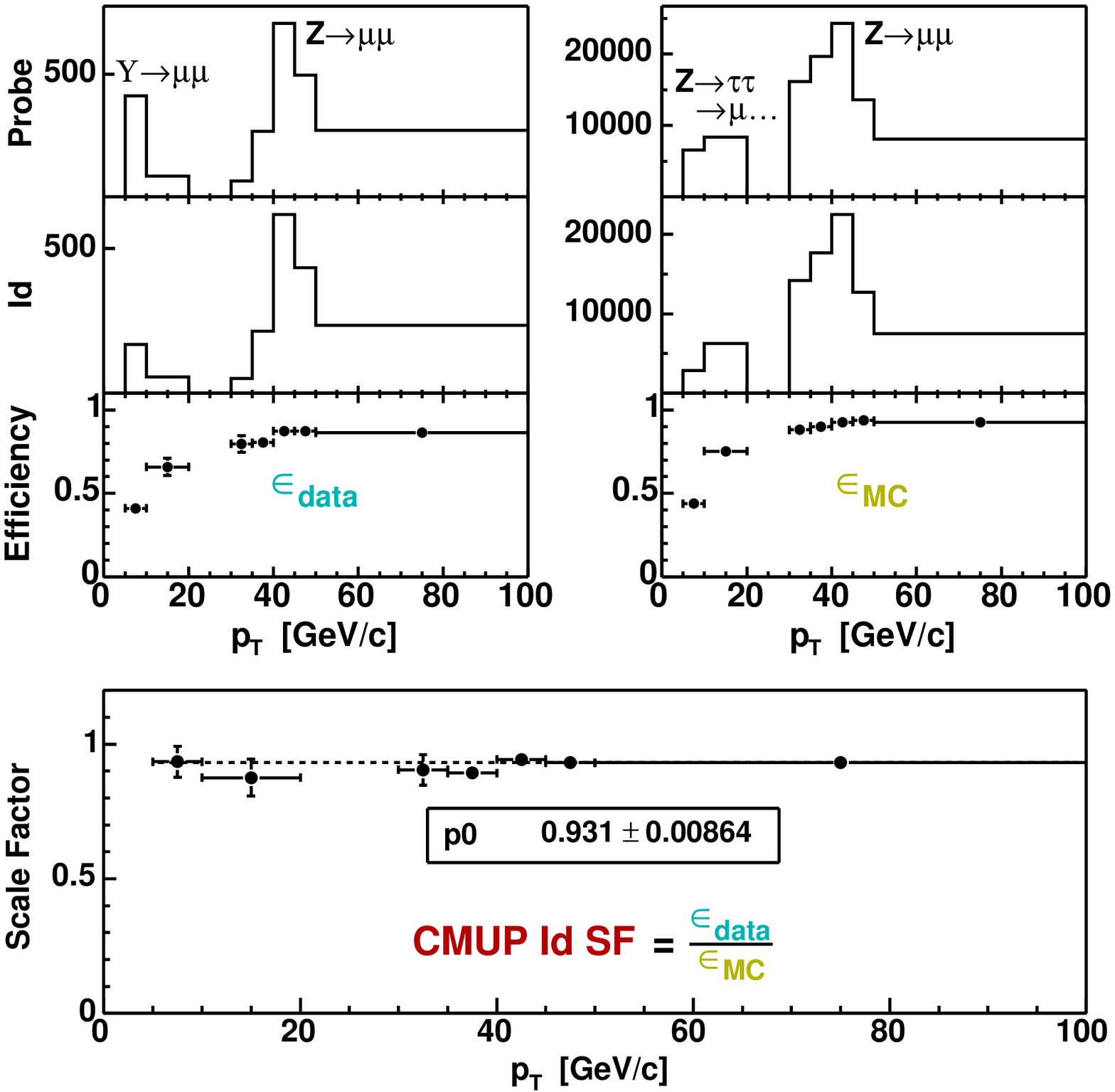}}
      \caption[CMUP muon identification scale factor vs. $\pt$]
              {CMUP muon identification scale factor vs. $\pt$. 
               This is obtained from dividing the efficiency 
               in data by the efficiency in MC.}
      \label{fig:MuSF_3}
   \end{center}
\end{figure}

Up to this point, all of the $\Upsilon\to\mu\mu$ candidates
in mass window (7, 13) GeV/$c^2$ are accepted. 
Now we break the probe into two $\pt$ 
bins 5 $<\pt<$ 10 GeV/$c$ and 10 $<\pt<$ 20 GeV/$c$.  
Fig.~\ref{fig:MuSF_1} shows the distributions
of the pair mass of the first leg and the second leg 
in each $\pt$ bin of the second leg, for CMUP probe.
We see three clear peaks at about 9.5, 10 and 10.3 
GeV/$c^2$.  This is the signature of $\Upsilon\to\mu\mu$.  
Now we subtract the linear background.  The signal mass window is 
defined as (9.2, 10.6) GeV/$c^2$.  We use a side-band 
method:
yield 
= entries in (9.2, 10.6) 
$-$ entries in (7.8, 8.5)
$-$ entries in (11.3, 12) GeV/$c^2$ 
mass windows.
         In $5<\pt<10$ ($10<\pt<20$) GeV/$c$ bin,
            N(muon + probe) = 410 (85), 
            N(muon + Id) = 168 (56), 
            we get efficiency = 41.0\% (65.9\%).
We put everything together to get the CMUP muon 
identification scale factor vs. $\pt$ and perfom 
a fit in $\pt$.  This is shown in Fig.~\ref{fig:MuSF_3}.

\begin{figure}
   \begin{center}
      \parbox{3.9in}{\epsfxsize=\hsize\epsffile{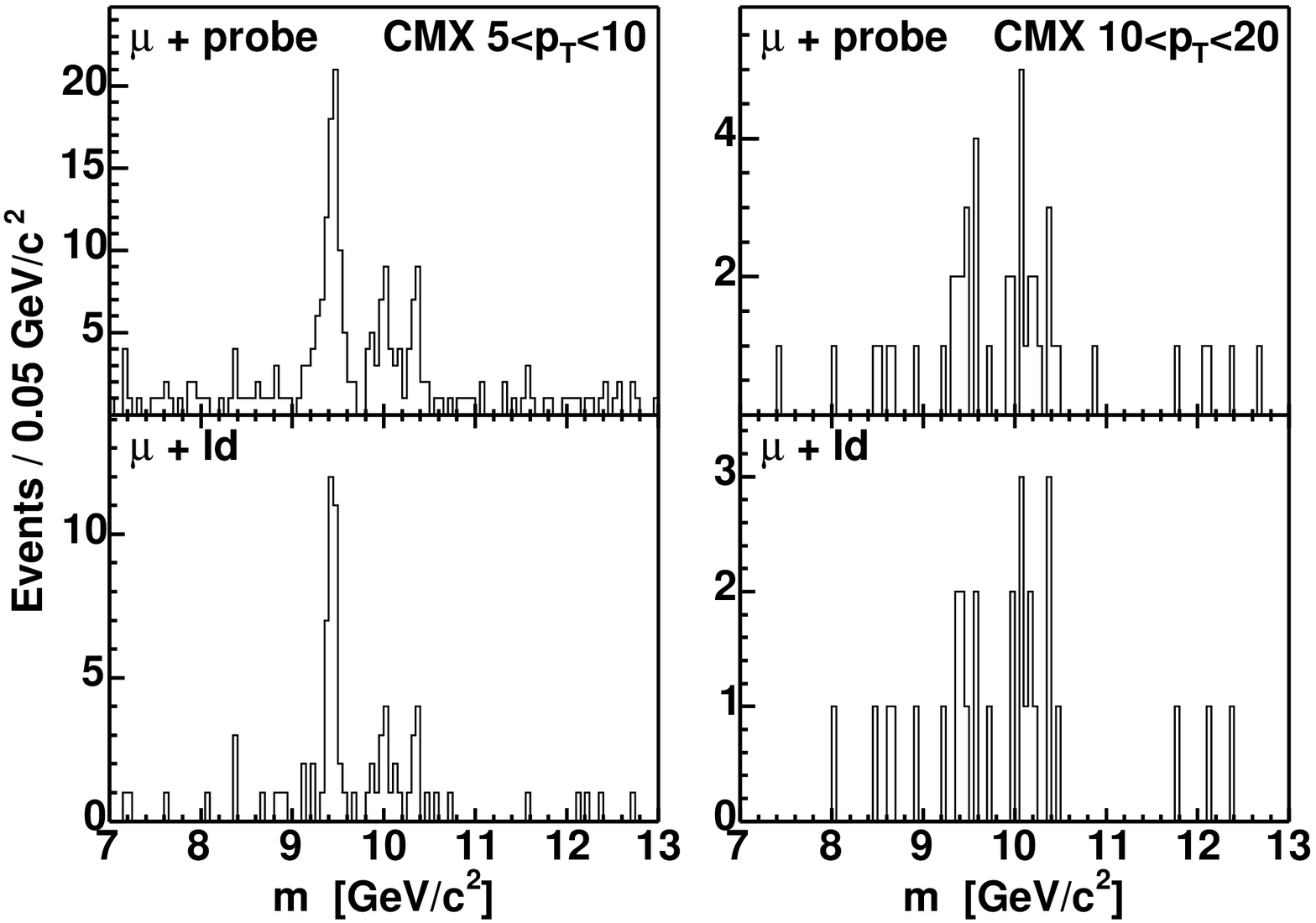}}
      \caption[$\Upsilon\to\mu\mu$ for CMX muon identification efficiency measurement]
              {Distributions of the invariant mass
               of $\Upsilon\to\mu\mu$ for medium $\pt$
               CMX muon identification 
               efficiency measurement in data.
               The three peaks are signature
               of $\Upsilon\to\mu\mu$.
               The left two plots are for CMX
               muons with $5<\pt<10$ GeV/$c$.
               The right two plots are for CMX
               muons with $10<\pt<20$ GeV/$c$.
               Side-band method is used for 
               background subtractions.}
      \label{fig:MuSF_2}
   \vspace{0.5in}
      \parbox{3.9in}{\epsfxsize=\hsize\epsffile{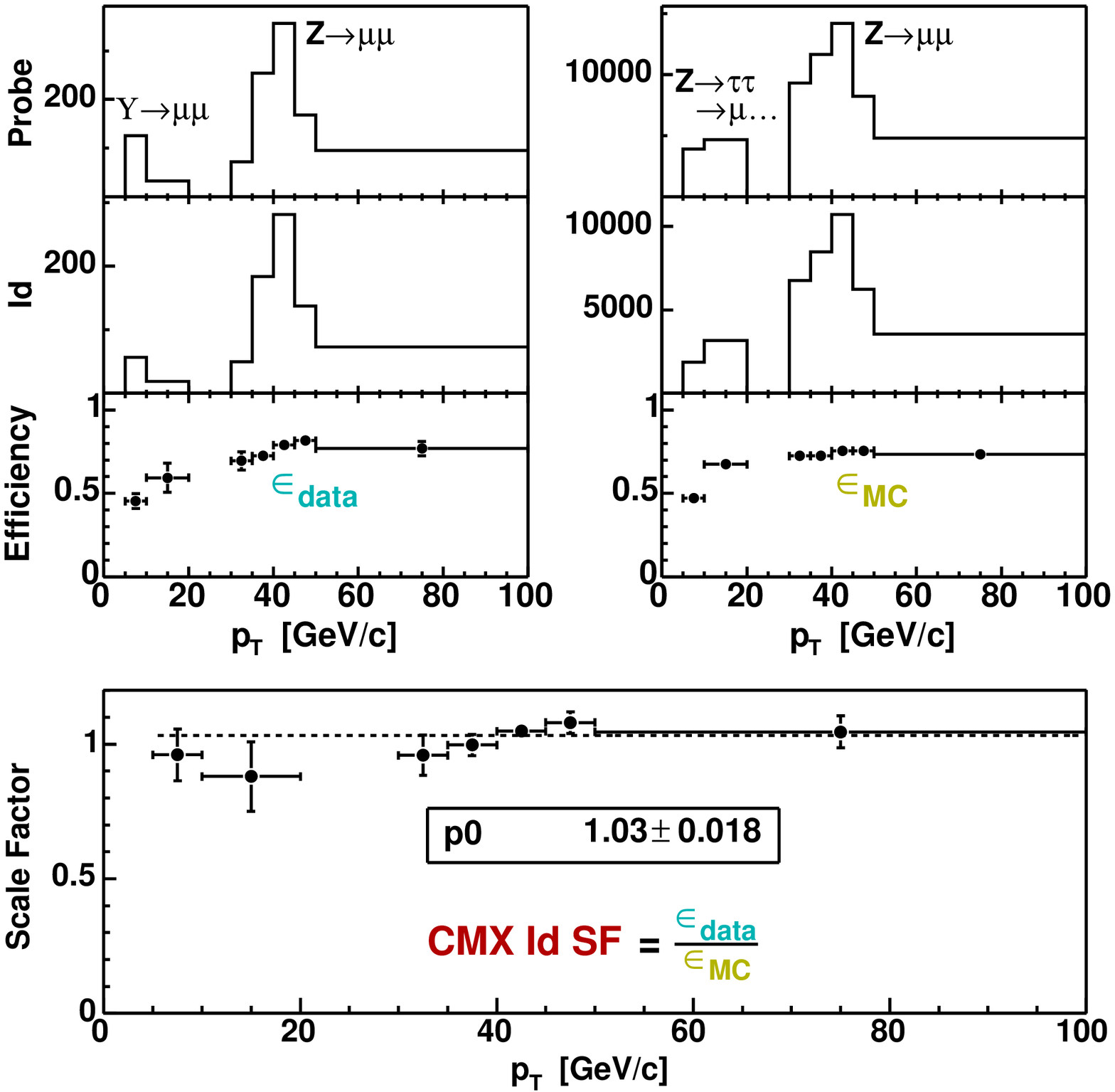}}
      \caption[CMX muon identification scale factor vs. $\pt$]
              {CMX muon identification scale factor vs. $\pt$. 
               This is obtained from dividing the efficiency 
               in data by the efficiency in MC.}
      \label{fig:MuSF_4}
   \end{center}
\end{figure}

Fig.~\ref{fig:MuSF_2} shows the 
mass distribution of muon pair in each $\pt$ bin 
of the second leg, for CMX probe.
         In $5<\pt<10$ ($10<\pt<20$) GeV/$c$ bin,
            N(muon + probe) = 126 (32), 
            N(muon + Id) = 57 (19), 
            we get efficiency = 45.2\% (59.4\%).
Fig.~\ref{fig:MuSF_4} shows
the procedure to get the CMX muon identification
scale factor. 

Analogous to the
electron scale factor study, we assign 
a conservative uncertainty of 4\%.  The resulting
identification scale factors are 
$0.93\pm0.04$ for CMUP muon, and
$1.03\pm0.04$ for CMX muon. 

\vspace{0.2in}
\noindent{\bf Muon Reconstruction Scale Factor}
\vspace{0.1in}

In the reconstruction scale factor study, the probe 
is a stubless muon which may or may not have a stub
in the muon chamber associated with the fiducial track.
It must have passed the vertex $z$ cut, calorimeter 
isolation, EM energy cut, and hadronic energy cut.
For such a probe, we check if it has a stub.
Table~\ref{tab:MuSF_3} shows the
procedure in data and Table~\ref{tab:MuSF_4}
shows the procedure in MC. 

\begin{table}
   \begin{center}
      \begin{tabular}{|l|c|c|c|c|l|} \hline
                        & \multicolumn{2}{c|}{$\Upsilon\to\mu\mu$}
                        & \multicolumn{2}{c|}{$Z\to\mu\mu$}
                        &
           \\ \cline{1-5}
         event          & \multicolumn{4}{c|}{11922805}
                        &
           \\ \cline{1-5}
         goodrun        & \multicolumn{4}{c|}{9103020}
                        &
           \\
         triggered      & \multicolumn{4}{c|}{1881529}
                        &
           \\
         unique         & \multicolumn{4}{c|}{1800059}
                        &
           \\ \cline{1-5}
         process        & \multicolumn{2}{c|}{691}
                        & \multicolumn{2}{c|}{1861}
                        &
           \\ \cline{1-5}
         second leg     & \multicolumn{2}{c|}{760}
                        & \multicolumn{2}{c|}{2583}
                        &
           \\ \cline{2-5}
                        & CMUP & CMX & CMUP & CMX &        \\ \cline{2-5}
                        &  570 & 190 & 1718 & 865 & Probe  \\ \hline \hline 
         has stub       &  474 & 170 & 1569 & 838 & Rec.   \\ \hline
      \end{tabular}
   \end{center}
   \caption[Muon reconstruction efficiency measurement in data]
           {Number of events for muon reconstruction efficiency
            measurement in data.}
   \label{tab:MuSF_3}
\end{table}

\begin{table}
   \begin{center}
      \begin{tabular}{|l|c|c|c|c|l|} \hline
                        & \multicolumn{2}{c|}{$Z\to\tau_{\mu}\tau_x$}
                        & \multicolumn{2}{c|}{$Z\to\mu\mu$}
                        &
           \\ \cline{1-5}
         event          & \multicolumn{2}{c|}{492000}
                        & \multicolumn{2}{c|}{405291}
                        &
           \\ \cline{1-5}
         muon           & \multicolumn{2}{c|}{170596}
                        & \multicolumn{2}{c|}{810582}
                        &
           \\ \cline{1-5}
         match          & \multicolumn{2}{c|}{27109}
                        & \multicolumn{2}{c|}{149411}
                        &
           \\ \cline{2-5}
                        &  CMUP  &  CMX  &  CMUP  &  CMX  &        \\ \cline{2-5}
                        &  17334 &  9775 &  95503 & 53908 & Probe  \\ \hline \hline 
         has stub       &  16672 &  9709 &  93044 & 53827 & Rec.   \\ \hline
      \end{tabular}
   \end{center}
   \caption[Muon reconstruction efficiency measurement in MC]
           {Number of events for muon reconstruction efficiency
            measurement in MC.}
   \label{tab:MuSF_4}
\end{table}

\begin{figure}
   \begin{center}
      \parbox{3.9in}{\epsfxsize=\hsize\epsffile{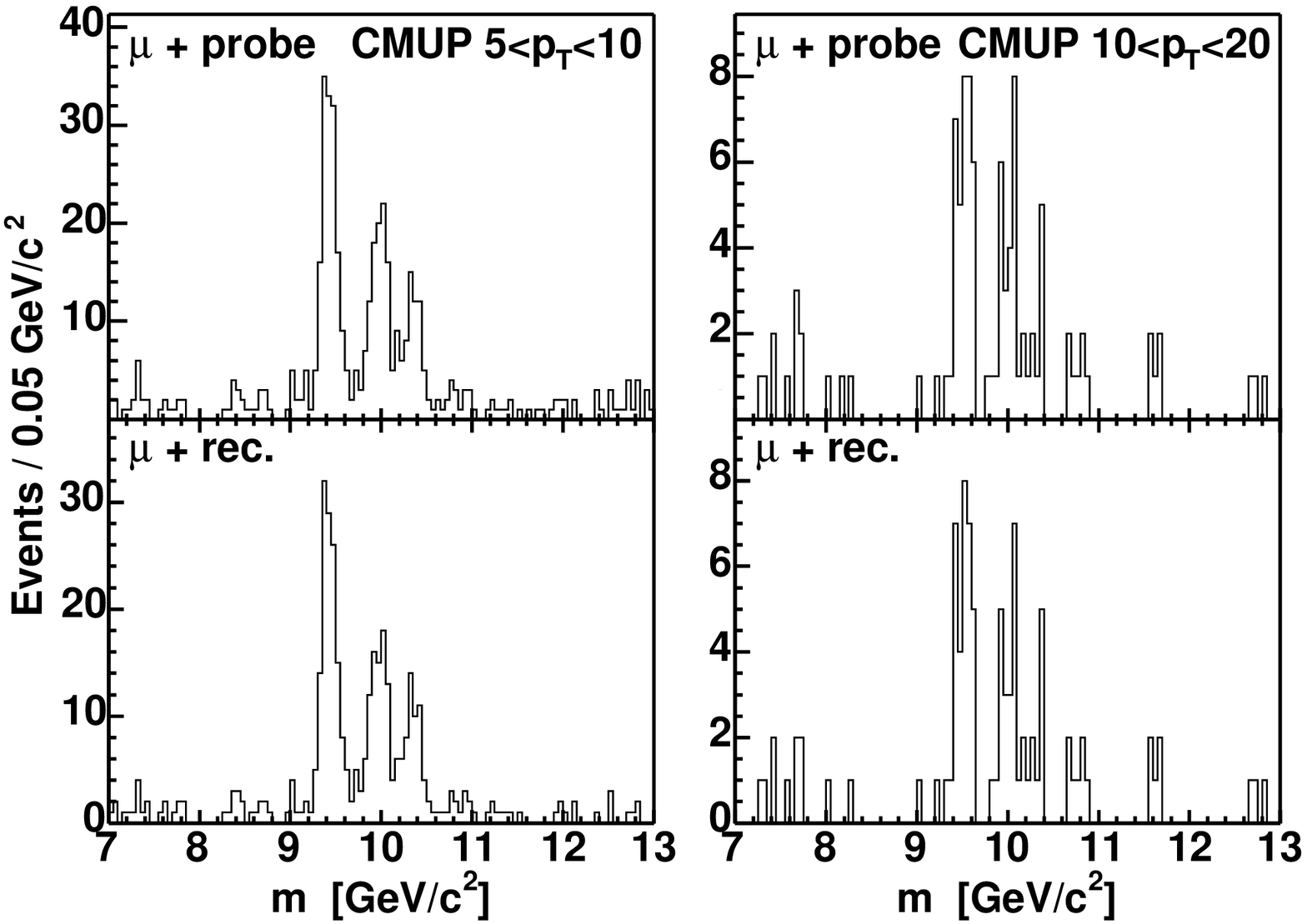}}
      \caption[$\Upsilon\to\mu\mu$ for CMUP muon reconstruction efficiency measurement]
              {Distributions of the invariant mass
               of $\Upsilon\to\mu\mu$ for medium $\pt$
               CMUP muon reconstruction
               efficiency measurement in data.
               The three peaks are signature
               of $\Upsilon\to\mu\mu$.
               The left two plots are for CMUP
               muons with $5<\pt<10$ GeV/$c$.
               The right two plots are for CMUP
               muons with $10<\pt<20$ GeV/$c$.
               Side-band method is used for 
               background subtractions.}
      \label{fig:MuSF_5}
   \vspace{0.5in}
      \parbox{3.9in}{\epsfxsize=\hsize\epsffile{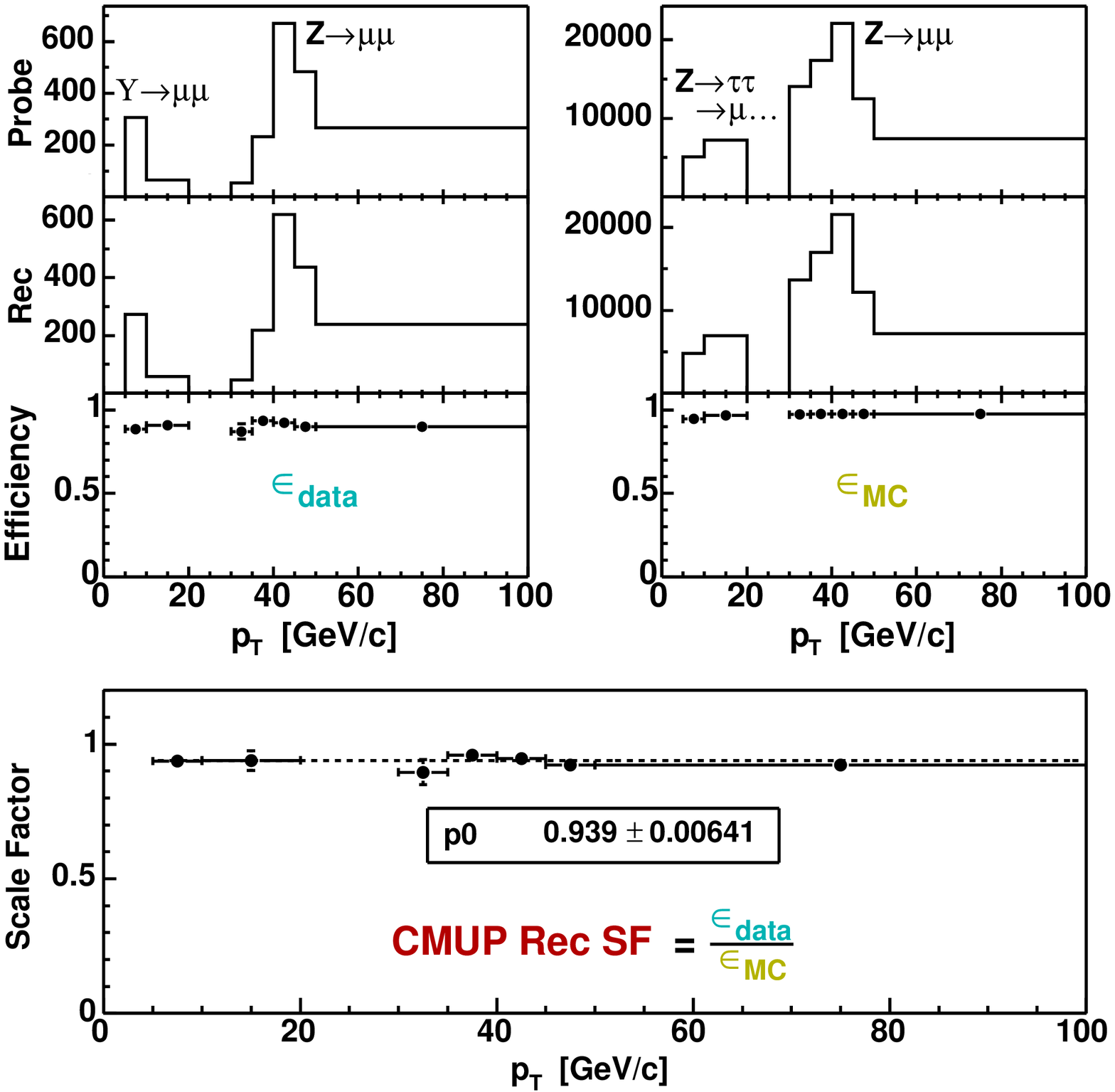}}
      \caption[CMUP muon reconstruction scale factor vs. $\pt$]
              {CMUP muon reconstruction scale factor vs. $\pt$. 
               This is obtained from dividing the efficiency 
               in data by the efficiency in MC.}
      \label{fig:MuSF_7}
   \end{center}
\end{figure}

We break the second leg muon into two 
$\pt$ bins: 5 $<\pt<$ 10 GeV/$c$ and 10 $<\pt<$ 20 GeV/$c$. 
Fig.~\ref{fig:MuSF_5} shows the distributions
of the pair mass of the first leg and the second leg 
in each $\pt$ bin of the second leg, for CMUP probe.
We use the side-band method to do background subtraction. 
         In $5<\pt<10$ ($10<\pt<20$) GeV/$c$ bin, 
            N(muon + probe) = 307 (65), 
            N(muon + stub) = 272 (59), 
            we get efficiency = 88.6\% (90.8\%).
We put everything together to get the muon 
reconstruction scale factor vs. $\pt$ and perform 
a fit in $\pt$.  This is shown in Fig.~\ref{fig:MuSF_7}.

\begin{figure}
   \begin{center}
      \parbox{3.9in}{\epsfxsize=\hsize\epsffile{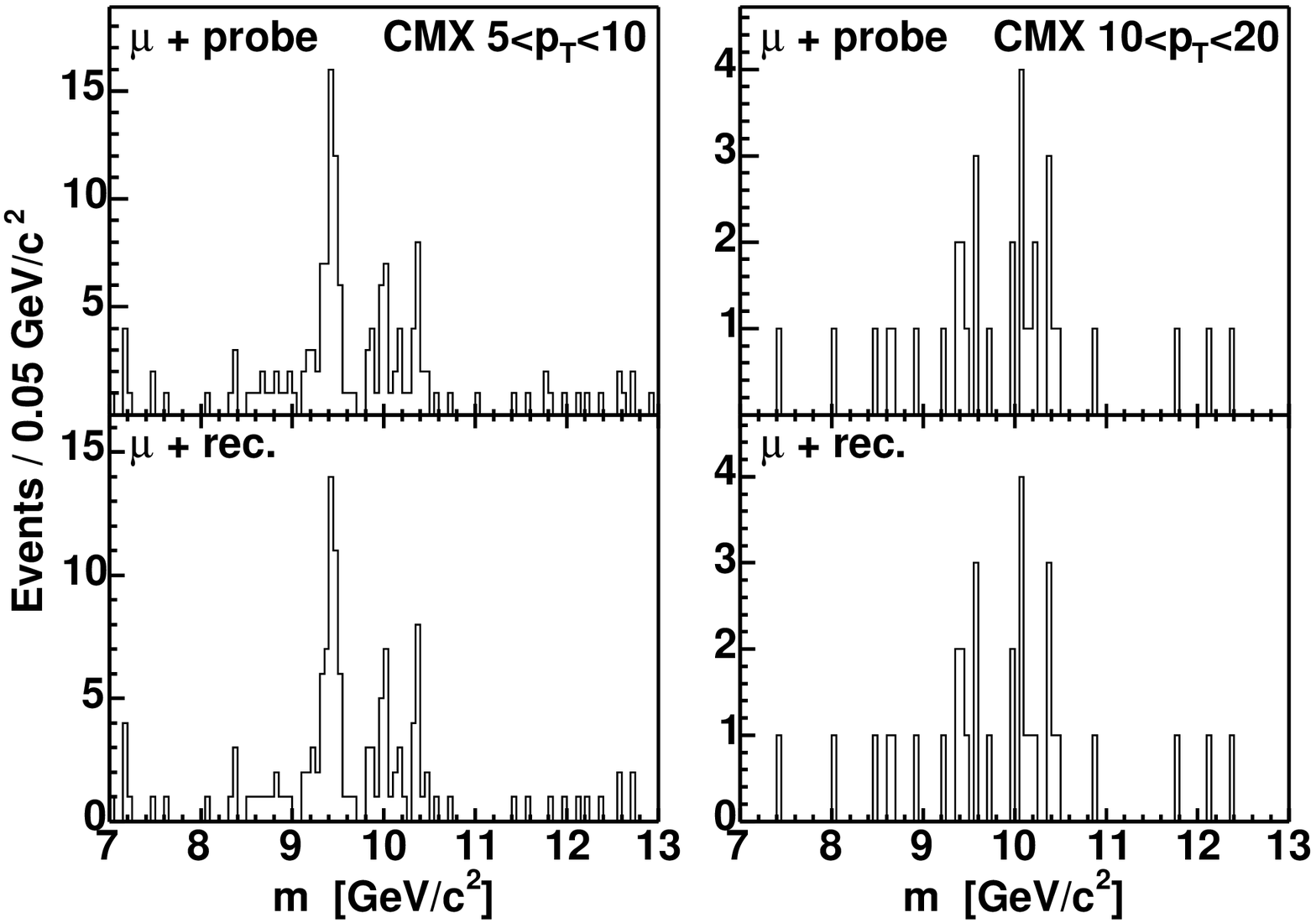}}
      \caption[$\Upsilon\to\mu\mu$ for CMX muon reconstruction efficiency measurement]
              {Distributions of the invariant mass
               of $\Upsilon\to\mu\mu$ for medium $\pt$
               CMX muon reconstruction
               efficiency measurement in data.
               The three peaks are signature
               of $\Upsilon\to\mu\mu$.
               The left two plots are for CMX
               muons with $5<\pt<10$ GeV/$c$.
               The right two plots are for CMX
               muons with $10<\pt<20$ GeV/$c$.
               Side-band method is used for 
               background subtractions.}
      \label{fig:MuSF_6}
   \vspace{0.5in}
      \parbox{3.9in}{\epsfxsize=\hsize\epsffile{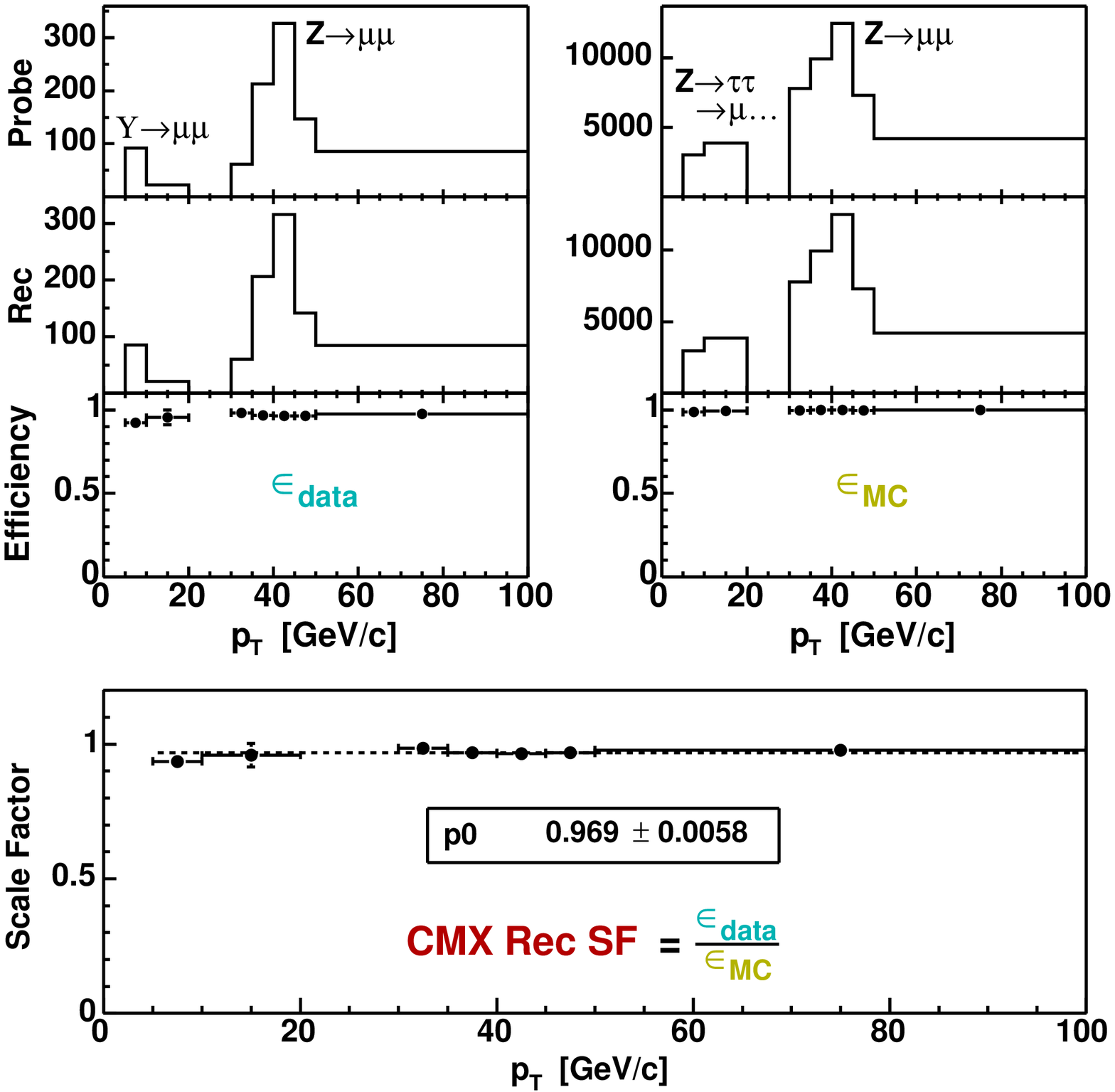}}
      \caption[CMX muon reconstruction scale factor vs. $\pt$]
              {CMX muon reconstruction scale factor vs. $\pt$. 
               This is obtained from dividing the efficiency 
               in data by the efficiency in MC.}
      \label{fig:MuSF_8}
   \end{center}
\end{figure}

Fig.~\ref{fig:MuSF_6} shows the mass distribution of muon pair
in each $\pt$ bin of the second leg, for CMX probe.
In $5<\pt<10$ ($10<\pt<20$) GeV/$c$ bin, 
            N(muon + probe) = 92 (22), 
            N(muon + stub) = 85 (21), 
            we get efficiency = 92.4\% (95.5\%).
Fig.~\ref{fig:MuSF_8} shows the procedure to get
the CMX muon reconstruction scale factor.

As in the
electron scale factor study, we assign 
a conservative systematic uncertainty of 4\%.  The results
of the reconstruction scale factors are 
$0.94\pm0.04$ for CMUP muon, and
$0.97\pm0.04$ for CMX muon. 

We summarize the muon reconstruction and identification 
scale factors with uncertainties:
\begin{eqnarray}
   f^{CMUP \;\; rec}_{data/MC} & = & 0.94\pm0.04 \\ 
   f^{CMUP \;\; Id}_{data/MC}  & = & 0.93\pm0.04 \\ 
   f^{CMX \;\; rec}_{data/MC}  & = & 0.97\pm0.04 \\ 
   f^{CMX \;\; Id}_{data/MC}   & = & 1.03\pm0.04 
\end{eqnarray}


\section{Missing Transverse Energy}
\label{sec:MET}

Weakly interacting particles such as neutrinos of
the Standard Model and the lightest supersymmetric 
particle (LSP) predicted in new physics, deposit 
no energy in the calorimeters.  Minimum 
ionizing particles such as muons leave little 
energy in the calorimeters.  When present these 
cause a significant vector sum of the transverse 
energy of all of the detected particles.  The
imbalance, i.e., the negative of the vector sum 
in the transverse plane corresponds to the missing 
transverse energy ($\met$).

Since $\met$ measures the vector sum of all of the
momentum of particles escaping detection in the calorimeters,  
there is no information on the energy and
direction of an individual particle or how many
particles escaped detection.  With many such
particles in an event there is also a chance
that their transverse momenta cancel each
other.

There is an instrumental source of $\met$
because the calorimeters are not perfect.  There
are crack regions due to the support structures,
and the transition regions between components,
for example from the central calorimeters to the 
plug calorimeters.
The probability that all the energy of a particle
is undetected is rather small.  But QCD processes
have a large production rate.  Some of the jets
can have a lot of energy undetected and make a
significant $\met$.  

In our high-mass tau pair analysis, we will use
an $\met$ cut and several other kinematic cuts 
related to $\met$.  To get the uncertainty 
due to the instrumental $\met$, we should get
the distributions in data and MC, and compare
the same variable.

In the real data, the physics processes
      $Z\to ee$ and $\gamma$ + jet, which have
      zero true missing energy, can be 
      used to study the effect of the 
      instrumental $\met$.  The latter is a 
      better choice for our purpose because 
      hadronic taus in the calorimeters are more 
      like jets than electrons.  The inclusive
      photon sample is used to select 
      $\gamma$ + jet events.  Jets are required 
      to be reconstructed as hadronic tau 
      objects.  The true $\met$ in this sample 
      should be zero.  The reconstructed 
      $\met$ corresponds to the instrumental 
      $\met$ in data.

The simulation uses $Z\to\tau_e\tau_h$ process and 
      requires a tight electron and a hadronic 
      tau object.  The difference between the 
      reconstructed $\met$ in the simulation minus
      the $\met$ from neutrinos corresponds to 
      the instrumental $\met$ in MC.  

\begin{figure}
   \begin{center}
      \parbox{3.0in}{\epsfxsize=\hsize\epsffile{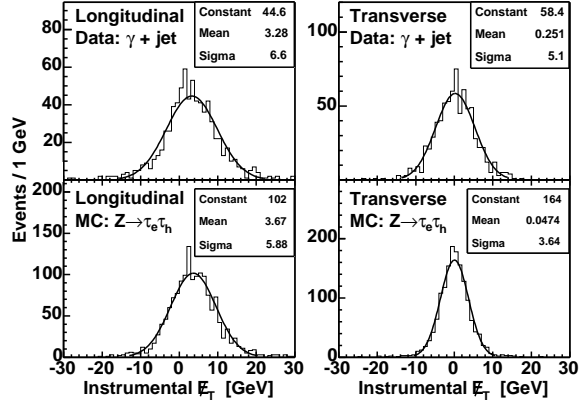}}
      \caption[Instrumental $\met$ in data and MC]
              {Distributions of the instrumental $\met$ in data 
               using $\gamma$+jet sample and MC using 
               $Z\to\tau_e\tau_h$ sample. 
               Instrumental $\met$ is projected to
               the direction of the reconstructed 
               leading tau object to get the longitudinal
               and transverse components.}
      \label{fig:MET}
   \end{center}
\end{figure}

Then the instrumental $\met$ is projected to the 
direction of hadronic tau object, shown in 
Fig.~\ref{fig:MET}.  The distributions in data
and MC are different for the longitudinal 
component and the transverse component, 
respectively.

To get the uncertainty due to the instrumental 
$\met$, we ``smear'' the longitudinal component and 
the transverse component of the instrumental 
$\met$ in MC according to their differences 
between data and MC, then add neutrinos back 
to get the smeared~$\met$.  

Now we can calculate the uncertainty of the cuts 
related to $\met$ by the effect with/without 
smearing the instrumental $\met$.  
Table~\ref{tab:MET} shows the effect in 
$\tau_e\tau_h$ channel of $\zprime(m=300\mbox{ GeV/}c^2)$ sample 
and $\zprime(m=600\mbox{ GeV/}c^2)$ sample.  The uncertainties 
are
(1191-1125)/1125 = 5.9\% in $\zprime(300)$ 
sample and
(1875-1814)/1814 = 3.4\% in $\zprime(600)$
sample.  Taking the larger value, we find that
the uncertainty in acceptance due~to~$\met \approx 6\%$.

\begin{table}
   \begin{center}
      \begin{tabular}{|l|c|c|c|c|} \hline
         sample & \multicolumn{2}{c|}{$\zprime(m=300\mbox{ GeV/}c^2)$}
                & \multicolumn{2}{c|}{$\zprime(m=600\mbox{ GeV/}c^2)$} \\ \hline
         $\tau\tau$ event  
                & \multicolumn{2}{c|}{100000}
                & \multicolumn{2}{c|}{100000} \\ 
         $\tau_e\tau_h$ decay mode   
                & \multicolumn{2}{c|}{23246}
                & \multicolumn{2}{c|}{23250} \\
         $\et^e>$ 10 GeV, $\pt^{\tau}>25$ GeV/$c$
                & \multicolumn{2}{c|}{2135}
                & \multicolumn{2}{c|}{3044} \\ \hline
         smear instrumental $\met$
                & $\;\;\;\;$  no $\;\;\;\;$
                & $\;\;\;\;$ yes $\;\;\;\;$
                & $\;\;\;\;$  no $\;\;\;\;$
                & $\;\;\;\;$ yes $\;\;\;\;$ \\ \hline
         $\met>$ 15 GeV                    & 1720 & 1801  &  2745 & 2829 \\
         $\Delta\phi(e - \met)<30^{\circ}$ & 1231 & 1299  &  1844 & 1907 \\
         $m_{vis}>$ 120 GeV/$c^2$          & 1125 & 1191  &  1814 & 1875 \\ \hline
         uncertainty  
                & \multicolumn{2}{c|}{5.9\%}
                & \multicolumn{2}{c|}{3.4\%} \\ \hline
      \end{tabular}
      \caption[The uncertainty in acceptance due to instrumental $\met$]
              {Number of $Z'\to\tau\tau$ events to study 
               the uncertainty in acceptance due to 
               the imperfect modeling of the instrumental 
               $\met$ in MC simulation. The uncertainty
               is obtained from the effect of with/without
               ``smearing'' the instrumental $\met$ in
               MC to that in data.}
      \label{tab:MET}
   \end{center}
\end{table}



\chapter{Event Kinematic Selection}
\label{cha:event}

In this chapter we first discuss the trigger paths.
Second, we discuss the good run selections and the 
integrated luminosities.  
Third, in addition to the particle identification, 
we add event kinematic cuts to further suppress 
backgrounds.  Since the kinematic cuts need to keep 
high efficiency for the signals, optimization on 
the event kinematic cuts is necessary.
Fourth, there are thresholds in the triggers.  The 
trigger primitives are not exactly the same as the 
offline variables we cut on, and so we need to evaluate
the marginal trigger efficiencies for selected events.


\section{Trigger Path}
\label{sec:event_trigger}

For the $\tau_e\tau_h$ selection, we use  
      the ``electron plus track'' trigger 
      called TAU\_ELE.  It requires an electron in 
      the CEM detector with $\et>8$ GeV, $\pt>8$ GeV/$c$ 
      and an isolated track with $\pt>5$ GeV/$c$. 

For the $\tau_{\mu}\tau_h$ selection, there are two 
      ``muon plus track'' triggers called TAU\_CMU (TAU\_CMX) 
      which requires a CMUP (CMX) muon with
      $\pt>8$~GeV/$c$ and an isolated track with 
      $\pt>5$~GeV/$c$.

For the $\tau_h\tau_h$ selection, we use 
      the ``$\met$ plus tau'' trigger
      called TAU\_MET.  It requires L1 $\met>25$ 
      GeV and an L3 tau object with $\et>20$ GeV, 
      track isolation and 
      $m$(tracks)~$<2.0$~GeV/$c^2$.

The TAU\_ELE, TAU\_CMU, and TAU\_CMX triggers 
are cleaned up by requiring an isolated track.  
The TAU\_MET trigger requires only one 
isolated tau object thus the other tau objects
in this trigger are not necessarily isolated.

The track isolation requirement in these triggers
is that there is no additional track in a
10$^o$ to 30$^o$ annulus.  This track isolation is 
looser than the offline tau track isolation with a 
shrinking inner cone.
The detailed descriptions of the tau triggers 
can be found in Ref.~\cite{Anastassov:2003vc}.

In addition to selecting the candidate events,
there is also an important issue regarding
the jet$\to\tau$ misidentification background.  
This fake background is not negligible because 
of the large production rate of jets.  
Using MC simulation to model all the processes 
of the fake background is not adequate.  
We estimate the contribution of these events 
directly from real data.  

For the purpose of estimating jet$\to\tau$ 
misidentification background, it is better to 
use those triggers without the isolation 
requirement in order to have a sample which 
has a larger statistics and is dominated by 
jet background.

There is an ELECTRON\_CENTRAL\_8 (abbreviated 
as CELE8) trigger which has the same 
requirement as TAU\_ELE but without the track 
isolation requirement.  There is also a 
MUON\_CMUP8 (abbreviated as CMUP8) trigger 
which has the same requirement as TAU\_CMU
but without the track isolation requirement.
The CELE8 and the CMUP8 triggers are
dynamically prescaled.  A prescale is imposed 
to reduce the rate of a trigger.  A fixed 
prescale under-utilizes the trigger bandwidth 
when the luminosity falls during a run.  A 
dynamic prescale is based on the availability 
of the trigger bandwidth, and automatically 
reduce the prescales as the luminosity falls.

There is not a corresponding trigger path
available for the TAU\_CMX trigger.  There 
is a prescaled trigger available for the 
TAU\_MET trigger but its prescale is 100 
which is too big.  Thus their jet$\to\tau$ 
fake background estimates have to be done 
with the trigger itself.


\section{Good Run Selection and Integrated Luminosity}
\label{sec:event_GTU}

We use the data samples collected in CDF from 
March 2002 to September 2003 for this analysis.  
The Good Run List~\cite{CDFnote:5613} used in 
this analysis is in the range of the run number 
141544$-$168889.  

We use the online initial filtering and the 
offline periodic classification to decide 
whether a run is good or bad.  The former gets 
rid of obviously bad runs where there are 
problems with the sub-detectors or the triggers.  
The latter is based on the classification using 
a large sample in a run, for example, of the 
$J/\Psi\to ee, \; \mu\mu$ events which is 
expected to have a very narrow peak,  
or the photon plus jet events
which is expected to have very good 
energy balancing, etc.

The status of a trigger or a sub-detector is a 
single bit 1 or 0, which means good or bad.
The bit 1 or 0 of a trigger is based on whether
the deadtime is less than 5\% and is set by the 
online run control shift crew.  The bit 1 or 0 
of a sub-detector at the online stage is based 
on the status of the high voltage, the calibration, 
the occupancy, etc. and is set by the monitoring 
operator.  The bit 1 or 0 of a sub-detector at the 
offline stage is based on, for example, the 
reconctructed $J/\Psi\to ee, \; \mu\mu$ mass which
can tell possible problems in the tracking system, 
the calorimeters or the muon chambers, and is set 
by the physics groups.

Here are the details of the requirements on a 
good run.  
There are several run configurations (trigger 
tables) when the CDF detector is taking data: 
test, calibration, cosmic, and physics.  A 
good run must be a physics run. 
At the online stage the losses of the beam should 
be low.  The ``on-tape'' luminosity should be 
greater than 10 nb$^{-1}$.  The bits of 
the L1, L2, L3 triggers, 
the calorimeters, the CMU detector, the CES 
detector should be 1.
At the offline stage the bits of the calorimeters, 
the COT detector, the CMU and CMP detectors 
should be 1.
The runs after 150145 when the CMX trigger
updated L1 hardware, in addition to the bits
above, are required to have the online and 
offline bits of the CMX detector set to 1.

The total integrated luminosity in the included 
good runs in the run number range 141544$-$168889 
is 195~pb$^{-1}$.  However, the good run of the 
data sample from the TAU\_CMX trigger starts from 
150145 and its integrated luminosity is 179~pb$^{-1}$;
the good run number of the data sample from the TAU\_MET 
trigger stops at 156487 and its integrated luminosity 
is 72~pb$^{-1}$.  The TAU\_MET trigger was changed 
after run 156487 to include L2 two-dimensional track 
isolation which needs further study.  The uncertainty 
in the luminosity measurements is about 
6\%~\cite{Klimenko:2003if}.

The integrated luminosity in the data sample
from the CELE8 trigger, which is 
dynamically prescaled, is 46~pb$^{-1}$.  It 
is calculated by adding the isolated track 
requirement and comparing its survived number 
of events with the total number of events in 
the data sample from the TAU\_ELE trigger 
whose luminosity is known.  Analogously, the 
integrated luminosity in the data sample from 
the CMUP8 trigger is found to be 38~pb$^{-1}$.

There were duplicate events incorrectly processed 
and put in the data samples that were later 
reprocessed.  We reprocessed all of the events.
To avoid the duplicate events, we pick one of 
them and require that it be a unique event.


\section{Selection Criteria}
\label{sec:event_cuts}

The event kinematic cuts are designed to further 
suppress background while to keep high signal
efficiency.  
Table~\ref{tab:event_cuts} shows the list of cuts 
for event selection.
We note several features of the requirements:
\begin{itemize}
\item The $\pt^{\tau}$ threshold is 25 GeV/$c$ because 
      tau identification is fully efficient at about
      25 GeV/$c$ and it is a high threshold to reduce
      background.

\item The $\et^e$, $\pt^e$, and $\pt^{\mu}$ thresholds 
      are 10 GeV, 10 GeV/$c$ and 10 GeV/$c$, respectively. 
      (The thresholds in the corresponding triggers are 
      8 GeV, 8 GeV/$c$ and 8 GeV/$c$.) For $\tau_h\tau_h$, 
      we require the second tau $\pt^{\tau_2}>10$ GeV/$c$.

\item The $\met$ cut and the angle cut 
      $\Delta\phi(l-\met)<30^o$ are designed to 
      remove hadronic jet backgrounds.  They are 
      explained below.

\item We use $m_{vis}>$ 120 GeV/$c^2$ cut to remove 
      the ``irreducible'' $Z/\gamma^*\to\tau\tau$ 
      background.  The low mass region with 
      $m_{vis}<$ 120 GeV/$c^2$ is our control region.

\item For the $\tau_{\mu}\tau_h$ selection, we have
      a cosmic veto~\cite{CDFnote:6089}.  

\item For the $\tau_h\tau_h$ selection, we require 
      the second tau has exactly one track to further
      clean up QCD backgrounds.  We will check tau
      signature by track multilicity on the leading
      tau side.
\end{itemize}
 
\begin{table}
   \begin{center}
      \begin{tabular}{|c|c|c|} \hline
         $\tau_e\tau_h$            & $\tau_{\mu}\tau_h$          & $\tau_h\tau_h$                 \\ \hline \hline
         $\pt^{\tau}>25$           & $\pt^{\tau}>25$             & $\pt^{\tau_1}>25$              \\
         $\et^e>10$, $\pt^e>10$    & $\pt^{\mu}>10$              & $\pt^{\tau_2}>10$              \\
         $\met>15$                 & $\met>15$                   & $\met>25$                      \\
         $\Delta\phi(e-\met)<30^o$ & $\Delta\phi(\mu-\met)<30^o$ & $\Delta\phi(\tau_2-\met)<30^o$ \\
         $m(e+\tau+\met)>120$      & $m(\mu+\tau+\met)>120$      & $m(\tau_1+\tau_2+\met)>120$    \\
                                   & cosmic veto                 &  $\tau_2$ num. track == 1      \\ \hline
      \end{tabular}
      \caption[Event kinematic cuts]
              {Event kinematic cuts.}
      \label{tab:event_cuts}
   \end{center}
\end{table}
     
The $\met$ measured in $\tau_{\mu}\tau_h$ channel
needs a muon correction since there is an effect 
of missing energy due to the fact that muons are 
minimum ionizing partilces.
The procedure 
of the muon correction is: first, we subtract the 
$\pt$ of a tight muon; second, we add muon energy 
deposits in the calorimeters to avoid counting the 
same energy twice.  

We require $\met>15$ GeV for the $\tau_e\tau_h$ 
and $\tau_{\mu}\tau_h$ selections.  For the 
$\tau_h\tau_h$ selection, we use data from the 
TAU\_MET trigger and we require $\met>25$ GeV to 
match the 25 GeV $\met$ trigger threshold.  We 
could suppress more backgrounds by requiring more 
significant $\met$.  However, for the signal 
processes, since there is at least one neutrino 
at each side, there is a chance that the transverse 
momentum of the neutrinos cancel each other, and 
hence raising $\met$ thresholds can reduce signal 
efficiency.  We found those $\met$ thresholds are 
at optimzed points.

The $\Delta\phi<30^o$ cut requires that the
significant $\met$ should follow the $e$ ($\mu$)
for the $\tau_e\tau_h$ ($\tau_{\mu}\tau_h$) 
channels and follow the lower $\pt$ tau object 
for the $\tau_h\tau_h$ channel.  The $\met$ 
measured is the vector sum of the neutrinos from 
the decays of the two taus.  Here is the example with the 
$\tau_e\tau_h$ channel which has one neutrino 
associated with $\tau_h$ and two neutrinos 
associated with $\tau_e$.  Thus this event 
topology cut is able to get the most of the 
signals, and to strongly suppress the backgrounds,
especially the jet$\to\tau$ misidentified fake 
backgrounds which mostly has a $\Delta\phi$ 
with a random topology.


\section{Marginal Efficiency Correction}
\label{sec:event_trigEff}

We need to include in our estimates of the signal 
and background rates the effect of the triggers.  
We are concerned, however, only with the effect 
of the triggers on those events passing the offline
cuts: the marginal efficiency.  

\begin{figure}
   \begin{center}
      \parbox{5.5in}{\epsfxsize=\hsize\epsffile{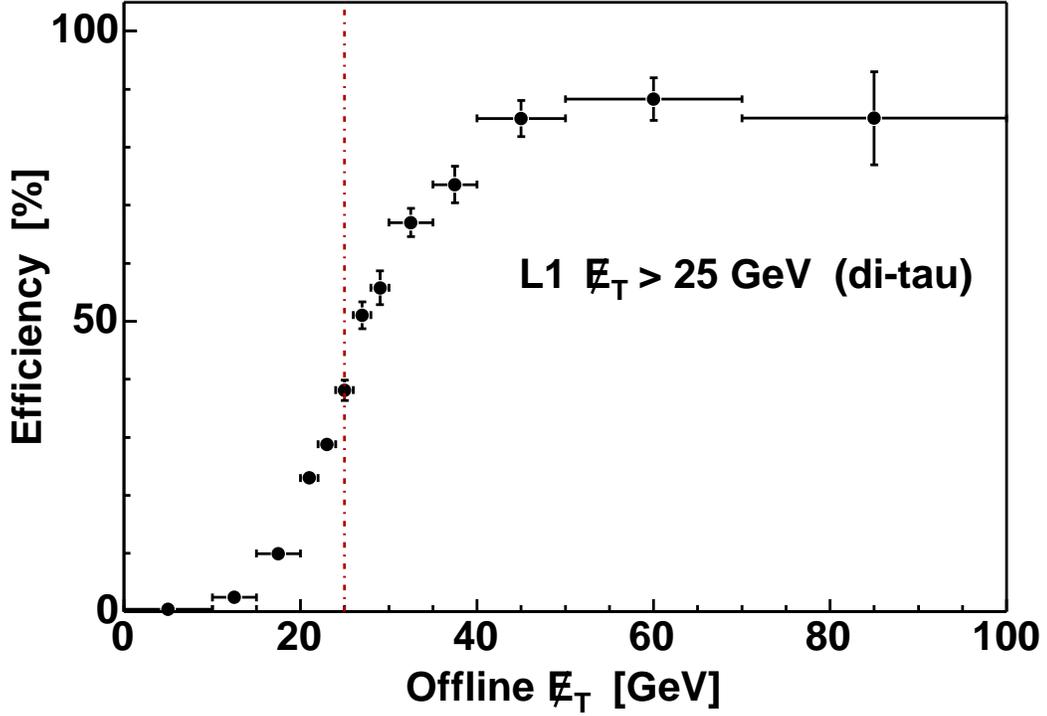}}
      \caption[L1 $\met>$ 25 GeV trigger efficiency vs. offline $\met$ for di-tau event]
              {L1 $\met>$ 25 GeV trigger efficiency vs. offline $\met$ for di-tau event.}
      \label{fig:event_trigEff}
   \end{center}
\end{figure}

The TAU\_MET trigger for the $\tau_h\tau_h$ 
analysis triggers directly on tau object, thus
there is no marginal trigger efficiency from 
the TAU side.  But there is a marginal trigger 
efficiency from the MET side which is based on a 
1 GeV tower threshold for a fast calculation
at L1 while the offline $\met$ is based on a 
0.1 GeV tower threshold.  We use the JET20 data 
sample and mimic the $\tau_h\tau_h$ event 
topology in the calorimeter by requiring one 
central jet with $\et>25$ GeV and at least 
another one central jet with $\et>10$ GeV.  
The L1 $\met>$ 25 GeV trigger efficiency 
vs. offline $\met$ for di-tau event is shown 
in Fig.~\ref{fig:event_trigEff}. 
The marginal trigger efficiency of the TAU\_MET 
trigger for the $\tau_h\tau_h$ analysis is a slow 
turn-on due to the large trigger tower threshold. 

The marginal efficiencies of the TAU\_ELE and
TAU\_CMU (TAU\_CMX) triggers for the $\tau_e\tau_h$ 
and $\tau_{\mu}\tau_h$ analyses are all at plateau,
\begin{eqnarray}
   \epsilon\mbox{(TAU\_ELE)} & = & 0.92\pm0.03 \\
   \epsilon\mbox{(TAU\_CMU)} & = & 0.85\pm0.03 \\
   \epsilon\mbox{(TAU\_CMX)} & = & 0.92\pm0.03         
\end{eqnarray}
The trigger efficiencies of the electron part,
the muon part and the isolated track part
are calculated by using conversion electrons 
from $\gamma\to ee$, muons from $\Upsilon/Z\to\mu\mu$,
and tracks from jet samples, respectively.
The details can be found in Ref.~\cite{CDFnote:6257}.
The biggest uncertainty is from the track provided 
by the XFT trigger, which uses the four axial 
$r-\phi$ superlayers (no stereo $r-z$ superlayers) 
of the COT detector with at least 10 hits (out of 
total 12 hits) in each axial superlayer.  In the
event reconstruction, we require at least 3 axial 
superlayers with at least 7 hits in each axial 
superlayer, and the same configuration for the 
stereo superlayers.  The marginal XFT track 
finding trigger efficiency is found to be a 
function of $\pt$, $\eta$, the number of prongs, 
and the different run ranges.  The overall 
uncertainty is about 3\%.



\chapter{Low Mass Control Region}
\label{cha:control}

The low-mass region with $m_{vis}<120$ GeV/$c^2$ is
used as the control region to test the event cuts
and background determination.  
If we find that the observed and predicted event
rates agree in the control region, we can proceed
to unblind the signal region.

The main source of events in the control region 
is from $Z/\gamma^*\to\tau\tau$.
The other backgrounds include $Z/\gamma^*\to ee$, 
$Z/\gamma^*\to\mu\mu$ and jet$\to\tau$ misidentified fake 
background.  Top background $t\bar{t}$ and 
di-boson backgrounds such as $WW$ and $WZ$ are 
negligible because their cross sections are two 
orders of magnitude smaller than Drell-Yan 
backgrounds and their event topology is the 
opposite of the requirement that a significant
$\met$ follows the lepton direction.  The 
jet$\to\tau$ misidentified fake background is not 
negligible because the dijet production cross section
is large.

For the jet$\to\tau$ misidentified fake background, 
rather than trying to model all the processes that 
could produce fake events, we estimate the 
contribution of these events from real data which 
includes any process contributing to the fake 
background.


\section{Drell-Yan Cross Section}
\label{sec:control_sigmaDY}

\begin{figure}
   \begin{center}
      \parbox{5.7in}{\epsfxsize=\hsize\epsffile{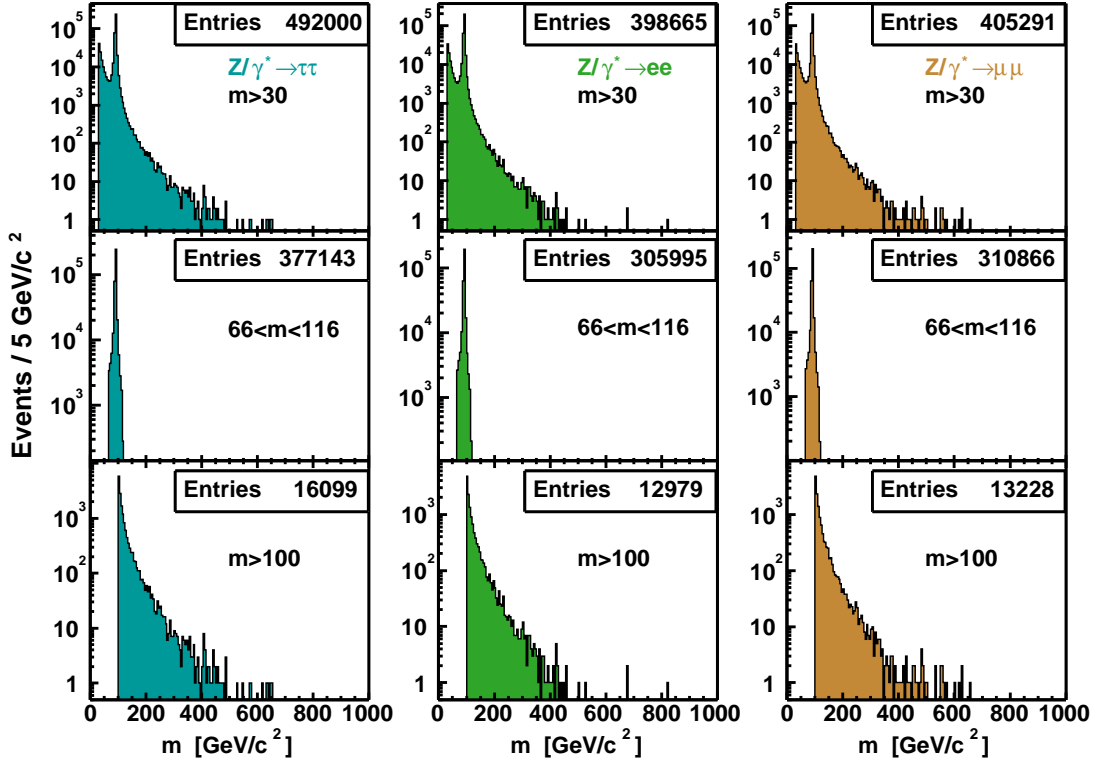}}
      \caption[Drell-Yan mass spectra in different mass regions]
              {Drell-Yan mass spectra in different mass regions.}
      \label{fig:control_sigmaDY}
   \end{center}
\end{figure}

The cross section times branching ratio of the
Drell-Yan processes in the mass window $66<m<116$ 
GeV/$c^2$ at $\sqrt{s}=1.96$ TeV is about 250 pb~\cite{Acosta:2004uq}.
Fig.~\ref{fig:control_sigmaDY} shows the mass 
spectrum and event counts in different 
mass regions. 
The $Z/\gamma^*\to\tau\tau$ 
sample has 377143 events in the mass window 
$66<m<116$~GeV/$c^2$ which corresponds to 
a 250~pb production cross section.  
The number of events in a mass 
window is proportional to the cross
section in that mass window.  For example,
the number of events 492000 in the mass window 
$m>30$~GeV/$c^2$ gives a cross section 
$250\times492000/377143\approx326$~pb.
By the same algebra, we get the cross 
sections in different mass windows:
\begin{eqnarray}
   \sigma\cdot B(Z/\gamma^*\to l^+l^-)_{66-116}      & \approx & 250 \mbox{ pb} \\
   \sigma\cdot B(Z/\gamma^*\to l^+l^-)_{>30\ \ \ \,} & \approx & 326 \mbox{ pb} \\
   \sigma\cdot B(Z/\gamma^*\to l^+l^-)_{30-100}      & \approx & 315 \mbox{ pb} \\
   \sigma\cdot B(Z/\gamma^*\to l^+l^-)_{>100\ \ \,}  & \approx & \ \, 11 \mbox{ pb} 
\end{eqnarray}

\section{Drell-Yan Background}
\label{sec:control_DY}

The Drell-Yan backgrounds can be estimated from 
MC simulation with three pieces:
\begin{equation}
  \mbox{Expected MC background} 
   =        \mbox{luminosity} 
     \times (\sigma\cdot B) 
     \times \mbox{acceptance}
\end{equation}
We just discussed the production cross section, and 
we have discussed the luminosity 
for each trigger path in Section~\ref{sec:event_GTU}.
Now we discuss the acceptance and the estimate of 
the Drell-Yan backgrounds.
Table~\ref{tab:control_DY} shows the Drell-Yan 
background acceptances,  the application of
the trigger efficiencies, the application of
the lepton data/MC scale factors, and the 
normalization to the integrated luminosities
of the data samples from the triggers.

\begin{table}
   \begin{center}
      \begin{tabular}{|l|r|r|r|} \hline
              source                               & $Z/\gamma^*\to\tau\tau$
                                                   & $Z/\gamma^*\to ee$
                                                   & $Z/\gamma^*\to\mu\mu$       \\
              mass window                          &  $m>30$ &  $m>30$ &  $m>30$ \\
              $\sigma\cdot B$ (pb)                 &     326 &     326 &     326 \\ 
              event                                &  492000 &  398665 &  405291 \\ \hline
         $\tau_e\tau_h$ (TAU\_ELE)                 &         &         &         \\
              $\tau(25)+e(10)$                     &    1528 &     272 &       1 \\
              $\met>15$                            &     514 &      29 &       1 \\
              $\Delta\phi(e-\met)<30^{\circ}$      &     415 &       2 &       0 \\
              $m_{vis}<120$                        &     405 &       1 &       0 \\ 
              trigger efficiency                   &  373.07 &    0.92 &    0.00 \\
              lepton scale factors                 &  351.03 &    0.87 &    0.00 \\
              normalized (195 pb$^{-1}$)           &   45.36 &    0.14 &    0.00 \\ \hline
         $\tau_{\mu}\tau_h$ (TAU\_CMU)             &         &         &         \\
              $\tau(25)+\mbox{CMUP }\mu(10)$       &     836 &       0 &     415 \\
              cosmic veto                          &     836 &       0 &     415 \\
              $\met>15$                            &     294 &       0 &     351 \\
              $\Delta\phi(\mu-\met)<30^{\circ}$    &     253 &       0 &       7 \\
              $m_{vis}<120$                        &     248 &       0 &       4 \\ 
              trigger efficiency                   &  226.06 &    0.00 &    3.65 \\
              lepton scale factors                 &  191.69 &    0.00 &    3.09 \\
              normalized (195 pb$^{-1}$)           &   24.77 &    0.00 &    0.48 \\ \hline
         $\tau_{\mu}\tau_h$ (TAU\_CMX)             &         &         &         \\
              $\tau(25)+\mbox{CMX }\mu(10)$        &     425 &       0 &     219 \\
              cosmic veto                          &     425 &       0 &     219 \\
              $\met>15$                            &     150 &       0 &     181 \\
              $\Delta\phi(\mu-\met)<30^{\circ}$    &     134 &       0 &       1 \\
              $m_{vis}<120$                        &     130 &       0 &       0 \\ 
              trigger efficiency                   &  118.50 &    0.00 &    0.00 \\
              lepton scale factors                 &  114.84 &    0.00 &    0.00 \\
              normalized (179 pb$^{-1}$)           &   13.62 &    0.00 &    0.00 \\ \hline
         $\tau_h\tau_h$ (TAU\_MET)                 &         &         &         \\
              $\tau_1(25)+\tau_2(10)$              &    4264 &       1 &       9 \\
              $\met>25$                            &     295 &       0 &       0 \\
              $\Delta\phi(\tau_2-\met)<30^{\circ}$ &     240 &       0 &       0 \\
              $\tau_2$ num. track == 1             &     185 &       0 &       0 \\
              $m_{vis}<120$                        &     169 &       0 &       0 \\ 
              trigger efficiency                   &   93.39 &    0.00 &    0.00 \\
              lepton scale factors                 &   87.87 &    0.00 &    0.00 \\
              normalized (72 pb$^{-1}$)            &    4.19 &    0.00 &    0.00 \\ \hline  
      \end{tabular}
      \caption[Drell-Yan background estimates in the control region]
              {Drell-Yan background estimates for each channel in
               the low mass control region.}
      \label{tab:control_DY}
   \end{center}
\end{table}


\section{Fake Background}
\label{sec:control_fake}

In a ``fake'' background event a jet is misidentified 
as a tau.  This background is not negligible 
because the dijet production cross section is large.
The relative jet$\to\tau$ misidentification rate and 
the relative tau identification efficiency corresponding 
to the denominator chosen is applied to the denominator 
tau objects to compute their weight for being a jet.
We sum up the weights of all the events to get 
jet$\to\tau$ misidentified fake background  
estimate in the sample, as described in 
Section~\ref{subsec:TauId_jetBg}.

There is also a probability that, for example, 
for $\tau_e\tau_h$ channel, a jet is misidentified 
as an electron.  But the jet$\to e$ misidentification 
rate is an order of magnitude smaller than the jet$\to\tau$ 
misidentification rate.  Electron identification
requires at most two calorimeter towers with EM 
energy fraction greater than 0.95 and other cuts.
Tau identification requires at most six calorimeter 
towers with EM energy fraction less than 0.8 
corresponding to $\xi$ greater than 0.2 and other 
cuts.  Naively assuming a flat distribution between
0 and 6 of the number of towers of jet, and a flat 
distribution between 0.0 and 1.0 of jet EM energy 
fraction, we have
\begin{equation}
   \frac{\mbox{jet}\to\tau}
        {\mbox{jet}\to e} 
   \approx \frac{(6-0)\times(0.8-0.0)} 
                {(2-0)\times(1.0-0.95)}
   = 48
\end{equation}
The electron side is much cleaner than the tau side.
It is a good approximation to estimate fakes
from the tau side.  The situation is the same for
$\tau_{\mu}\tau_h$ channel.

There is a subtlety in the fake estimate for 
$\tau_h\tau_h$ channel.  
In the data sample from the TAU\_MET trigger,
we order the tau objects in each event by
their $\pt$.  To illustrate the subtlety, here we  
temporately call the leading tau object with 
the highest $\pt$ as $\tau_1$ in the case it is 
a true tau or jet$_1$ in the case it is a true 
jet, and the second tau object with a lower 
$\pt$ as $\tau_2$ or jet$_2$.
The trigger only requires one isolated tau 
object.  We estimate the fake background from
the second tau object side which is not 
necessarily isolated.  This is able to cover the two 
cases (a) and (b) out of the total three cases of the
fake background sources: 
(a)~$\tau_1$~+~jet$_2$, (b)~jet$_1$~+~jet$_2$, and
(c)~jet$_1$~+~$\tau_2$.  Jet$_1$ has a lower 
misidentification rate than jet$_2$ because of 
its higher $\pt$, so we get $\mbox{c}<\mbox{a}$ 
and $\mbox{a}+\mbox{b}\approx\mbox{a}+\mbox{b}+\mbox{c}$.  
The fake estimate from the second tau object 
side is an approximation.

The procedure to estimate the jet fake background 
in the various channels is
shown in Table~\ref{tab:control_fake_1}.  We need to
define a specific denominator according to data sample
from the trigger path available, and we need to find out 
the normalization factors of the dynamically prescaled 
trigger paths.  

The denominators $D_{\xi}$ which is up 
to the electron removal cut $\xi>0.2$ and $D_{trkIso10Deg}$ 
which is up to the 10$^{\circ}$ track isolation cut are 
explained in Table~\ref{tab:TauId_cuts} in 
Section~\ref{subsec:TauId_cuts}.  Note that the relative
jet$\to\tau$ misidentification rate and the relative tau 
identification efficiency for different denominator 
samples are different.

The available dynamically prescaled triggers 
are discussed in Section~\ref{sec:event_trigger},
and their integrated luminosities are discussed
in Section~\ref{sec:event_GTU}.

\begin{itemize}
\item The $\tau_e\tau_h$ channel has a dynamically
      prescaled data sample from the CELE8
      trigger path available.  There is no trigger 
      cut on the tau objects, so it is ideal for the 
      fake background estimate.  We apply the cuts 
      up to the electron removal cut $\xi>0.2$ listed
      in Table~\ref{tab:TauId_cuts} on the tau objects
      and use the denominator $D_{\xi}$ to estimate
      the fakes.  The integrated luminosity
      of this trigger path is 46~pb$^{-1}$, thus
      the normalization factor to the integrated
      luminosity 195~pb$^{-1}$ of the data sample 
      from the TAU\_ELE trigger is 195/46 = 4.239.

\item The $\tau_{\mu}\tau_h$ with CMUP muon channel
      has a dynamically prescaled data sample from
      the CMUP8 trigger path available.  There is 
      no trigger cut on the tau objects, and we use
      the denominator $D_{\xi}$ to estimate the fakes.  
      The normalization factor is 195/38 = 5.132.

\item The $\tau_{\mu}\tau_h$ with CMX muon channel
      has to use the TAU\_CMX trigger itself for
      the fake background estimate.  The tau objects
      have already been cleaned up by the 10$^{\circ}$ 
      track isolation cut in the trigger.  We apply 
      the cuts up to the 10$^{\circ}$ track isolation 
      cut listed in Table~\ref{tab:TauId_cuts} on the 
      tau objects and use the denominator 
      $D_{trkIso10Deg}$ to estimate the fakes. 

\item The $\tau_h\tau_h$ channel has to use the
      TAU\_MET trigger itself.  The leading tau 
      object is cleaned up by track isolation, 
      but the second tau object is not.  We estimate 
      the fake background from the second tau object
      side, and use the denominator~$D_{\xi}$.
\end{itemize}

For each event, we substitute the relative tau identification
efficiency and the relative jet$\to\tau$ misidentification
rate corresponding to the defined denominator into 
Eq.~(\ref{eq:weight_jet_notpassing}) if the tau object does 
not pass the full set  of the tau identification cuts, or 
into Eq.~(\ref{eq:weight_jet_passing}) if it does, to 
calculate the weight to be a jet.

We sum up the weights of all the events 
in the sample to estimate the jet background,
using Eq.~(\ref{eq:sum_weights_jet}).
We then apply the event kinematic cuts
and normalize the numbers to the luminosities 
of the data samples of the tau trigger paths.

\begin{table}
   \begin{center}
      \begin{tabular}{|l|r|r|r|r|r|r|r|r|}  \hline
         channel                     & \multicolumn{2}{c|}{$\tau_e\tau_h$} 
                                     & \multicolumn{2}{c|}{CMUP $\tau_{\mu}\tau_h$} 
                                     & \multicolumn{2}{c|}{CMX $\tau_{\mu}\tau_h$} 
                                     & \multicolumn{2}{c|}{$\tau_h\tau_h$} \\ \hline
         trigger path                & \multicolumn{2}{c|}{CELE8}    
                                     & \multicolumn{2}{c|}{CMUP8}             
                                     & \multicolumn{2}{c|}{TAU\_CMX}            
                                     & \multicolumn{2}{c|}{TAU\_MET}       \\
         denominator                 & \multicolumn{2}{c|}{$D_{\xi}$}          
                                     & \multicolumn{2}{c|}{$D_{\xi}$}             
                                     & \multicolumn{2}{c|}{$D_{trkIso10Deg}$}      
                                     & \multicolumn{2}{c|}{$D_{\xi}$}          \\ 
         norm. factor                & \multicolumn{2}{c|}{4.239} 
                                     & \multicolumn{2}{c|}{5.132} 
                                     & \multicolumn{2}{c|}{1}
                                     & \multicolumn{2}{c|}{1}              \\ \hline
                                     & $\sum\omega^{jet}$ & event  
                                     & $\sum\omega^{jet}$ & event
                                     & $\sum\omega^{jet}$ & event
                                     & $\sum\omega^{jet}$ & event          \\ \cline{2-9} 
         $\sum\omega^{jet}$ or event &             92.1   & 2292     
                                     &             12.4   &  362     
                                     &             64.4   &  379     
                                     &            106.8   & 2778           \\
         kinematic cuts              &              0.903 &   56   
                                     &              0.403 &   12  
                                     &              1.649 &   30    
                                     &              3.163 &   43           \\ \hline
         normalized                  & \multicolumn{2}{c|}{$3.83\pm0.51$}   
                                     & \multicolumn{2}{c|}{$2.07\pm0.60$}   
                                     & \multicolumn{2}{c|}{$1.65\pm0.30$}   
                                     & \multicolumn{2}{c|}{$3.16\pm0.48$}  \\ \hline
      \end{tabular}
      \caption[Fake background estimates in the control region]
              {Fake background estimates in the low mass control region.
               Uncertainties are statistical.}
      \label{tab:control_fake_1}
   \end{center}
\end{table}

The event entries which are integers corresponding 
to the sum of weights which are real numbers are
also shown in Table~\ref{tab:control_fake_1}.  The 
event entries are used to estimate the statistical 
uncertainties.  

There is a systematic uncertainty due to the uncertainty 
in the jet$\to\tau$ misidentification fake rate.  The 
rate used is the average fake rate of the JET samples.  
We use the individual fake rate of the JET20, JET50, 
JET70, and JET100 samples to estimate this uncertainty, 
shown in Table~\ref{tab:control_fake_2}.  For example, 
for the $\tau_e\tau_h$ channel, using the average fake 
rate we get an estimate of 3.83; while using the
individual fake rates from the JET20, JET50, JET70, 
and JET100 samples, we get estimates
4.43, 3.25, 3.03, and 2.94, respectively. 
We take the biggest difference, i.e. $3.83-2.94=0.89$ as 
the systematic uncertainty.  The fractional  systematic 
uncertainty for this channel is $0.89/3.83\approx20\%$.  
The fractional systematic uncertainties of other channels 
are about 20\% too.

\begin{table}
   \begin{center}
      \begin{tabular}{|l|c|c|c|c|}  \hline
         channel    & $\tau_e\tau_h$ 
                    & CMUP $\tau_{\mu}\tau_h$
                    & CMX $\tau_{\mu}\tau_h$ 
                    & $\tau_h\tau_h$            \\ \hline
         average    & 3.83 & 2.07 & 1.65 & 3.16 \\ \hline
         JET20      & 4.43 & 2.23 & 1.83 & 3.26 \\
         JET50      & 3.25 & 1.61 & 1.35 & 2.90 \\
         JET70      & 3.03 & 1.62 & 1.40 & 3.01 \\
         JET100     & 2.94 & 1.77 & 1.32 & 3.20 \\ \hline
         syst. err. & 0.89 & 0.46 & 0.33 & 0.26 \\ \hline
      \end{tabular}
      \caption[Systematic uncertainties on fake backgrounds in the control region]
              {Systematic uncertainties of fake background estimates
               in the low mass control region.}
      \label{tab:control_fake_2}
   \end{center}
\end{table}

Combining in quadrature the statistical 
uncertainties in
Table~\ref{tab:control_fake_1} and the 
systematic uncertainties in
Table~\ref{tab:control_fake_2},
we get
\begin{eqnarray}
                \tau_e\tau_h     \mbox{ fake} & = & 3.83\pm1.03 \\
   \mbox{CMUP } \tau_{\mu}\tau_h \mbox{ fake} & = & 2.07\pm0.76 \\
   \mbox{CMX }  \tau_{\mu}\tau_h \mbox{ fake} & = & 1.65\pm0.45 \\
                \tau_h\tau_h     \mbox{ fake} & = & 3.16\pm0.55 
\end{eqnarray}


\section{Cross Check Fake Background}
\label{sec:control_crossCheck}

\begin{figure}
   \begin{center}
      \parbox{4.4in}{\epsfxsize=\hsize\epsffile{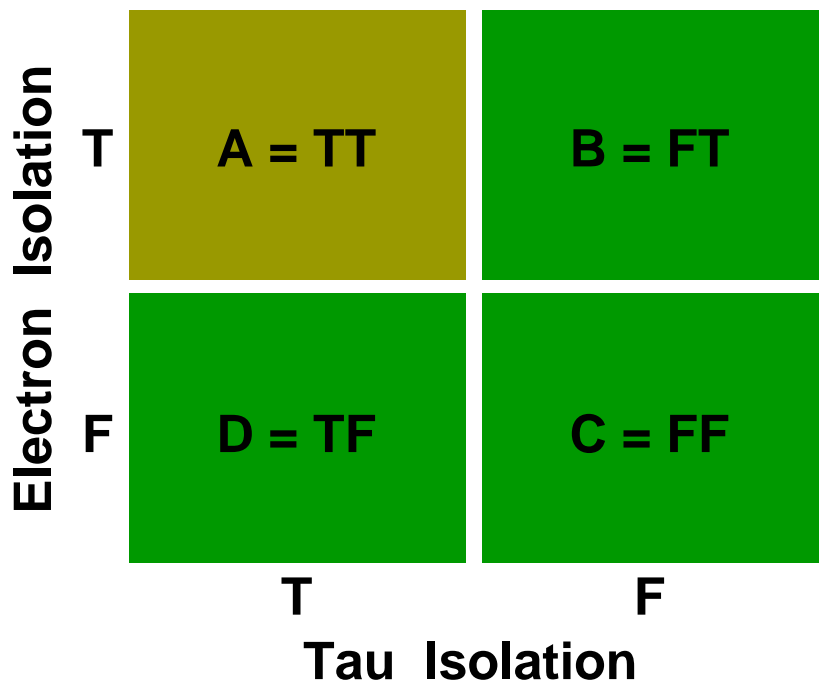}}
      \caption[Cross check fake background estimate]
              {Using the uncorrelated tau isolation
               and electron isolation to estimate fake background
               for $\tau_e\tau_h$ channel.}
      \label{fig:control_crossCheck}
   \end{center}
\end{figure}

We perform a cross check on the fake background 
estimate as follows: relax the tau isolation and 
the lepton isolation, and apply all of the other 
cuts.  The tau isolation and the lepton isolation 
are uncorrelated, thus we can extrapolate from the 
fake regions into the signal region.  For 
example, for $\tau_e\tau_h$ channel, the signal
region A and the background regions B, C and D 
are defined as in 
Fig.~\ref{fig:control_crossCheck}, 
and the fake backgrounds in A extrapolated = 
$\mbox{B}\times\mbox{D}/\mbox{C}$.

Unfortunately, we can only cross check the fake
background for $\tau_e\tau_h$ channel using the 
data sample from the CELE8 trigger path
and possibly for $\tau_{\mu}\tau_h$ with CMUP 
muon channel using the data sample from the CMUP8
trigger path.  Neither sample has isolation in 
the trigger.  There is no such sample for 
$\tau_{\mu}\tau_h$ with CMX muon channel.  There 
is a sample without isolation for $\tau_h\tau_h$ 
channel, but its prescale is 100 which is too large 
for this exercise.

Due to the statistics in the region B, C and D, 
this cross check can only be done for the 
$\tau_e\tau_h$ channel in the low mass control 
region.  The numbers in region B, C and D are 12, 
142 and 13, respectively.  When we extrapolate to 
region A, we find that 
\begin{equation}
   \mbox{A}
   = \mbox{B}\times\mbox{D}/\mbox{C}
   = 12\times13/142 
   = 1.099
\end{equation}
The normalization factor is 4.239, 
thus we get
   $\tau_e\tau_h$ fake extroplated
   = $1.099\times4.239$ 
   = 4.66.
This is in good agreement with $3.83\pm1.03$ 
obtained by summing up the weights of tau object 
being a jet.  It does give us confidence in the 
method of the jet$\to\tau$ misidentified fake 
background estimate.


\section{Uncertainties in Control Region}
\label{sec:control_err}

The statistical uncertainty and the systematic uncertainty 
of Drell-Yan background estimate include
\begin{itemize}
\item statistical uncertainty,
\item $\sigma\cdot B$ uncertainty, 2\%, aside from luminosity 
      uncertainty (see Ref.~\cite{Acosta:2004uq}),
\item trigger efficiencies (see Section~\ref{sec:event_trigEff}),
\item lepton scale factors (see Section~\ref{subsec:TauSF_sf} for $\tau$ scale factor,
                                Section~\ref{subsec:EleSF} for $e$ scale factor, 
                            and Section~\ref{subsec:MuSF} for $\mu$ scale factors),
\item $\met$ uncertainty, 6\% (see Section~\ref{sec:MET}), and
\item luminosity, 6\% (see Ref.~\cite{Klimenko:2003if}).
\end{itemize}

The statistical uncertainty and systematic uncertainty
of the jet$\to\tau$ misidentified fake background estimate 
are discussed in Section~\ref{sec:control_fake}.

We combine the $\tau_{\mu}\tau_h$ CMUP muon channel 
with a luminosity 195 pb$^{-1}$ and the 
$\tau_{\mu}\tau_h$ CMX muon channel with a luminosity
179 pb$^{-1}$ into one channel, simply called the
$\tau_{\mu}\tau_h$ channel.
The observed events in 
$\tau_e\tau_h$, 
$\tau_{\mu}\tau_h$ and
$\tau_h\tau_h$ channels are 
46, 
36 and 
8, respectively.

\begin{figure}
   \begin{center}
      \parbox{5.4in}{\epsfxsize=\hsize\epsffile{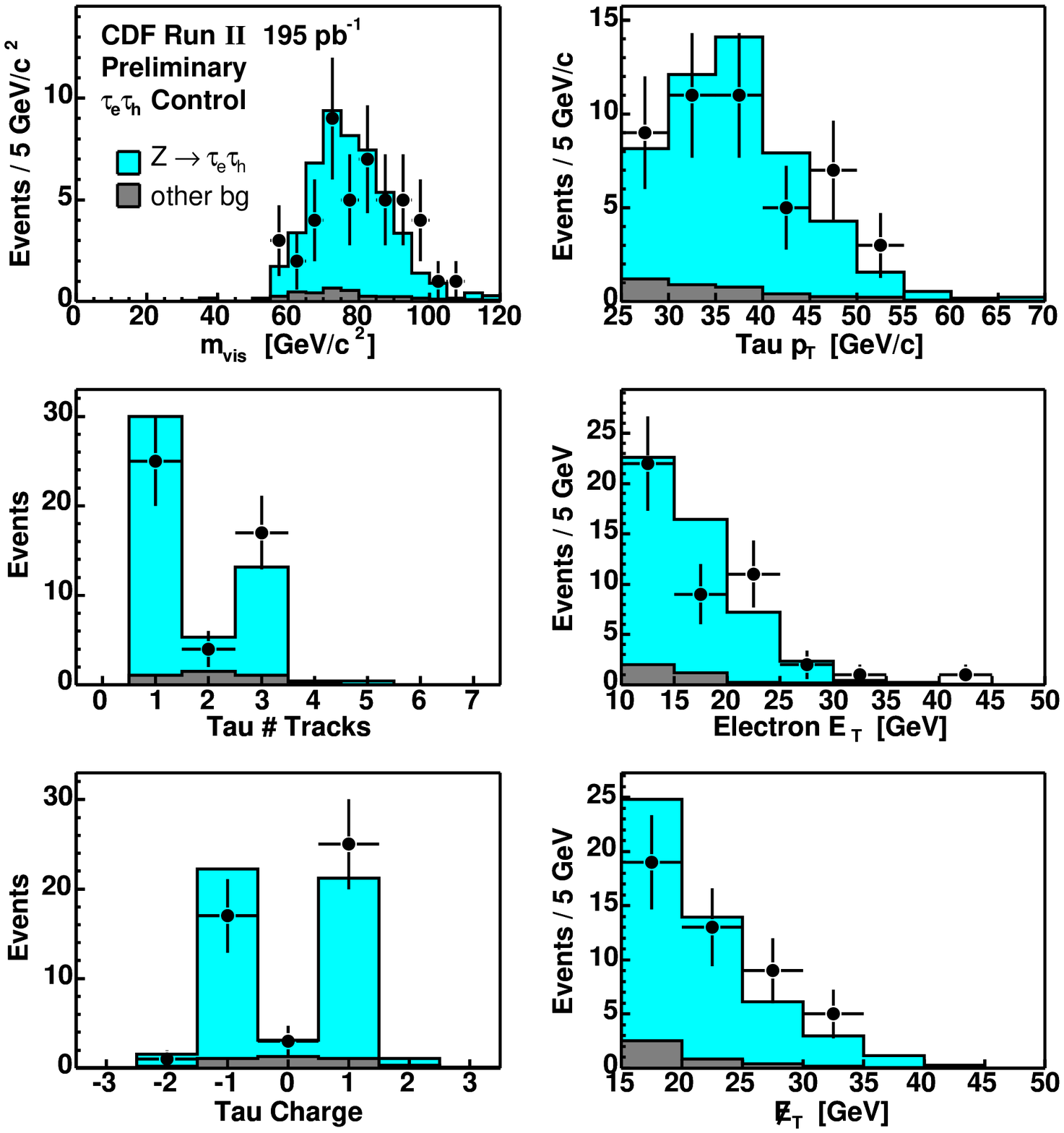}}
      \caption[Distributions of the $\tau_e\tau_h$ channel in the control region]
              {Distributions of the $\tau_e\tau_h$ channel in the control region
               for data (points) and predicted backgrounds (histograms).}
      \label{fig:control_ZTauTau_1}
   \end{center}
\end{figure}

\begin{figure}
   \begin{center}
      \parbox{5.4in}{\epsfxsize=\hsize\epsffile{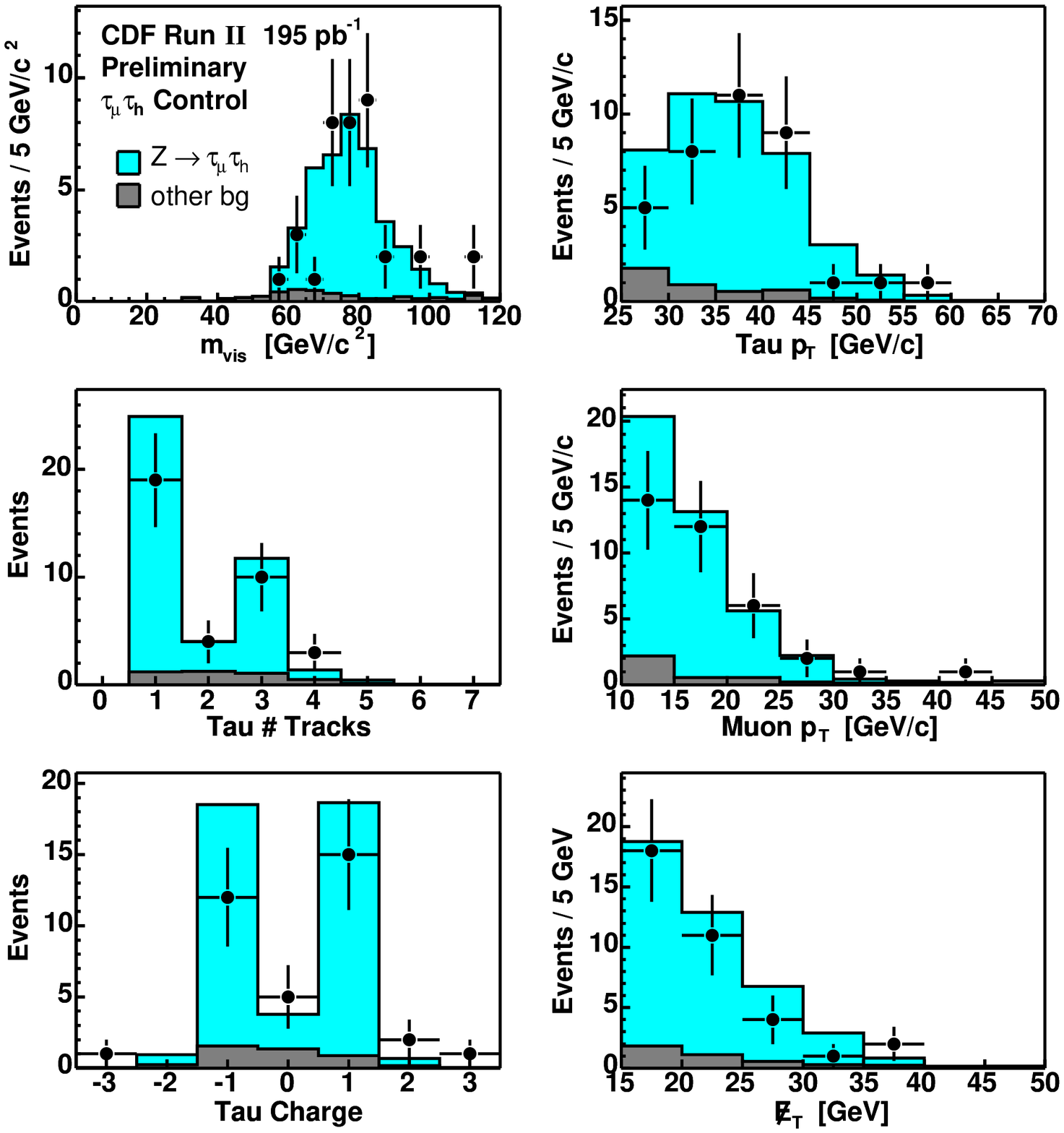}}
      \caption[Distributions of the $\tau_{\mu}\tau_h$ channel in the control region]
              {Distributions of the $\tau_{\mu}\tau_h$ channel in the control region
               for data (points) and predicted backgrounds (histograms).}
      \label{fig:control_ZTauTau_2}
   \end{center}
\end{figure}

\begin{figure}
   \begin{center}
      \parbox{5.4in}{\epsfxsize=\hsize\epsffile{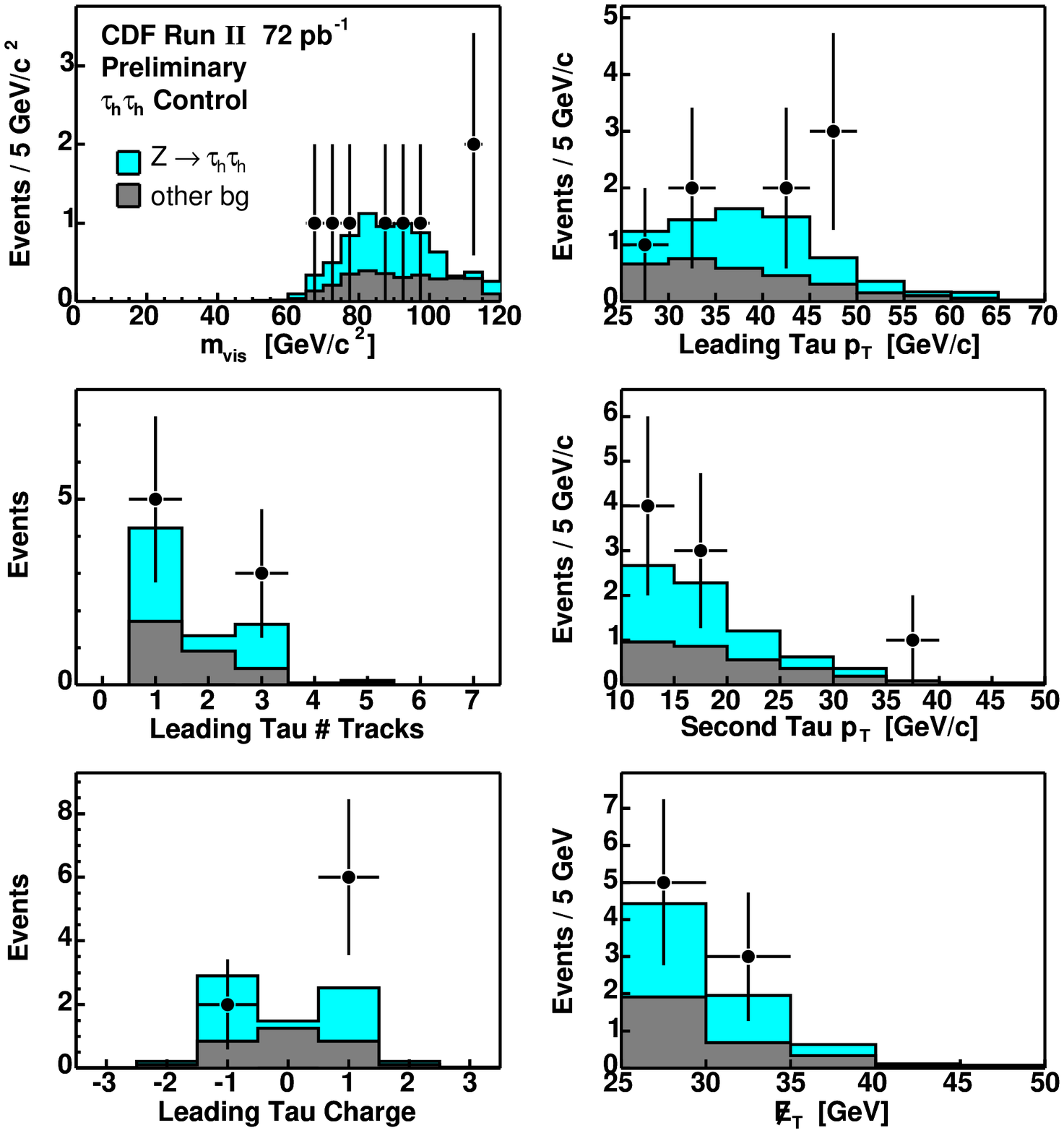}}
      \caption[Distributions of the $\tau_h\tau_h$ channel in the control region]
              {Distributions of the $\tau_h\tau_h$ channel in the control region
               for data (points) and predicted backgrounds (histograms).}
      \label{fig:control_ZTauTau_3}
   \end{center}
\end{figure}

Table~\ref{tab:control_err}
shows the summary of the control sample in low mass 
region for 195 pb$^{-1}$ (72 pb$^{-1}$ for 
$\tau_h\tau_h$).
The total background estimate is $99.27\pm12.55$,
dominated by the source from $Z/\gamma^*\to\tau\tau$ 
as expected.
The observed number of events, 90, in the control
region is in good agreement with this prediction.
Fig.~\ref{fig:control_ZTauTau_1}$-$\ref{fig:control_ZTauTau_3}
show the distributions of each channel in the low mass
control region. The observed distributions
in the data are in good agreement with the predicted
distributions.

\begin{table}
   \begin{center}
      \begin{tabular}{|c|c|c|c|c|} \hline
         Source                  & $\tau_e\tau_h$ & $\tau_{\mu}\tau_h$ & $\tau_h\tau_h$ &           Total \\ \hline
         $Z/\gamma^*\to\tau\tau$ & $45.36\pm6.84$ &     $38.39\pm5.72$ &  $4.19\pm0.77$ & $87.94\pm12.38$ \\
         $Z/\gamma^*\to ee$      &  $0.14\pm0.14$ &                  0 &              0 &  $0.14\pm0.14$  \\
         $Z/\gamma^*\to\mu\mu$   &              0 &      $0.48\pm0.25$ &              0 &  $0.48\pm0.25$  \\
         Jet$\to\tau$            &  $3.83\pm1.03$ &      $3.72\pm0.88$ &  $3.16\pm0.55$ & $10.71\pm1.46$  \\ \hline
         Expected                & $49.32\pm6.94$ &     $42.59\pm5.85$ &  $7.35\pm0.95$ & $99.27\pm12.55$ \\ \hline
         Observed                &             46 &                 36 &              8 &              90 \\ \hline
      \end{tabular}
      \caption[Expected and observed events in the control region]
              {Number of expected events for each channel
               and each source, compared with the number
               observed, in the control region $m_{vis}<120$ GeV/$c^2$.}
      \label{tab:control_err}
   \end{center}
\end{table}


%

\chapter{High Mass Signal Region}
\label{cha:signal}

The high mass region with $m_{vis}>120$ GeV/$c^2$ 
is the signal region.
First we calculate signal acceptance, 
then we estimate the backgrounds.  The main 
backgrounds are $Z/\gamma^*\to\tau\tau$,
$Z/\gamma^*\to ee$, $Z/\gamma^*\to\mu\mu$ which 
can be estimated from MC simulation, and 
the jet$\to\tau$ misidentified fake background 
which can be estimated from data, as in the
control region.


\section{Signal Acceptance}
\label{sec:signal_acc}

\begin{figure}
   \begin{center}
      \parbox{5.1in}{\epsfxsize=\hsize\epsffile{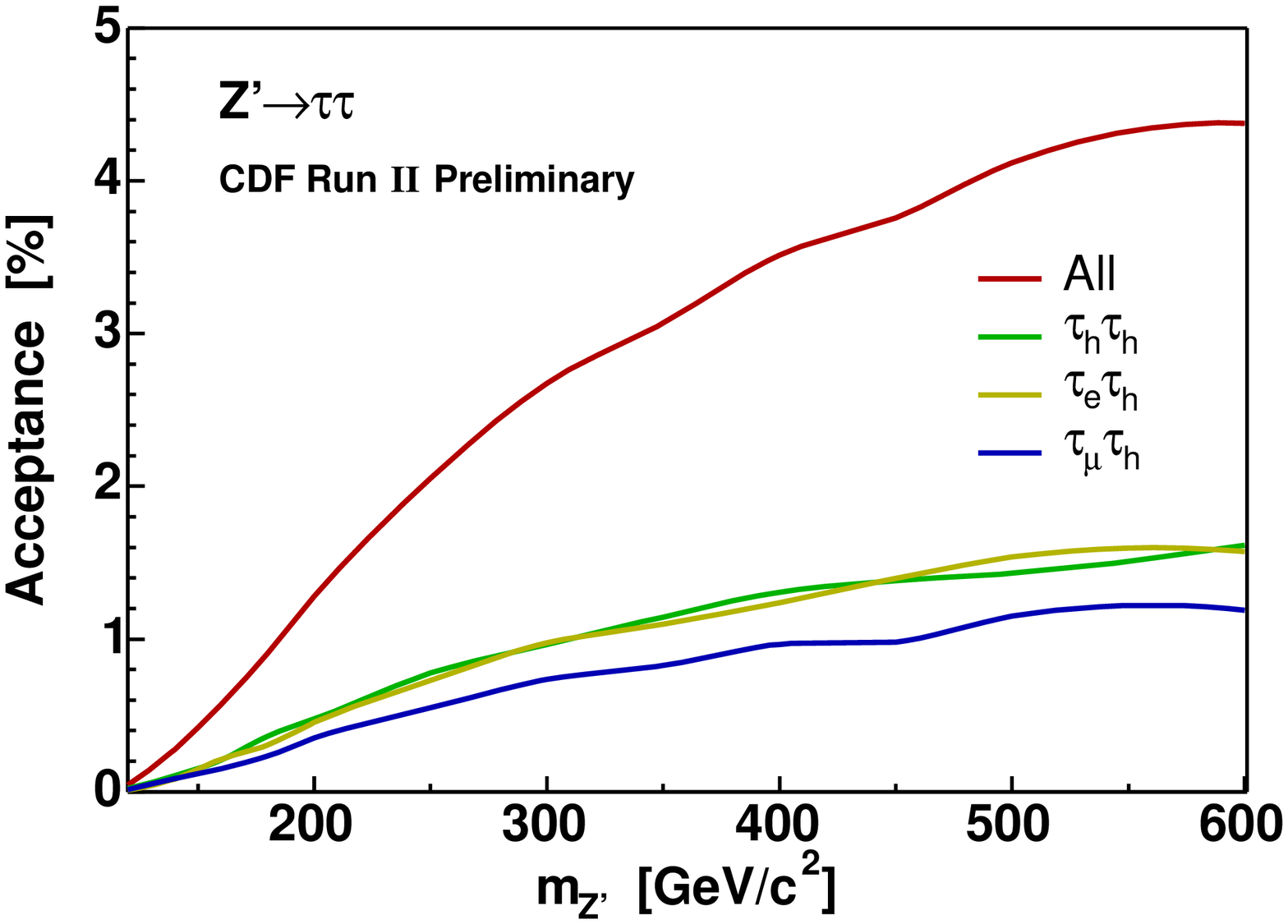}}
      \caption[Signal acceptance of $\zprime\to\tau\tau$]
              {Signal acceptance of a new vector particle 
               $\zprime\to\tau\tau$ in each channel,
               as a function of the $\zprime$ mass.}
      \label{fig:signal_ZprimeAcc}
   \vspace{0.5in}
      \parbox{5.1in}{\epsfxsize=\hsize\epsffile{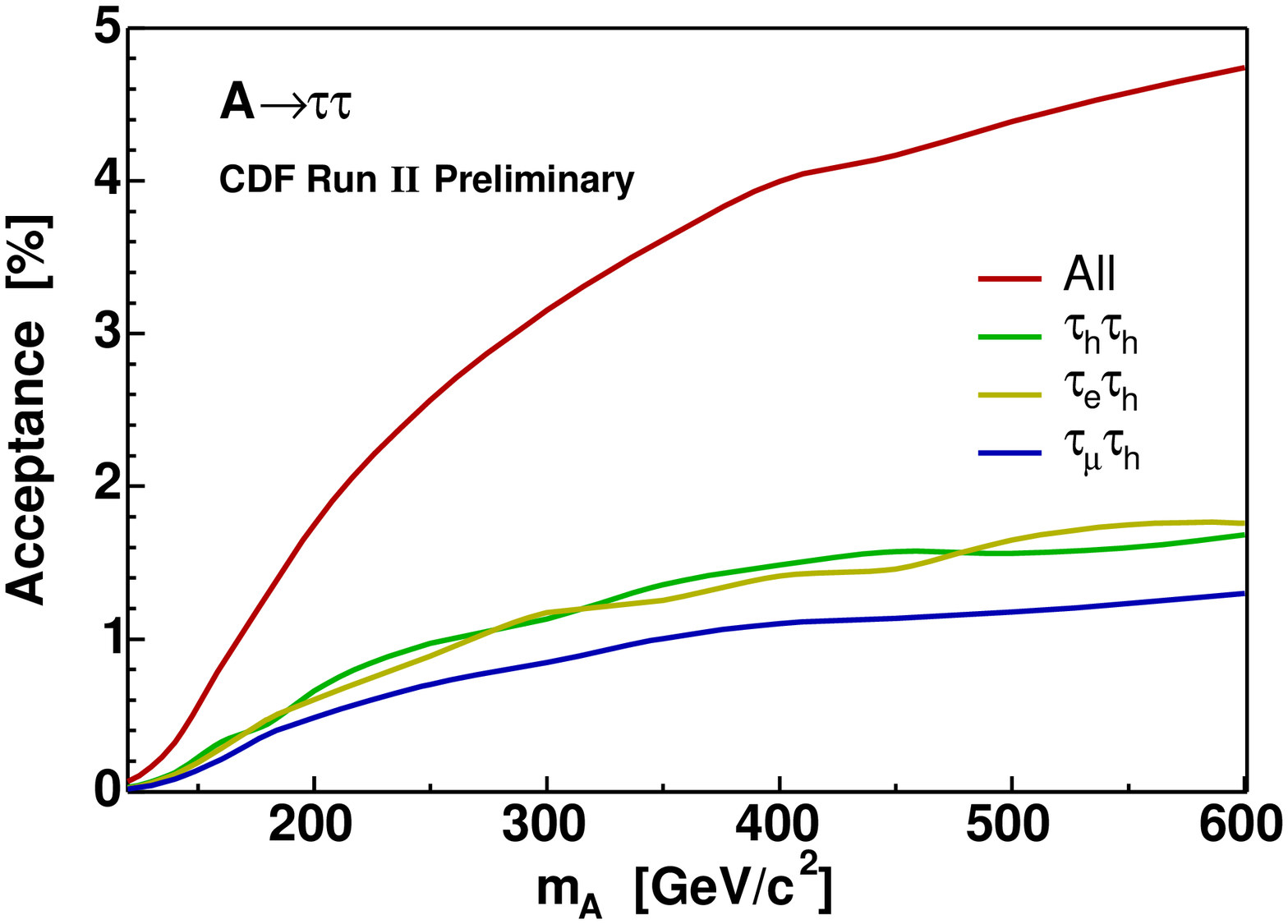}}
      \caption[Signal acceptance of $A\to\tau\tau$]
              {Signal acceptance of a new scalar particle 
               $A\to\tau\tau$ in each channel,
               as a function of the $A$ mass.}
      \label{fig:signal_AAcc}
   \end{center}
\end{figure}

Table~\ref{tab:signal_ZprimeAcc}
shows the procedure to measure the signal acceptances
in each channel for the new vector particle decaying to
two taus, using $\zprime\to\tau\tau$ events.  For example, 
for the $\tau_e\tau_h$ channel, we match the offline 
tau object and electron object with the $\tau_h$ and $\tau_e$
by requiring the separation angle be less than 0.2 radian,
apply the event kinematic cuts, multiply the number of accepted 
events by the trigger efficiency and the lepton scale factors,
and calculate the overall acceptance.  Since the mass of the
$\zprime$ is unknown, we calculate the signal acceptance as a 
function of its mass.  Only five 
mass points (120, 180, 300, 450, 600) GeV/$c^2$ out of total
twelve mass points (120, 140, 160, 180, 200, 250, 300, 
350, 400, 450, 500, 600) GeV/$c^2$ are shown in
Table~\ref{tab:signal_ZprimeAcc}.
The signal acceptances of the $\tau_{\mu}\tau_h$ channel with 
a CMUP muon and of the $\tau_{\mu}\tau_h$ with a CMX muon are 
combined into one signal acceptance for the $\tau_{\mu}\tau_h$ 
channel.  The total acceptance is a combination of the acceptance
of the $\tau_e\tau_h$, the $\tau_{\mu}\tau_h$, and
the $\tau_h\tau_h$ channels. 
The signal acceptances are shown in 
in Fig.~\ref{fig:signal_ZprimeAcc}.

Table~\ref{tab:signal_AAcc}
shows the procedure to measure the signal acceptances
in each channel for the new scalar particle decaying to two 
taus, using $A\to\tau\tau$.  We set $\tan\beta$ = 20 as a 
representative value of $\tan\beta$.  Similarly,
since the mass of $A$ is unknown, we calculate the signal 
acceptances as a function of mass, as shown in
Fig.~\ref{fig:signal_AAcc}.

\begin{table}
   \begin{center}
      \begin{tabular}{|l|r|r|r|r|r|} \hline
              $\zprime\to\tau\tau$                 &   m=120 &   m=180 &   m=300 &   m=450 &   m=600 \\
              event                                &  100000 &  100000 &  100000 &  100000 &  100000 \\ \hline
         $\tau_e\tau_h$ (TAU\_ELE)                 &         &         &         &         &         \\
              $\tau_e\tau_h$ decay                 &   23527 &   23209 &   23246 &   23345 &   23250 \\
              match $\tau(25)+e(10)$               &     761 &    1256 &    2135 &    2816 &    3044 \\
              $\met>15$                            &     380 &     797 &    1720 &    2416 &    2745 \\
              $\Delta\phi(e-\met)<30^{\circ}$      &     296 &     583 &    1231 &    1655 &    1844 \\
              $m_{vis}>120$                        &      14 &     355 &    1125 &    1610 &    1814 \\
              trigger efficiency                   &    12.9 &   327.0 &  1036.3 &  1483.1 &  1671.0 \\
              lepton scale factors                 &    12.1 &   307.7 &   975.1 &  1395.4 &  1572.2 \\
              acceptance (\%)                      &   0.012 &   0.308 &   0.975 &   1.395 &   1.572 \\ \hline
         $\tau_{\mu}\tau_h$ (TAU\_CMU)             &         &         &         &         &         \\
              $\tau_{\mu}\tau_h$  decay            &   22540 &   22500 &   22437 &   22358 &   22463 \\
              match $\tau(25)+\mbox{CMUP }\mu(10)$ &     418 &     698 &    1121 &    1492 &    1775 \\
              cosmic veto                          &     418 &     698 &    1121 &    1491 &    1775 \\
              $\met>15$                            &     198 &     460 &     894 &    1313 &    1615 \\
              $\Delta\phi(\mu-\met)<30^{\circ}$    &     169 &     348 &     677 &     919 &    1134 \\
              $m_{vis}>120$                        &      14 &     208 &     632 &     882 &    1114 \\
              trigger efficiency                   &    12.8 &   189.6 &   576.1 &   804.0 &  1015.4 \\
              lepton scale factors                 &    10.8 &   160.8 &   488.5 &   681.7 &   861.1 \\
              acceptance (\%)                      &   0.011 &   0.161 &   0.489 &   0.682 &   0.861 \\ \hline
         $\tau_{\mu}\tau_h$ (TAU\_CMX)             &         &         &         &         &         \\
              $\tau_{\mu}\tau_h$  decay            &   22540 &   22500 &   22437 &   22358 &   22463 \\
              match $\tau(25)+\mbox{CMX }\mu(10)$  &     196 &     322 &     505 &     551 &     605 \\
              cosmic veto                          &     196 &     322 &     505 &     551 &     605 \\
              $\met>15$                            &      99 &     200 &     408 &     473 &     535 \\
              $\Delta\phi(\mu-\met)<30^{\circ}$    &      88 &     140 &     301 &     345 &     379 \\
              $m_{vis}>120$                        &       2 &      83 &     279 &     336 &     372 \\
              trigger efficiency                   &     1.8 &    75.7 &   254.3 &   306.3 &   339.1 \\
              lepton scale factors                 &     1.8 &    73.3 &   246.4 &   296.8 &   328.6 \\
              acceptance (\%)                      &   0.002 &   0.073 &   0.246 &   0.297 &   0.329 \\ \hline
         $\tau_h\tau_h$ (TAU\_MET)                 &         &         &         &         &         \\
              $\tau_h\tau_h$ decay                 &   41677 &   41880 &   41934 &   41772 &   42027 \\
              match $\tau_1(25)+\tau_2(10)$        &    1662 &    2449 &    3415 &    3932 &    4257 \\
              $\met>25$                            &     277 &     940 &    2037 &    2888 &    3383 \\
              $\Delta\phi(\tau_2-\met)<30^{\circ}$ &     242 &     832 &    1679 &    2244 &    2459 \\
              $\tau_2$ num. track == 1             &     185 &     653 &    1335 &    1789 &    2043 \\
              $m_{vis}>120$                        &      31 &     526 &    1282 &    1768 &    2028 \\
              trigger efficiency                   &    21.2 &   388.3 &  1023.7 &  1469.1 &  1716.9 \\
              lepton scale factors                 &    19.9 &   365.3 &   963.2 &  1382.3 &  1615.4 \\
              acceptance (\%)                      &   0.020 &   0.365 &   0.963 &   1.382 &   1.615 \\ \hline
         channels combined                         &         &         &         &         &         \\
              acceptance (\%)                      &   0.045 &   0.907 &   2.673 &   3.756 &   4.377 \\ \hline
      \end{tabular}
      \caption[$\zprime\to\tau\tau$ signal acceptance]
              {New vector particle $\zprime\to\tau\tau$ signal acceptance,
               for each channel, as a function of the $\zprime$ mass.} 
      \label{tab:signal_ZprimeAcc}
   \end{center}
\end{table}

\begin{table}
   \begin{center}
      \begin{tabular}{|l|r|r|r|r|r|} \hline
              $A\to\tau\tau$                       &   m=120 &   m=180 &   m=300 &   m=450 &   m=600 \\
              event                                &  100000 &  100000 &  100000 &  100000 &  100000 \\ \hline
         $\tau_e\tau_h$ (TAU\_ELE)                 &         &         &         &         &         \\
              $\tau_e\tau_h$ decay                 &   23427 &   23391 &   23364 &   23051 &   23242 \\
              match $\tau(25)+e(10)$               &    1063 &    1806 &    2556 &    2991 &    3375 \\
              $\met>15$                            &     539 &    1237 &    2098 &    2665 &    3142 \\
              $\Delta\phi(e-\met)<30^{\circ}$      &     396 &     870 &    1445 &    1723 &    2047 \\
              $m_{vis}>120$                        &      23 &     547 &    1354 &    1684 &    2028 \\
              trigger efficiency                   &    21.2 &   503.9 &  1247.3 &  1551.3 &  1868.1 \\
              lepton scale factors                 &    19.9 &   474.1 &  1173.6 &  1459.6 &  1757.7 \\
              acceptance (\%)                      &   0.020 &   0.474 &   1.174 &   1.460 &   1.758 \\ \hline
         $\tau_{\mu}\tau_h$ (TAU\_CMU)             &         &         &         &         &         \\
              $\tau_{\mu}\tau_h$  decay            &   22649 &   22759 &   22344 &   22472 &   22398 \\
              match $\tau(25)+\mbox{CMUP }\mu(10)$ &     650 &    1001 &    1454 &    1832 &    2076 \\
              cosmic veto                          &     650 &    1000 &    1454 &    1832 &    2076 \\
              $\met>15$                            &     353 &     671 &    1198 &    1634 &    1923 \\
              $\Delta\phi(\mu-\met)<30^{\circ}$    &     286 &     492 &     855 &    1088 &    1272 \\
              $m_{vis}>120$                        &      17 &     329 &     790 &    1063 &    1265 \\
              trigger efficiency                   &    15.5 &   299.9 &   720.1 &   969.0 &  1153.1 \\
              lepton scale factors                 &    13.1 &   254.3 &   610.6 &   821.6 &   977.8 \\
              acceptance (\%)                      &   0.013 &   0.254 &   0.611 &   0.822 &   0.978 \\ \hline
         $\tau_{\mu}\tau_h$ (TAU\_CMX)             &         &         &         &         &         \\
              $\tau_{\mu}\tau_h$  decay            &   22649 &   22759 &   22344 &   22472 &   22398 \\
              match $\tau(25)+\mbox{CMX }\mu(10)$  &     239 &     407 &     552 &     601 &     612 \\
              cosmic veto                          &     239 &     406 &     552 &     601 &     612 \\
              $\met>15$                            &     120 &     297 &     449 &     522 &     553 \\
              $\Delta\phi(\mu-\met)<30^{\circ}$    &      88 &     214 &     291 &     363 &     370 \\
              $m_{vis}>120$                        &       6 &     138 &     266 &     355 &     365 \\
              trigger efficiency                   &     5.5 &   125.8 &   242.5 &   323.6 &   332.7 \\
              lepton scale factors                 &     5.3 &   121.9 &   235.0 &   313.6 &   322.4 \\
              acceptance (\%)                      &   0.005 &   0.122 &   0.235 &   0.314 &   0.322 \\ \hline
         $\tau_h\tau_h$ (TAU\_MET)                 &         &         &         &         &         \\
              $\tau_h\tau_h$ decay                 &   41813 &   41837 &   42008 &   42104 &   41891 \\
              match $\tau_1(25)+\tau_2(10)$        &    2325 &    3117 &    3951 &    4333 &    4348 \\
              $\met>25$                            &     495 &    1322 &    2534 &    3316 &    3653 \\
              $\Delta\phi(\tau_2-\met)<30^{\circ}$ &     400 &    1072 &    1969 &    2467 &    2579 \\
              $\tau_2$ num. track == 1             &     293 &     821 &    1531 &    2005 &    2106 \\
              $m_{vis}>120$                        &      46 &     630 &    1483 &    1985 &    2101 \\
              trigger efficiency                   &    30.8 &   472.8 &  1202.9 &  1672.1 &  1789.5 \\
              lepton scale factors                 &    29.0 &   444.9 &  1131.8 &  1573.3 &  1683.8 \\
              acceptance (\%)                      &   0.029 &   0.445 &   1.132 &   1.573 &   1.684 \\ \hline
         channels combined                         &         &         &         &         &         \\
              acceptance (\%)                      &   0.067 &   1.295 &   3.151 &   4.168 &   4.742 \\ \hline
      \end{tabular}
      \caption[$A\to\tau\tau$ signal acceptance]
              {New scalar particle $A\to\tau\tau$ signal acceptance,
               for each channel, as a function of the $A$ mass.} 
      \label{tab:signal_AAcc}
   \end{center}
\end{table}

\section{Drell-Yan Background}
\label{sec:signal_DY}

The largest portion of the production cross section 
for the Drell-Yan backgrounds is at the $Z$ boson 
resonance peak, about 91 GeV/$c^2$.  However the 
events in the high mass signal region are mostly 
from the high mass Drell-Yan tail.  To model 
the high mass tail better, we need more statistics
in MC simulation at that region.  To achieve this,
we break the generation level mass into two 
exclusively separated regions: $30<m<100$ GeV/$c^2$ 
and $m>100$ GeV/$c^2$, and simulate them separately.
The production cross sections in these two regions
are about 315 pb and 11 pb, respectively (see
Section~\ref{sec:control_sigmaDY}).
Therefore we have a low-mass sample and a high-mass 
sample for each $Z/\gamma^*\to l^+l^-$ source. 

Table~\ref{tab:signal_DY}
shows the procedure to estimate Drell-Yan backgrounds.
We apply the event kinematic cuts on the MC samples, 
multiply the number of surviving events by the trigger 
efficiencies and the lepton scale factors, normalize 
to the integrated luminosity 195 pb$^{-1}$ (179 pb$^{-1}$ 
for the TAU\_CMX trigger, 72 pb$^{-1}$ for the TAU\_MET
trigger), and combine the estimate for the low-mass 
Drell-Yan sample and the estimate for the high-mass 
Drell-Yan sample.

\begin{table}
   \begin{center}
      \begin{tabular}{|l|r|r|r|r|r|r|} \hline
              source                               & \multicolumn{2}{c|}{$Z/\gamma^{*}\to\tau\tau$}
                                                   & \multicolumn{2}{c|}{$Z/\gamma^{*}\to ee$}
                                                   & \multicolumn{2}{c|}{$Z/\gamma^{*}\to\mu\mu$}              \\ \cline{2-7}
              mass window                          &30$-$100 &  $>$100 &30$-$100 &  $>$100 &30$-$100 &  $>$100 \\
              $\sigma\cdot B$ (pb)                 &     315 &      11 &     315 &      11 &     315 &      11 \\
              event                                &  475901 &  160000 &  385686 &  160000 &  392063 &  160000 \\ \hline
         $\tau_e\tau_h$ (TAU\_ELE)                 &         &         &         &         &         &         \\
              $\tau(25)+e(10)$                     &    1405 &    1062 &     257 &     190 &       1 &       1 \\
              $\met>15$                            &     456 &     472 &      28 &      20 &       1 &       0 \\
              $\Delta\phi(e-\met)<30^{\circ}$      &     381 &     364 &       2 &       2 &       0 &       0 \\
              $m_{vis}>120$                        &       0 &      48 &       1 &       2 &       0 &       0 \\
              trigger efficiency                   &   0.000 &  44.216 &   0.921 &   1.842 &   0.000 &   0.000 \\
              lepton scale factors                 &   0.000 &  41.603 &   0.867 &   1.733 &   0.000 &   0.000 \\
              normalized (195 pb$^{-1}$)           &   0.000 &   0.558 &   0.138 &   0.023 &   0.000 &   0.000 \\ \cline{2-7}
              combined                             & \multicolumn{2}{c|}{0.56}
                                                   & \multicolumn{2}{c|}{0.16}
                                                   & \multicolumn{2}{c|}{0.00}                                 \\ \hline
         $\tau_{\mu}\tau_h$ (TAU\_CMU)             &         &         &         &         &         &         \\
              $\tau(25)+\mbox{CMUP }\mu(10)$       &     783 &     554 &       0 &       0 &     408 &     139 \\
              cosmic veto                          &     783 &     554 &       0 &       0 &     408 &     139 \\
              $\met>15$                            &     272 &     233 &       0 &       0 &     346 &     124 \\
              $\Delta\phi(\mu-\met)<30^{\circ}$    &     238 &     179 &       0 &       0 &       7 &       0 \\
              $m_{vis}>120$                        &       0 &      24 &       0 &       0 &       3 &       0 \\
              trigger efficiency                   &   0.000 &  21.877 &   0.000 &   0.000 &   2.735 &   0.000 \\
              lepton scale factors                 &   0.000 &  18.551 &   0.000 &   0.000 &   2.319 &   0.000 \\
              normalized (195 pb$^{-1}$)           &   0.000 &   0.249 &   0.000 &   0.000 &   0.363 &   0.000 \\ \cline{2-7}
              combined                             & \multicolumn{2}{c|}{0.25}
                                                   & \multicolumn{2}{c|}{0.00}
                                                   & \multicolumn{2}{c|}{0.36}                                 \\ \hline
         $\tau_{\mu}\tau_h$ (TAU\_CMX)             &         &         &         &         &         &         \\
              $\tau(25)+\mbox{CMX }\mu(10)$        &     384 &     284 &       0 &       0 &     212 &      49 \\
              cosmic veto                          &     384 &     284 &       0 &       0 &     212 &      49 \\
              $\met>15$                            &     127 &     129 &       0 &       0 &     174 &      41 \\
              $\Delta\phi(\mu-\met)<30^{\circ}$    &     114 &     107 &       0 &       0 &       1 &       1 \\
              $m_{vis}>120$                        &       0 &      23 &       0 &       0 &       1 &       1 \\
              trigger efficiency                   &   0.000 &  20.965 &   0.000 &   0.000 &   0.912 &   0.912 \\
              lepton scale factors                 &   0.000 &  20.318 &   0.000 &   0.000 &   0.883 &   0.883 \\
              normalized (179 pb$^{-1}$)           &   0.000 &   0.250 &   0.000 &   0.000 &   0.127 &   0.011 \\ \cline{2-7}
              combined                             & \multicolumn{2}{c|}{0.25}
                                                   & \multicolumn{2}{c|}{0.00}
                                                   & \multicolumn{2}{c|}{0.14}                                 \\ \hline
         $\tau_h\tau_h$ (TAU\_MET)                 &         &         &         &         &         &         \\
              $\tau_1(25)+\tau_2(10)$              &    4023 &    2524 &       1 &       3 &       8 &       3 \\
              $\met>25$                            &     249 &     428 &       0 &       0 &       0 &       2 \\
              $\Delta\phi(\tau_2-\met)<30^{\circ}$ &     202 &     361 &       0 &       0 &       0 &       0 \\
              $\tau_2$ num. track == 1             &     158 &     269 &       0 &       0 &       0 &       0 \\
              $m_{vis}>120$                        &       2 &      84 &       0 &       0 &       0 &       0 \\
              trigger efficiency                   &   1.547 &  63.373 &   0.000 &   0.000 &   0.000 &   0.000 \\
              lepton scale factors                 &   1.455 &  59.627 &   0.000 &   0.000 &   0.000 &   0.000 \\
              normalized (72 pb$^{-1}$)            &   0.069 &   0.295 &   0.000 &   0.000 &   0.000 &   0.000 \\ \cline{2-7}
              combined                             & \multicolumn{2}{c|}{0.36}
                                                   & \multicolumn{2}{c|}{0.00}
                                                   & \multicolumn{2}{c|}{0.00}                                 \\ \hline
      \end{tabular}
      \caption[Drell-Yan background estimates in the signal region]
              {Drell-Yan background estimates for each channel in
               the high mass signal region.}
      \label{tab:signal_DY}
   \end{center}
\end{table}


\section{Fake Background}
\label{sec:signal_fake}

The procedure to estimate the jet$\to\tau$ 
fake background is similar to what we have done for
low mass control region in Section~\ref{sec:control_fake}.  
The trigger path, the luminosity normalization factor, 
the denominator tau object definition, and the sum
of the weights of tau objects being a jet in the high 
mass signal region are exactly the same as those in the
low mass control region.  The only one difference 
is this cut: $m_{vis}<120$ GeV/$c^2$ for the low mass 
control region, while $m_{vis}>120$ GeV/$c^2$ for the 
high mass signal region.  
Now we repeat the same procedure, as shown in
Table~\ref{tab:signal_fake_1}.  
The event entries which are integers corresponding to the sum 
of weights which are real numbers are also shown.  
The event entries are used to estimate the statistical 
uncertainties.  

\begin{table}
   \begin{center}
      \begin{tabular}{|l|r|r|r|r|r|r|r|r|}  \hline
         channel                     & \multicolumn{2}{c|}{$\tau_e\tau_h$} 
                                     & \multicolumn{2}{c|}{CMUP $\tau_{\mu}\tau_h$} 
                                     & \multicolumn{2}{c|}{CMX $\tau_{\mu}\tau_h$} 
                                     & \multicolumn{2}{c|}{$\tau_h\tau_h$} \\ \hline
         trigger path                & \multicolumn{2}{c|}{CELE8}    
                                     & \multicolumn{2}{c|}{CMUP8}             
                                     & \multicolumn{2}{c|}{TAU\_CMX}            
                                     & \multicolumn{2}{c|}{TAU\_MET}       \\
         denominator                 & \multicolumn{2}{c|}{D\_xi}          
                                     & \multicolumn{2}{c|}{D\_xi}             
                                     & \multicolumn{2}{c|}{D\_trkIso10Deg}      
                                     & \multicolumn{2}{c|}{D\_xi}          \\ 
         norm. factor                & \multicolumn{2}{c|}{4.239} 
                                     & \multicolumn{2}{c|}{5.132} 
                                     & \multicolumn{2}{c|}{1}
                                     & \multicolumn{2}{c|}{1}              \\  \hline
                                     & $\sum\omega^{jet}$ & event  
                                     & $\sum\omega^{jet}$ & event
                                     & $\sum\omega^{jet}$ & event
                                     & $\sum\omega^{jet}$ & event          \\ \cline{2-9} 
         $\sum\omega^{jet}$ or event &             92.1   & 2292     
                                     &             12.4   &  362     
                                     &             64.4   &  379     
                                     &            106.8   & 2778           \\
         kinematic cuts              &              0.068 &   13
                                     &              0.006 &    1
                                     &              0.152 &    4
                                     &              0.282 &   12           \\ \hline
         normalized                  & \multicolumn{2}{c|}{$0.29\pm0.08$}   
                                     & \multicolumn{2}{c|}{$0.03\pm0.03$}   
                                     & \multicolumn{2}{c|}{$0.15\pm0.08$}   
                                     & \multicolumn{2}{c|}{$0.28\pm0.08$}  \\ \hline
      \end{tabular}
      \caption[Fake background estimates in the signal region]
              {Fake background estimates in the signal region.
               Uncertainties are statistical.}
      \label{tab:signal_fake_1}
   \end{center}
\end{table}

There is a systematic uncertainty due to the uncertainty 
in the jet$\to\tau$ fake rate.  The 
rate used is the average fake rate of the JET samples.  
We use the individual fake rate of the JET20, JET50, 
JET70, and JET100 samples to estimate this uncertainty, 
as shown in Table~\ref{tab:signal_fake_2}. 

\begin{table}
   \begin{center}
      \begin{tabular}{|l|c|c|c|c|}  \hline
         channel   & $\tau_e\tau_h$ 
                   & CMUP $\tau_{\mu}\tau_h$
                   & CMX $\tau_{\mu}\tau_h$ 
                   & $\tau_h\tau_h$            \\ \hline
         average   & 0.29 & 0.03 & 0.15 & 0.28 \\ \hline
         JET20     & 0.18 & 0.03 & 0.16 & 0.31 \\
         JET50     & 0.23 & 0.04 & 0.15 & 0.23 \\
         JET70     & 0.31 & 0.03 & 0.14 & 0.25 \\
         JET100    & 0.28 & 0.03 & 0.13 & 0.22 \\ \hline
         syst. err & 0.11 & 0.01 & 0.02 & 0.06 \\ \hline
      \end{tabular}
      \caption[Systematic uncertainties on fake backgrounds in the signal region]
              {Systematic uncertainties of fake background estimates
               in the signal region.}
      \label{tab:signal_fake_2}
   \end{center}
\end{table}

Combining in quadrature the statistical 
uncertainties in
Table~\ref{tab:signal_fake_1} and the 
systematic uncertainties in
Table~\ref{tab:signal_fake_2},
we get
\begin{eqnarray}
                \tau_e\tau_h     \mbox{ fake} & = & 0.29\pm0.14 \\
   \mbox{CMUP } \tau_{\mu}\tau_h \mbox{ fake} & = & 0.03\pm0.03 \\
   \mbox{CMX }  \tau_{\mu}\tau_h \mbox{ fake} & = & 0.15\pm0.08 \\
                \tau_h\tau_h     \mbox{ fake} & = & 0.28\pm0.10
\end{eqnarray}


\section{Uncertainties in Signal Region}
\label{sec:signal_err}

We summarize all of the systematic uncertainties in
the high mass signal region in this section.  Some of 
these are due to statistical uncertainties on the 
various backgrounds due to limited Monte Carlo or 
other statistics.  Others come from separate 
external studies as indicated.  And in this
section, we combine the $\tau_{\mu}\tau_h$ with
CMUP muon channel and the $\tau_{\mu}\tau_h$ with
CMX muon channel into one single $\tau_{\mu}\tau_h$
channel.

The systematic uncertainty in the Drell-Yan and 
new particle signal rates due to the imperfect
knowledge of the parton density functions 
(PDF's)~\cite{Lai:1999wy} is calculated by 
comparing the acceptance change ratio for 
various PDF's. The CTEQ5L is used in PYTHIA.
We add in quadrature the difference between
MRST72  to CTEQ5L, 
MRST75  to MRST72,
CTEQ6L1 to CTEQ6L, and
CTEQ6M  to CTEQ5L PDF's.
The MRST72 and MRST75 compare the effect of varying 
$\alpha_s$ on the PDF.  The CTEQ5L set is leading 
order, and the CTEQ6M sets are next to leading 
order but at the same value of $\alpha_s$.  Using 
$\zprime\to\tau\tau$, this is shown in 
Table~\ref{tab:signal_err_1}.  
We take 8\% as a conservative number.

\begin{table}
   \begin{center}
      \begin{tabular}{|l|c|c|c|c|c|} \hline
         $\zprime\to\tau\tau$ & $m=120$ & $m=180$ & $m=300$ & $m=400$ & $m=600$ \\ \hline
         MRST72  / CTEQ5L     &   1.047 &   1.029 &   1.021 &   1.006 &   1.002 \\
         MRST75  / MRST72     &   0.951 &   0.980 &   0.983 &   0.995 &   0.993 \\
         CTEQ6L1 / CTEQ6L     &   1.006 &   1.006 &   1.003 &   0.999 &   1.002 \\
         CTEQ6M  / CTEQ5L     &   1.035 &   1.023 &   1.021 &   1.008 &   1.004 \\ \hline
         PDF uncertainty      &   7.7\% &   4.2\% &   3.4\% &   1.1\% &   0.8\% \\ \hline
      \end{tabular}
      \caption[PDF uncertainty]
              {PDF uncertainty.}
      \label{tab:signal_err_1}
   \end{center}
\end{table}

We are careful to identify the correlated and 
the uncorrelated systematic uncertainties.
The correlated uncertainties include 
the uncertainties of the PDF, the integrated 
luminosity, the $e$, $\mu$, $\tau$ scale 
factors, the $\met$, and the jet$\to\tau$ fake rate.
Table~\ref{tab:signal_err_2} lists the 
uncertainties, their magnitude, and the affected
channels. (When uncertainties are correlated we 
assume a 100\% correlation.)

\begin{table}
   \begin{center}
      \begin{tabular}{|c|c|c|} \hline
         Uncertainty            & Magnitude (\%) & Affected Channels  \\ \hline \hline
         PDF                    &        8       &        all         \\
         integrated luminosity  &        6       &        all         \\
         $e$ scale factor       &        4       & $\tau_e\tau_h$     \\
         $\mu$ scale factor     &      5.5       & $\tau_{\mu}\tau_h$ \\
         $\tau$ scale factor    &       10       &        all         \\
         $\met$                 &        6       &        all         \\
         jet$\to\tau$ fake rate &       20       &        all         \\ \hline
      \end{tabular}
      \caption[Systematic uncertainties and the affected channels]
              {Systematic uncertainties, in percent, 
               and the affected channels.}
      \label{tab:signal_err_2}
   \end{center}
\end{table}

The $\zprime\to\tau\tau$ and $A\to\tau\tau$ signal 
acceptances and the systematic uncertainties are listed in 
Table~\ref{tab:signal_err_3}$-$\ref{tab:signal_err_4}.
The acceptance itself reflects the effects of trigger 
efficiency and the lepton scale factors.
The uncertainties include the contributions from
\begin{itemize}
\item statistical uncertainty (MC statistics),
\item PDF uncertainty (this Section),
\item trigger efficiencies (see Section~\ref{sec:event_trigEff}),
\item lepton scale factors (see Section~\ref{subsec:TauSF_sf} for $\tau$ scale factor,
                                Section~\ref{subsec:EleSF} for $e$ scale factor, 
                            and Section~\ref{subsec:MuSF} for $\mu$ scale factors), and
\item $\met$ uncertainty (see Section~\ref{sec:MET}).
\end{itemize}

\begin{table}
   \begin{center}
      \begin{tabular}{|c|c|c|c|c|} \hline
         $m(\zprime)$
              & $\tau_e\tau_h$ (\%) & $\tau_{\mu}\tau_h$ (\%) & $\tau_h\tau_h$ (\%) & combined (\%)   \\ \hline \hline
         120  & $0.012\pm0.004$     & $0.013\pm0.004$         & $0.020\pm0.005$     & $0.045\pm0.009$ \\ 
         140  & $0.084\pm0.015$     & $0.088\pm0.015$         & $0.105\pm0.020$     & $0.278\pm0.043$ \\ 
         160  & $0.213\pm0.035$     & $0.151\pm0.025$         & $0.206\pm0.038$     & $0.571\pm0.086$ \\ 
         180  & $0.308\pm0.049$     & $0.234\pm0.037$         & $0.365\pm0.066$     & $0.907\pm0.136$ \\ 
         200  & $0.453\pm0.070$     & $0.351\pm0.054$         & $0.476\pm0.085$     & $1.280\pm0.190$ \\ 
         250  & $0.727\pm0.111$     & $0.548\pm0.083$         & $0.776\pm0.137$     & $2.052\pm0.30$3 \\ 
         300  & $0.975\pm0.148$     & $0.735\pm0.110$         & $0.963\pm0.170$     & $2.673\pm0.394$ \\ 
         350  & $1.098\pm0.167$     & $0.826\pm0.124$         & $1.144\pm0.202$     & $3.068\pm0.452$ \\ 
         400  & $1.239\pm0.188$     & $0.966\pm0.144$         & $1.308\pm0.230$     & $3.512\pm0.517$ \\ 
         450  & $1.395\pm0.211$     & $0.979\pm0.146$         & $1.382\pm0.243$     & $3.756\pm0.553$ \\ 
         500  & $1.537\pm0.232$     & $1.148\pm0.172$         & $1.431\pm0.252$     & $4.116\pm0.604$ \\ 
         600  & $1.572\pm0.237$     & $1.190\pm0.178$         & $1.615\pm0.284$     & $4.377\pm0.644$ \\ \hline
      \end{tabular}
      \caption[Uncertainties of $\zprime\to\tau\tau$ signal acceptance]
              {Uncertainties of $f\bar{f}\to\zprime\to\tau\tau$
               signal acceptance (SM coupling).}
      \label{tab:signal_err_3} 
   \end{center}
\end{table}

\begin{table}
   \begin{center}
      \begin{tabular}{|c|c|c|c|c|} \hline
         $m(A)$
              & $\tau_e\tau_h$ (\%) & $\tau_{\mu}\tau_h$ (\%) & $\tau_h\tau_h$ (\%) & combined (\%)   \\ \hline \hline
         120  & $0.020\pm0.005$     & $0.018\pm0.005$         & $0.029\pm0.007$     & $0.067\pm0.012$ \\ 
         140  & $0.113\pm0.019$     & $0.082\pm0.015$         & $0.126\pm0.024$     & $0.321\pm0.050$ \\ 
         160  & $0.284\pm0.045$     & $0.213\pm0.034$         & $0.324\pm0.058$     & $0.822\pm0.123$ \\ 
         180  & $0.474\pm0.074$     & $0.376\pm0.058$         & $0.445\pm0.079$     & $1.295\pm0.191$ \\ 
         200  & $0.603\pm0.093$     & $0.485\pm0.074$         & $0.660\pm0.117$     & $1.748\pm0.259$ \\ 
         250  & $0.889\pm0.135$     & $0.703\pm0.106$         & $0.972\pm0.172$     & $2.564\pm0.379$ \\ 
         300  & $1.174\pm0.178$     & $0.846\pm0.127$         & $1.132\pm0.199$     & $3.151\pm0.463$ \\ 
         350  & $1.254\pm0.190$     & $1.004\pm0.150$         & $1.356\pm0.239$     & $3.614\pm0.532$ \\ 
         400  & $1.411\pm0.213$     & $1.101\pm0.165$         & $1.485\pm0.261$     & $3.996\pm0.588$ \\ 
         450  & $1.460\pm0.220$     & $1.135\pm0.170$         & $1.573\pm0.277$     & $4.168\pm0.614$ \\ 
         500  & $1.649\pm0.249$     & $1.177\pm0.176$         & $1.561\pm0.275$     & $4.386\pm0.644$ \\ 
         600  & $1.758\pm0.265$     & $1.300\pm0.194$         & $1.684\pm0.296$     & $4.742\pm0.696$ \\ \hline 
      \end{tabular}
      \caption[Uncertainties of $A\to\tau\tau$ signal acceptance]
              {Uncertainties of $gg\to A\to\tau\tau$ 
               signal acceptance ($\tan\beta = 20$).}
      \label{tab:signal_err_4} 
   \end{center}
\end{table}

The systematic uncertainties on the Drell-Yan 
backgrounds and the jet$\to\tau$ misidentified
fake backgrounds are listed in
Table~\ref{tab:signal_err_5}.
The systematic uncertainties on the Drell-Yan
backgrounds incorporate the effects of
\begin{itemize}
\item statistical uncertainty (MC statistics),
\item PDF uncertainty (this Section),
\item $\sigma\cdot B$ uncertainty, 2\%, aside 
      from luminosity uncertainty (see Ref.~\cite{Acosta:2004uq}),
\item trigger efficiencies (see Section~\ref{sec:event_trigEff}),
\item lepton scale factors (see Section~\ref{subsec:TauSF_sf} for $\tau$ scale factor,
                                Section~\ref{subsec:EleSF} for $e$ scale factor, 
                            and Section~\ref{subsec:MuSF} for $\mu$ scale factors),
\item $\met$ uncertainty (see Section~\ref{sec:MET}), and
\item luminosity, 6\% (see Ref.~\cite{Klimenko:2003if}).
\end{itemize}
The systematic uncertainties on the jet$\to\tau$ 
misidentified fake background incorporates the 
effects of
\begin{itemize}
\item statistical uncertainty (see Section~\ref{sec:signal_fake}), and
\item systematic uncertainty due to jet$\to\tau$ 
      misidentification rate (see Section~\ref{sec:signal_fake}).
\end{itemize}

\begin{table}
   \begin{center}
      \begin{tabular}{|c|c|c|c|c|} \hline
         Source                  & $\tau_e\tau_h$ & $\tau_{\mu}\tau_h$ & $\tau_h\tau_h$ &         Total \\ \hline
         $Z/\gamma^*\to\tau\tau$ &  $0.56\pm0.11$ &      $0.50\pm0.10$ &  $0.36\pm0.08$ & $1.42\pm0.23$ \\
         $Z/\gamma^*\to ee$      &  $0.16\pm0.14$ &                  0 &              0 & $0.16\pm0.14$ \\
         $Z/\gamma^*\to\mu\mu$   &              0 &      $0.50\pm0.26$ &              0 & $0.50\pm0.26$ \\
         Jet$\to\tau$            &  $0.29\pm0.14$ &      $0.18\pm0.09$ &  $0.28\pm0.10$ & $0.75\pm0.19$ \\ \hline
         Expected                &  $1.01\pm0.24$ &      $1.18\pm0.30$ &  $0.64\pm0.13$ & $2.83\pm0.46$ \\ \hline
      \end{tabular}
      \caption[Uncertainties of backgrounds in the signal region]
              {Uncertainties of backgrounds in signal region, 
               195 pb$^{-1}$ (72 pb$^{-1}$ for $\tau_h\tau_h$).}
      \label{tab:signal_err_5}
   \end{center}
\end{table}



\chapter{Results}
\label{cha:results}


\section{Observed Events}
\label{sec:box}

\begin{figure}
   \begin{center}
      \parbox{5.7in}{\epsfxsize=\hsize\epsffile{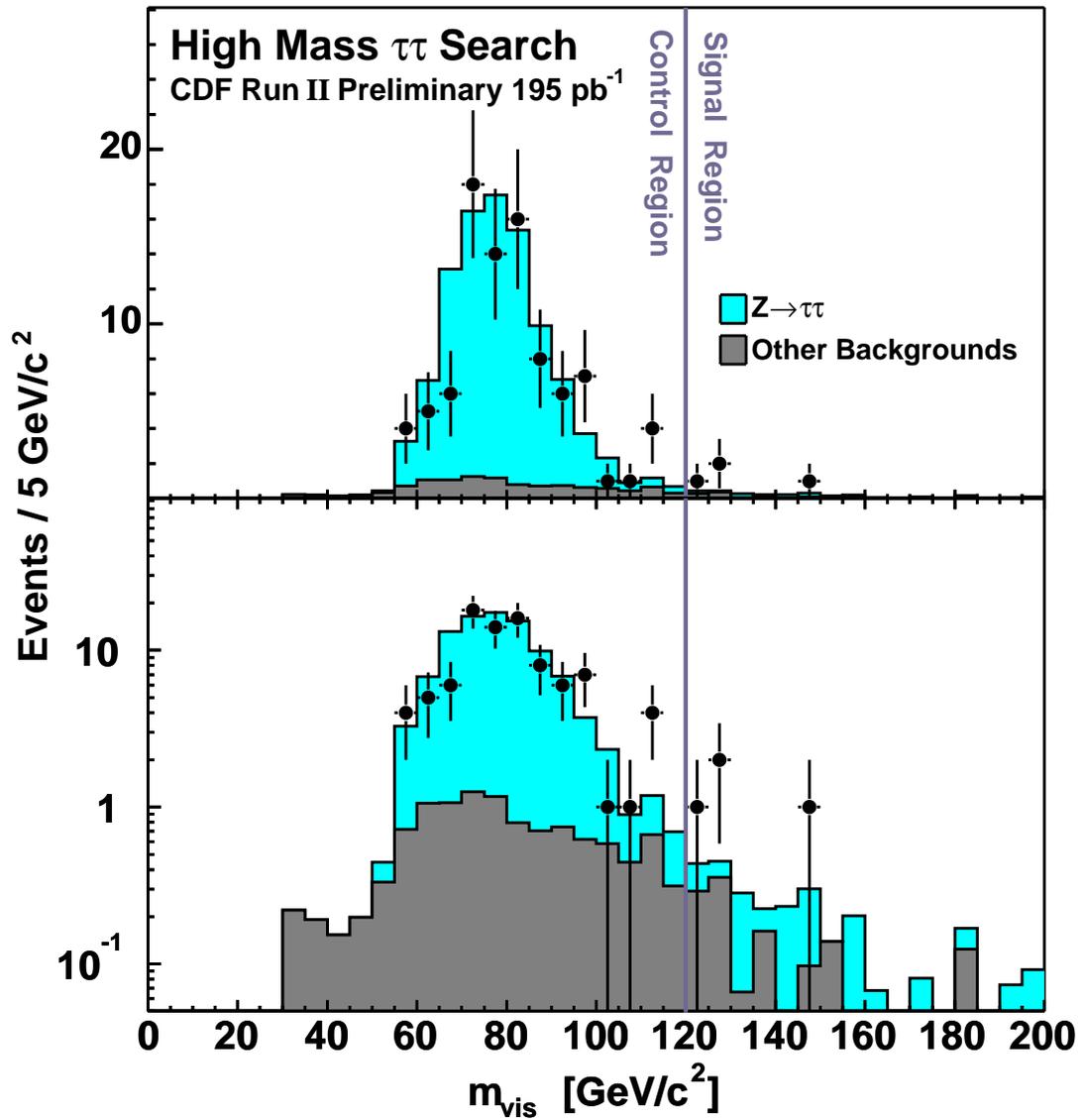}}
      \caption[Distribution of visible mass in the signal and control regions]
              {Distribution of visible mass ($m_{vis}$) 
               for data (points) and predicted backgrounds (histograms)
               in the signal and control regions. The upper plot
               is in linear scale. The lower plot is in log scale.}
      \label{fig:results_1}
   \end{center}
\end{figure}

\begin{figure}
   \begin{center}
      \parbox{5.7in}{\epsfxsize=\hsize\epsffile{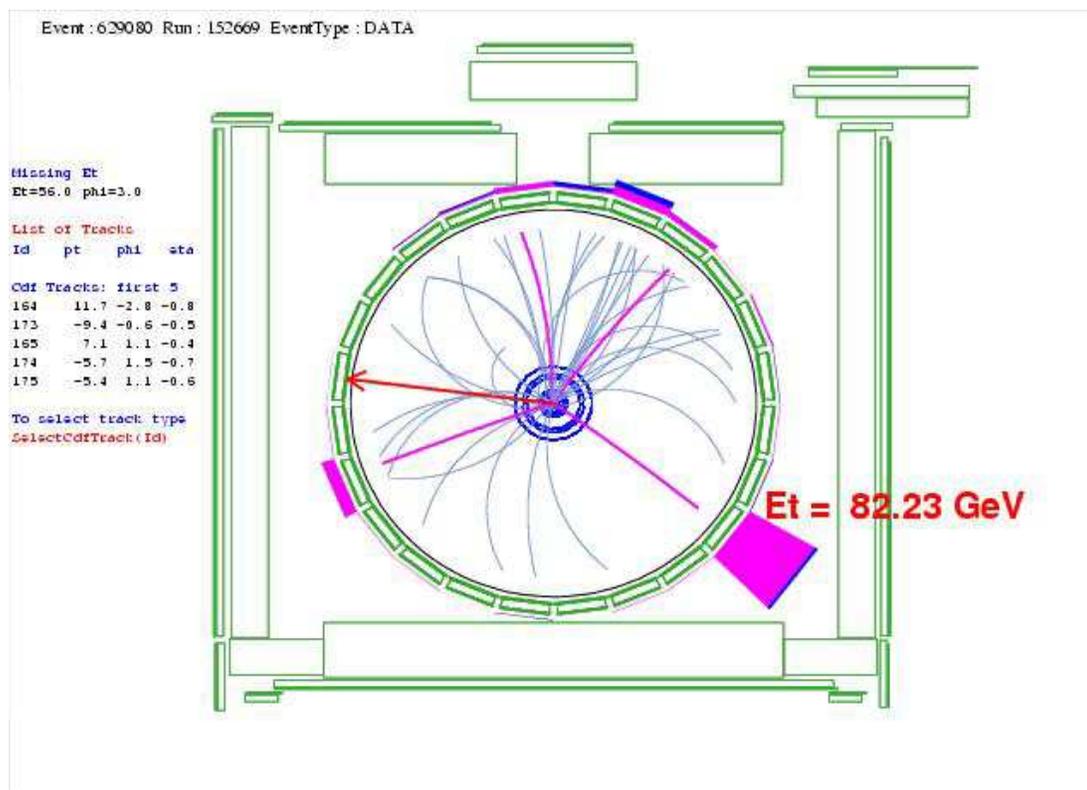}}
   \end{center}
   \begin{center}
      \parbox{5.7in}{\epsfxsize=\hsize\epsffile{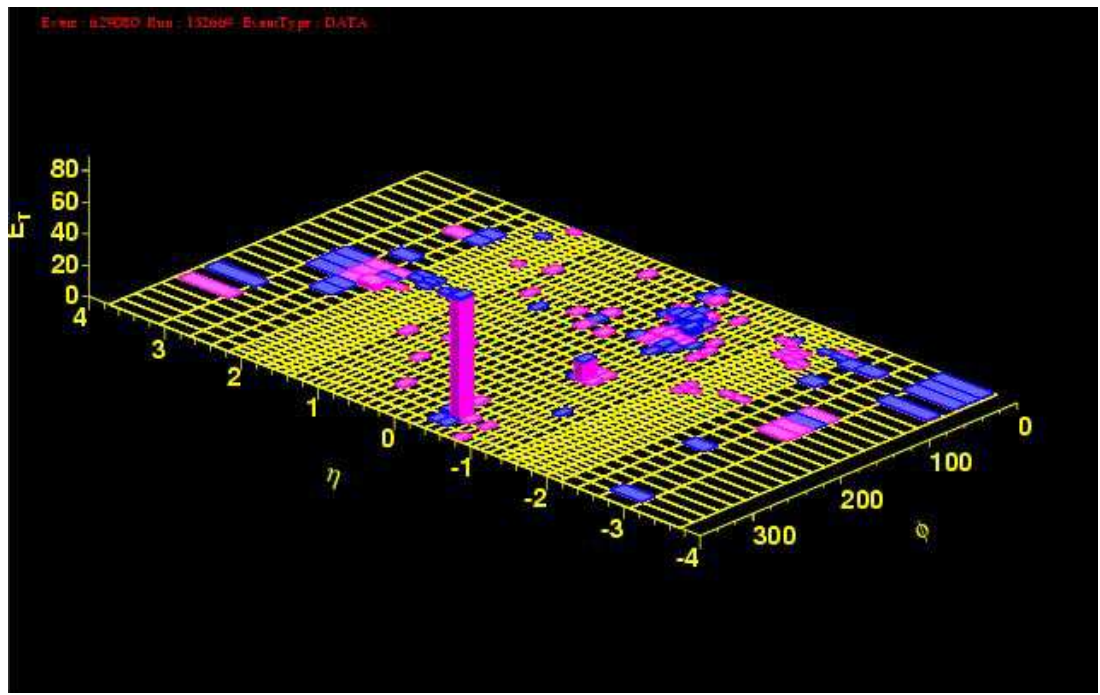}}
   \end{center}
   \begin{center}
      \caption[$\tau_e\tau_h$ candidate run=152669 event=629080 $m_{vis}$=148 GeV/$c^2$]
              {Event display $\tau_e\tau_h$ candidate 
               run=152669 
               event=629080 
               $m_{vis}$=148 GeV/$c^2$.}
      \label{fig:results_2}
   \end{center}
\end{figure}

\begin{figure}
   \begin{center}
      \parbox{5.7in}{\epsfxsize=\hsize\epsffile{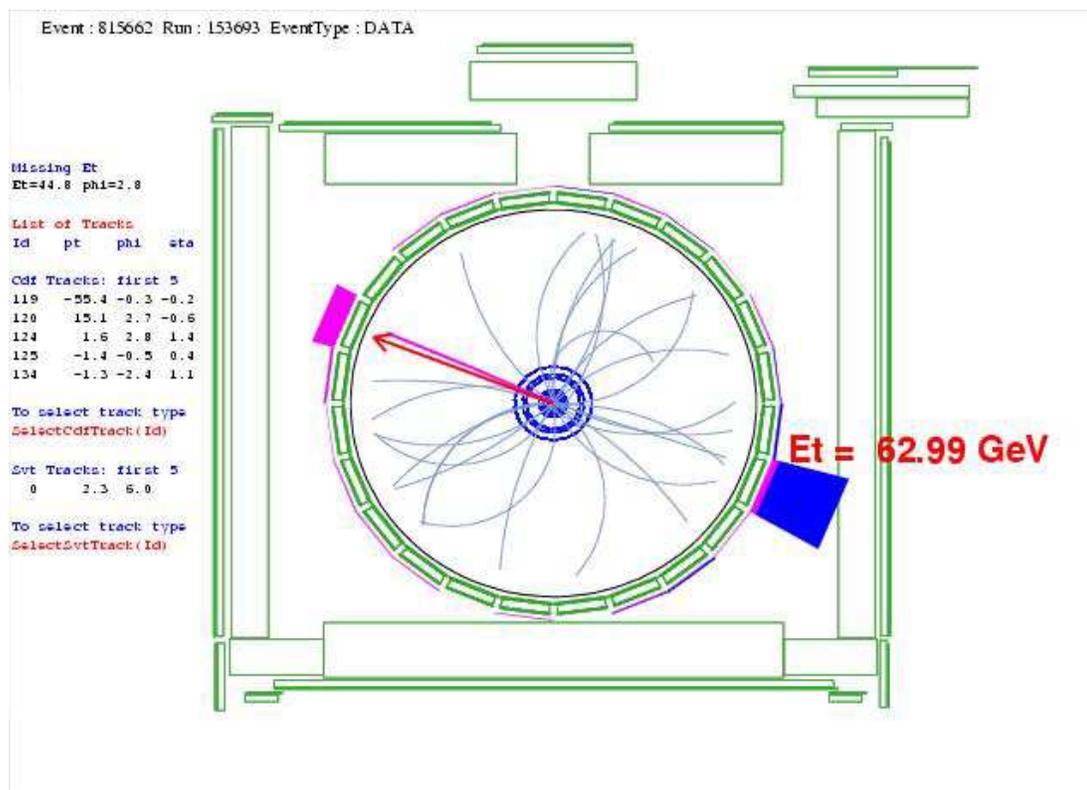}}
   \end{center}
   \begin{center}
      \parbox{5.7in}{\epsfxsize=\hsize\epsffile{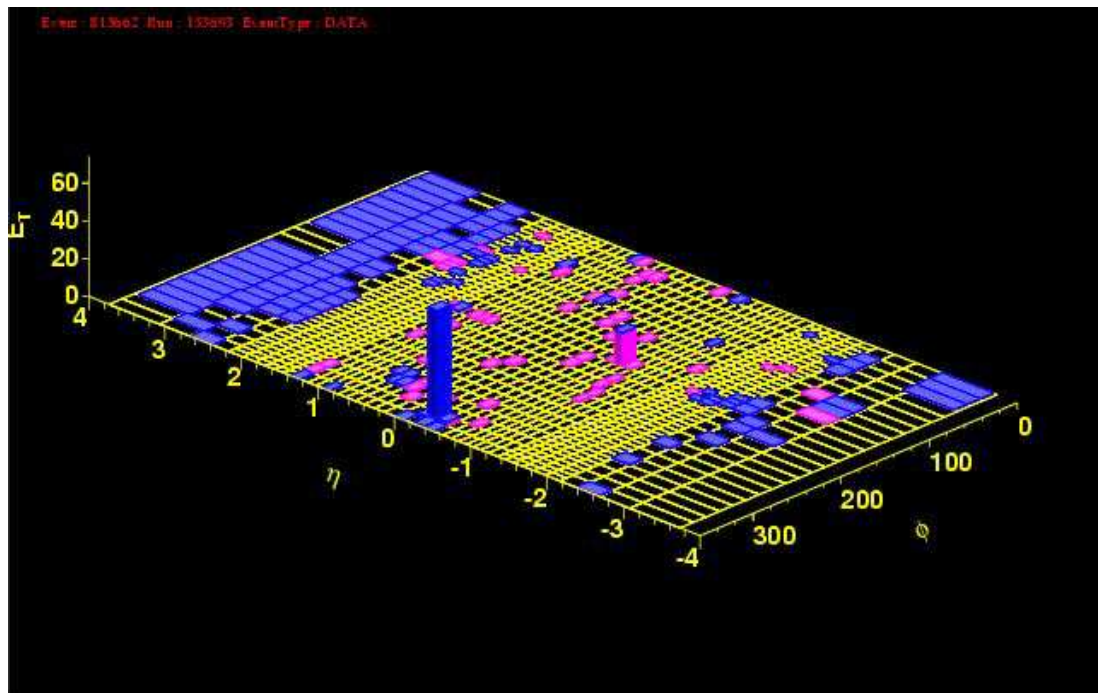}}
   \end{center}
   \begin{center}
      \caption[$\tau_e\tau_h$ candidate run=153693 event=815662 $m_{vis}$=129 GeV/$c^2$]
              {Event display $\tau_e\tau_h$ candidate 
               run=153693 
               event=815662 
               $m_{vis}$=129 GeV/$c^2$.}
      \label{fig:results_3}
   \end{center}
\end{figure}

\begin{figure}
   \begin{center}
      \parbox{5.7in}{\epsfxsize=\hsize\epsffile{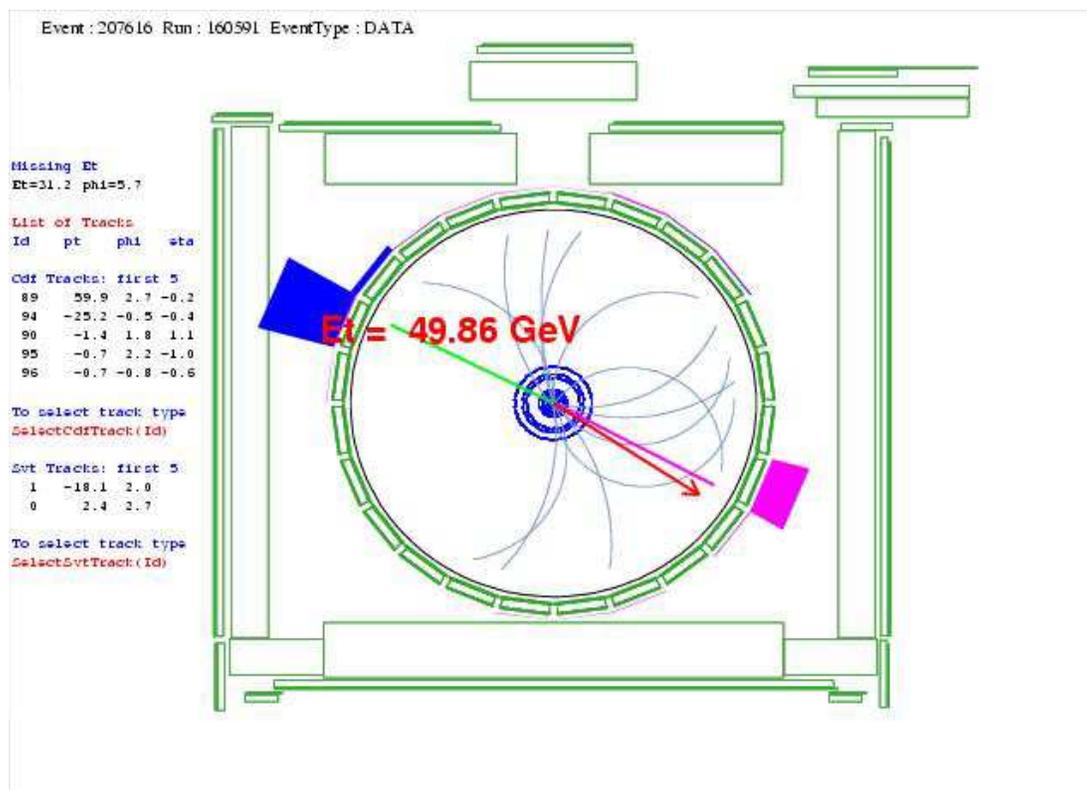}}
   \end{center}
   \begin{center}
      \parbox{5.7in}{\epsfxsize=\hsize\epsffile{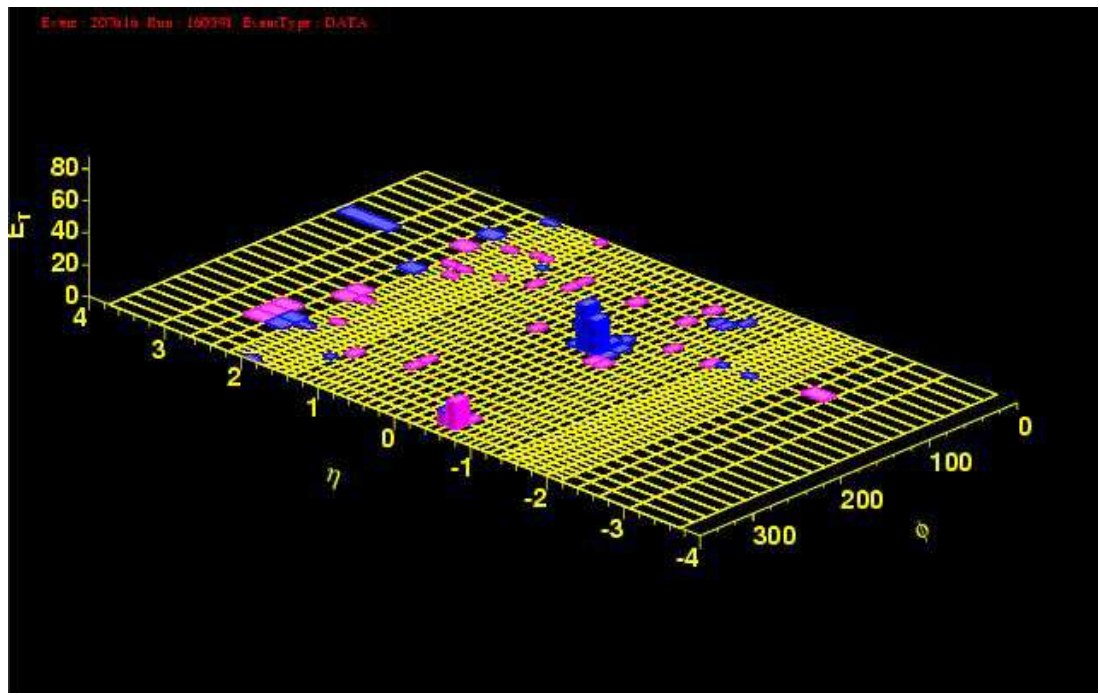}}
   \end{center}
   \begin{center}
      \caption[$\tau_e\tau_h$ candidate run=160591 event=207616 $m_{vis}$=125 GeV/$c^2$]
              {Event display $\tau_e\tau_h$ candidate 
               run=160591 
               event=207616 
               $m_{vis}$=125 GeV/$c^2$.}
      \label{fig:results_4}
   \end{center}
\end{figure}

\begin{figure}
   \begin{center}
      \parbox{5.7in}{\epsfxsize=\hsize\epsffile{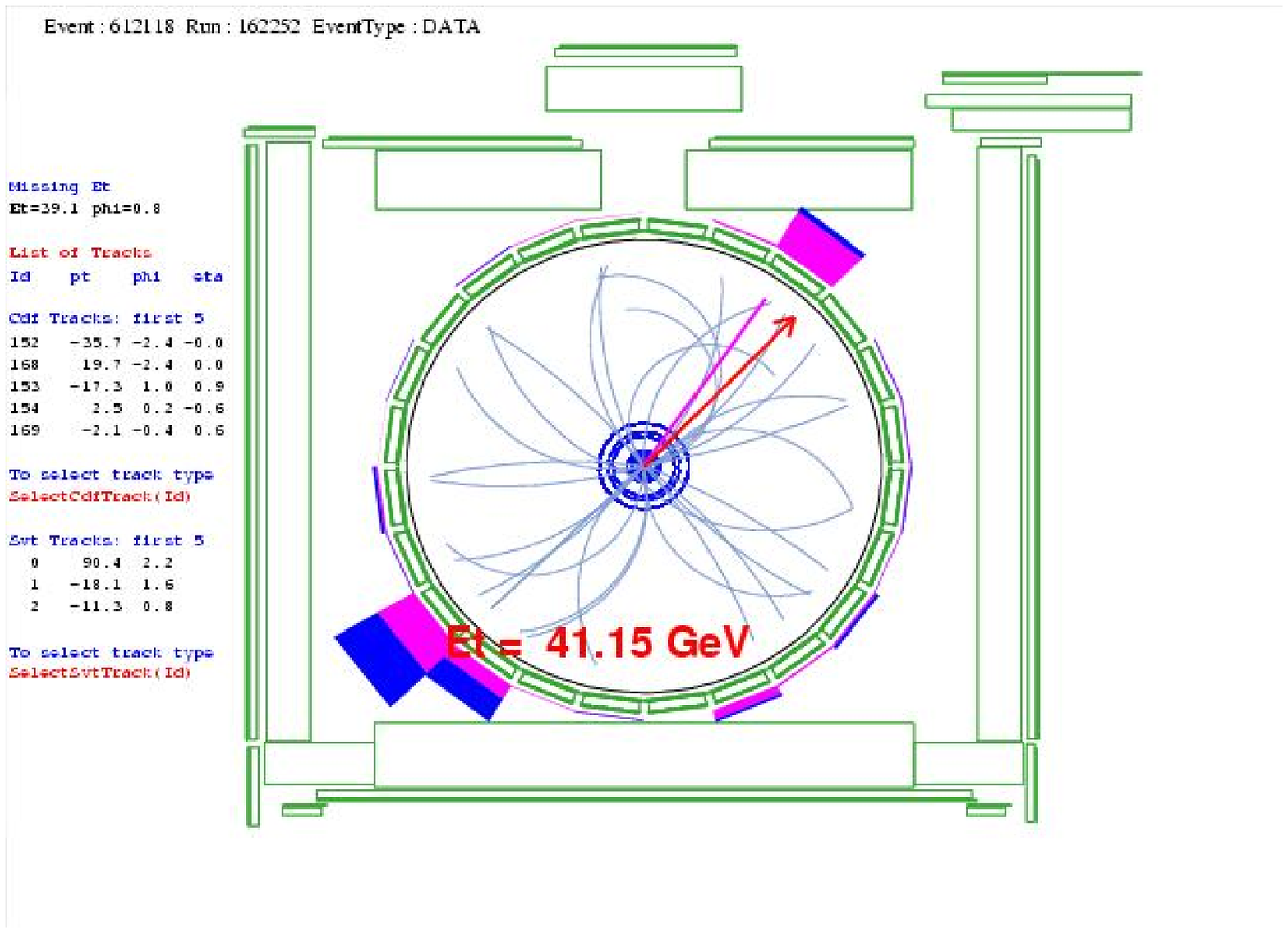}}
   \end{center}
   \begin{center}
      \parbox{5.7in}{\epsfxsize=\hsize\epsffile{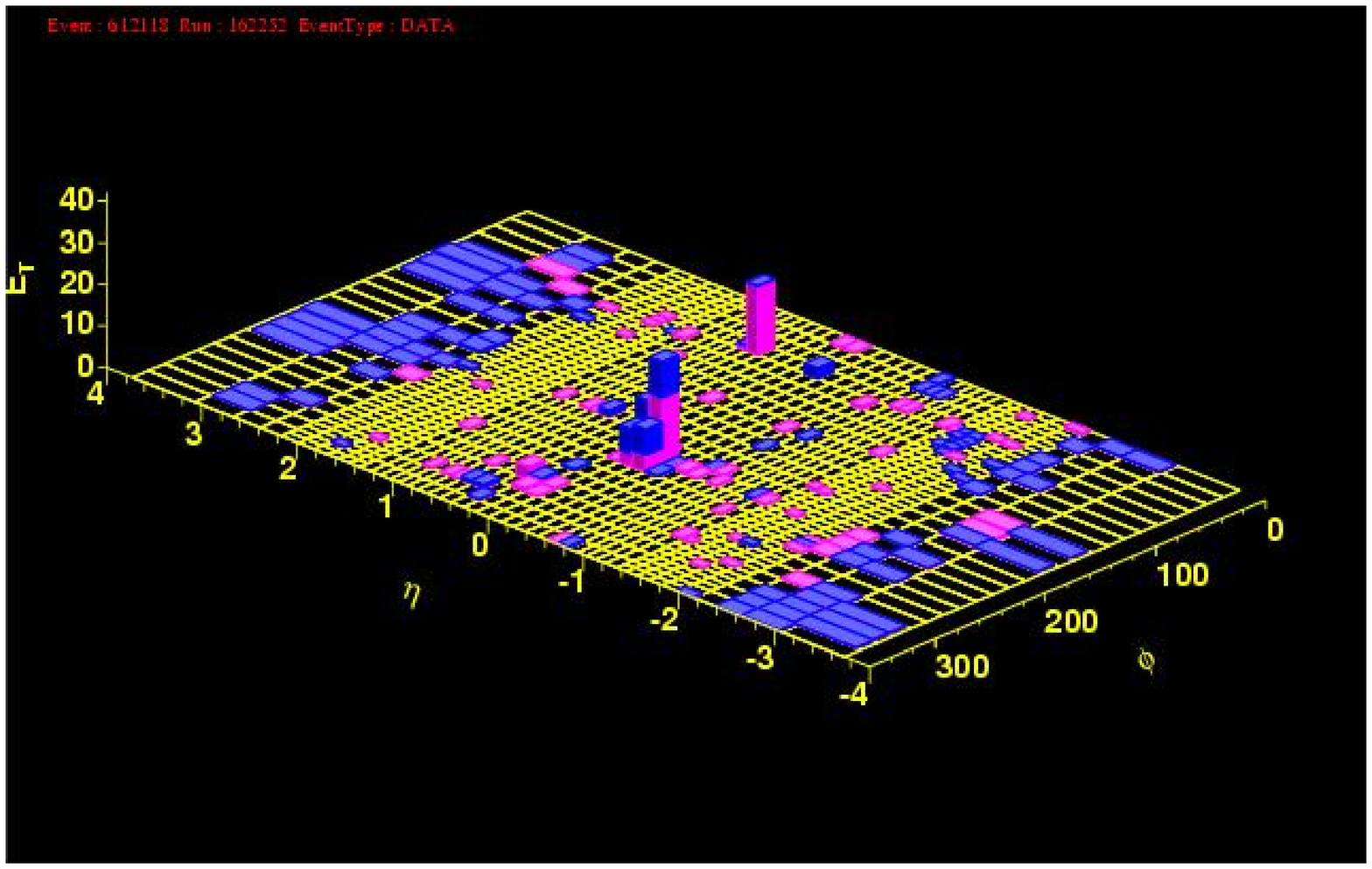}}
   \end{center}
   \begin{center}
      \caption[$\tau_e\tau_h$ candidate run=162252 event=612118 $m_{vis}$=124 GeV/$c^2$]
              {Event display $\tau_e\tau_h$ candidate 
               run=162252 
               event=612118 
               $m_{vis}$=124 GeV/$c^2$.}
      \label{fig:results_5}
   \end{center}
\end{figure}

After unblinding the signal region, we observe four 
events in $\tau_e\tau_h$ channel, zero events in 
$\tau_{\mu}\tau_h$ channel, and zero events in
$\tau_h\tau_h$ channel.  The numbers of background 
events estimated and observed are in 
Table~\ref{tab:results_1}.
Fig.~\ref{fig:results_1} shows the $m_{vis}$
distribution.
Fig.~\ref{fig:results_2}$-$\ref{fig:results_5}
shows the event displays of the four events 
observed in $\tau_e\tau_h$ channel. 

\begin{table}
   \begin{center}
      \begin{tabular}{|c|c|c|c|c|} \hline
         Source                  & $\tau_e\tau_h$ & $\tau_{\mu}\tau_h$ & $\tau_h\tau_h$ &         Total \\ \hline
         $Z/\gamma^*\to\tau\tau$ &  $0.56\pm0.11$ &      $0.50\pm0.10$ &  $0.36\pm0.08$ & $1.42\pm0.23$ \\
         $Z/\gamma^*\to ee$      &  $0.16\pm0.14$ &                  0 &              0 & $0.16\pm0.14$ \\
         $Z/\gamma^*\to\mu\mu$   &              0 &      $0.50\pm0.26$ &              0 & $0.50\pm0.26$ \\
         Jet$\to\tau$            &  $0.29\pm0.14$ &      $0.18\pm0.09$ &  $0.28\pm0.10$ & $0.75\pm0.19$ \\ \hline
         Expected                &  $1.01\pm0.24$ &      $1.18\pm0.30$ &  $0.64\pm0.13$ & $2.83\pm0.46$ \\ \hline
         Observed                &              4 &                  0 &              0 &             4 \\ \hline
      \end{tabular}
      \caption[Expected and observed events in the signal region]
              {Number of expected events for each channel
               and each source, and number of
               observed events, in the signal region.}
      \label{tab:results_1}
   \end{center}
\end{table}

\section{Experimental Limits}
\label{sec:limits}

Since we observe no excess, we proceed to calculate 
the 95\% confidence level (CL) upper limit on the 
cross section times branching ratio for new particle
production using a Bayesian procedure described in 
Ref.~\cite{CDFnote:6428}.

We need to combine multiple search channels and 
incorporate both uncorrelated and correlated
systematic uncertainties.
For each channel $i$, 
the integrated luminosity,
the signal acceptance,
the expected background events, and 
the observed events are denoted as
$L_i$,
$\epsilon_i$,
$b_i$, and
$n_i$,
respectively;
the uncorrelated uncertainties of 
the signal acceptance and
the expected background events
are denoted as
$f_{\epsilon i}$ and 
$f_{bi}$, 
respectively.
The correlated uncertainties of
the integrated luminosity,
the signal acceptance, and
the expected background events
are denoted as
$g_L$,
$g_{\epsilon}$, and
$g_b$,
respectively.
(Note that the $f$ factors carry $i$ indices and 
the $g$ factors do not.)
With a signal cross section $\sigma_{sig}$, the expected 
number of event $\mu_i$ in each channel can be written as
\begin{equation}
   \mu_i
    = (1+g_L)L_i\sigma_{sig}(1+f_{\epsilon i})(1+g_{\epsilon})\epsilon_i
      +(1+f_{bi})(1+g_b)b_i
\end{equation}
where the $f$ and $g$ factors are in a form 
$1+x$ thus \emph{relative} systematic uncertainties. 
We define a likelihood which is the product of the 
Poisson probabilities of observing $n_i$ events 
in each channel, 
\begin{equation}
   \mathcal{L}(\bar{n}|\sigma_{sig},\bar{b},\bar{\epsilon}) =
   \prod_i\mathcal{L}(n_i|\mu_i) = 
   \prod_i\frac{\mu_i^{n_i}e^{-\mu_i}}{n_i!} 
\end{equation}
where the overbars indicate that the variables
are arrays carrying an $i$ index.
We use a Monte Carlo method to
convolute the effects of the systematic 
uncertainties using Gaussian prior 
probability density functions 
for the $f$ and $g$ factors.
For an evaluating point of the $\sigma_{sig}$,
we sample the $f$ and $g$ factors 
within
their Gaussian widths around a central value of
zero, calculate the $\mu_i$ 
and 
the 
$\mathcal{L}(n_i|\mu_i)$ for each channel, 
and average the resulting likelihood
$\mathcal{L}(\bar{n}|\sigma_{sig},\bar{b},\bar{\epsilon})$. 
Using Bayes' Theorem,
we then construct a probability density function
for the signal cross section,
\begin{equation}
   \mathcal{P}(\sigma_{sig}|\bar{n},\bar{b},\bar{\epsilon}) 
   = \frac{                \mathcal{L}(n|\sigma_{sig}, \bar{b},\bar{\epsilon})P(\sigma_{sig})                } 
          {\int_0^{\infty} \mathcal{L}(n|\sigma_{sig}',\bar{b},\bar{\epsilon})P(\sigma_{sig}') d\sigma_{sig}'}
\end{equation}
with 
a prior 
probability density function 
$P(\sigma_{sig})$ which expresses the subjective
``degree of belief'' for the value of the 
signal cross section.
The 95\% CL upper limit $\sigma_{95}$ is obtained
by solving this integral equation
\begin{equation}
   \int_0^{\sigma_{95}} 
   \mathcal{P}(\sigma_{sig}|\bar{n},\bar{b},\bar{\epsilon}) 
   d\sigma_{sig} 
   = 0.95
   \label{eqn:sigma95}
\end{equation}
We assume a uniform prior in the signal cross section
up to some high cutoff; the value of the cutoff has
no significant influence on the 95\% CL upper limit.

\begin{figure}
   \begin{center}
      \parbox{5.5in}{\epsfxsize=\hsize\epsffile{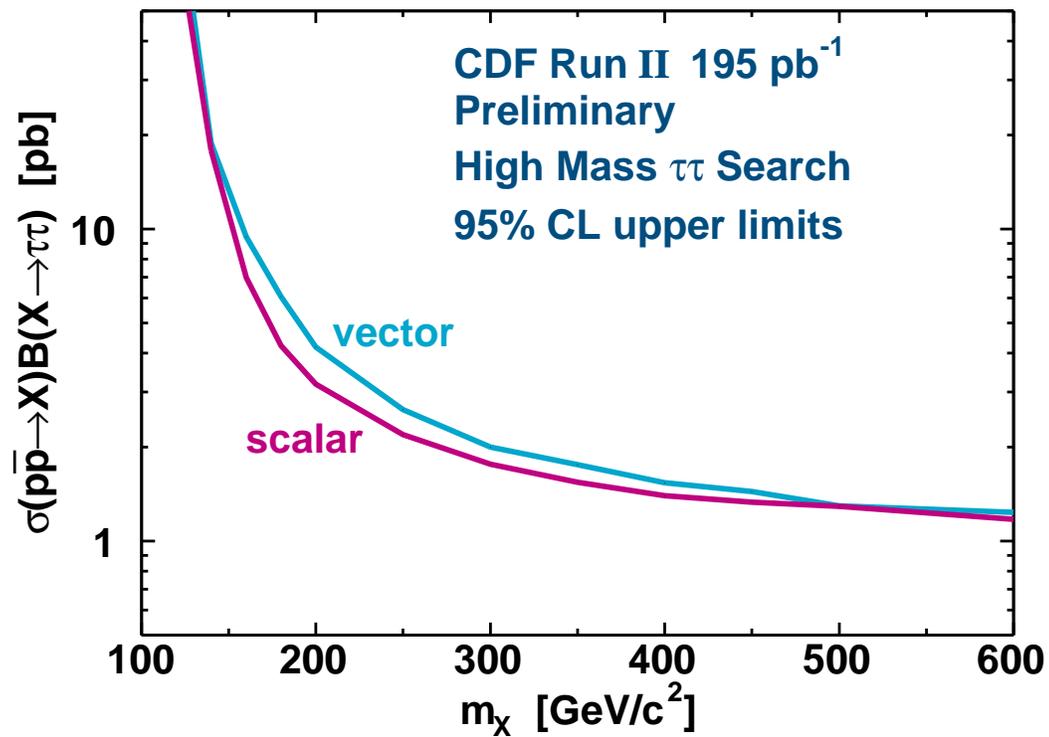}}
      \caption[Upper limits at 95\% CL for vector and scalar bosons]
              {Upper limits at 95\% CL on
               $\sigma(p\bar{p}\to X)B(X\to\tau\tau)$ 
               for vector and scalar boson,
               as a function of mass.}
      \label{fig:results_6}
   \end{center}
\end{figure}

We thereby extract the experimental 95\% CL
upper limit of $\sigma\cdot B$ for models 
using vector boson and scalar boson, 
respectively.
The results are listed in 
Table~\ref{tab:results_2} 
and shown in
Fig.~\ref{fig:results_6}.  
These are the generic limits for 
$gg\to X_{\mbox{scalar}}\to\tau\tau$
and
$f\bar{f}\to X_{\mbox{vector}}\to\tau\tau$
which can be interpreted in various models.

\begin{table}
   \begin{center}
      \begin{tabular}{|c|c|c|} \hline
            mass    &   vector    &  scalar         \\
        (GeV/$c^2$) &  limit (pb) & limit (pb)      \\ \hline
            120     &  122.294    &  87.338         \\
            140     &   18.884    &  17.899         \\
            160     &    9.446    &   6.996         \\
            180     &    6.066    &   4.229         \\
            200     &    4.185    &   3.187         \\
            250     &    2.637    &   2.192         \\
            300     &    1.999    &   1.764         \\
            350     &    1.757    &   1.540         \\
            400     &    1.537    &   1.396         \\
            450     &    1.441    &   1.330         \\
            500     &    1.296    &   1.290         \\
            600     &    1.237    &   1.174         \\ \hline
      \end{tabular}
      \caption[The 95\% CL upper limits]
              {The 95\% CL upper limits on vector
               and scalar particle production and
               decay to tau pairs.}
      \label{tab:results_2}
   \end{center}
\end{table}

\section{Exclusion Regions}
\label{sec:exclusion}

Now we can put the theoretical predictions on high mass
tau pair production discussed in Section~\ref{sec:theory_tt}
and the experimental 95\% CL upper limits together. 
We take the region where the theoretical prediction is bigger than
the upper limit to be excluded~at~95\%~CL.

\begin{figure}
   \begin{center}
      \parbox{5.5in}{\epsfxsize=\hsize\epsffile{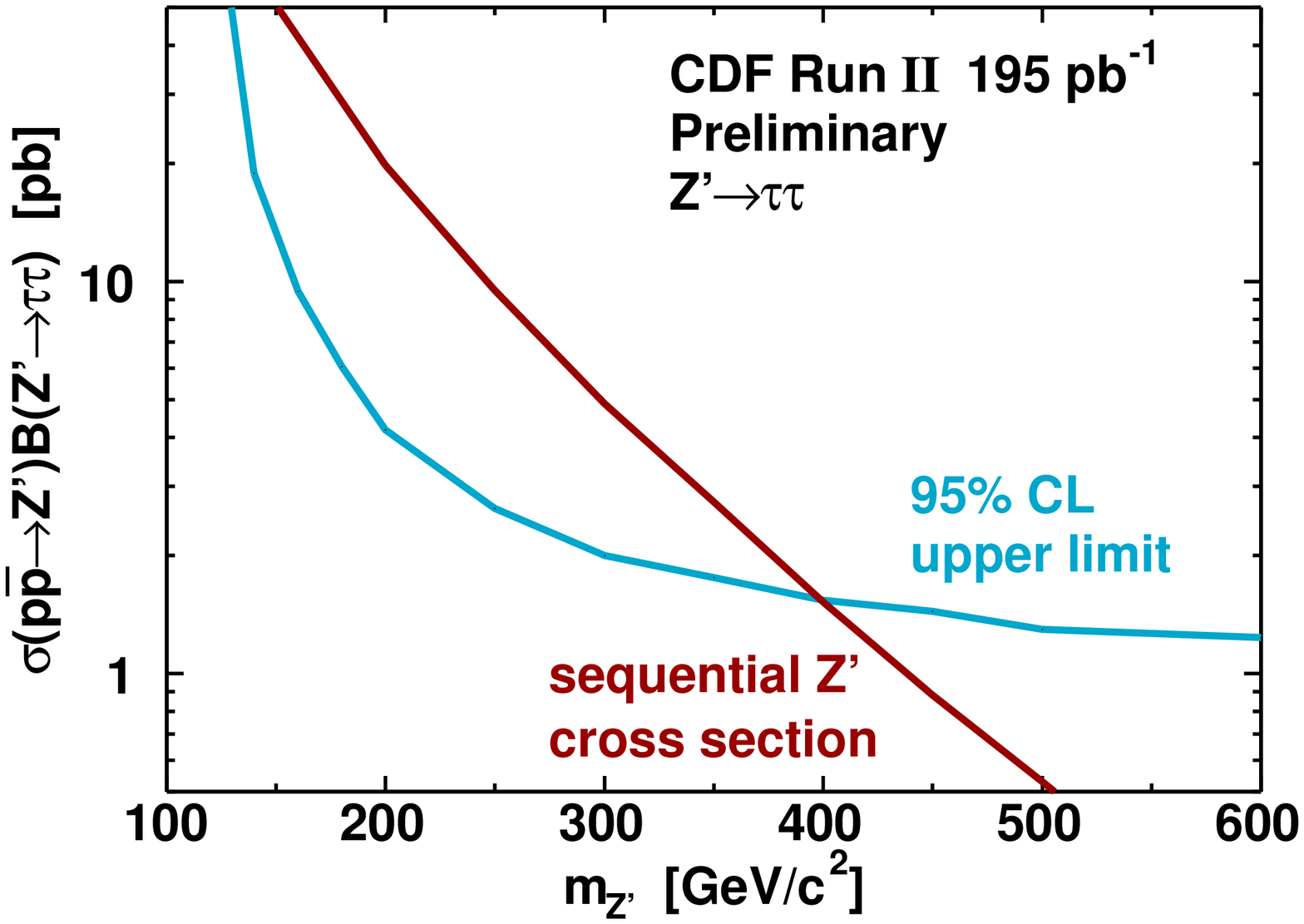}}
      \caption[Exclusion region for $\zprime$]
              {Upper limits at 95\% CL and theoretical predictions of
               $\sigma(p\bar{p}\to\zprime)B(\zprime\to\tau\tau)$.
               The excluded region is the region with $m(Z')<394$ GeV/$c^2$.} 
      \label{fig:results_7}
   \end{center}
\end{figure}

For reference, this analysis would thus exclude at 95\%
CL a $\zprime$ with standard model couplings having a
mass of less than 394 GeV/$c^2$, as shown in
Fig.~\ref{fig:results_7}.  For the MSSM pseudoscalar
Higgs boson $A$, this analysis is not sensitive to exclude
a region yet.

\chapter{Conclusions}
\label{cha:conclusions}

We have performed a blind search for high mass
tau pairs using data corresponding to 195
pb$^{-1}$ of integrated luminosity from Run II
of the Tevatron, using the CDF detector.  In 
the high-mass region with $m_{vis}>120$ GeV/$c^2$, 
we expect $2.8\pm0.5$ events from known background
sources, and observe $4$ events in the data
sample.  Thus no significant excess is observed,
and we use the result to set upper limits on 
the cross section times branching ratio to tau 
pairs of scalar and vector particles as a 
function of mass, 
shown in
Table~\ref{tab:results_2} 
and ploted in
Fig.~\ref{fig:results_6}.

\appendix


\chapter{The Structure of the Standard Model}
\label{cha:app_structure}

      The fundamental constituents of matter in Nature are 
      fermions: leptons and quarks, with interactions specified by
      the gauge symmetries
      SU(3)$_C$$\times$SU(2)$_L$$\times$U(1)$_Y$
      in the framework of the Standard Model (SM).
 
      Why are the fermions in an electroweak doublet?
      Why are left-handed fermions in a doublet, and 
             right-handed fermions in a singlet?
      What tells us that quarks have color degrees-of-freedom?
      Why must quark doublets be paired with lepton doublets? 
      Here we come to a brief review of how the structure of 
      the SM emerged.  A good introduction can be found in
      Ref.~\cite{Field:1994}.

      The relationships among the fermions are interpreted from 
      the interactions they experience, namely the cross sections and 
      decay widths measured, calculated, and measured ...  an 
      interplay of experimental inputs and theoretical constraints.  
      The objective is to unify the different interactions. 


      \vspace{0.2in}
      \noindent{\bf Charged Current}
      \vspace{0.1in}

      Let us recall Fermi's theory~\cite{Fermi:1934sk}
      of charged current (CC) 
      weak interaction for four fermions, e.g. the 
      crossed $\beta$-decay, $e p \to n\nu_e$,
      \begin{center}
         \parbox{2.0in}{\epsfxsize=\hsize\epsffile{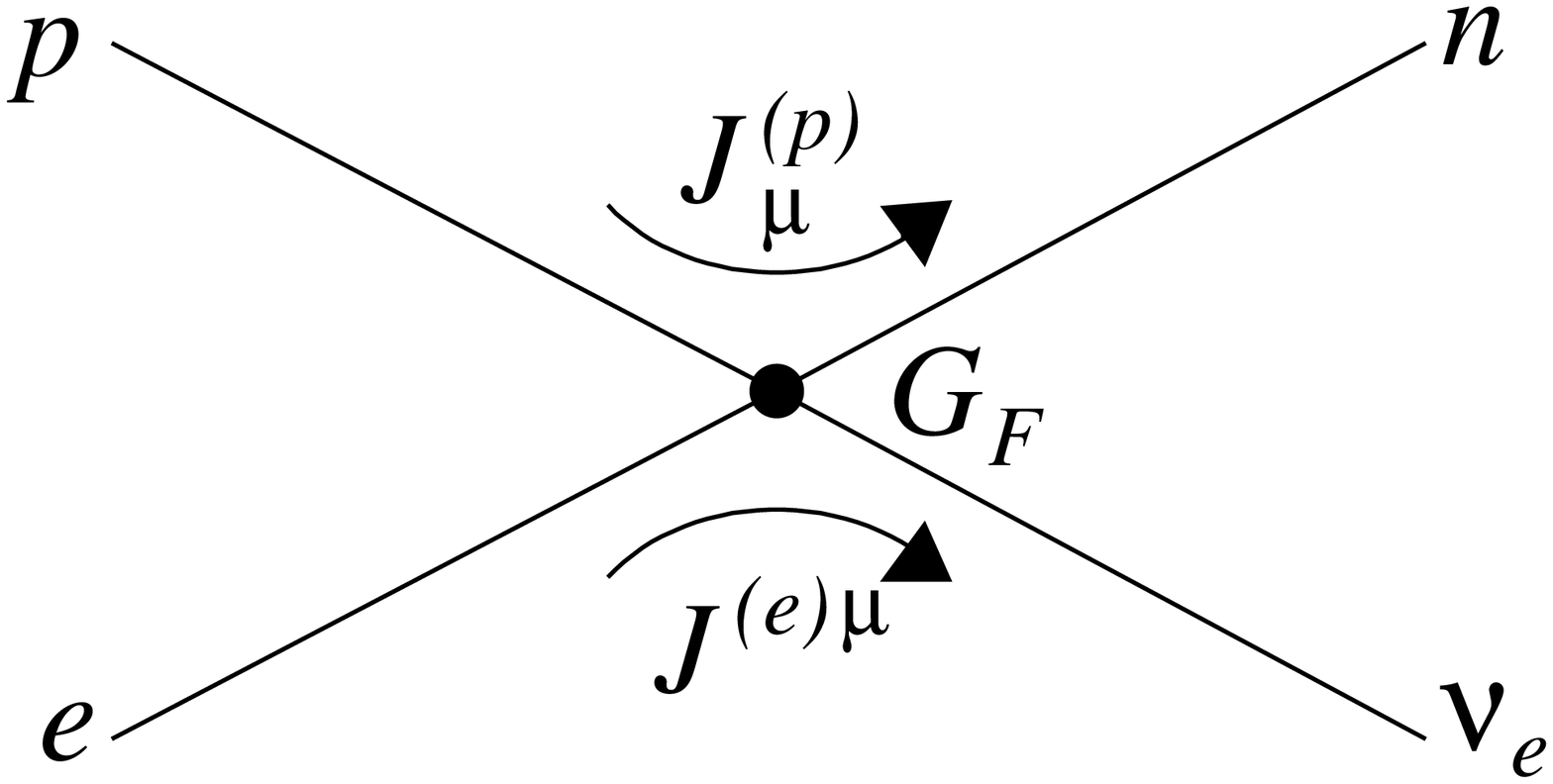}}
      \end{center}
      The amplitude (matrix element) of this process can be written as
      \begin{equation}
         \mathcal{M} = G_F J^{(e)\mu} J^{(p)}_{\mu} 
      \end{equation}
      where $G_F$ is Fermi's constant and the charged
      currents for the fermion fields are
      \begin{equation}
         J^{(e)\mu}    = \bar{u}_e \gamma^{\mu} u_{\nu_e}, \;\;\;\;\;\;
         J^{(p)}_{\mu} = \bar{u}_p \gamma_{\mu} u_n
      \end{equation}

      The next advance came after the discovery that 
      CC violates parity maximally~\cite{Lee:1956qn}
      and the V-A theory of the weak interaction~\cite{Feynman:1958ty}
      was proposed.  Only 
      left-handed fermions, which are projected by a 
      V-A operator $\frac{1}{2}(1 - \gamma_5)$, 
      appears in CC.  
      \begin{center}
         \parbox{2.0in}{\epsfxsize=\hsize\epsffile{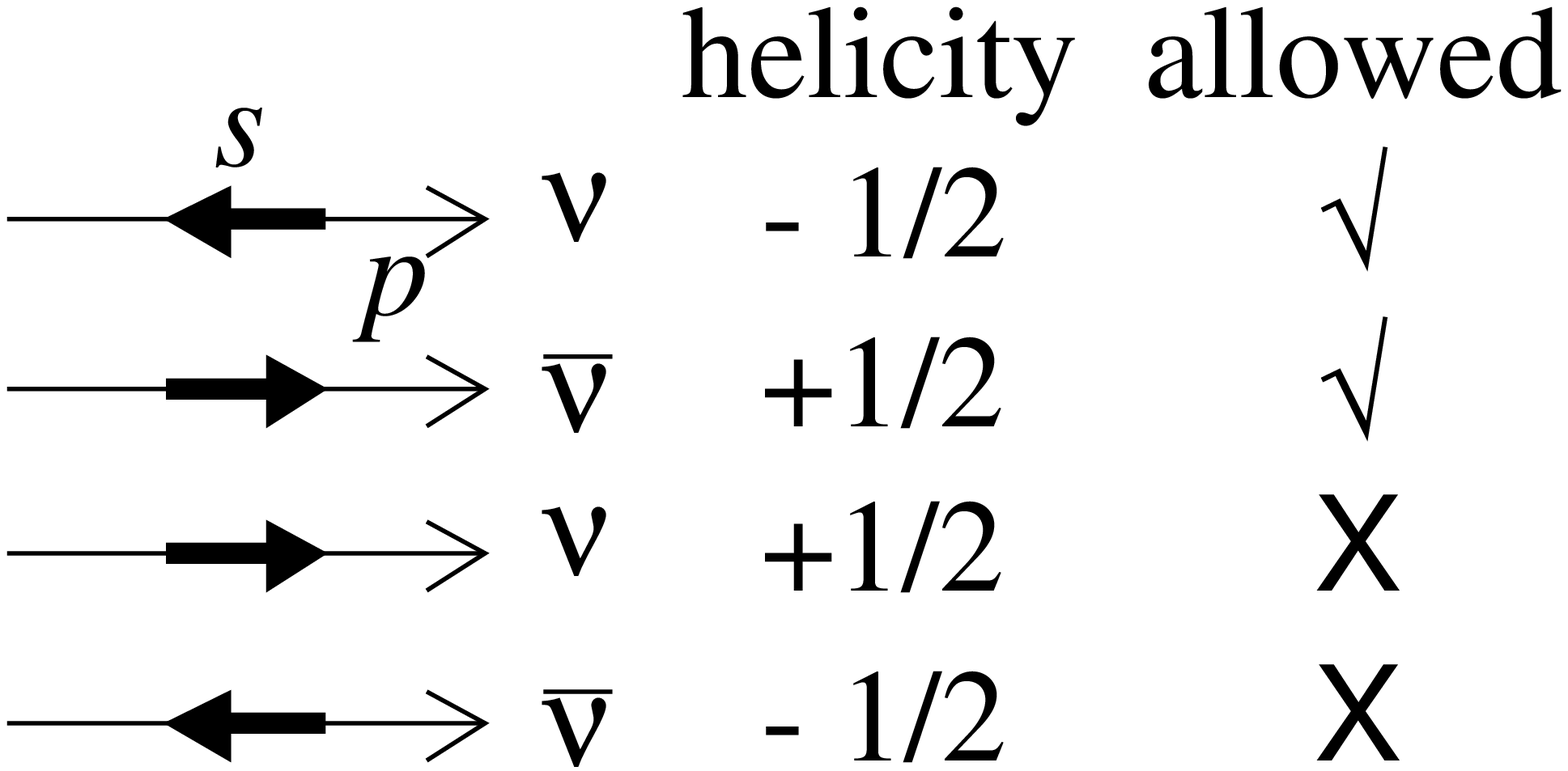}}
      \end{center}
      \begin{equation}
         \mathcal{M} = \frac{4G}{\sqrt{2}} J_{\mu}^{\dagger} J^{\mu} 
      \end{equation}
      \begin{equation}
         J^{\mu} = \bar{u}_e \gamma^{\mu} \frac{1}{2}(1-\gamma_5) u_{\nu_e}
                  +\bar{u}_p \gamma^{\mu} \frac{1}{2}(1-\gamma_5) u_n
      \end{equation}

      After the introduction of quarks~\cite{Gell-Mann:1964nj}
      for understanding the classification
      of the hadrons, it was natural to re-write the hadronic part
      of CC in terms of quark fields.  The transition 
      $u \to d$ occurs via CC, with the other two quarks 
      in the nucleon being spectators. 
       \begin{center}
         \parbox{2.0in}{\epsfxsize=\hsize\epsffile{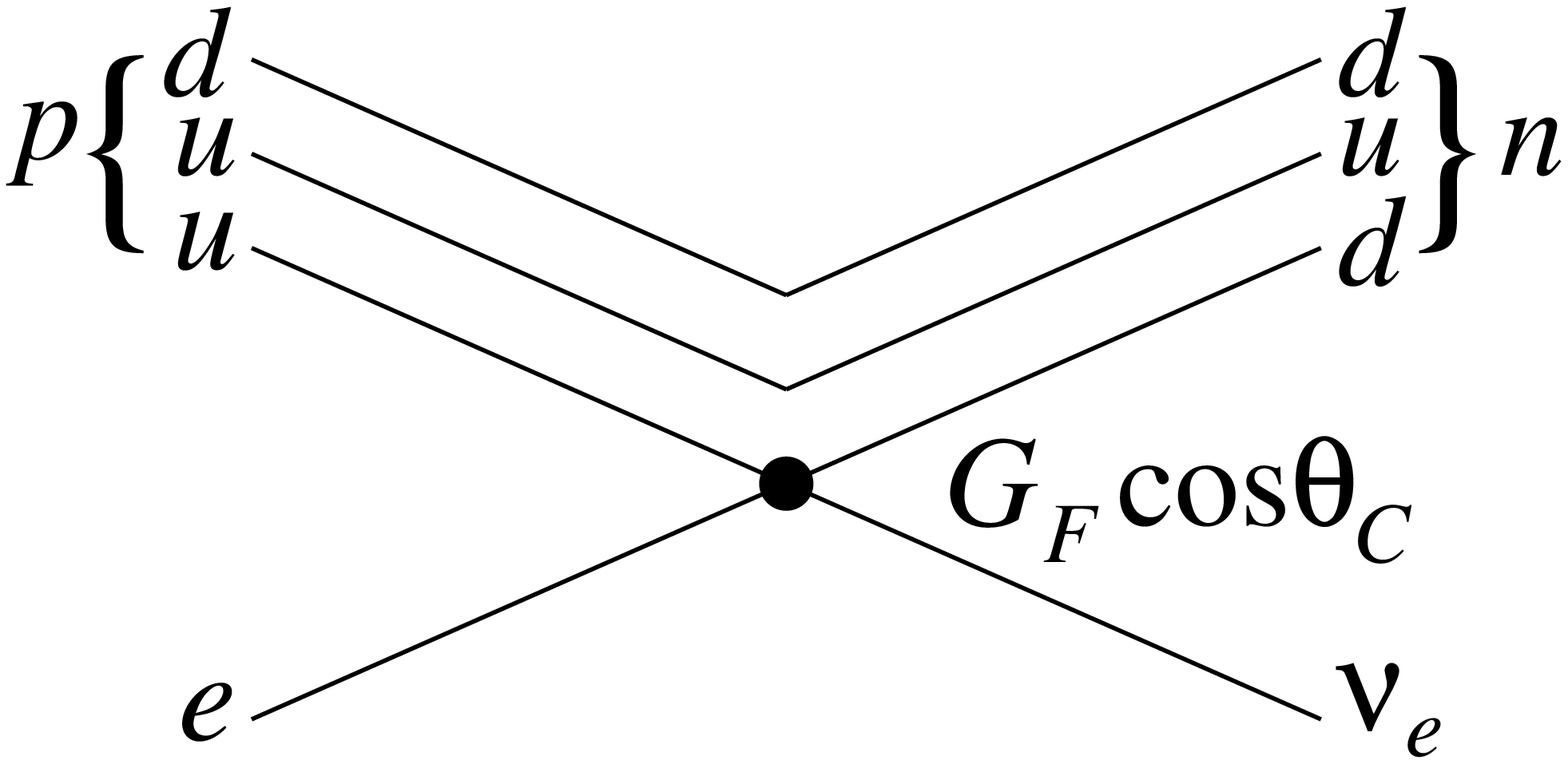}}
      \end{center}
      \begin{equation}
         J^{\mu} = \bar{e} \gamma^{\mu} \frac{1}{2}(1-\gamma_5) \nu_e
                  +\bar{u} \gamma^{\mu} \frac{1}{2}(1-\gamma_5) d'        
      \end{equation}

      There was an inconsistency found
      in the value of the Fermi constant $G_F$ as determined 
      from $\beta$-deay and the purely leptonic muon decay.
      This lead Cabibbo to the hypothesis that the quark states 
      in CC are not the physical states (eigenstates of mass), 
      but rather a quantum superposition of the physical 
      states.  
      \begin{equation}
         \left( \begin{array}{c}
                   d \\
                   s
                \end{array}
         \right)_{\mbox{weak}}
         = 
         \left( \begin{array}{cc}
                    \cos\theta_C & \sin\theta_C \\
                   -\sin\theta_C & \cos\theta_C
                \end{array}
         \right)
         \left( \begin{array}{c}
                   d \\
                   s
                \end{array}
         \right)_{\mbox{mass}}
      \end{equation}
      where $\theta_C$ is Cabibbo angle, thus the Fermi constant
      is replaced by $G_F\cos\theta_C$.  This idea was generalized
      to the case of three quark generations in terms of
      the CKM (Cabbibo-Kobayashi-Maskawa) matrix~\cite{Cabibbo:1963yz},
      \begin{equation}
         \left( \begin{array}{c}
                   d \\
                   s \\
                   b
                \end{array}
         \right)_{\mbox{weak}}
         = 
         \left( \begin{array}{ccc}
                   V_{ud} & V_{us} & V_{ub} \\
                   V_{cd} & V_{cs} & V_{cb} \\
                   V_{td} & V_{ts} & V_{tb} 
                \end{array}
         \right)
         \left( \begin{array}{c}
                   d \\
                   s \\
                   b
                \end{array}
         \right)_{\mbox{mass}}
      \end{equation}

      Glashow proposed the intermediate vector 
      boson model (IVB) in 1961~\cite{Glashow:1961tr} 
      and the form has 
      been incorportated in the SM.  The 
      basic idea is to replace the four fermion 
      interaction by the exchange of a massive
      charged boson $W^{\pm}$, e.g.
      $\nu_{\mu} e^- \to \nu_e \mu^-$,
      (a) four fermion interaction,
      (b) the IVB model:
      \begin{center}
         \parbox{4.0in}{\epsfxsize=\hsize\epsffile{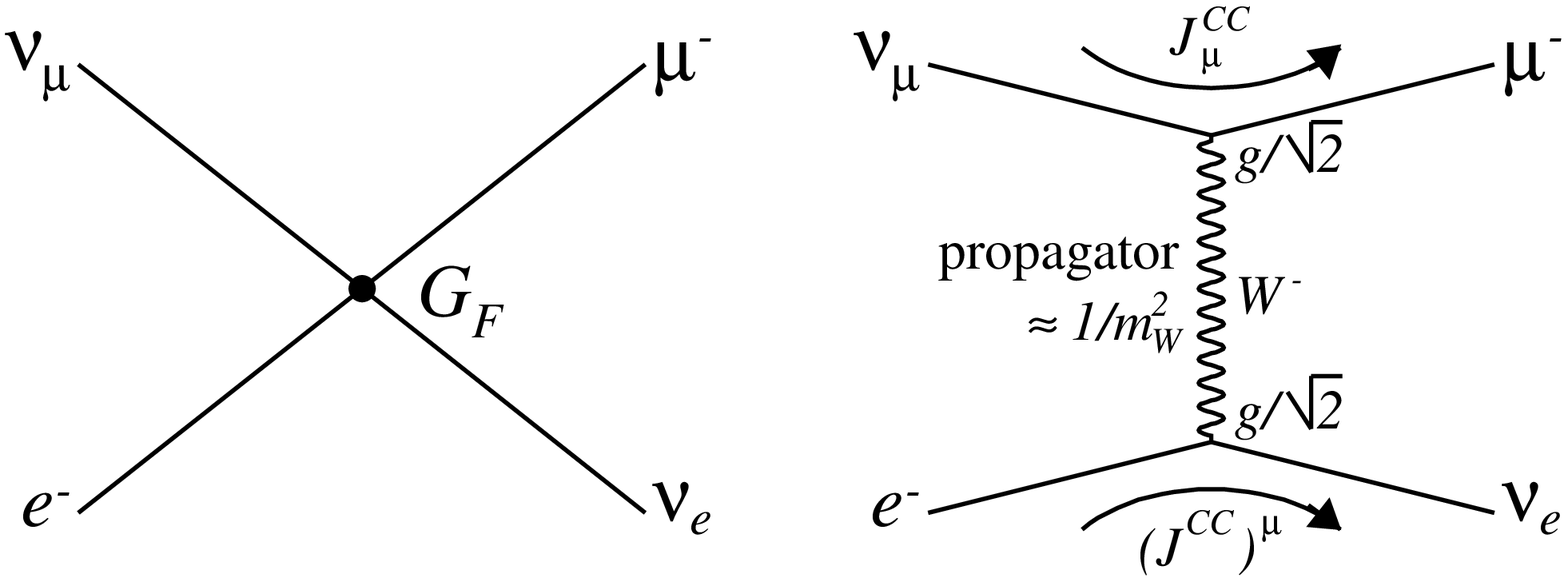}}
      \end{center}
      The matrix element can be written as
      \begin{equation}
         \mathcal{M}_{Fermi}^{CC} =     \frac{4G_F}{\sqrt{2}}
                                        J_{\mu}^{CC}
                                        \left(J^{CC}\right)^{\mu}
         \label{eq:CC_Fermi}
      \end{equation}
      \begin{equation}
         \mathcal{M}_{IVB}^{CC} \approx \frac{g}{\sqrt{2}}
                                        J_{\mu}^{CC}
                                        \left(\frac{1}{m_W^2}\right)
                                        \frac{g}{\sqrt{2}}
                                        \left(J^{CC}\right)^{\mu}
         \label{eq:CC_IVB}
      \end{equation}
      Comparing Eq.~(\ref{eq:CC_Fermi}) with Eq.~(\ref{eq:CC_IVB}),
      substituting $g = e/\sin\theta_W$
      and $\alpha = e^2/(4\pi)$,
      and using experimental values:
      $\alpha = 1/137$, 
      $G_F = 1.166\times10^{-5}$ GeV$^{-2}$,
      $\sin^2\theta_W = 0.22$  
      ($\sin^2\theta_W$ was first determined 
      from the NC/CC cross section ratio in
      neutrino scattering where NC is the neutral
      current interaction explained below),
      this leads to the prediction for the W mass:
      \begin{equation}
         m_W = \left(\frac{\sqrt{2}g^2}{8G_F}\right)^{1/2}
             = \frac{37.3}{\sin\theta_W}
             = 79.5 \; \mbox{GeV}/c^2
         \label{eq:mW_g2_GF}
      \end{equation}
      This may be compared with the experimental
      value~\cite{Eidelman:2004wy}:
      \begin{equation}
         m_W = 80.425 \pm 0.038 \; \mbox{GeV}/c^2
      \end{equation}
      The interdediate $W^{\pm}$ bosons, along with the $Z^0$ bosons
      explained below, were discovered
      at CERN in 1983~\cite{Arnison:1983rp}.

      \vspace{0.2in}
      \noindent{\bf A Doublet in Weak Isospin Space}
      \vspace{0.1in}

      We write the left-handed leptons in a weak isospin SU(2)$_L$ doublet
      and the right-handed leptons in a singlet, for example,
      \begin{equation}
         L =
         \left( 
            \begin{array}{c}
               \nu_e \\
               e_L^-
            \end{array}
         \right),
         \;\;\;
         e_R 
      \end{equation}
      The generators of the SU(2)$_L$ transformations are
      $T_L^i = \frac{1}{2}\tau^i$,
      where $\tau^i$ are Pauli matrices.  The charge
      raising opertator $\tau^+$, the charge lowering operator
      $\tau^-$, and the original $\tau^3$ are 
      \begin{eqnarray}
         \tau^+ & = &   \frac{1}{2}(\tau_1 + i\tau_2)
                      = \left( \begin{array}{cc}
                                  0 & 1 \\
                                  0 & 0
                               \end{array}
                        \right) \\
         \tau^- & = &   \frac{1}{2}(\tau_1 - i\tau_2)
                      = \left( \begin{array}{cc}
                                  0 & 0 \\
                                  1 & 0
                               \end{array}
                        \right) \\
         \tau^3 & = &   \left( \begin{array}{cc}
                                  1 & 0 \\
                                  0 & -1
                               \end{array}
                        \right)
      \end{eqnarray}
      The currents can be written as
      \begin{eqnarray}
         J_{\mu}^+    & = &        \bar{\nu}_e 
                                   \gamma_{\mu} 
                                   \frac{1}{2}(1 - \gamma_5)
                                   e
                            \equiv \bar{\nu}_e 
                                   \gamma_{\mu} 
                                   e_L
                            =      \bar{L}
                                   \gamma_{\mu}
                                   \tau^+
                                   L \\
         J_{\mu}^-    & = &        \bar{e}
                                   \gamma_{\mu} 
                                   \frac{1}{2}(1 - \gamma_5)
                                   \nu_e
                            \equiv \bar{e}_L 
                                   \gamma_{\mu} 
                                   \nu_e
                            =      \bar{L}
                                   \gamma_{\mu}
                                   \tau^-
                                   L \\
         J_{\mu}^3    & = &        \frac{1}{2}
                                   \left[ \bar{\nu}_L
                                          \gamma_{\mu} 
                                          \nu_L
                                         -\bar{e}_L
                                          \gamma_{\mu} 
                                          e_L
                                   \right]
                            =      \bar{L}
                                   \gamma_{\mu}
                                   \frac{1}{2}\tau^3
                                   L 
      \end{eqnarray}
      These can be combined into an isospin triplet of 
      currents
      \begin{equation}
         \mathbf{J}_{\mu} = \bar{L} 
                            \gamma_{\mu}
                            \mathbf{T}
                            L
      \end{equation}
      The weak isospin invariance implies that the SU(2)$_L$ 
      invariant Lagrangian to describe the interaction
      between the $W$ bosons and the current $\mathbf{J}$
      with a coupling $g$ is of the form
      \begin{equation}
         \mathcal{L} = g \mathbf{J}^{\mu} \cdot \mathbf{W}_{\mu}
      \end{equation}
      Hence a neutral IVB W$^3$ should exist, coupling 
      to $J^3$. 
      Since the electromagnetic current $J_{\mu}^{em}$ 
      is parity conserving,
      \begin{equation}
         J_{\mu}^{em} = e( \bar{e}_R
                           \gamma_{\mu}
                           e_R
                          +\bar{e}_L
                           \gamma_{\mu}
                           e_L
                         )
      \end{equation}
      Whereas $J_{\mu}^3$ has a V-A structure. 
      $J_{\mu}^3$ cannot be directly identified with 
      the electromagnetic current, nor W$^3$ with the 
      photon.


      \vspace{0.2in}
      \noindent{\bf Neutral Current}
      \vspace{0.1in}

      Next came the inputs from the neutral current 
      (NC) interactions.  NC were discovered by the 
      Gargamelle Collaboration at CERN in 1973~\cite{Hasert:1973ff}, 
      $\nu_{\mu} q \to \nu_{\mu} q$.
      \begin{center}
         \parbox{4.0in}{\epsfxsize=\hsize\epsffile{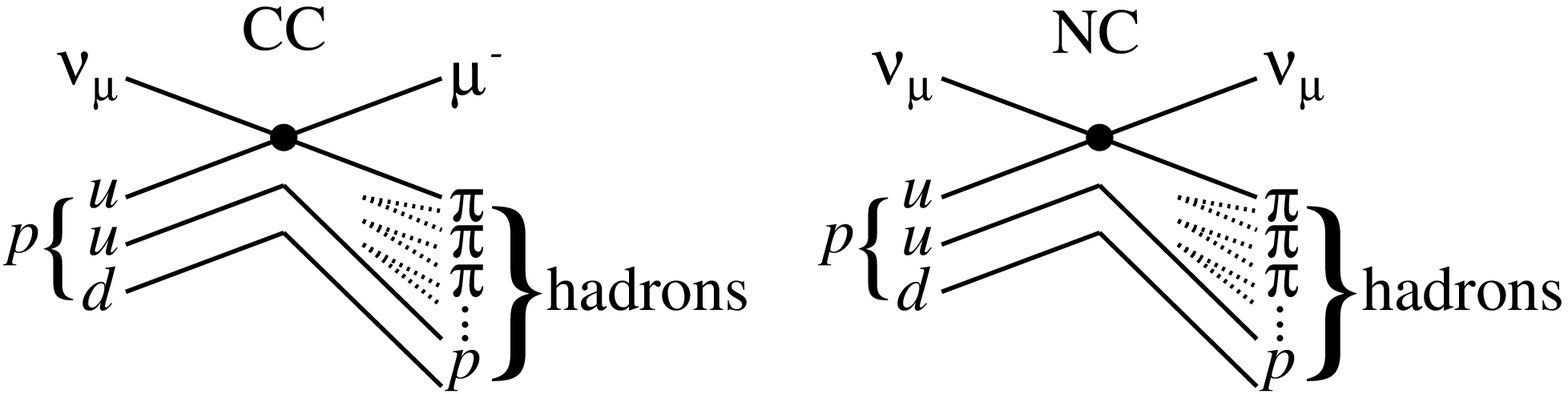}}
      \end{center}
      The matrix element can be written as
      \begin{equation}
         \mathcal{M} = \frac{8G_F\rho}{\sqrt{2}}
                       \left(J^{NC}\right)^{\mu}
                       J^{NC}_{\mu}
      \end{equation}
      with NC in the form
      \begin{equation}
         \left(J^{NC}\right)^{\mu}
         = \sum_l\left[ \bar{\nu}_l 
                        \gamma^{\mu} 
                        \frac{1}{2}(1 - \gamma_5)
                        \nu_l
                 \right]
          +\sum_f\left[ \bar{f} 
                        \gamma^{\mu} 
                        \frac{1}{2}(C_V^f - C_A^f\gamma_5)
                        f
                 \right] 
      \end{equation}
      \[
         l = e, \mu, \tau;
         \;\;\;
         f = l, q;
         \;\;\;
         q = u, d, s, c, b, t 
      \]
      The neutrino part has a V-A structure.
      The lepton/quark part has 
      parity violation ($C_A^f \neq 0$),
      but not maximally ($C_A^f \neq C_V^f$).
      Universality of NC and 
      CC requires $\rho = 1$, 
      later predicted in the SM.

      We can write the NC interactions in terms of IVB, e.g.
      $\nu_{\mu} e^- \to \nu_{\mu} e^-$,
      (a) four fermion interaction,
      (b) the IVB model:
      \begin{center}
         \parbox{4.0in}{\epsfxsize=\hsize\epsffile{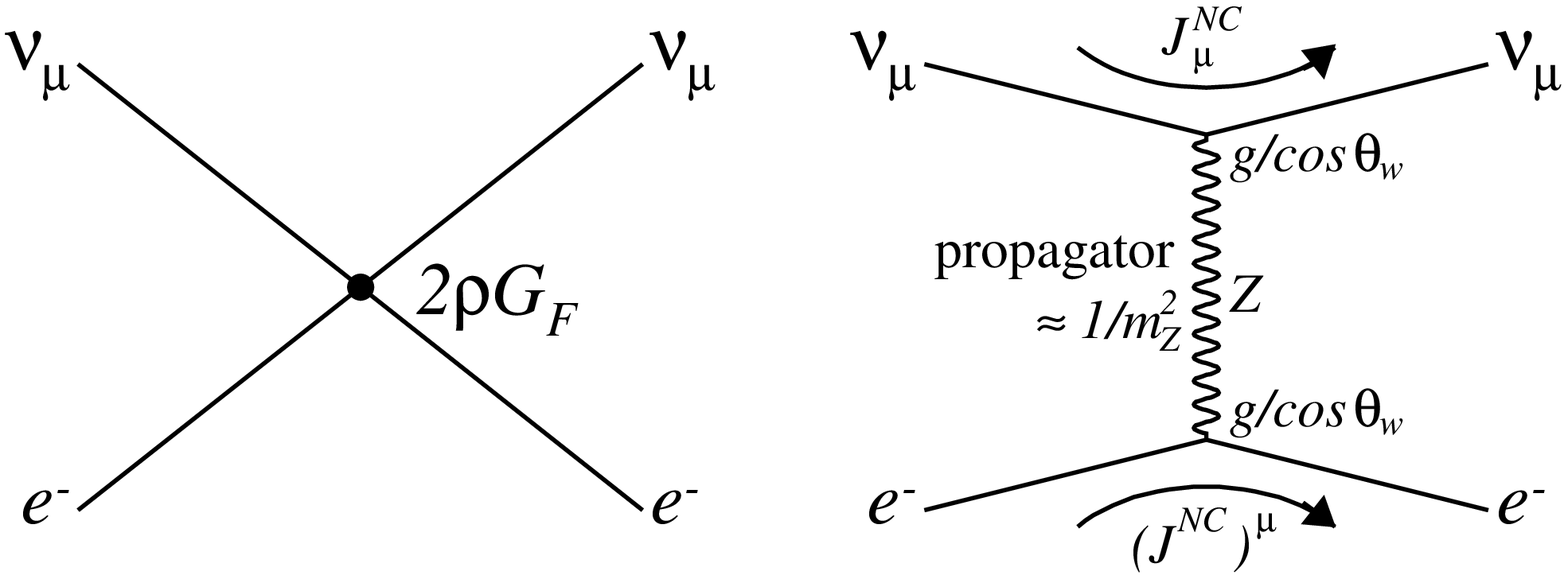}}
      \end{center}
      The matrix element can be written as
      \begin{equation}
         \mathcal{M}_{Fermi}^{NC} =     \frac{8 \rho G_F}{\sqrt{2}}
                                        J_{\mu}^{NC}
                                        \left(J^{NC}\right)^{\mu}
         \label{eq:NC_Fermi}
      \end{equation}
      \begin{equation}
         \mathcal{M}_{IVB}^{NC} \approx \frac{g}{\cos\theta_W}
                                        J_{\mu}^{NC}
                                        \left(\frac{1}{m_Z^2}\right)
                                        \frac{g}{\cos\theta_W}
                                        \left(J^{NC}\right)^{\mu}
         \label{eq:NC_IVB}
      \end{equation}
      Comparing Eq.~(\ref{eq:NC_Fermi}) with Eq.~(\ref{eq:NC_IVB}),
      and assuming universality of the charged and
      neutral currents ($\rho = 1$), this gives the 
      prediction for Z mass:
      \begin{equation}
         m_Z = \left(\frac{\sqrt{2}g^2}{8G_F}\right)^{1/2}
               \frac{1}{\sqrt{\rho}\cos\theta_W}
             = \frac{m_W}{\sqrt{\rho}\cos\theta_W}
             = \frac{79.5}{\cos\theta_W}
             = 90 \; \mbox{GeV}/c^2
      \end{equation}
      This may be compared with the experimental
      value~\cite{Eidelman:2004wy}:
      \begin{equation}
         m_Z = 91.1876 \pm 0.0021 \; \mbox{GeV}/c^2
      \end{equation}


      \vspace{0.2in}
      \noindent{\bf Flavor Changing Neutral Current}
      \vspace{0.1in}

      The flavor changing neutral current (FCNC) 
      interaction is strongly suppressed~\cite{Eidelman:2004wy}, 
      (a) CC, (b) FCNC:
      \begin{center}
         \parbox{4.0in}{\epsfxsize=\hsize\epsffile{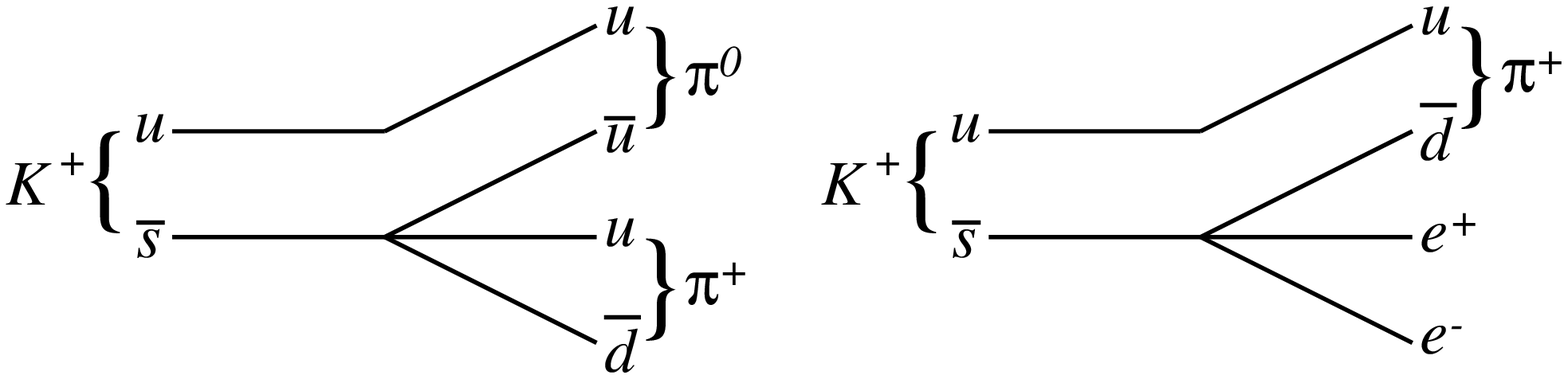}}
      \end{center}
      \begin{equation}
         J^{NC} = \bar{u}u + \bar{d}'d'
                = \bar{u}u + \bar{d}d\cos^2\theta_C + \bar{s}s\sin^2\theta_C
                           + \underbrace{
                                (\bar{s}d + \bar{d}s)
                             }_{\mbox{FCNC}}
                             \sin\theta_C\cos\theta_C
      \end{equation}
      \begin{eqnarray}
         \mbox{BR}(K^+ \to \pi^+ \pi^0)   & = & (21.13 \pm 0.14)\% \\
         \mbox{BR}(K^+ \to \pi^+ e^+ e^-) & = & (2.88  \pm 0.13)\times10^{-7}
      \end{eqnarray}
      The GIM (Glashow-Iliopoulos-Maiani) mechanism~\cite{Glashow:1970gm} 
      proposed that
      quarks must be paired in doublets.  This naturally solved 
      FCNC.  In addition, $c$ quark was predicted and later 
      discovered~\cite{Aubert:1974js}.
      \begin{equation}
         \left( \begin{array}{c}
                   u \\
                   d'
                \end{array}
         \right)
         \;\;\;
         \left( \begin{array}{c}
                   c \\
                   s'
                \end{array}
         \right)
      \end{equation}
      \begin{equation}
         J^{NC} = \bar{u}u + \bar{c}c + \bar{d}'d' + \bar{s}'s'
                = \bar{u}u + \bar{c}c + \bar{d}d + \bar{s}s
      \end{equation}


      \vspace{0.2in}
      \noindent{\bf A Triplet in Quark Color Space}
      \vspace{0.1in}

      The quarks in the spin-$\frac{3}{2}$ baryons 
      are in a symmetrical state of space, spin 
      and flavor degrees of freedom, e.g. 
      \begin{equation}
         \Delta^{++} = uuu,
         \;\;\;\;\;\;
         \Omega^- = sss
      \end{equation}
      However the requirements of Fermi-Dirac 
      statistics imply the total antisymmetry of 
      the wave function.  The solution was the 
      introduction of the color degree of freedom, 
      with indices as red ($r$), green ($g$), and 
      blue ($b$).
      \begin{equation}
         q = \left( \begin{array}{c}
                       q_r \\
                       q_g \\
                       q_b
                    \end{array}
             \right)
      \end{equation}

      One of the tests of the number of charged 
      fundamental constituents is provided by 
      \begin{equation}
         R = \frac{\sigma(e^+e^- \to \mbox{hadrons})}
                  {\sigma(e^+e^- \to \mu^+\mu^-)}
      \end{equation}
      The virtual photon emitted by the $e^+e^-$ 
      annihilation will excite all kinematically
      accessible $q\bar{q}$ pairs from the vacuum. 
      \begin{equation}
         R = \sum_q e_q^2
      \end{equation}
      At low energy where only the $u$, $d$ and $s$ 
      quarks are available, in the absense of color 
      degree of freedom, we expect
      \begin{equation}
         R = e_u^2 + e_d^2 + e_s^2 
           = \left(\frac{2}{3}\right)^2
            +\left(-\frac{1}{3}\right)^2
            +\left(-\frac{1}{3}\right)^2   
           = \frac{2}{3}
      \end{equation}
      If quarks have three colors,
      \begin{equation}
         R = 3(e_u^2 + e_d^2 + e_s^2)
           = 2
      \end{equation}
      For energies above 10 GeV, $c$ and $b$ quarks 
      are available,
      \begin{equation}
         R = 3(e_u^2 + e_d^2 + e_s^2 + e_c^2 + e_b^2)
           = \frac{11}{3}
      \end{equation}
      The color triplet model is excellently supported 
      by data, see the ``$\sigma$ and $R$ in $e^+e^-$ Collisions'' plots
      in the Section ``Plots of cross sections and related quantities (Rev.)''
      in PDG~\cite{Eidelman:2004wy}.


      \vspace{0.2in}
      \noindent{\bf Pair Quarks with Leptons}
      \vspace{0.1in}

      Some classical symmetries, known as anomalous
      symmetries~\cite{Adler:1969gk}
      are broken by quantum effects.
      The requirement for an anomaly-free theory~\cite{Adler:1969er}
      is that:
      \begin{equation}
         \sum Q_f = 0
      \end{equation}
      where the sum is over all quarks and leptons. For
      example consider the two doublets,
      \begin{eqnarray}
         \left(\begin{array}{c}
                  \nu_e \\
                  e
               \end{array}
         \right)
         \nonumber \\
         \left(\begin{array}{c}
                  u \\
                  d
               \end{array}
         \right)
         \nonumber
      \end{eqnarray}
      \begin{equation}
         \sum Q_f = (0 - 1)
                   +3\times(\frac{2}{3} - \frac{1}{3})
                  = 0
      \end{equation}
      Cancellation of anomalies requires that quark
      doublets must be paired with lepton doublets.
      The SM identifies a generation in a natural 
      way by identifying the doublet containing the 
      heaviest charged lepton with the doublet 
      containing the heaviest quarks (and so on), 
      but one could in principle associate any quark 
      doublet with any lepton doublet and call that 
      a generation, because there are no interactions
      between quarks and leptons in the SM.  What  
      needs to be guaranteed is that the number of 
      quark and lepton generations must be equal.  



\chapter{Gauge Symmetry \& Spontaneous Symmetry Breaking}
\label{cha:app_gs_ssb}

The interactions between the fermions and the vector bosons
in the Standard Model (SM) are uniquely specified by requiring 
the theory, i.e. the SM Lagrangian, invariant under gauge 
transformations which are local and involve transformations 
varying from point to point.  Some of the standard texts 
are listed in Ref.~\cite{Quigg:1983}.

A symmetry indicates a deeper relationship
among the elementary particles with a further 
unification of the interactions and makes the form 
of a Lagrangian more compact.  Symmetry dictates 
design and plays the central role in the direction 
to \emph{find the simplest model}. 

\vspace{0.2in}
\noindent{\bf Gauge Symmetry}
\vspace{0.1in}

\begin{figure}
   \begin{center}
      \parbox{5.5in}{\epsfxsize=\hsize\epsffile{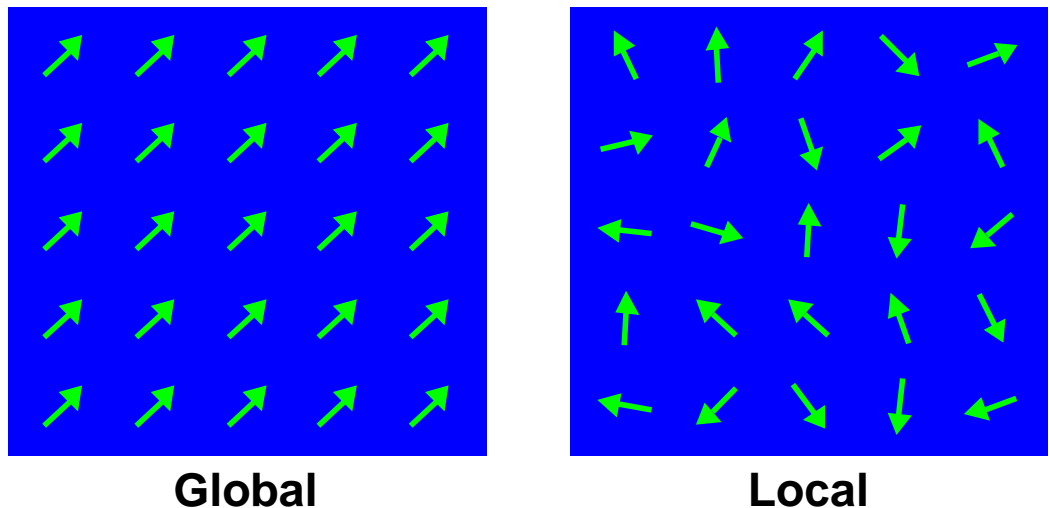}}
      \caption[Global and local transformations]
              {Global and local transformations.}
      \label{fig:Global_Local}
   \end{center}
\end{figure}

Let us take electromagnetism as an example and consider 
the Lagrangian for a free fermion field $\Psi(x)$.
\begin{equation}
   \mathcal{L}_0 = \bar{\Psi}(x)
                   (i\gamma^{\mu}\partial_{\mu} - m)
                   \Psi(x) 
   \label{eq:FreeFermion_a}
\end{equation}
This is invariant under a global 
U(1) phase transformation which is space-time
independent and is illustrated in
the left plot in Fig.~\ref{fig:Global_Local},
\begin{equation}
   \Psi(x) \to \Psi'(x) = e^{-iQ\theta}\Psi(x)
\end{equation}
where $Q$ is the charge or the U(1) quantum number
of the fermion. For example, the charge assignment 
for $u$ quark, $d$ quark, $\nu_e$, and $e$ are
+2/3, -1/3, 0, and -1, respectively. 

We are going to construct an invariant Lagrangian
under a local, i.e., gauge, U(1) phase transformation 
which is space-time dependent and is illustrated in 
the right plot in Fig.~\ref{fig:Global_Local}.
\begin{equation}
   \Psi(x) \to \Psi'(x) = e^{-iQ\theta(x)}\Psi(x)
\end{equation}
The partial derivative $\partial_{\mu}$ 
in Eq.~(\ref{eq:FreeFermion_a}) spoils the invariance.  
We need to form a gauge-covariant derivative 
$D_{\mu}$ which will have the simple 
transformation property,
\begin{equation}
   D_{\mu}\Psi(x) \to e^{-iQ\theta(x)}D_{\mu}\Psi(x)
   \label{eq:CovariantDerivative_a1}
\end{equation}
so that the combination $\bar{\Psi}D_{\mu}\Psi$ is 
gauge invariant.  To achieve this, we enlarge the 
Lagrangian with a new vector gauge field $A_{\mu}(x)$ 
and form the covariant form as
\begin{equation}
   D_{\mu}\Psi = (\partial_{\mu} + ieQA_{\mu})\Psi
   \label{eq:CovariantDerivative_a2}
\end{equation}
where $e$ is a free parameter which eventually
will be identified as the coupling of the gauge 
field to the fermion field.  The transformation 
property in Eq.~(\ref{eq:CovariantDerivative_a1}) will 
be satisfied if the gauge field $A_{\mu}(x)$ has 
the transformation property
\begin{equation}
   A_{\mu}(x) \to A'_{\mu}(x) 
                = A_{\mu}(x) 
                 +\frac{1}{e}\partial_{\mu}\theta(x)
   \label{eq:GaugeFieldTransformation_a}
\end{equation}
Note that the coupling of the gauge field (photon) 
to any fermion field is determined by its 
transformation property under the symmetry group.  
This is usually referred to as \emph{universality}. 
Also note that photon is massless because an 
$A_{\mu}A^{\mu}$ term is not gauge invariant under 
this transformation.  

To make the photon field a 
truely dynamical variable we need to add a kinetic 
term to the Lagrangian involving its derivatives.  
The simplest gauge-invariant term with a 
conventional normalization is
\begin{equation}
   \mathcal{L}_A = -\frac{1}{4}F_{\mu\nu}F^{\mu\nu}
   \label{eq:GaugeFieldKineticTerm_a}
\end{equation}
where
\begin{equation}
   F_{\mu\nu} = \partial_{\mu}A_{\nu}
               -\partial_{\nu}A_{\mu}
   \label{eq:GaugeFieldDerivative_a}
\end{equation}
Terms with higher powers are omitted in order that 
the theory be renormalizable. 
We notice that photon does not have self-coupling 
because it does not carry a charge.

Now we arrive at the 
gauge-invariant QED Lagrangian
\begin{equation}
   \mathcal{L}_{QED} = \bar{\Psi}
                       i\gamma^{\mu}(\partial_{\mu}+ieQA_{\mu})
                       \Psi
                      -m\bar{\Psi}\Psi
                      -\frac{1}{4}F_{\mu\nu}F^{\mu\nu}
   \label{eq:L_QED}
\end{equation}

Most remarkably, if one demands the symmetry
be local, one is forced to include the 
electromagnetic field, and hence, light.  Recall
that there are four Maxwell equations.  
While here we just require
``gauge symmetry'' and electromagnetism
is determined.  This 
illustrates how 
physics becomes simpler.

\vspace{0.2in}
\noindent{\bf Non-Abelian Gauge Symmetry}
\vspace{0.1in}

Yang and Mills extended the gauge principle to 
non-Abelian symmetry~\cite{Yang:1954ek}.
Consider the simplest case 
isospin SU(2). Let the fermion field be an isospin 
doublet,
\begin{equation}
   \Psi = \left(\begin{array}{c}
                   \Psi_1 \\
                   \Psi_2
                \end{array}
          \right)
\end{equation}
The free Lagragian 
\begin{equation}
   \mathcal{L}_0 = \bar{\Psi}(x)
                   (i\gamma^{\mu}\partial_{\mu} - m)
                   \Psi(x) 
   \label{eq:FreeFermion_b}
\end{equation}
is invariant under the global SU(2) transformation
\begin{equation}
    \Psi(x) \to \Psi'(x) 
               = e^{-i\mathbf{T} \cdot \mathbf{\theta}}
                 \Psi(x)
\end{equation}
where $\mathbf{\theta} = (\theta_1, \theta_2, \theta_3)$
are the SU(2) transformation parameters and
$\mathbf{T} = \frac{\mathbf{\tau}}{2}$ 
are the SU(2) generators
with $\mathbf{\tau} = (\tau_1, \tau_2, \tau_3)$
the Pauli matrices satisfying
\begin{equation}
   [T_i, T_j] = i\epsilon_{ijk}T_k
                \;\;\;\;\;\;
                i,j,k = 1,2,3
\end{equation}
with $\epsilon_{ijk}$ the structure constants for SU(2).

It is easy to check that two successive SU(2) 
transformations do not commute because the generators 
do not commute and this is why SU(2) is called 
a non-Abelian symmetry, in contrast to an Abelian 
symmetry such as U(1) where two successive U(1) 
transformations commute.  

Under the local symmetry 
transformation
\begin{equation}
    \Psi(x) \to \Psi'(x) 
               = e^{-i\mathbf{T} \cdot \mathbf{\theta}(x)}
                 \Psi(x)
\end{equation}
the partial derivative $\partial_{\mu}$ in Eq.~(\ref{eq:FreeFermion_b})
spoils the invariance.  To construct a 
gauge-invariant Lagrangian we follow a procedure
similar to that of the Abelian case: 

\begin{itemize}
\item We form a gauge-covariant derivative
      \begin{equation}
         D_{\mu}\Psi(x) \to e^{-i\mathbf{T} \cdot \mathbf{\theta}(x)}
                            D_{\mu}\Psi(x)
         \label{eq:CovariantDerivative_b1}
      \end{equation}
      by introducing vector gauge fields 
      $A_{\mu}^i$, $i = 1, 2, 3$ (one 
      for each group generator) and
      a coupling $g$
      \begin{equation}
         D_{\mu}\Psi = (\partial_{\mu} + ig\mathbf{T} \cdot \mathbf{A}_{\mu})
                       \Psi
         \label{eq:CovariantDerivative_b2}
      \end{equation}
      and defining the transformation property
      for the vector gauge fields as,
      \begin{equation}
         A_{\mu}^i \to A_{\mu}^{i'} 
                      = A_{\mu}^i
                       -\epsilon^{ijk}\theta^jA_{\mu}^k
                       +\frac{1}{g}\partial_{\mu}\theta^i
         \label{eq:GaugeFieldTransformation_b}
      \end{equation}
      The gauge fields are massless 
      because an $A_{\mu}^iA^{i\mu}$ term is not 
      gauge invariant, similar to an Abelian field.
      But, the second term is clearly the transformation
      for a triplet representation under SU(2), 
      thus the $A_{\mu}^i$ fields carry charges.

\item Then we add a gauge invariant kinetic term
      for the gauge fields
      \begin{equation}
         \mathcal{L}_A = -\frac{1}{4}F_{\mu\nu}^iF^{i\mu\nu}
         \label{eq:GaugeFieldKineticTerm_b}
      \end{equation}
      where
      \begin{equation}
         F_{\mu\nu}^i = \partial_{\mu}A_{\nu}^i
                       -\partial_{\nu}A_{\mu}^i
                       -g\epsilon^{ijk}A_{\mu}^jA_{\nu}^k
         \label{eq:GaugeFieldDerivative_b}
      \end{equation}
      The third term shows that the gauge fields 
      have self-coupling because they carry 
      charge, in contrast to an Abelian 
      field.
\end{itemize}

We arrive at the complete gauge-invariant Lagrangian
which describes the interaction between the gauge 
fields $A_{\mu}^i$ and the SU(2) doublet fields,
\begin{equation}
   \mathcal{L} =  \bar{\Psi}
                  i\gamma^{\mu}
                  (\partial_{\mu} + ig\mathbf{T} \cdot \mathbf{A}_{\mu})
                  \Psi
                 -m\bar{\Psi}\Psi
                 -\frac{1}{4}F_{\mu\nu}^iF^{i\mu\nu}
\end{equation}

Generalization of the Yang-Mills theory to a higher 
group SU(N) with $N\geq3$ 
is straightforward.  

\vspace{0.2in}
\noindent{\bf SU(3)$_C$$\times$SU(2)$_L$$\times$U(1)$_Y$}
\vspace{0.1in}

The structure of the gauge symmetries in the SM is
SU(3)$_C$$\times$SU(2)$_L$$\times$U(1)$_Y$.
For a particular fermion $\Psi$, its quantum field 
is a product of factors,
\begin{equation}
   \Psi = \left(\begin{array}{c}
                   \mbox{space-time} \\
                   \mbox{factor}
                \end{array}
          \right)
          \times
          \left(\begin{array}{c}
                   \mbox{spin} \\
                   \mbox{factor}
                \end{array}
          \right)
          \times
          \left(\begin{array}{c}
                   \mbox{U(1)}_Y \\
                   \mbox{factor}
                \end{array}
          \right)
          \times
          \left(\begin{array}{c}
                   \mbox{SU(2)}_L \\
                   \mbox{factor}
                \end{array}
          \right)
          \times
          \left(\begin{array}{c}
                   \mbox{SU(3)}_C \\
                   \mbox{factor}
                \end{array}
          \right)
\end{equation}
Each factor has some labels, coordinates, or indices.
The orthonormality of the quantum field holds separately
for each factor.  
Since the gauge bosons of one 
of the symmetry groups do not transform under the other 
gauge symmetries in the product of groups, the gauge 
invariant Lagrangian may be simply written as a sum of 
the terms of individual groups.  
The gauge 
symmetric Lagrangian in the framework of Yang-Mills 
theory is
\begin{eqnarray}
   \mathcal{L}_{symmetric} 
   & = & \bar{\Psi}
         i\gamma^{\mu}
         \left(\partial_{\mu} 
                + i g_1 \frac{Y}{2}B_{\mu}
                + i g_2 T^j W_{\mu}^j
                + i g_3 \lambda^a G_{\mu}^a
         \right)\Psi
         \label{eq:L_symmetric}
         \\
   &   & - \frac{1}{4}B_{\mu\nu}B^{\mu\nu}
         - \frac{1}{4}W_{\mu\nu}^iW^{i\mu\nu}
         - \frac{1}{4}G_{\mu\nu}^aG^{a\mu\nu}
         \nonumber
\end{eqnarray}
where 
the eight $G_{\mu}^a$ and $\lambda^a$, 
the three $W_{\mu}^i$ and $T^i$, 
the one   B$_{\mu}$   and $Y$
are the gauge bosons and generators corresponding to 
the SU(3)$_C$ color, 
the SU(2)$_L$ weak isospin, and
the U(1)$_Y$  hypercharge gauge symmetries, 
respectively;
$g_i$ are the gauge couplings;
and
\begin{eqnarray}
   B_{\mu\nu}   & = &  \partial_{\mu}B_{\nu}
                     - \partial_{\nu}B_{\mu} 
                       \\
   W_{\mu\nu}^i & = &  \partial_{\mu}W_{\nu}^i
                     - \partial_{\nu}W_{\mu}^i
                     + g_2\epsilon^{ijk}W_{\mu}^jW_{\nu}^k
                      \\
   G_{\mu\nu}^a & = &  \partial_{\mu}G_{\nu}^a
                     - \partial_{\nu}G_{\mu}^a
                     + g_3f^{abc}G_{\mu}^bG_{\nu}^c
\end{eqnarray}
with $\epsilon^{ijk}$ and $f^{abc}$ the structure 
constants for SU(2) and SU(3).

At this stage, all of the gauge bosons and 
fermions are massless.  The explicit mass terms 
break gauge invariance.  
For gauge bosons, the 
exptected mass terms 
\begin{equation}
   m_W^2W_{\mu}W^{\mu}
\end{equation}
plus similar terms for the others, are clearly 
not invariant under gauge transformations
  $W_{\mu}^i \to W_{\mu}^{i'} 
   = W_{\mu}^i
    -\epsilon^{ijk}\theta^jW_{\mu}^k
    +\frac{1}{g_2}\partial_{\mu}\theta^i$.
This is true for any gauge theory.
For fermions, using the left- and right-handed 
projection operator $P_L$ and $P_R$, the mass 
term can be written as
\begin{eqnarray}
   m\bar{\Psi}\Psi
   & = & m\bar{\Psi}(P_L + P_R)\Psi \nonumber \\
   & = & m\bar{\Psi}P_LP_L\Psi +
         m\bar{\Psi}P_RP_R\Psi \nonumber \\
   & = & m(\bar{\Psi}_R\Psi_L +
           \bar{\Psi}_L\Psi_R)
\end{eqnarray}
In the SM, left-handed fermions are in SU(2) 
doublets and the right-handed fermions are in SU(2) 
singlets, thus they transform differently.  The 
$\bar{\Psi}_R\Psi_L$ and $\bar{\Psi}_L\Psi_R$ 
terms are not SU(2) singlets and would not give
an SU(2) invariant Lagrangian.  

However the description that all of the gauge 
bosons and fermions are massless is not true in 
Nature.  We need to 
\begin{itemize}
\item[(a)] generate the masses of the leptons
           and quarks; 
\item[(b)] generate the masses of the $W^+$,
           $W^-$, and $Z^0$ weak vector bosons; 
\item[(c)] but also keep the photon and gluon
           massless.  
\end{itemize}
In other words, the SU(3)$_C$ 
will be kept precise, and the gluon will remain massless. 
We need to break SU(2)$_L$$\times$U(1)$_Y$ down to 
U(1)$_{EM}$, resulting in mixing between the 
$B_{\mu}$ and $W_{\mu}^3$ fields, and non-zero masses 
for three of the gauge bosons ($W^{\pm}$ and $Z^0$).  
The photon ($A$) remain massless, due to a residual
U(1)$_{EM}$ gauge symmetry that remains 
unbroken.

\vspace{0.2in}
\noindent{\bf Spontaneous Symmetry Breaking}
\vspace{0.1in}

The solution in the SM is to add a spontaneous
symmetry breaking (SSB) term into the symmetric
Lagrangian ``by hand''.  The Lagrangian will 
remain symmetric but the physical vacuum does not 
respect the symmetry. In this case, the symmetry 
of the Lagrangian is said to be spontaneously 
broken. 
\begin{equation}
   \mathcal{L} = \mathcal{L}_{symmetric}
                +\mathcal{L}_{SSB}
   \label{eq:L}
\end{equation}

The assumption to construct $\mathcal{L}_{SSB}$ 
is that the universe is filled with a scalar 
field, called Higgs field.  One real scalar field 
could solve (a).  One complex field could solve (a) 
and create one massive vector boson.  To achieve (a), 
(b) and (c), the minimum requirement of the Higgs 
field is two complex fields arranged in a doublet
in the SU(2) space and carries U(1) hypercharge +1
(electric charge $Q = T_L^3 + \frac{Y}{2}$ is +1
and 0 for the upper and lower component, respectively),
but is a singlet in color space. 
\begin{equation}
   \phi = \left(\begin{array}{c}
                   \phi^+ \\
                   \phi^0
                \end{array}
          \right)
        = \frac{1}{\sqrt{2}}
          \left(\begin{array}{c}
                   \phi_1 + i\phi_2 \\
                   \phi_3 + i\phi_4
                \end{array}
          \right)
\end{equation}  
Under a SU(2)$_L$$\times$U(1)$_Y$ gauge
transformation, the doublet transforms as
\begin{equation}
   \phi \to e^{-i\frac{1}{2}\alpha(x)}
            e^{-iT^i\beta^i(x)}
            \phi
   \label{eq:HiggsTransformation}
\end{equation} 

The scalar field can be given gauge invariant
terms in $\mathcal{L}_{SSB}$: 
the kinetic term required by gauge invariance, 
the Higgs potential including a mass-like term 
      and a self-interaction term, and
the Yukawa coupling between the doublet and
      a particular fermion $\Psi$.
\begin{equation}
   \mathcal{L}_{SSB}
   = (D_{\mu}\phi)^{\dagger}(D^{\mu}\phi)
     - V(\phi)
     - \mathcal{L}_{Yukawa}
   \label{eq:L_SSB}
\end{equation}
with
\begin{eqnarray}
   D_{\mu} 
   & = & \partial_{\mu}
         +ig_1\frac{1}{2}B_{\mu}
         +ig_2T^jW_{\mu}^j
         \\
   V(\phi)
   & = & \mu^2          \phi^{\dagger}\phi
         +\lambda \left(\phi^{\dagger}\phi\right)^2 
         \\
   \mathcal{L}_{Yukawa}
   & = & g_f\bar{\Psi}\phi\Psi
   \label{eq:YukawaCoupling}
\end{eqnarray}

\begin{figure}
   \begin{center}
      \parbox{5.5in}{\epsfxsize=\hsize\epsffile{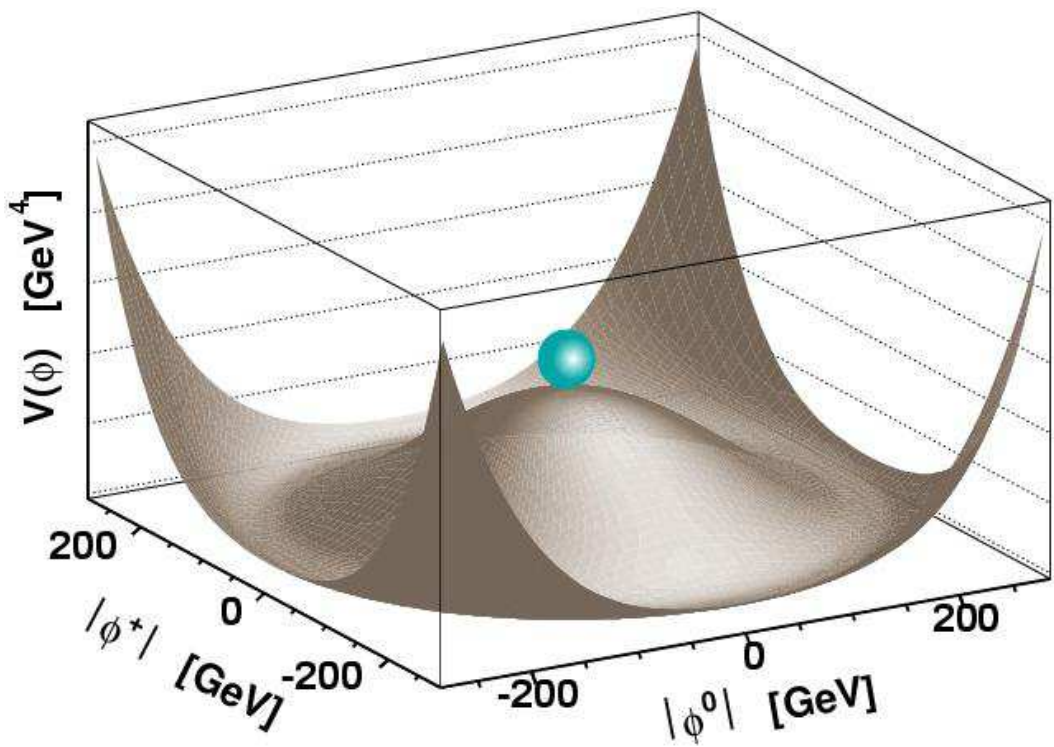}}
      \caption[Spontaneous symmetry breaking of Higgs potential]
              {Spontaneous symmetry breaking of Higgs potential.}
      \label{fig:SSB}
   \end{center}
\end{figure}

Spontaneous symmetry breaking of the Higgs potential~\cite{Higgs:1964ia}
is possible by assuming $\mu^2<0$ (also a positive 
$\lambda$ to possess a stable vacuum).
This is shown in Fig.~\ref{fig:SSB}.
The minimum of the Higgs potential shifts 
(in field space) from $\phi = 0$ to
\begin{equation}
   \phi^{\dagger}\phi
   = \frac{1}{2}
     \left(
        \phi_1^2 + \phi_2^2 + \phi_3^2 + \phi_4^2
     \right)
   = \frac{-\mu^2}{\lambda}
   = v^2
\end{equation}
The field thus acquires a non-zero vacuum expectation
value (VEV).  Choosing $\langle\phi^3\rangle = v$, we 
expand about $v$,
\begin{equation}
   \phi = \frac{1}{\sqrt{2}}
          \left(\begin{array}{c}
                   \phi_1 + i\phi_2 \\
                   v + H  + i\phi_4
                \end{array}
          \right)
\end{equation}  
with $\phi^3 = v + H$.  Any SU(2) doublet can be
written as 
\begin{equation}
   \phi = \left(e^{-iT^i\theta^i(x)}\right)^{\dagger}
          \left(\begin{array}{c}
                   0 \\
                   \sigma(x)
                \end{array}
          \right)
\end{equation} 
By applying the gauge 
symmetry of 
$\mathcal{L}_{SSB}$ under the transformation 
of the Higgs doublet in Eq.~(\ref{eq:HiggsTransformation}), 
the algebra can be simplied by ``gauging away'' 
three of the four real degrees of freedom of
the Higgs doublet with $\phi^1, \phi^2, \phi^4 = 0$, 
\begin{equation}
   \phi
        = \frac{1}{\sqrt{2}}
          \left(\begin{array}{c}
                   0 \\
                   v + H(x)
                \end{array}
          \right)
   \label{eq:UnitaryGauge}
\end{equation}
This is called the unitary gauge.  On the other 
hand, the physical quantities are independent of 
the choice of gauge.  This indicates these degrees 
of freedom are unphysical.  

\vspace{0.2in}
\noindent{\bf Gauge Boson Mass}
\vspace{0.1in}

The generators of the 
SU(2)$_L$ transformations are 
$T_L^i = \frac{1}{2}\tau^i$, where $\tau^i$ are 
Pauli matrices.  
\begin{eqnarray}
   T^1 & = & \frac{1}{2}\tau^1
         =   \frac{1}{2}\left( \begin{array}{cc}
                                  0 & 1 \\
                                  1 & 0
                               \end{array}
                        \right) \\
   T^2 & = & \frac{1}{2}\tau^2
         =   \frac{1}{2}\left( \begin{array}{cc}
                                  0 & -i \\
                                  i & 0
                               \end{array}
                        \right) \\
   T^3 & = & \frac{1}{2}\tau^3
         =   \frac{1}{2}\left( \begin{array}{cc}
                                  1 & 0 \\
                                  0 & -1
                               \end{array}
                        \right)
\end{eqnarray}
We write explicitly
\begin{equation}
   g_1\frac{1}{2}B_{\mu} + 
   g_2T^jW_{\mu}^j
   =
   g_1\frac{1}{2}B_{\mu} + 
   g_2\frac{1}{2}
   \left( \begin{array}{cc}
             W_{\mu}^3 &
             W_{\mu}^1 - iW_{\mu}^2 \\
             W_{\mu}^1 + iW_{\mu}^2 &
             - W_{\mu}^3
          \end{array}
   \right)
   \label{eq:g1B_g2W}
\end{equation}
We then substitute Eq.~(\ref{eq:UnitaryGauge}) and 
Eq.~(\ref{eq:g1B_g2W}) into the kinetic term and the 
Higgs potential of $\mathcal{L}_{SSB}$ in 
Eq.~(\ref{eq:L_SSB}).  After some algebra the 
tree-level mass terms for the $H$ field and the gauge 
bosons are present.  The unphysical scalars reappear 
as the longitudinal polarizations of the weak bosons.  
\begin{eqnarray}
   (D_{\mu}\phi)^{\dagger}(D^{\mu}\phi)
   - V(\phi)
   & = & - \frac{1}{2}\left( 2\lambda v^2 \right) 
           H^2
           \\
   &   & + \left( \frac{g_2\nu}{2} \right)^2 
           W^{+\mu}W^-_{\mu}
           \\
   &   & + \left( \frac{\nu}{2} \right)^2
           \left( g_2W^3_{\mu} - g_1B_{\mu} \right)
           \left( g_2W^{3\mu}  - g_1B^{\mu} \right)
           \\
   &   & + \cdots
           \nonumber
\end{eqnarray}
and many other interaction terms.  The fields 
$W_{\mu}^{\pm}$ are defined as the electric charge 
eigenstates.  The SSB has mixed the $B_{\mu}$ and 
$W_{\mu}^3$ gauge bosons with the weak mixing angle 
$\theta_W$.  
\begin{equation}
   W_{\mu}^{\pm} = \frac{1}{\sqrt{2}}
                   \left(W_{\mu}^1 \mp W_{\mu}^2\right) 
   \label{eq:Wpm}
\end{equation}
\begin{equation}
   \left( \begin{array}{c}
             Z_{\mu} \\
             A_{\mu}
          \end{array}
   \right)      
   = 
   \left( \begin{array}{cc}
             \cos\theta_W & -\sin\theta_W \\
             \sin\theta_W &  \cos\theta_W
          \end{array}
   \right)
   \left( \begin{array}{c}
             W_{\mu}^3 \\
             B_{\mu} 
          \end{array}
   \right)
   \label{eq:ZA}
\end{equation}
\begin{equation}
   \tan\theta_W = \frac{g_1}{g_2} 
\end{equation}
Now we can read out the tree-level masses,
\begin{eqnarray}
   m_H        & = & v \sqrt{2\lambda} \\
   m_W        & = & v \frac{g_2}{2} \\
   m_Z        & = & v \frac{g_2}{2\cos\theta_W} \\
   m_{\gamma} & = & 0
\end{eqnarray}
Using 
$m_W = (\frac{\sqrt{2}g_2^2}{8G_F})^{1/2}$
in Eq.~(\ref{eq:mW_g2_GF})
with Fermi's constant $G_F = 1.166\times10^{-5}$ GeV$^{-2}$,
we can estimate the VEV of the 
Higgs field:
\begin{equation}
   v = \frac{2m_W}{g_2}
     = (\sqrt{2}G_F)^{-1/2}
     \approx 246 \; \mbox{GeV}
\end{equation}
The quantity
\begin{equation}
   \rho = \frac{m_W}{m_Z\cos\theta_W}
\end{equation}
is the universality parameter of the neutral current
interactions and the charged current interactions.  It
is predicted to be one at tree level in the SM, 
thus provides a test of the SM realization of SSB
compared to other models.  Any deviation from
$\rho = 1$ would be an important signal of new 
physics.

\vspace{0.2in}
\noindent{\bf Eletroweak Unification}
\vspace{0.1in}

Substituting the physical state of 
$W_{\mu}^{\pm}$ in Eq.~(\ref{eq:Wpm}) and 
$Z_{\mu}$, $A_{\mu}$ in Eq.~(\ref{eq:ZA})
into the electroweak interaction in the covariant 
derivative term in Eq.~(\ref{eq:L_symmetric}), 
and using $Y = 2(Q - T^3)$,
we can identify the weak CC, weak NC, and 
electromagnetic interactions. 
\begin{eqnarray}
   &   & \bar{\Psi}
         i\gamma^{\mu}
         \left(
            i g_1 \frac{Y}{2}B_{\mu} +
            i g_2 T^j W_{\mu}^j
         \right)
         \Psi
         \nonumber \\
   & = & - \bar{\Psi}\gamma^{\mu}
         \left[
            g_1 \left(Q - T^3\right) B_{\mu} + 
            g_2 \left(T^1W_{\mu}^1 + T^2W_{\mu}^2 + T^3W_{\mu}^3\right)
         \right]
         \Psi
         \nonumber \\
   & = & \begin{array}[t]{ll}
         - \bar{\Psi}\gamma^{\mu}
         \left[
            \frac{g_2}{\sqrt{2}}\left(T^-W_{\mu}^+ + T^+W_{\mu}^-\right)
         \right]
         \Psi
       & \mbox{(weak CC)}
         \\
         - \bar{\Psi}\gamma^{\mu}
         \left[
            \frac{g_2}{\cos\theta_W}\left(T^3-\sin^2\theta_WQ\right)Z_{\mu} 
         \right]
         \Psi
       & \mbox{(weak NC)}
         \\
         - \bar{\Psi}\gamma^{\mu}
            g_2\sin\theta_WQA_{\mu}   
         \Psi
       & \mbox{(electromagnetic)}
         \end{array}
   \label{eq:WZA}
\end{eqnarray}
Comparing the electromagnetic part with 
the $-\bar{\Psi}\gamma^{\mu}eQA_{\mu}\Psi$
term of $\mathcal{L}_{QED}$ in Eq.~(\ref{eq:L_QED}),
this implies the unification relation:
\begin{equation}
   e = g_2\sin\theta_w = g_1\cos\theta_w
   \label{eq:EWKunification}
\end{equation}

\vspace{0.2in}
\noindent{\bf Yukawa Coupling}
\vspace{0.1in}

Now we check the fermion masses.  The structure of
the lepton fields, for example, of the first generation is
\begin{equation}
   \begin{array}{cc}
      L = \left(\begin{array}{c} \nu_e \\ e_L \end{array}\right),
      & e_R
   \end{array}
\end{equation}
The Higgs field is an SU(2) doublet.  This makes
it possible to write an SU(2)-invariant interaction
of the fermions with the Higgs field, i.e., the Yukawa
coupling term in Eq.~(\ref{eq:YukawaCoupling}), which
can be written as
\begin{eqnarray}
   \mathcal{L}_{Yukawa}
   & = & g_e \bar{L} \phi e_R +
         g_e \bar{e}_R \phi^{\dagger} L 
         \label{eq:YukawaCoupling_LR_1}\\
   & = & g_e\left(\begin{array}{cc}
                     \times & \times
                  \end{array}
            \right)
            \left(\begin{array}{c}
                     \times \\
                     \times
                  \end{array}
            \right)
            \left(\begin{array}{c}
                     \times
                  \end{array}
            \right)
            +
         g_e\left(\begin{array}{c}
                     \times
                  \end{array}
            \right)
            \left(\begin{array}{cc}
                     \times & \times
                  \end{array}
            \right)
            \left(\begin{array}{c}
                     \times \\
                     \times
                  \end{array}
            \right) \nonumber
\end{eqnarray}
Here $\bar{L}\phi$ is an SU(2) invariant.
Multiplying by the $e_R$ does not change the
SU(2) invariance.  The second term is the Hermitian
conjugate of the first.  The coupling $g_e$ is 
arbitrary because it is not specified by the gauge 
symmetry principle of the theory.  After SSB by 
substituting $\phi$ with Eq.~(\ref{eq:UnitaryGauge}), 
and using $\bar{e}_Le_R + \bar{e}_Re_L = \bar{e}{e}$,
we get
\begin{eqnarray}
   \mathcal{L}_{Yukawa}
   & = & \frac{g_ev}{\sqrt{2}}
         \left(\bar{e}_Le_R + \bar{e}_Re_L\right) 
        +\frac{g_e}{\sqrt{2}}
         \left(\bar{e}_Le_R + \bar{e}_Re_L\right) 
         H
         \nonumber \\
   & = & m_e\bar{e}e
        +\frac{m_e}{v}\bar{e}eH
        \label{eq:YukawaCoupling_SSB_1}
\end{eqnarray}
We have identified the fermion mass as
$m_e = \frac{g_ev}{\sqrt{2}}$.  Thus the theory can 
now accommodate a non-zero fermion mass.  The second 
term says that there is a lepton-Higgs coupling 
$\frac{m_e}{v}$.
We notice that there is no mass term occured for neutrinos,
$m_{\nu} = 0$.  By assumption the theory contains no 
right-handed neutrino state $\nu_R$, therefore 
a term analogous to Eq.~(\ref{eq:YukawaCoupling_LR_1})
cannot be written that will lead to a mass term
$\bar{\nu}_R\nu_L$.  And this implies neutrinos do
not interact with $H$.

The structure of the quark fields, for example, of 
the first generation is
\begin{equation}
   \begin{array}{ccc}
      Q_L = \left(\begin{array}{c} u_L \\ d_L \end{array}\right),
      & u_R,
      & d_R
   \end{array}
\end{equation}
Since the structure of the right-handed quark is
different from the lepton case, there is a subtlety 
in writing down the Yukawa coupling term.  We know 
$\phi$ is an SU(2) doublet, then so is 
\begin{equation}
   \tilde{\phi}
    = i\tau^2\phi^*
    = \left(
         \begin{array}{c} 
            \phi^{0*} \\ 
            -\phi^- 
         \end{array}
      \right)
\end{equation}
This is true for any SU(2) doublet.  Since $\phi$ has 
hypercharge $Y = +1$, $\tilde{\phi}$ has $Y = -1$, 
and for each state, $Q = T^3 + Y/2$ is still satisfied.  
After SSB, $\tilde{\phi}$ becomes
\begin{equation}
   \tilde{\phi}
   \to \frac{1}{\sqrt{2}}
       \left(
          \begin{array}{c} 
             v + H \\
             0
          \end{array}
       \right)
   \label{eq:UnitaryGauge_tilde}
\end{equation}
The SU(2)-invariant Yukawa coupling for the quarks can be 
written as
\begin{equation}
   \mathcal{L}_{Yukawa}
   = g_d \bar{Q}_L \phi d_R +
     g_u \bar{Q}_L \tilde{\phi} u_R +
     \mbox{h.c.} 
     \label{eq:YukawaCoupling_LR_2}
\end{equation}
After SSB by substituting $\phi$ with Eq.~(\ref{eq:UnitaryGauge}), 
$\tilde{\phi}$ with Eq.~(\ref{eq:UnitaryGauge_tilde}), and using 
$\bar{q}_Lq_R + \bar{q}_Rq_L = \bar{q}{q}$, we get
\begin{eqnarray}
   \mathcal{L}_{Yukawa}
   & = &
     \frac{g_uv}{\sqrt{2}}\bar{u}u
    +\frac{g_dv}{\sqrt{2}}\bar{d}d
    +\frac{g_u}{\sqrt{2}}\bar{u}uH
    +\frac{g_d}{\sqrt{2}}\bar{d}dH
   \nonumber \\
   & = &
     m_u\bar{u}u
    +m_d\bar{d}d
    +\frac{m_u}{v}\bar{u}uH
    +\frac{m_d}{v}\bar{d}dH
    \label{eq:YukawaCoupling_SSB_2}
\end{eqnarray}
Again the quark masses can be accommodated, but are 
arbitrary parameters. They have to be provided by
experiment.  The last two terms describe the 
interaction of $u$ and $d$ quarks with $H$.

The procedure can be copied for the second and third
generations with $e \to \mu, \tau$ and with $u \to c, t$
and $d \to s, b$.  Since $H$ interacts with a coupling
proportional to $m_f$, it couples most strongly to
the heaviest generation.

\vspace{0.2in}
\noindent{\bf CKM Matrix}
\vspace{0.1in}

The spaces we have been working on are an internal quantum phase
space called gauge space of fermions, and an internal field space
of the Higgs potential.  The logic line is gauge symmetry $+$ SSB.  
Let us write down the SM Lagrangian~(\ref{eq:L}) explicitly by 
combining Eq.~(\ref{eq:L_symmetric}) and Eq.~(\ref{eq:L_SSB}).  This time 
is not for a particular fermion $\Psi_f$ 
only.  There are three generations of fermions 
in the SM.  We will sum up all of them.  Once we do that, there is 
a new internal space: generation space.  The eigenstates of the 
fermions in gauge space could be \emph{not} the eigenstates of the 
fermions in generation space which are the physical mass eigenstates 
we observe in experiment.
\begin{eqnarray}
   \mathcal{L} 
   & = & \mathcal{L}_{symmetric}
        +\mathcal{L}_{SSB}
         \\
   & = & \begin{array}[t]{ll}
         \sum_f
         \bar{\Psi}_f
         i\gamma^{\mu}
         \left(\partial_{\mu} 
                + i g_1 \frac{Y}{2}B_{\mu}
                + i g_2 T^j W_{\mu}^j
                + i g_3 \lambda^a G_{\mu}^a
         \right)\Psi_f
       & (L_{symm, \; covariant})
         \\
         - \frac{1}{4}B_{\mu\nu}B^{\mu\nu}
         - \frac{1}{4}W_{\mu\nu}^iW^{i\mu\nu}
         - \frac{1}{4}G_{\mu\nu}^aG^{a\mu\nu}
       & (L_{symm, \; GK})
         \\
         +
         \left|
            \left(
              \partial_{\mu}
              +ig_1\frac{1}{2}B_{\mu}
              +ig_2T^jW_{\mu}^j
            \right)
            \phi
         \right|^2
       & (L_{SSB, \;\;\; kinetic})
         \\
         -
         \left[
            \mu^2          \phi^{\dagger}\phi
            +\lambda \left(\phi^{\dagger}\phi\right)^2 
         \right]
       & (L_{SSB, \;\;\; V(\phi)})
         \\
         -
         \sum_f
         g_f\bar{\Psi}_f\phi\Psi_f
       & (L_{SSB, \;\;\; Yukawa})
         \end{array}
         \nonumber
\end{eqnarray}
We collect all of the terms for the fermions after SSB:
the kinetic and QCD terms in Eq.~(\ref{eq:L_symmetric}),
the mass and Higgs coupling terms in Eq.~(\ref{eq:YukawaCoupling_SSB_1}) 
for the leptons and in Eq.~(\ref{eq:YukawaCoupling_SSB_2}) for the quarks,
and the weak CC, weak NC and electromagnetic terms in Eq.~(\ref{eq:WZA}).
We simplify the notation for a fermion field $\Psi_f$ as $f$.  
The part of the SM Lagrangian for fermions is given by
\begin{equation}
   \mathcal{L}_F   =  \begin{array}[t]{ll}
                       \sum_f 
                       \bar{f}
                       \left(  i /\!\!\!\partial
                             - m_f
                             - \frac{m_f}{v}H
                       \right)
                       f
                     & (\mbox{Higgs})
                       \\
                       - \;
                       \frac{g_3}{2}
                       \sum_q
                       \bar{q}_{\alpha} 
                       \gamma^{\mu} 
                       \lambda^a_{\alpha\beta}
                       q_{\beta}
                       G_{\mu}^a
                     & (\mbox{QCD})
                       \\
                       - \; 
                       e
                       \sum_f
                       Q_f
                       \bar{f} 
                       \gamma^{\mu} 
                       f
                       A_{\mu}
                     & (\mbox{QED})
                       \\
                       - \;
                       \frac{g_2}{\cos\theta_w}
                       \sum_f
                       \bar{f} 
                       \gamma^{\mu} 
                       \left(T^3-\sin^2\theta_WQ\right)
                       f
                       Z_{\mu}
                     & (\mbox{weak NC})
                       \\
                       - \;
                       \frac{g_2}{\sqrt{2}}
                       \sum_f
                       \bar{f} 
                       \gamma^{\mu} 
                       (T^+W^+_{\mu} + T^-W^-_{\mu})
                       f
                     & (\mbox{weak CC})
                       \end{array}
   \label{eq:L_F}
\end{equation}
We denote the gauge eigenstate triplets in the generation space as
\begin{eqnarray}
   \begin{array}{cc}
      \mathbf{e}_L = \left(\begin{array}{ccc}
                              e_L \\
                              \mu_L \\
                              \tau_L
                           \end{array}
                     \right),
      &
      \mathbf{e}_R = \left(\begin{array}{ccc}
                              e_R \\
                              \mu_R \\
                              \tau_R
                           \end{array}
                     \right)
   \end{array}
   \nonumber \\
   \begin{array}{cc}
      \mathbf{u}_L = \left(\begin{array}{ccc}
                              u_L \\
                              c_L \\
                              t_L
                           \end{array}
                     \right),
      &
      \mathbf{u}_R = \left(\begin{array}{ccc}
                              u_R \\
                              c_R \\
                              t_R
                           \end{array}
                     \right)
   \end{array}
   \\
   \begin{array}{cc}
      \mathbf{d}_L = \left(\begin{array}{ccc}
                              d_L \\
                              s_L \\
                              b_L
                           \end{array}
                     \right),
      &
      \mathbf{d}_R = \left(\begin{array}{ccc}
                              d_R \\
                              s_R \\
                              b_R
                           \end{array}
                     \right)

   \end{array}
   \nonumber 
\end{eqnarray}
and denote the rotations from the gauge eigenstates to 
the mass eigenstates as unitary matrices 
$L_e$, $R_e$, $L_u$, $R_u$, $L_d$, and $R_d$
such that
\begin{eqnarray}
   \begin{array}{cc}
      \mathbf{e}_L \to L_e\mathbf{e}_L,
      &
      \mathbf{e}_R \to R_e\mathbf{e}_R
   \end{array}
   \nonumber \\
   \begin{array}{cc}
      \mathbf{u}_L \to L_u\mathbf{u}_L,
      &
      \mathbf{u}_R \to R_u\mathbf{u}_R
   \end{array}
   \\
   \begin{array}{cc}
      \mathbf{d}_L \to L_d\mathbf{d}_L,
      &
      \mathbf{d}_R \to R_d\mathbf{d}_R
   \end{array}
   \nonumber 
\end{eqnarray}
Because all of the neutrinos in the SM are massless,
they are degenerate in the mass eigenstates, namely we 
cannot tell the differences among the mass eigenstates.  
We set the rotation for neutrinos as a unit matrix 
denoted as $I_{\nu}$.  

First we check the QED part 
in Eq.~(\ref{eq:L_F}) to see if there is any change 
under the rotations,
\begin{eqnarray}
   \mathcal{L}_{F}^{QED} 
                 & = &
                       - \;
                       e
                       \sum_f 
                       Q_f
                       \bar{f} 
                       \gamma^{\mu} 
                       f
                       A_{\mu}
                       \nonumber \\
                 & = &
                       - \;
                       e
                       Q_{\mathbf{f}}
                       \left(
                          \bar{\mathbf{f}}_L 
                          \gamma^{\mu} 
                          \mathbf{f}_L
                          +
                          \bar{\mathbf{f}}_R 
                          \gamma^{\mu} 
                          \mathbf{f}_R
                       \right)
                       A_{\mu},
                       \;\;\;\;\;\;
                       \mbox{ with }
                       \mathbf{f} = \mathbf{e}, \mathbf{u}, \mathbf{d}
                       \nonumber \\
                & \to &
                       - \;
                       e
                       Q_{\mathbf{f}}
                       \left(
                          \bar{\mathbf{f}}_L 
                          \gamma^{\mu} 
                          L_{\mathbf{f}}^{\dagger}L_{\mathbf{f}}
                          \mathbf{f}_L
                          +
                          \bar{\mathbf{f}}_R 
                          \gamma^{\mu} 
                          R_{\mathbf{f}}^{\dagger}R_{\mathbf{f}}
                          \mathbf{f}_R
                       \right)
                       A_{\mu}
                 \label{eq:L_F_QED_rotated}
                       \\
                 & = &
                       - \;
                       e
                       Q_{\mathbf{f}}
                       \left(
                          \bar{\mathbf{f}}_L 
                          \gamma^{\mu} 
                          \mathbf{f}_L
                          +
                          \bar{\mathbf{f}}_R 
                          \gamma^{\mu} 
                          \mathbf{f}_R
                       \right)
                       A_{\mu}
                       \nonumber
\end{eqnarray}
where we have let $L_f^{\dagger}$ ($R_f^{\dagger}$) 
pass $\gamma^{\mu}$ forward in Eq.~(\ref{eq:L_F_QED_rotated})
because the former rotates in the generation space and 
the latter is in the spinor space.  Since the unitary
rotation matrices give $L_f^{\dagger}L_f = I$
and $R_f^{\dagger}R_f = I$, the electromagnetic 
interaction is diagonized in both the gauge eigenstates
and the mass eigenstates.

The same result holds for the Higgs, QCD, and weak NC 
parts in Eq.~(\ref{eq:L_F}) for the same reason.  For 
the weak NC, this is called the GIM mechanism~\cite{Glashow:1970gm}.  
The flavor changing neutral currents (FCNC), e.g. 
$s \to d$ decay ``off-diagonal'' in the generation 
space, are strongly suppressed.  On the other hand, 
the FCNC rare decays are very interesting because 
they are possible probes for new interactions.

Now we check the weak CC in Eq.~(\ref{eq:L_F}).
For leptons, we have 
\begin{eqnarray}
   \mathcal{L}_l^{CC}
                 & = &
                       - \;
                       \frac{g_2}{\sqrt{2}}
                       \sum_l 
                       \bar{l} 
                       \gamma^{\mu} 
                       (T^+W^+_{\mu} + T^-W^-_{\mu})
                       l
                       \nonumber \\
                 & = &
                       - \;
                       \frac{g_2}{\sqrt{2}}
                       \left(
                          \bar{\mathbf{\nu}}
                          \gamma^{\mu} 
                          \mathbf{e}_L
                          W^-_{\mu}
                          + h.c.
                       \right)
                       \nonumber \\
                 & \to &
                       - \;
                       \frac{g_2}{\sqrt{2}}
                       \left(
                          \bar{\mathbf{\nu}}
                          I_{\nu}^{\dagger}L_e
                          \gamma^{\mu} 
                          \mathbf{e}_L
                          W^-_{\mu}
                          + h.c.
                       \right)
                 \label{eq:L_l_CC_rotated}
                       \\
                 & = &
                       - \;
                       \frac{g_2}{\sqrt{2}}
                       \left(
                          \bar{\mathbf{\nu}}
                          \gamma^{\mu} 
                          \mathbf{e}_L
                          W^-_{\mu}
                          + h.c.
                       \right)
                       \nonumber
\end{eqnarray}
where we have let $L_e$ pass $\gamma^{\mu}$ backward
in~(\ref{eq:L_l_CC_rotated}).  With $I_{\nu}^{\dagger}L_e$ 
acting backward on the vector of the degenerated 
neutrino mass eigenstates, we just go back to the 
original form, and the leptonic weak CC interactions
are diagonized in both kinds of the eigenstates.

So far, the distinction between the gauge eigenstates
and the mass eigenstates has been seen to have no apparent
effect.  However, mixing between generations does 
manifest itself in the system of the weak CC for quarks.  
By convention, the quark mixing is assigned to the 
down-type quarks,
\begin{eqnarray}
   \mathcal{L}_q^{CC}
                 & = &
                       - \;
                       \frac{g_2}{\sqrt{2}}
                       \sum_q 
                       \bar{q} 
                       \gamma^{\mu} 
                       (T^+W^+_{\mu} + T^-W^-_{\mu})
                       q
                       \nonumber \\
                 & = &
                       - \;
                       \frac{g_2}{\sqrt{2}}
                       \left(
                          \bar{\mathbf{u}}_L
                          \gamma^{\mu} 
                          \mathbf{d}_L
                          W^-_{\mu}
                          + h.c.
                       \right)
                       \nonumber \\
                 & \to &
                       - \;
                       \frac{g_2}{\sqrt{2}}
                       \left(
                          \bar{\mathbf{u}}_L
                          \gamma^{\mu} 
                          L_u^{\dagger}L_d
                          \mathbf{d}_L
                          W^-_{\mu}
                          + h.c.
                       \right)
                 \label{eq:L_q_CC_rotated}
                       \\
                 & = &
                       - \;
                       \frac{g_2}{\sqrt{2}}
                       \left(
                          \bar{\mathbf{u}}_L
                          \gamma^{\mu} 
                          V
                          \mathbf{d}_L
                          W^-_{\mu}
                          + h.c.
                       \right)
                       \nonumber
\end{eqnarray}
where
\begin{equation}
   V = L_u^{\dagger}L_d
\end{equation}
Thus the down-type quark gauge states participating in
the transitions of the weak CC are linear combinations of
their mass eigenstates.  For three generations, it is called
the CKM (Cabibbo-Kobayashi-Maskawa) matrix~\cite{Cabibbo:1963yz}.
The SM does 
not predict the content of $V$.  Rather its matrix 
elements must be extracted from experiment.
\begin{equation}
   \left( \begin{array}{c}
             d \\
             s \\
             b
          \end{array}
   \right)_{\mbox{weak}}
   = 
   \left( \begin{array}{ccc}
             V_{ud} & V_{us} & V_{ub} \\
             V_{cd} & V_{cs} & V_{cb} \\
             V_{td} & V_{ts} & V_{tb} 
          \end{array}
   \right)
   \left( \begin{array}{c}
             d \\
             s \\
             b
          \end{array}
   \right)_{\mbox{mass}}
   \label{eq:CKM_Matrix}
\end{equation}

Any $3\times3$ complex matrix has 18 paramters. 
The quark mixing matrix $V$, being the product of two 
unitary matrices, is itself unitary, $V^{\dagger}V = 1$,
and this eliminates 9 paramters.  The rest of 9 parameters
can be identified with 3 rotation angle, and 6
phase angles with 5 of them eliminated by rephasing the 
relative quark phase angles in Eq.~(\ref{eq:CKM_Matrix}) and 
leaving 1 global phase angle.  So the actual total number 
of free parameters is $18 - 9 - 5 = 4$, which includes 3 
rotation angle and 1 phase angle.  

The ``standard'' 
parametrization of the CKM matrix advocated in 
PDG~\cite{Eidelman:2004wy} is
\begin{equation}
   V =
   \left(
      \begin{array}{ccc}
         c_{12}c_{13} & 
         s_{12}c_{13} & 
         s_{13}e^{-i\delta_{13}} \\
         -s_{12}c_{23}-c_{12}s_{23}s_{13}e^{i\delta_{13}} & 
         c_{12}c_{23}-s_{12}s_{23}s_{13}e^{i\delta_{13}} & 
         s_{23}c_{13} \\
         s_{12}s_{23}-c_{12}c_{23}s_{13}e^{i\delta_{13}} & 
         -c_{12}s_{23}-s_{12}c_{23}s_{13}e^{i\delta_{13}} & 
         c_{23}c_{13} 
   \end{array}
   \right)
   \label{eq:CKM_StandardParametrization}
\end{equation} 

In this equation, $c_{ij} = \cos\theta_{ij}$ and
$s_{ij} = \sin\theta_{ij}$, with $i$ and $j$ labeling
the generations.  The interpretation is that if $\theta_{ij}$
vanishes, so does the mixing between those two generations.
For example, in the limit $\theta_{23} = \theta_{13} = 0$,
the third generation decouples and it reduces to two generations 
with $\theta_{12}$ identified as the Cabibbo angle.

The complex parameter in phase angle goes into the weak 
charged interaction terms 
$\bar{\mathbf{u}}\gamma^{\mu}P_LV\mathbf{d}W_{\mu}$,
and from quantum theory we know that the Hamiltonian will not
be invariant under time reversal, or equivalently, CP.
So this induces CP violation.  

The magnitude of the complex
matrix element in the CKM matrix presently measured is
\begin{equation}
   \left(
      \begin{array}{ccc}
         0.9739-0.9751  &  0.221-0.227  & 0.0029-0.0045 \\
          0.221-0.227   & 0.9730-0.9744 &  0.039-0.044  \\
         0.0048-0.0140  &  0.037-0.043  & 0.9990-0.9992 
   \end{array}
   \right)
   \label{eq:CKM_Measurement}
\end{equation} 
Here we discuss some of the immediate consequences.  
For top quark, with $V_{tb}\approx0.999$, we have
\begin{equation}
   \mbox{BR}(t \to W b) \approx 100\%
\end{equation}
For bottom quark, with $V_{cb}\approx0.04$ ten times 
larger than $V_{ub}\approx0.004$, it mostly decays by
$b \to W c$.  Then $W$ can decay to $e\nu$, $\mu\nu$,
$\tau\nu$, $u\bar{d}$, and $c\bar{s}$ with a color factor
3 for each quark decaying mode.  The width of $b$ decays is
$\Gamma_b\approx(9V_{cb}^2G_F^2m_b^5)/(192\pi^3)$.
This gives 
\begin{equation}
   \frac{\Gamma_b}{\Gamma_{\tau}}
   \approx
      \frac{9}{5}
      V_{cb}^2
      \left(
         \frac{m_b}{m_{\tau}}
      \right)^5
   \approx
      0.4
\end{equation}
So the life time of $b$ is about two and half times
longer than the life time of $\tau$ because its 
decay can only happen by the rotation from the
mass eigenstates to the weak eigenstates and 
the magnitudes of the matrix elements for this 
rotation are small.

\vspace{0.2in}
\noindent{\bf Couplings to Fermions}
\vspace{0.1in}

For convenience, we repeat Eq.~(\ref{eq:L_F}) here.
\begin{equation}
   \mathcal{L}_F   =  \begin{array}[t]{ll}
                       \sum_f 
                       \bar{f}
                       \left(  i /\!\!\!\partial
                             - m_f
                             - \frac{m_f}{v}H
                       \right)
                       f
                     & (\mbox{Higgs})
                       \\
                       - \;
                       \frac{g_3}{2}
                       \sum_q
                       \bar{q}_{\alpha} 
                       \gamma^{\mu} 
                       \lambda^a_{\alpha\beta}
                       q_{\beta}
                       G_{\mu}^a
                     & (\mbox{QCD})
                       \\
                       - \; 
                       e
                       \sum_f
                       Q_f
                       \bar{f} 
                       \gamma^{\mu} 
                       f
                       A_{\mu}
                     & (\mbox{QED})
                       \\
                       - \;
                       \frac{g_2}{\cos\theta_w}
                       \sum_f
                       \bar{f} 
                       \gamma^{\mu} 
                       \left(T^3-\sin^2\theta_WQ\right)
                       f
                       Z_{\mu}
                     & (\mbox{weak NC})
                       \\
                       - \;
                       \frac{g_2}{\sqrt{2}}
                       \sum_f
                       \bar{f} 
                       \gamma^{\mu} 
                       (T^+W^+_{\mu} + T^-W^-_{\mu})
                       f
                     & (\mbox{weak CC})
                       \end{array}
   \label{eq:repeat_L_F}
\end{equation}
We can read out the couplings to the fermions in the SM 
as follows:
\begin{itemize}
\item The Higgs coupling for $H \to f\bar{f}$ 
      is $\frac{m_f}{v}$.

\item The QCD coupling for $g \to q\bar{q}$ is 
      $\frac{g_3}{2}\lambda^a$. 
      (For the electroweak interactions of the quarks
       $\gamma/Z/W/H \to \bar{q}_cq_c$, 
       the effect of the color charge is that
       the probabilities, i.e., the decay widths
       are multiplied by a constant color factor 
       $N_c = 3$, rather than that the couplings
       appearing in the amplitudes are multiplied
       by the color generator $\lambda^a$.
       This is because that $\gamma/Z/W/H$ are 
       colorless and the number of color 
       combinations of $\bar{q}_cq_c$ is fixed 
       to be three.)

\item The electromagnetic coupling for 
      $\gamma \to f\bar{f}$ is $eQ_f$.

\item The neutral weak coupling for 
      $Z^0 \to f\bar{f}$ is 
      $\frac{g_2}{\cos\theta_W}(T^3 - \sin^2\theta_W Q_f)$
      for left-handed fermions and
      $\left(-\frac{g_2Q_f\sin^2\theta_W}{\cos\theta_W}\right)$
      for right-handed fermions.

\item The charged weak coupling is 
      $\frac{g_2}{\sqrt{2}}$, 
      and this only applies to left-handed fermions.
      We notice that the coupling for 
      $W^{\pm} \to l \nu_l$ is $\frac{g_2}{\sqrt{2}}$, 
      while the coupling for $W^{\pm} \to qq'$ 
      should be multiplied by a quark mixing element 
      in the CKM matrix and it becomes
      $V_{qq'}\frac{g_2}{\sqrt{2}}$.
\end{itemize}
These results are summarized in 
Table~\ref{tab:CouplingsToFermions}
in Section~\ref{sec:theory_SM}.



\chapter{How to Calculate Cross Section}
\label{cha:HowToCalculateXSec}

We are concerned about the resonance production of tau pairs in the
SM i.e. $p\bar{p}\to\gamma^*/Z\to\tau\tau$. 
This is a good example to see how event 
generator~\cite{Barger:1997}~\cite{Richardson:2003}
works by using Monte Carlo simulation.

\begin{figure}
   \begin{center}
      \parbox{5.5in}{\epsfxsize=\hsize\epsffile{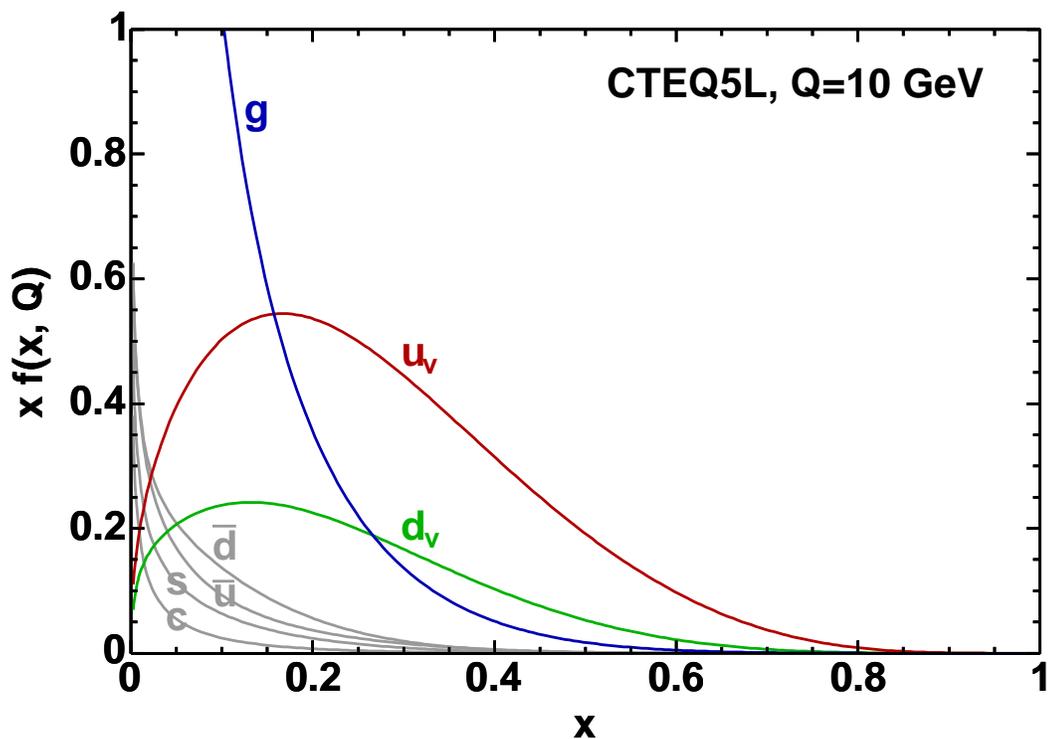}}
      \caption[Proton's parton density functions]
              {Proton's parton density functions.}
      \label{fig:PDF}
   \end{center}
\end{figure}

At $p\bar{p}$ collider, the production of any process starts 
from parton interaction.  A proton is made of quarks and gluons 
and can be written as
\begin{equation}
   \mbox{proton} =
   \underbrace{uud}_{valence} + \;\;
   \underbrace{u\bar{u}+d\bar{d}+\cdots}_{sea} \;\; + \;\;
   \underbrace{g+g+\cdots}_{gluons}
\end{equation}
The probability density for a given parton $i$ in a proton
carrying momentum fraction $x$ and being ``seen'' in an interaction
by an intermediate boson with energy scale $Q$ is characterized
by a function $f_i(x, Q)$, called the Parton Density Function
(PDF)~\cite{Lai:1999wy}. 
The momentum density of a parton is its PDF multiplied 
by its momentum fraction and is expressed as $x f_i(x, Q)$.
An example of parametrization is shown in Fig.~\ref{fig:PDF}.

The differential cross section for $12\to34$ can be written as
\begin{equation}
   d\sigma =
   \frac{\left(2\pi\right)^4}{2\hat{s}}
   \frac{d^3p_3}{\left(2\pi\right)^32E_3}
   \frac{d^3p_4}{\left(2\pi\right)^32E_4}
   \delta^4(p_1+p_2-p_3-p_4)
   dx_1dx_2f_1(x_1)f_2(x_2)
   \sum_{spins}\left|\mathcal{M}\right|^2_{12\to34}
\end{equation}
where $\hat{s}$ is the parton center-of-mass energy squared, 
$p_i$ ($E_i$) is the momentum (energy) of the $i$th particle,
$x_{1,2}$ are the fractions of the momenta of the incoming
beam particles carried by the incoming partons, $f_i(x_i)$
are the PDF's with an implicit dependence on the energy scale of the
interaction, and $\sum_{spins}\left|\mathcal{M}\right|^2_{12\to34}$
is the matrix element squared for the process averaged over
the spins and colors of the incoming particles and summed 
over the spins and colors of the outgoing particles.

First we consider the phase space. We perform the integral over 
the three-momentum of $p_4$, and reexpress the integral over the
momentum of $p_3$ in terms of the magnitude of the 
three-momentum $p$ in the parton center-of-mass frame and the 
angle with respect to the beam $\theta$ and the azimuthal angle 
$\phi$.  Then we make a transformation using $\hat{s}=x_1x_2s$
with $s$ the $p\bar{p}$ center-of-mass energy squared and we
get $dx_2=d\hat{s}/(sx_1)$. After some algebra, the differential 
cross section becomes
\begin{equation}
   d\sigma =
   \frac{p}{32\pi^2\hat{s}^{5/2}}
   d\cos\theta
   d\phi
   d\hat{s}
   \frac{dx_1}{x_1}
   x_1f_1(x_1)x_2f_2(x_2)
   \sum_{spins}\left|\mathcal{M}\right|^2_{12\to34}
\end{equation}
The angular part can be uniformly generated with $0<\phi<2\pi$ 
and $-1<\cos\theta<1$.  The momentum fraction $dx_1/x_1$ part 
can be transformed to $\ln x_1$ and then uniformly generated.  
For the distribution over $\hat{s}$, we impose a minimum value 
of $\hat{s}$.  There are two types 
of distributions to be smoothed in order to converge faster
for the Monte Carlo simulation.  One type is a power law 
distribution $1/\hat{s}^{\alpha}$ with $\alpha>1$ which is the rise 
in the cross section due to the photon exchange at small 
center-of-mass energies.  The other type is the Breit-Wigner 
resonance due to the $Z$ boson exchange with a mass $m$ and 
a width~$\Gamma$,
\begin{equation}
\begin{array}[c]{lllll}
   \int_{\hat{s}/s}^1\frac{dx_1}{x_1} 
 & \rho\equiv\ln x_1
 & \to
 & \int d\rho 
 & \mbox{uniformly} \\
   \int_{\hat{s}_{min}}^s\frac{d\hat{s}}{\hat{s}^{\alpha}} 
 & \rho\equiv\hat{s}^{(1-\alpha)}
 & \to
 & \int d\rho 
 & \mbox{uniformly} \\
   \int_{\hat{s}_{min}}^s\frac{d\hat{s}}{\left(\hat{s}-m^2\right)^2 + m^2\Gamma^2}
 & \rho\equiv\tan^{-1}\left(\frac{\hat{s}-m^2}{m\Gamma}\right)
 & \to
 & \int d\rho
 & \mbox{uniformly}
\end{array}
\end{equation}

Second we consider the matrix element which is the interesting part.
With a non-constant matrix element, the distribution is expected
to deviate from the pure phase space distribution. Further,
compared with the distributions described by the SM, there are 
probably deviations in the distributions in real data due to some 
unknown matrix elements of new physics. 
The effects shown in cross section could be an enhancement or 
a suppression, a new resonance, changes in the angular distributions, 
a divergence or a cancellation by interference, etc.  A good deal of 
particle physics consists of the measurements and the interpretations 
of such effects in cross section. 
For the SM process $q\bar{q}\to\gamma^*/Z\to\tau\tau$, we have
\begin{equation}
   \sum_{spins}\left|\mathcal{M}\right|^2_{12\to34} =
   \begin{array}[t]{l}
      (\hat{t}-m_3^2)
      (\hat{t}-m_4^2)
      (|g^{RL}|^2+|g^{LR}|^2)
      \\
      +
      (\hat{u}-m_3^2)
      (\hat{u}-m_4^2)
      (|g^{LL}|^2+|g^{RR}|^2)
      \\
      +
      2m_3m_4
      \mathcal{R}e
      \{g^{RL}g^{RR*}+g^{LR}g^{LL*}\}
   \end{array}
\end{equation}
where 
$\hat{t} = (p_1-p_3)^2$,
$\hat{u} = (p_1-p_4)^2$,
$m_{3,4}$ are the masses of the outgoing tau particles.
In the center-of-mass frame using
$p_{cm}^2 = \frac{1}{4\hat{s}}[\hat{s}-(m_3+m_4)^2][\hat{s}-(m_3-m_4)^2]$,
the value of $\hat{t}$ can be expressed as
$\hat{t} = m_3^2-\hat{s}^{1/2}(E_3-p_{cm}\cos\theta)$,
and the value of $\hat{u}$ can be expressed as
$\hat{u} = m_4^2-\hat{s}^{1/2}(E_4-p_{cm}\cos\theta)$.
The couplings are defined to be
\begin{equation}
   g^{ab}
 = \sum_{i=\gamma*/Z}
   \frac{g_{in}^ag_{out}^b}
        {(\hat{s}-m_i^2)^2+m_i^2\Gamma_i^2}
\end{equation}
where the sum runs over $\gamma^*/Z$ the intermediate gauge
bosons with mass $0/m_Z$ and width $0/\Gamma_Z$,  and
$g_{in}^{L,R}$ is the coupling of the gauge boson to the incoming
partons and $g_{out}^{L,R}$ is the coupling of the gauge boson to
the outgoing tau particles.  The couplings to the fermions in the 
SM are listed in Table~\ref{tab:CouplingsToFermions}. 

To summarize, the major parts for generating an event include: 
(a) generating randomly the incoming partons and incorporating the PDF's, 
(b) generating randomly the kinematic variables which describe the event 
    in the phase space of the final particles, and 
(c) calculating the matrix element.
Now we can put the parts together and get the weight for an event by 
multiplying all of the factors.  The weight is in GeV$^{-2}$ and we 
need to convert to picobarn with a conversion constant $3.89379\times10^8$ 
GeV$^2$ pb.  

After generating a large sample of events, we can fill the weights of 
the events into a histogram, for example, a one-dimensional histogram 
of $\hat{s}$ which is the invariant mass of the tau pairs.  The 
differential cross section versus the invariant mass of the tau pairs 
can be obtained by dividing the histogram by the number of events 
generated and the size of the bins.  The result is shown in 
Fig.~\ref{fig:TauTau_3} in Section~\ref{sec:theory_tt}.



\chapter{Separation Angle under Boost}
\label{cha:Alpha_Gamma}

\begin{figure}
   \begin{center}
      \parbox{5.5in}{\epsfxsize=\hsize\epsffile{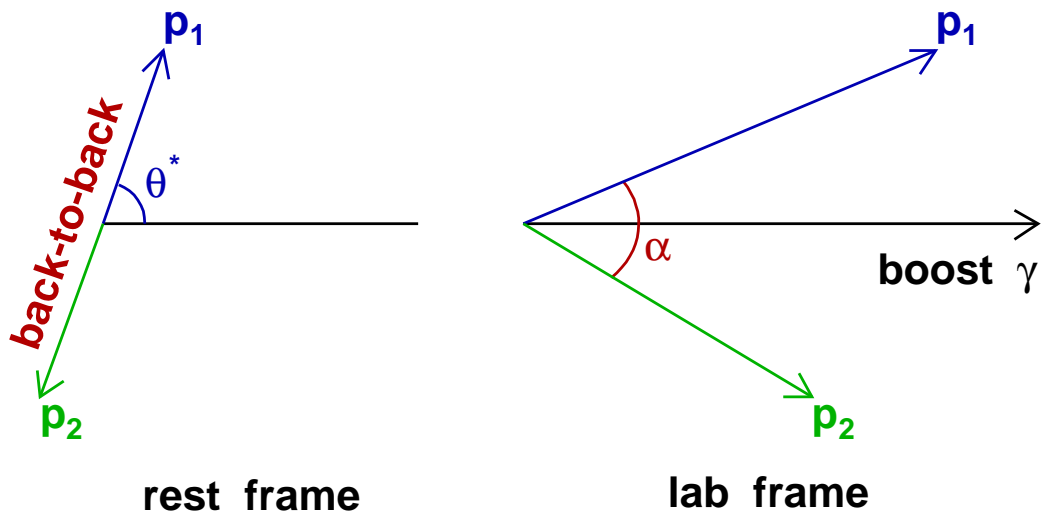}}
      \caption[Boost two-body decay from rest frame to lab frame]
              {Boost two-body decay from rest frame to lab frame.}
      \label{fig:Boost_1}
   \end{center}
\end{figure}

The calculable case is to boost the simplest
phase space, i.e. a two-body decay, from the
rest frame to the lab frame, as shown in 
Fig.~\ref{fig:Boost_1}.  The two final 
particles are back-to-back in the rest frame.  
The separation angle $\alpha$ of the two 
final particles in the lab frame can be 
parametrized as a function of $\theta^*$ the 
polar angle in the rest frame, which has an 
equal probability to be any value between 
$0^o$ and $90^o$, and the boost~$\gamma$.  

\begin{figure}
   \begin{center}
      \parbox{5.3in}{\epsfxsize=\hsize\epsffile{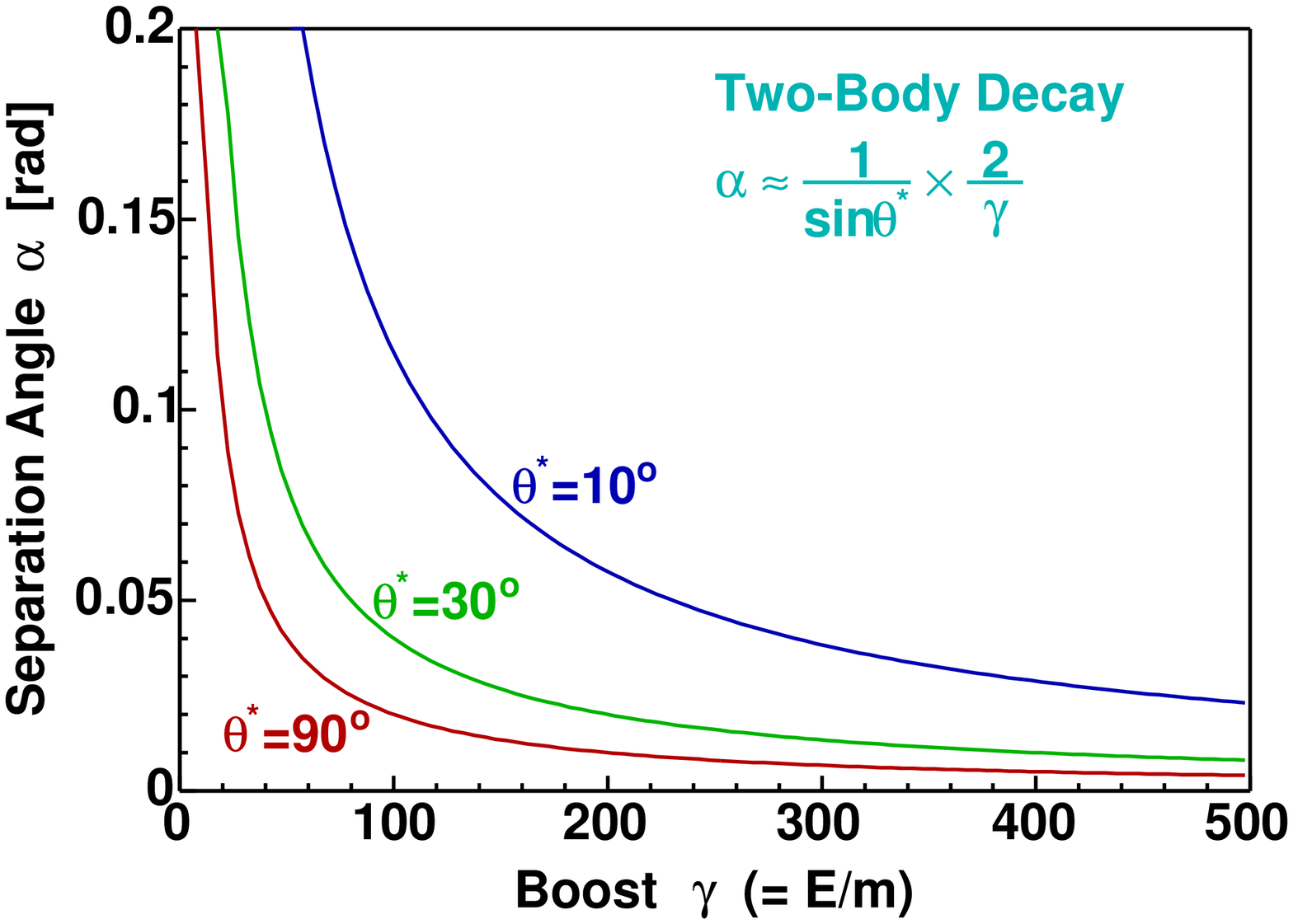}}
      \caption[Separation angle vs. boost, calculated in $\theta^*$ slices]
              {Separation angle vs. boost, calculated in $\theta^*$ slices.}
      \label{fig:Boost_2}
   \vspace{0.5in}
      \parbox{5.3in}{\epsfxsize=\hsize\epsffile{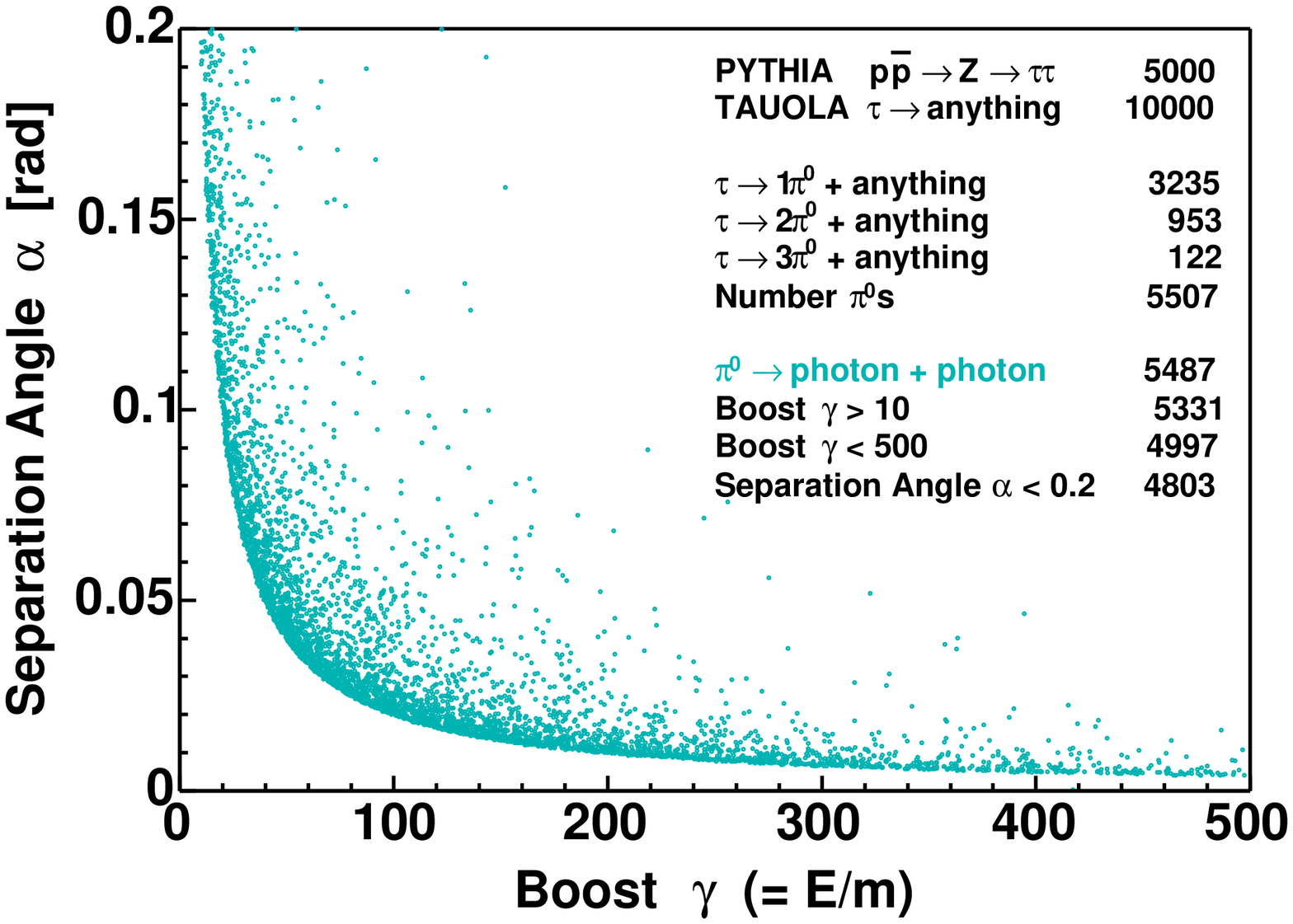}}
      \caption[Separation angle vs. boost, Monte Carlo distribution]
              {Separation angle vs. boost, Monte Carlo distribution.}
      \label{fig:Boost_3}
   \end{center}
\end{figure}

Let us consider two massless final particles, e.g.
two photons from a $\pi^0$ decay with a mass $m$, 
an energy $E$, and a boost $\gamma = E/m$.  We boost 
the four-momentum of $p_1$ from the rest frame
to the lab frame,
\begin{equation}
   \left(
      \begin{array}{cccc}
         1   &   0   &          0        &        0          \\
         0   &   1   &          0        &        0          \\
         0   &   0   &       \gamma      & \sqrt{\gamma^2-1} \\
         0   &   0   & \sqrt{\gamma^2-1} &     \gamma   
      \end{array}      
   \right)
   \left(        
      \begin{array}{c}
         0                       \\
         \frac{m}{2}\sin\theta^* \\
         \frac{m}{2}\cos\theta^* \\
         \frac{m}{2}
      \end{array}      
   \right)
   =
   \left(        
      \begin{array}{c}
         0                                                 \\
         \frac{m}{2}\sin\theta^*                           \\
         \frac{m}{2}(\gamma\cos\theta^*+\sqrt{\gamma^2-1}) \\
         \frac{m}{2}(\sqrt{\gamma^2-1}\cos\theta^*+\gamma) \\
      \end{array}      
   \right)      
\end{equation}
We denote the angle between $p_1$ in the lab frame 
and the direction of the boost as $\theta_1$.  
We have
\begin{equation}
   \sin\theta_1 =
   \frac{\sin\theta^*}
        {\sqrt{\sin^2\theta^*+(\gamma\cos\theta^*+\sqrt{\gamma^2-1})^2}}
\end{equation}
We denote the angle between $p_2$ in the lab frame 
and the direction of the boost as $\theta_2$.
By substituting $\theta^*$ with $\theta^*+\pi$, 
we have
\begin{equation}
   \sin\theta_2 =
   \frac{-\sin\theta^*}
        {\sqrt{\sin^2\theta^*+(-\gamma\cos\theta^*+\sqrt{\gamma^2-1})^2}}
\end{equation}
Now we can calculate the separation angle $\alpha$,
\begin{equation}
   \sin\alpha = 
   \sin[\theta_1 + (2\pi - \theta_2)] = 
   \frac{2\sin\theta^*\sqrt{\gamma^2-1}}
        {\sin^2\theta^*(\gamma^2-1)+1}
\end{equation}
For $\theta^*$ not too small and $\gamma\gg1$,
we get an approximation for small $\alpha$, 
\begin{equation}
   \alpha \approx 
   \frac{1}{\sin\theta^*}
   \times
   \frac{2}{\gamma} 
   \label{eq:Alpha_Gamma}
\end{equation}
For fixed $\theta^*$ (not too small) values, 
the functions are shown in Fig.~\ref{fig:Boost_2}.
For large boosts, the smearing by $\theta^*$ is 
small, thus the correlation between the separation 
angle and the boost (energy) is very strong.  

Since $\theta^*$ has an equal probability 
to be any value between $0^o$ and $90^o$, 
the probability that the separation angle stays 
between the curve for $\theta^* = 30^o$ and the 
curve for $\theta^* = 90^o$ is three times 
larger than the probability that the separation
angle stays between the curve for 
$\theta^* = 10^0$ and the curve for 
$\theta^* = 30^o$.  The effect is very obvious.  
We use Monte Carlo simulation to check the same 
plot, as shown in Fig.~\ref{fig:Boost_3}.
It confirms that the simplest case of two-body 
decay is indeed calculable and the correlation 
between the separation angle and the boost 
(energy) is very strong.  
 
For the more complicated phase spaces such as 
those of tau's hadronic decays, the calculation 
is very hard.  But Eq.~(\ref{eq:Alpha_Gamma}) is 
still a good hint.  We need to use
Monte Carlo simulation to get the distribution,  
which is shown in Fig.~\ref{fig:TauId_shr1} in 
Section~\ref{subsec:TauId_shrinking}.

\bibliographystyle{h-physrev3}
\bibliography{thesis}

\vita

\vspace{0.1in}
\begin{center}
   {\bf Zongru Wan}
\end{center}
\vspace{0.1in}

\begin{center}
\begin{tabular}{ll}
{\bf 1973}      & Born November 5 in Anhui, China. \\
                & \\
{\bf 1991}      & Graduated from Taihu High School, Anqing, Anhui, China. \\
                & \\
{\bf 1991-1995} & Attended Beijing Institute of Technology, Beijing, China. \\
                & Majored in Mechanical \& Electrical Engineering. \\
                & \\
{\bf 1995}      & B.E. in Mechanical \& Electrical Engineering. \\
                & Beijing Institute of Technology, Beijing, China. \\
                & \\
{\bf 1995-1998} & Research Staff, BES Collaboration \\
                & at Institute of High Energy Physics, Beijing, China. \\
                & \\
{\bf 1999-2005} & Graduate Studies in Physics \\
                & at Rutgers, the State University of New Jersey, USA. \\
                & \\
{\bf 1999}      & Teaching Assistant, Department of Physics and Astronomy. \\
                & Rutgers, the State University of New Jersey, USA. \\
                & \\
{\bf 1999-2005} & Research Assistant, Department of Physics and Astronomy. \\
                & Rutgers, the State University of New Jersey, USA. \\
                & \\
{\bf 2005}      & Ph.D. in Physics. \\
                & Rutgers, the State University of New Jersey, USA. \\
\end{tabular}
\end{center}

\end{document}